\documentclass[structabstract]{aa}  
\usepackage{graphicx}
\usepackage{txfonts}
\usepackage{natbib}
\usepackage{multirow}
\usepackage{longtable}

\def\4he{$^4$He}

\def\kms{\mathrm{km\,s}^{-1}}
\def\e#1{\times 10^{#1}}

\def\msol{\mathrm{M}_\odot}
\def\lsol{L_\odot}

\def\eqn#1{Eq. \ref{#1}}
\def\up#1{$^{#1}$}
\def\down#1{$_{#1}$}

\def\h2{$\mathrm{H}_2$}
\def\mic{\mathrm{ }\mu\mathrm{m}}
\def\spy{\;\msol~\mathrm{ yr}^{-1}}

\begin{document}
   \title{New observations and models of circumstellar CO line emission of AGB stars in the \textit{Herschel}\thanks{\textit{Herschel} is an ESA space observatory with science instruments provided by European-led Principal Investigator consortia and with important participation from NASA.} SUCCESS programme\thanks{Based on observations carried out with the IRAM 30m Telescope. IRAM is supported by INSU/CNRS (France), MPG (Germany) and IGN (Spain).}}
  \titlerunning{Observations and models of circumstellar CO line emission of AGB stars in the SUCCESS programme}
   \author{T. Danilovich
          \inst{1}
          \and
          D. Teyssier\inst{2}
          \and
          K. Justtanont\inst{1}
          \and
          H. Olofsson \inst{1}
          \and L. Cerrigone \inst{3}
          \and V. Bujarrabal \inst{4}
          \and J. Alcolea \inst{5}
          \and J. Cernicharo \inst{6}
          \and A. Castro-Carrizo \inst{7}
          \and P. Garc\'ia-Lario \inst{2}
          \and A. Marston \inst{2}
          }

   \institute{Onsala Space Observatory, Department of Earth and Space Sciences, Chalmers University of Technology, 439 92 Onsala, Sweden   
         \and
             European Space Astronomy Centre, Urb. Villafranca del Castillo, PO Box 50727, 28080, Madrid, Spain
         \and ASTRON, the Netherlands Institute for Radioastronomy, PO Box 2, 7990 AA, Dwingeloo, The Netherlands
         \and Observatorio Astron\'omico Nacional (IGN), PO Box 112, 28803 Alcal\'a de Henares, Spain
         \and Observatorio Astron\'omico Nacional (IGN), Alfonso XII, 3 y 5, 28014 Madrid, Spain
         \and Group of Molecular Astrophysics. ICMM. CSIC. C/ Sor Juana In\'es de La Cruz N3. E-28049, Madrid. Spain
         \and Institut de Radioastronomie Millim\'etrique, 300 rue de la Piscine, F-38406, St-Martin d'H\`eres, France
        \\     \email{taissa@chalmers.se}
             }

   \date{Received 9 June 2015 / Accepted 28 June 2015}

 
  \abstract
   {Asymptotic giant branch (AGB) stars are in one of the latest evolutionary stages of low to intermediate-mass stars. Their vigorous mass loss has a significant effect on the stellar evolution, and is a significant source of heavy elements and dust grains for the interstellar medium. The mass-loss rate can be well traced by carbon monoxide (CO) line emission.}
   {We present new \textit{Herschel} HIFI and IRAM 30m telescope CO line data for a sample of 53 galactic AGB stars. The lines cover a fairly large range of excitation energy from the $J=1\to0$ line to the $J=9\to8$ line, and even the $J=14\to13$ line in a few cases. We perform radiative transfer modelling for 38 of these sources to estimate their mass-loss rates.}
   {We used a radiative transfer code based on the Monte Carlo method to model the CO line emission. We assume spherically symmetric circumstellar envelopes that are formed by a constant mass-loss rate through a smoothly accelerating wind.}
   {We find models that are consistent across a broad range of CO lines for most of the stars in our sample, i.e., a large number of the circumstellar envelopes can be described with a constant mass-loss rate. We also find that an accelerating wind is required to fit, in particular, the higher-J lines and that a velocity law will have a significant effect on the model line intensities. The results cover a wide range of mass-loss rates ($\sim 10^{-8}$ to $2\e{-5}\spy$) and gas expansion velocities (2 to $21.5\;\kms$) , and include M-, S-, and C-type AGB stars. Our results generally agree with those of earlier studies, although we tend to find slightly lower mass-loss rates by about 40\%, on average. We also present ``bonus" lines detected during our CO observations.}
  {}
   
   \keywords{Stars: AGB and post-AGB -- circumstellar matter -- stars: mass loss -- stars: evolution}

   \maketitle
%

\section{Introduction}

Towards the end of their lives, low and intermediate mass stars (with masses $\sim$ 0.8--8 $\msol$) will exhaust their supply of He and cease fusion reactions in their cores, leaving a quiescent C/O core with H and He fusion reactions only taking place in thin shells surrounding the core. This evolutionary phase is known as the asymptotic giant branch (AGB) \citep{Herwig2005}.

AGB stars are also a significant source of heavy elements in the universe. It is thought that about half of all elements heavier than Fe originate in AGB stars through the $s$-process of slow neutron capture \citep{Herwig2005}. It is during the AGB phase that this enriched material is brought to the surface. At the same time, the star experiences vigorous mass loss, ejecting matter to form a circumstellar envelope (CSE) around the star. Molecules and dust grains form in the CSE, and will eventually chemically enrich the interstellar medium (ISM). 

It is believed that AGB stars begin their lives on the AGB as oxygen-rich M-type stars, and eventually some of these, those with masses in the range $\sim 1.5$--$4\;\msol$ \citep{Herwig2005}, will transition into carbon-rich C stars. With a C/O ratio close to 1, S stars are believed to occupy the evolutionary phase between M and C stars. The lowest mass AGB stars ($\lesssim 1 ~\msol$) do not transform into C stars because they do not undergo a third dredge-up event. The highest mass AGB stars ($\gtrsim 4~\msol$) also do not end their lives as C stars due to hot bottom burning (HBB), unless the mass loss quenches the HBB process, leaving time for the star to evolve into a C star before leaving the AGB.

Radiative transfer modelling of circumstellar CO radio lines has long been used to derive the mass-loss rates of AGB stars \citep{Morris1987,Kastner1992,Justtanont1994,Groenewegen1998,Schoier2001,Olofsson2002,Decin2006,Ramstedt2009,De-Beck2010}. These data were almost exclusively obtained with ground-based telescopes. The \textit{Herschel} Space Observatory allowed observations of higher energy lines than possible from ground-based telescopes. This has led to studies that model molecular emission (not only that of CO) in more detail and over a wide range of energies as in \citet{Schoier2011}, \citet{Khouri2014}, and \citet{Danilovich2014}. However, each of those papers deals with only one or a handful of stars. The new data have not yet been applied to mass-loss rate determinations for larger samples of stars.

In this paper we present new data obtained as part of the SUCCESS project \citep[SUbmilimetre Catalogue of Circumstellar EnvelopeS with \textit{HerSchel}/HIFI,][]{Teyssier2011}, a CO survey of a large sample of AGB and post-AGB stars across all three chemical types. \textit{Herschel}/HIFI was used to observe the CO $J=5\rightarrow4$ and $J=9\rightarrow8$ lines in 53 AGB stars. Of these, six stars were also observed in the CO $J=14\rightarrow13$ line and 29 stars were also observed in the CO $J=1\rightarrow0$ and $J=2\rightarrow1$ lines using the IRAM 30m telescope. Of these stars we present radiative transfer models for 38 objects, supplementing our new high-resolution observations with archival data where possible. Inclusion of the high-$J$ lines from \textit{Herschel}/HIFI allows us to better constrain our models over a broad range of temperatures, and hence produce models that better represent the CSE over a large radial range.


\section{Sample and observations}

\begin{table*}
\caption{Basic stellar parameters for the AGB stars in the SUCCESS sample}             
\label{stars}      
\centering                          
\begin{tabular}{l c l l c c c}
\hline \hline					
Star	&	Alternate name	&	RA	&	Dec	&	Variability	&	Period	&	$\upsilon_\mathrm{lsr}$	\\
	&		&	\,\,(J2000)	&	\,\,(J2000)	&		&	[days]	&	[km s$^{-1}$]	\\
\hline			
\quad \it C stars	\\										
R Scl	&	IRC$-$30015	&	01 26 58.09	&	$-$32 32 35.5	&	SRb	&	370	&	-19	\\
V701 Cas	&	AFGL 482	&	03 23 36.57	&	+70 27 07.5	&	M	&	567	&	-14	\\
V384 Per	&	IRC+50096	&	03 26 29.53	&	+47 31 50.2	&	M	&	535	&	-16	\\
GY Cam	&	IRC+60144	&	04 35 17.54	&	+62 16 23.8	&	SR	&	...	&	-47	\\
R Lep	&	IRC$-$10080 	&	04 59 36.35	&	$-$14 48 22.5	&	M	&	445	&	11	\\
V1259 Ori	&	AFGL 865	&	06 03 59.84	&	+07 25 54.4	&	M	&	696	&	42	\\
IRAS 06192+4657	&		&	06 22 58.52	&	+46 55 34.9	&	?	&	...	&	-23	\\
UU Aur	&	IRC +40158	&	06 36 32.84	&	+38 26 43.8	&	SRb	&	235	&	8	\\
V688 Mon	&	AFGL 971	&	06 36 54.24	&	+03 25 28.7	&	M	&	653	&	3	\\
AI Vol	&	IRAS 07454 $-$7112	&	07 45 02.80	&	$-$71 19 43.2	&	M	&	511	&	-39	\\
U Hya	&	IRC$-$10242	&	10 37 33.27	&	$-$13 23 04.4	&	SRb	&	183	&	-32	\\
X TrA	&	IRAS 15094 $-$6953	&	15 14 19.18	&	$-$70 04 46.1	&	SR	&	361	&	-2	\\
II Lup	&	IRAS 15194 $-$5115	&	15 23 04.91	&	$-$51 25 59.0	&	M	&	575	&	-15.5	\\
V CrB	&	IRC+40273	&	15 49 31.31	&	+39 34 17.9	&	M	&	358	&	-99	\\
V821 Her	&	IRC+20370	&	18 41 54.39	&	+17 41 08.5	&	M	&	524	&	-0.5	\\
V Aql	&	IRC$-$10486	&	19 04 24.16	&	$-$05 41 05.4	&	SRb	&	407	&	54	\\
V1968 Cyg	&	AFGL 2494	&	20 01 08.51	&	+40 55 40.2	&	M	&	783	&	28	\\
RV Aqr	&	IRC+00499	&	21 05 51.68	&	$-$00 12 40.3	&	M	&	453	&	1	\\
TX Psc	&	IRC+00532	&	23 46 23.52	&	+03 29 12.5	&	LB	&	...	&	13	\\
\quad \it S stars\\
R And	&	IRC+40009	&	00 24 01.95	&	+38 34 37.3	&	M	&	409	&	-16	\\
S Cas	&	IRC+70024	&	01 19 41.97	&	+72 36 39.3	&	M	&	612	&	-30	\\
W And	&	IRC+40037	&	02 17 32.96	&	+44 18 17.8	&	M	&	397	&	-35	\\
R Gem	&	IRC+20171	&	07 07 21.27	&	+22 42 12.7	&	M	&	370	&	-60	\\
Y Lyn	&	IRC+50180	&	07 28 11.61	&	+45 59 26.2	&	SRc	&	110	&	-0.5	\\
RS Cnc	&	IRC+30209	&	09 10 38.80	&	+30 57 47.3	&	SRb	&	242	&	6.5	\\
R Cyg	&	IRC+50301	&	19 36 49.38	&	+50 11 59.5	&	M	&	426	&	-17	\\
$\pi$ Gru	&	AFGL 4289	&	22 22 44.21	&	-45 56 52.6	&	SRb	&	196	&	-12	\\
\quad \it M stars\\
KU And	&	IRC+40004	&	00 06 52.94	&	+43 05 00.0	&	M	&	720	&	-22	\\
V370 And	&	IRC+50049	&	01 58 44.33	&	+45 26 06.9	&	SRb	&	228	&	-2	\\
AFGL 292	&	IRC+10025	&	02 02 38.63	&	+07 40 36.5	&	?	&	...	&	23	\\
R Hor	&	IRAS 02522 $-$5005	&	02 53 52.77	&	$-$49 53 22.7	&	M	&	408	&	37	\\
NV Aur	&	IRC+50137	&	05 11 19.44	&	+52 52 33.2	&	M	&	635	&	2	\\
BX Cam	&	IRC+70066	&	05 46 44.10	&	+69 58 25.2	&	M	&	...	&	-2	\\
GX Mon	&	IRC+10143 	&	06 52 46.91	&	+08 25 19.0	&	M	&	527	&	-9	\\
L$_2$ Pup	&	IRAS 07120 $-$4433	&	07 13 32.32	&	$-$44 38 23.1	&	SRb	&	141	&	33.5	\\
S CMi	&	IRC+10167	&	07 32 43.07	&	+08 19 05.2	&	M	&	333	&	51	\\
R LMi	&	IRC+30215	&	09 45 34.28	&	+34 30 42.8	&	M	&	372	&	0	\\
R Leo	&	IRC+10215	&	09 47 33.49	&	+11 25 43.7	&	M	&	310	&	0	\\
R Crt	&	IRC$-$20222	&	11 00 33.85	&	$-$18 19 29.6	&	SRb	&	160	&	11	\\
BK Vir	&	IRC+00220	&	12 30 21.01	&	+04 24 59.2	&	SRb	&	140	&	17	\\
Y UMa	&	IRC+60220	&	12 40 21.28	&	+55 50 47.6	&	SRb	&	168	&	19	\\
RT Vir	&	IRC+10262 	&	13 02 37.98	&	+05 11 08.4	&	SRb	&	158	&	18	\\
SW Vir	&	IRC+00230	&	13 14 04.39	&	$-$02 48 25.2	&	SRb	&	146	&	-11	\\
R Hya	&	IRC$-$20254	&	13 29 42.78	&	$-$23 16 52.8	&	M	&	380	&	-10	\\
RX Boo	&	IRC+30257	&	14 24 11.63	&	+25 42 13.4	&	SRb	&	158	&	2	\\
S CrB	&	IRC+30272	&	15 21 23.96	&	+31 22 02.6	&	M	&	360	&	0	\\
X Her	&	IRC+50248	&	16 02 39.17	&	+47 14 25.3	&	SRb	&	102	&	-73	\\
V1111 Oph	&	IRC+10365	&	18 37 19.26	&	+10 25 42.2	&	M	&	...	&	-32	\\
RR Aql	&	IRC+00458	&	19 57 36.06	&	$-$01 53 11.3	&	M	&	395	&	28	\\
V1943 Sgr	&	IRC$-$30425	&	20 06 55.24	&	$-$27 13 29.8	&	SRb	&	330	&	-15	\\
V1300 Aql	&	IRC$-$10529	&	20 10 27.87	&	$-$06 16 13.6	&	M	&	680	&	-18	\\
T Cep	&	IRC+70168	&	21 09 31.78	&	+68 29 27.2	&	M	&	388	&	-2	\\
EP Aqr	&	IRC+00509	&	21 46 31.85	&	$-$02 12 45.9	&	SRb	&	55	&	-34	\\
\hline
\end{tabular}
\tablefoot{An ellipsis (...) indicates an unknown property. Variability and period information was obtained from the International Variable Star Index (VSX) database.}
\end{table*}

SUCCESS is a \textit{Herschel}/HIFI Guaranteed Time project \citep{Teyssier2011} which observed 74 AGB and post-AGB stars in the CO $J=5\rightarrow4$ and $J=9\rightarrow8$ lines, and 10 of those objects also in the CO $J=14\rightarrow13$ line. Excluding post-AGB objects and extreme OH/IR stars \citep[the latter presented in ][]{Justtanont2013}, the sample comprised of 53 AGB stars observed in the former two lines and six observed in all three lines. The sample was primarily built based on the evolved star lists considered in the studies
by \cite{Knapp1998}, \cite{Schoier2001}, and \cite{Castro-Carrizo2010}, complemented by some of the brightest line 
calibrators regularly used at the APEX telescope \citep{Gusten2008}. 
The selection criterion was related to the intensity of the CO $J=2\to1$ as observed at the IRAM 30m telescope, and aimed at sources exhibiting peak intensities in excess of 1.5\,K ($T_{\rm mb}$) in this line.

A reduced sample of 29 objects was also observed with the IRAM 30m telescope in the CO $J = 1\to0$ and $J =2\to1$ lines.
The objective was to obtain well-calibrated data observed with state-of-the-art receivers, especially for those sources where the literature spectra dated from several decades ago and/or had been obtained from 
a varied set of facilities.

The full sample is summarised in Table \ref{stars}, including the pulsation periods and the systemic velocities. A summary of all our new observations and integrated line intensities is given in Table \ref{obstbl}. We have indicated with an * those lines which have some ISM contamination that was corrected for when calculating the integrated intensity. In these cases, the contamination tended to occur only on one side of the line and we calculated the intensity by integrating from the centre to the edge of the non-contaminated side and then doubling this result. The beam-widths for each line and telescope are given in Table \ref{scopedat}.

The new data for stars we have modelled are shown in Figures \ref{Cmods}, \ref{Smods}, and \ref{Mmods}. The new observations for stars we have not modelled are shown in Figures \ref{Cplots}, \ref{Splots}, and \ref{Mplots}.

\subsection{IRAM observations}

The complementary observations at the IRAM 30m telescope were obtained in December 2013 respectively using the E090 and E230 EMIR receivers and the FTS backend to cover the CO $J=1\to0$ and $J=2\to1$ lines. We used the Wobbler Switching mode with a throw of 4\arcmin.
Owing to the large instantaneous bandwidth offered by the EMIR receivers, several bonus lines from  $^{13}$CO, CN, SiS, and HC\down{3}N were simultaneously observed and are presented in Section~\ref{sec:bonus}. Data were processed with the GILDAS/Class\footnote{http://www.iram.fr/IRAMFR/GILDAS} software and converted into the $T_{\rm mb}$ scale assuming main beam efficiencies of 0.78 and 0.58 at 3 and 1\,mm respectively. We have corrected for the antenna elevation gain, which accounts for up to 25\% line intensity loss in our data. Finally, we re-adjusted our line intensities based on the monitoring of reference spectra of the CO lines in IRC+10216 (with fiducial line peak intensities of 24.3 K and 54.5 K for the CO $J=1\to0$ and $J=2\to1$ lines, respectively), resulting in corrections between 0 and 20\% depending on the line and the observing day. The achieved noise rms are in the range 30--150\,mK and 55--360\,mK at 3 and 1\,mm, respectively, and per native velocity resolution element (0.5 and 0.25\,$\kms$, respectively).

\subsection{HIFI observations}

The SUCCESS sample was observed between July 2010 and April 2012 using the HIFI instrument \citep{de-Graauw2010} aboard the \textit{Herschel} Space Observatory \citep{Pilbratt2010}. The data were obtained with the double beam switching mode, using reference positions separated by 3\arcmin{} from the target position \citep{Roelfsema2012}. The spectra were sampled on the Wide Band Acousto-Optical Spectrometer (WBS), offering a native resolution of 1.1 MHz (0.6 $\kms$ at the lowest observed frequency). The selected frequency tunings were optimised for the targeted CO lines, but bonus lines of CN and SiO were also obtained in the instantaneous  bandwidth of 4 GHz (2.4 GHz for the CO $J=14\to13$ line) provided by the WBS (see Sect. \ref{sec:bonus}). The full list of {\it Herschel} observation identifiers (ObsIDs) is given in Table \ref{obsids}.

The HIFI data have been processed using HIPE 12\footnote{HIPE is a joint development by the \textit{Herschel} Science Ground Segment Consortium, consisting of ESA, the NASA \textit{Herschel} Science Center, and the HIFI, PACS, and SPIRE consortia.} and calibrated in the $T^*_{\rm A}$ scale. On top of that, the CO $J=14\to13$ data have been corrected for the so-called electrical standing waves using the {\it doHebCorrection} task \citep{Kester2014}. A sideband gain ratio different from unity has been used for the CO $J=5\to4$ observations \citep{Higgins2014}. Finally, all our data have been converted to the $T_{\rm mb}$ using the revised main beam efficiencies (M\"ueller et al., 2014\footnote{{http://herschel.esac.esa.int/twiki/pub/Public/HifiCalibrationWeb/\\HifiBeamReleaseNote\_Sep2014.pdf}}). Likewise, we refer to this technical note for the beam size assumed in our modelling (Table \ref{scopedat}). The achieved 1-$\sigma$ noise rms was 15 mK ($T_{\rm mb}$) for a smoothed resolution channel of 3 $\kms$ for both CO $J=5\to4$ and $J=9\to8$, and 44 mK for CO $J=14\to13$.

\begin{table*}
\caption{HIFI and IRAM CO line observations}             
\label{obstbl}      
\centering                          
\begin{tabular}{l c c c c c c}
\hline \hline
 & & \multicolumn{2}{c}{IRAM} & \multicolumn{3}{c}{HIFI}\\
Star	&	$\upsilon_\infty$	&	CO ($1\rightarrow0$)	&	CO ($2\rightarrow1$)	&	CO ($5\rightarrow4$)	&	CO ($9\rightarrow8$)	&	CO ($14\rightarrow13$)	\\
	&	[km s$^{-1}$]	&	[K km s$^{-1}$]	&	[K km s$^{-1}$]	&	[K km s$^{-1}$]	&	[K km s$^{-1}$]	&	[K km s$^{-1}$]	\\
\hline	
\quad \it C stars	\\											
R Scl	&	16.5	&	81.9	\,\,(1.0)	&	103.3	\,\,(0.9)\phantom{1}	&	5.30	\,\,(0.08)	&	1.57	\,\,(0.07)	&	1.33	\,\,(0.38)	\\
V701 Cas	&	11.5	&*	36.4	\,\,(0.3)\phantom{ *}	&*	44.4	\,\,(0.4)\phantom{ *}	&	1.67	\,\,(0.22)	&	1.93	\,\,(0.07)	&	...		\\
V384 Per	&	15	&	63.6	\,\,(0.2)	&	90.1	\,\,(0.3)	&	5.52	\,\,(0.18)	&	6.25	\,\,(0.09)	&	...		\\
GY Cam	&	20	&	43.8	\,\,(0.2)	&	70.1	\,\,(0.3)	&	3.14	\,\,(0.20)	&	2.90	\,\,(0.08)	&	...		\\
R Lep	&	18	&	32.5	\,\,(0.4)	&	100.3	\,\,(0.6)\phantom{1}	&	4.23	\,\,(0.13)	&	4.57	\,\,(0.08)	&	...		\\
V1259 Ori	&	16	&	63.8	\,\,(0.2)	&	85.7	\,\,(0.3)	&	3.93	\,\,(0.21)	&	3.23	\,\,(0.07)	&	...		\\
IRAS 06192+4657	&	6	&	...				&	...				&	$<$0.22	\,\,(0.16)\phantom{$<$}	&	0.290	\,\,(0.171)	&	...		\\
UU Aur	&	12	&	...				&	...				&	2.47	\,\,(0.15)	&	2.31	\,\,(0.08)	&	...		\\
V688 Mon	&	13.5	&*	30.8	\,\,(0.3)\phantom{ *}	&*	67.4	\,\,(0.2)\phantom{ *}	&	1.96	\,\,(0.19)	&	2.31	\,\,(0.09)	&	...		\\
AI Vol	&	12	&	...	&	...				&	6.87	\,\,(0.13)	&	8.12	\,\,(0.08)	&	7.62	\,\,(0.23)	\\
U Hya	&	6.5	&	8.87	\,\,(0.37)	&	53.7	\,\,(1.0)	&	2.98	\,\,(0.16)	&	2.07	\,\,(0.11)	&	...		\\
X TrA	&	6.5	&	...				&	...				&	1.36	\,\,(0.18)	&	1.76	\,\,(0.08)	&	...		\\
II Lup	&	21.5	&	...				&	...				&	16.8	\,\,(0.1)\phantom{1}	&	17.6	\,\,(0.07)	&	...		\\
V CrB	&	7.5	&	...				&	...				&	0.541	\,\,(0.159)	&	1.07	\,\,(0.08)	&	...		\\
V821 Her	&	13.5	&	82.8	\,\,(0.5)	&	136.5	\,\,(0.7)\phantom{1}	&	6.32	\,\,(0.15)	&	7.46	\,\,(0.08)	&	...		\\
V Aql	&	11	&	12.6	\,\,(0.2)	&	23.1	\,\,(0.3)	&	1.36	\,\,(0.13)	&	1.30	\,\,(0.07)	&	...		\\
V1968 Cyg	&	20	&	37.2	\,\,(0.3)	&	!				&	5.58	\,\,(0.18)	&	5.79	\,\,(0.07)	&	...		\\
RV Aqr	&	15	&	46.0	\,\,(0.2)	&	73.9	\,\,(0.4)	&	3.64	\,\,(0.13)	&	4.16	\,\,(0.07)	&	...		\\
TX Psc	&	4	&	...				&	...				&	0.887	\,\,(0.141)	&	1.24	\,\,(0.08)	&	...		\\
\quad \it S stars \\
R And	&	8	&	27.0	\,\,(0.4)	&	56.9	\,\,(0.4)	&	3.72	\,\,(0.16)	&	5.58	\,\,(0.08)	&	...		\\
S Cas	&	19	&	37.6	\,\,(0.3)	&	90.9	\,\,(0.5)	&	4.27	\,\,(0.18)	&	6.08	\,\,(0.08)	&	...		\\
W And	&	6	&	8.14	\,\,(0.1)	&	24.5	\,\,(0.4)	&	1.07	\,\,(0.13)	&	1.34	\,\,(0.09)	&	...		\\
R Gem	&	5	&	7.01	\,\,(0.3)	&	10.9	\,\,(0.6)	&	0.565	\,\,(0.230)	&	0.471	\,\,(0.156)	&	...		\\
Y Lyn	&	8	&	11.1	\,\,(0.4)	&	29.6	\,\,(0.6)	&	1.52	\,\,(0.19)	&	1.28	\,\,(0.09)	&	...		\\
RS Cnc	&	2.5	&	22.9	\,\,(0.2)	&	63.4	\,\,(0.5)	&	3.60	\,\,(0.19)	&	3.17	\,\,(0.08)	&	...		\\
R Cyg	&	9	&	15.6	\,\,(0.3)	&	43.2	\,\,(0.5)	&	2.21	\,\,(0.16)	&	1.42	\,\,(0.08)	&	...		\\
$\pi$ Gru	&	10	&	...				&	...				&	13.8	\,\,(0.15)	&	13.9	\,\,(0.04)	&	10.0	\,\,(0.1)\phantom{1}	\\
\quad \it M stars \\
KU And	&	20	&	48.9	\,\,(0.3)	&	83.5	\,\,(0.4)	&	3.15	\,\,(0.16)	&	4.41	\,\,(0.07)	&	...		\\
V370 And	&	9	&	...				&	...				&	2.96	\,\,(0.22)	&	2.35	\,\,(0.09)	&	...		\\
AFGL 292	&	8.5	&	4.56	\,\,(0.23)	&	10.1	\,\,(0.2)	&	0.492	\,\,(0.229)	&	0.481	\,\,(0.127)	&	...		\\
R Hor	&	4	&	...				&	...				&	4.23	\,\,(0.14)	&	3.31	\,\,(0.08)	&	...		\\
NV Aur	&	18	&	43.3	\,\,(0.2)	&	59.1	\,\,(0.2)	&	2.54	\,\,(0.13)	&	1.84	\,\,(0.07)	&	...		\\
BX Cam	&	19	&	33.9	\,\,(0.2)	&	50.8	\,\,(0.2)	&	2.96	\,\,(0.19)	&	2.58	\,\,(0.08)	&	...		\\
GX Mon	&	19	&	64.2	\,\,(0.1)	&	128.1	\,\,(0.2)\phantom{1}	&	5.83	\,\,(0.18)	&	4.07	\,\,(0.07)	&	...		\\
L$_2$ Pup	&	2	&	...				&	...				&	1.82	\,\,(0.14)	&	2.56	\,\,(0.09)	&	3.41	\,\,(0.24)	\\
S CMi	&	2	&	\phantom{0}0.608	\,\,(0.358)	&	\phantom{1}3.84	\,\,(0.60)	&	0.271	\,\,(0.108)	&	0.436	\,\,(0.084)	&	...		\\
R LMi	&	7.5	&	...				&	...				&	1.51	\,\,(0.13)	&	1.27	\,\,(0.08)	&	...		\\
R Leo	&	8.5	&	...				&	...				&	5.86	\,\,(0.15)	&	8.35	\,\,(0.08)	&	...		\\
R Crt	&	12	&	...				&	...				&	3.57	\,\,(0.18)	&	3.26	\,\,(0.08)	&	...		\\
BK Vir	&	6	&	\phantom{1}3.73	\,\,(0.16)	&	13.1	\,\,(0.2)	&	0.474	\,\,(0.101)	&	1.03	\,\,(0.08)	&	...		\\
Y UMa	&	7.5	&	...				&	...				&	0.735	\,\,(0.088)	&	0.807	\,\,(0.090)	&	...		\\
RT Vir	&	8.5	&	...				&	...				&	1.90	\,\,(0.14)	&	1.53	\,\,(0.10)	&	...		\\
SW Vir	&	8.5	&	...				&	...				&	4.34	\,\,(0.23)	&	4.14	\,\,(0.09)	&	...		\\
R Hya	&	10	&	...				&	...				&	6.94	\,\,(0.22)	&	7.85	\,\,(0.09)	&	3.98	\,\,(0.17)	\\
RX Boo	&	10	&	...				&	...				&	4.52	\,\,(0.23)	&	3.79	\,\,(0.09)	&	2.92	\,\,(0.24)	\\
S CrB	&	7	&	\phantom{1}3.19	\,\,(0.33)	&	!	&	1.19	\,\,(0.11)	&	0.910	\,\,(0.092)	&	...		\\
X Her	&	8.5	&	...				&	...				&	2.37	\,\,(0.18)	&	1.98	\,\,(0.08)	&	...		\\
V1111 Oph	&	17	&	47.2	\,\,(0.3)	&	82.1	\,\,(0.4)	&	3.13	\,\,(0.19)	&	2.50	\,\,(0.07)	&	...		\\
RR Aql	&	9	&	...				&	...				&	1.50	\,\,(0.11)	&	1.14	\,\,(0.08)	&	...		\\
V1943 Sgr	&	6.5	&	...				&	...				&	0.862	\,\,(0.168)	&	0.754	\,\,(0.082)	&	...		\\
V1300 Aql	&	18	&	68.9	\,\,(0.4)	&	102.7	\,\,(0.7)\phantom{1}	&	3.66	\,\,(0.15)	&	3.12	\,\,(0.06)	&	...		\\
T Cep	&	5.5	&	...				&	...				&	1.44	\,\,(0.10)	&	1.90	\,\,(0.09)	&	...		\\
EP Aqr	&	12	&	...				&	...				&	3.73	\,\,(0.16)	&	3.35	\,\,(0.08)	&	...		\\
\hline
\end{tabular}
\tablefoot{The value in brackets after the flux gives the integrated noise RMS. An ellipsis (...) indicates that the line was not observed for the indicated star; * indicates that flux has been corrected for ISM emission; ! indicates the line was observed but with unreliable flux calibration.}
\end{table*}

\begin{table}
\caption{Telescope parameters for all lines referred to in this paper}             
\label{scopedat}      
\centering                          
\begin{tabular}{l c c c}
\hline \hline
Transition & Frequency &	Telescope&	$\theta$ \\
     &       [GHz]     & 		&[$\arcsec$]   \\
\hline
CO ($1\rightarrow0$) & \phantom{0}115.271& IRAM & 21.4\\
 & & NRAO & 55\phantom{.0}\\
 & & OSO & 33\phantom{.0}\\
 & & SEST & 45\phantom{.0}\\
CO ($2\rightarrow1$) & \phantom{0}230.538 & APEX & 27\phantom{.0}\\
 & & CSO & 30\phantom{.0}\\
 & & IRAM & 10.7\\
 & & JCMT & 21\phantom{.0}\\
 & & SEST & 23\phantom{.0}\\
CO ($3\rightarrow2$) & \phantom{0}345.796& APEX & 18\phantom{.0}\\
 & & CSO & 20\phantom{.0}\\
 & & JCMT & 14\phantom{.0}\\
 & & SEST & 15\phantom{.0}\\
CO ($4\rightarrow3$) & \phantom{0}461.041& APEX & 14\phantom{.0} \\
 & & CSO & 15.5\\
 & & JCMT & 12\phantom{.0}\\
CO ($5\rightarrow4$) & \phantom{0}576.268 & HIFI & 36.1\\
CO ($6\rightarrow5$) & \phantom{0}691.473 & CSO & 10.3\\
 & & HIFI & 30.4\\
 & & JCMT & 8\\
CO ($7\rightarrow6$) & \phantom{0}806.652 & APEX & \phantom{1}7.7\\
CO ($9\rightarrow8$) & 1036.912 & HIFI & 20.1\\
CO ($10\rightarrow9$)& 1151.985 & HIFI & 18.2\\
CO ($14\rightarrow13$)& 1611.794 & HIFI & 12.9\\
CO ($16\rightarrow15$)& 1841.345 & HIFI & 11.5\\
\hline
\end{tabular}
\tablefoot{APEX is the Atacama Pathfinder Experiment; CSO is the Caltech Submillimeter Observatory; IRAM refers to the 30m telescope at the Institut de Radioastronomie Millim\'etrique; JCMT is the James Clerk Maxwell Telescope; HIFI is the Heterodyne Instrument for the Far-Infrared aboard {\it Herschel}; NRAO refers to the 12 m telescope at the (US) National Radio Astronomy Observatory; OSO is the 20 m telescope at the Onsala Space Observatory; SEST is the Swedish-ESO submillimetre telescope. }
\end{table}

\subsection{Supplementary data}\label{suppobs}

To better constrain our models, we used previously observed low- and intermediate-$J$ CO lines from a variety of sources. A summary of the supplementary observations is given in Table \ref{lowjobs}.

The telescopes and their corresponding beam widths for different frequencies are listed Table \ref{scopedat}, covering all the new and archival observations used in this paper.

As well as archival line data, we have used results from the APEX Pointing Catalogue, which can be found online\footnote{{http://www.apex-telescope.org/observing/pointing/spectra/}}.

For our SED models we primarily used photometry from IRAS and 2MASS \citep{hog2000}. See Sect. \ref{SED} for more details. 

\subsection{Bonus lines}\label{sec:bonus}

In the course of our new HIFI and IRAM observations, we also acquired some ``bonus" line spectra for molecules that were observable within our target frequency ranges. Our HIFI observations covered the CN $(5_{9/2}\to4_{7/2})$ and $(5_{11/2}\to4_{9/2})$ line groups, which were detected in a handful of C stars, and SiO $(13\to12)$, which was detected mostly in M stars. Our IRAM observations covered the CN $N=1\to0$ and CN $N=2\to1$ line groups, which were detected in most C stars and one S star (S Cas). Also covered by IRAM were the \up{13}CO ($1\to0$) line, which was detected in most observed sources, the SiS ($6\to5$) line, which was detected in higher mass-loss rate sources, and the HC$_3$N ($12\to11$) line, which was detected in the higher mass-loss rate C stars.

We will not be modelling these additional species but we discuss the bonus detections in greater detail in Sect. \ref{a:bonus}. The observations are plotted in Figures \ref{13COlines}, \ref{SiOlines}, \ref{SiSlines}, and \ref{CNlines}. The integrated intensities for the detected lines are listed in Table \ref{bonustbl}.

\section{Radiative transfer modelling}\label{modelling}

\subsection{SED modelling}\label{SED}

We begin our radiative transfer modelling by estimating some key dust properties of each star in the sample. The SED modelling is performed using DUSTY\footnote{http://www.pa.uky.edu/$\sim$moshe/dusty/}, a publicly available radiative transfer code \citep{Ivezic1997}. We found the best fit for each star using primarily 2MASS and IRAS photometric observations. The distances taken from the literature and the luminosities were calculated from the period-luminosity relation of \cite{Glass1981}:
\begin{eqnarray}
M_{bol} =& 0.76 -2.09 \log P
\end{eqnarray}
The resulting { effective temperature of the central black body}, dust optical depth (given at $10~\mic$) and the inner radius of our model, based on the dust condensation temperature, are listed in Table \ref{SEDtbl}.

As the DUSTY code has been widely used, some of the stars in our sample have already been modelled using the same methods and in these cases we simply use the earlier results from \citet{Schoier2007}, \citet{Schoier2013}, and \cite{Ramstedt2014}. These are indicated with a $\dagger$ in Table \ref{SEDtbl}.

\begin{table*}
\caption{SED parameters}             
\label{SEDtbl}      
\centering                          
\begin{tabular}{l r r r l c l}
\hline \hline
Star	&	$D\;\,$	&	$L_*\;$	&	$T_*\,\;$	&	$\;\tau_{10}$	&	$R_\mathrm{in}$	&	Reference	\\
	&	[pc]	&	[$\lsol$]	&	[K]\,\,	&		&	[$\times 10^{14}$ cm]	&		\\
\hline	
\quad \it C stars\\
V701 Cas	&	1720	&	7800	&	2800	&	0.85	&	2.1	&		\citet{Menzies2006}	\\
V384 Per	&	560	&	8100	&	2000	&	0.25	&	2.1	&	$\dagger$	\citet{Schoier2013}	\\
GY Cam	& 1030	&	7800	&	2000	&	0.3	&	2.0	&		\citet{Groenewegen2002}	\\
R Lep	&	432	&	5500	&	2200	&	0.06	&	1.7	&	$\dagger$	\cite{Ramstedt2014}	\\
V1259 Ori	&	1600	&	9300	&	2200	&	1.7	&	2.2	&		\citet{Menzies2006}	\\
UU Aur	&	260	&	6900	&	2800	&	0.017	&	1.9	&	$\dagger$	\citet{Schoier2013}	\\
V688 Mon	&	1770	&	8800	&	2800	&	0.6	&	2.2	&		\citet{Menzies2006}	\\
AI Vol	&	710	&	9000	&	2100	&	0.45	&	2.2	&	$\dagger$	\citet{Schoier2007}	\\
U Hya	&	208	&	*4000	&	2400	&	0.012	&	1.5	&	$\dagger$	\cite{Ramstedt2014}	\\
X TrA	&	360	&	5400	&	2200	&	0.024	&	1.7	&		\citet{Cox2012}	\\
II Lup	&	500	&	8800	&	2400	&	0.55	&	2.2	&	$\dagger$	\citet{Schoier2013}	\\
V CrB	&	630	&	5300	&	1800	&	0.035	&	1.7	&		\citet{Cox2012}	\\
V821 Her	&	600	&	7900	&	2200	&	0.45	&	2.1	&	$\dagger$	\citet{Schoier2013}	\\
V Aql	&	330	&	6500	&	2800	&	0.02	&	1.9	&	$\dagger$	\citet{Schoier2013}	\\
V1968 Cyg	&	1480	&	10200	&	2400	&	0.85	&	2.3	&		\citet{Menzies2006}	\\
RV Aqr	&	670	&	6800	&	2200	&	0.27	&	1.9	&	$\dagger$	\citet{Schoier2013}	\\
\quad \it S stars\\
R And	&	350	&	6300	&	1900	&	0.05	&	1.8	&	$\dagger$	\cite{Ramstedt2014}		\\
S Cas	&	570	&	8000	&	1800	&	0.5	&	2.1	&	$\dagger$	\cite{Ramstedt2014}		\\
W And	&	450	&	5800	&	2400	&	0.1	&	1.8	&		\cite{van-Leeuwen2007}		\\
R Gem	&	820	&	5500	&	2400	&	0.035	&	1.7	&		\cite{Whitelock2008}	$\ddagger$	\\
Y Lyn	&	253	&	*4000	&	2400	&	0.02	&	1.5	&		\cite{Ramstedt2014}		\\
R Cyg	&	690	&	6200	&	2600	&	0.14	&	1.8	&		\cite{Whitelock2008}	$\ddagger$	\\
\quad \it M stars\\
KU And	&	680	&	11800	&	2000	&	0.90	&	2.5	&	$\dagger$	\citet{Schoier2013}		\\
AFGL 292	&	319	&	*6000	&	2200	&	0.02	&	1.8	&		\cite{Winters2003}		\\
R Hor	&	310	&	8500	&	2200	&	0.30	&	2.1	&	$\dagger$	\citet{Schoier2013}		\\
NV Aur	&	1200	&	9800	&	2000	&	3.50	&	2.3	&	$\dagger$	\citet{Schoier2013}		\\
BX Cam	&	500	&	7500	&	2800	&	1.30	&	2.0	&	$\dagger$	\citet{Schoier2013}		\\
GX Mon	&	550	&	8200	&	2600	&	2.00	&	2.1	&	$\dagger$	\citet{Schoier2013}		\\
L$_2$ Pup	&	86	&	*4000	&	2800	&	0.07	&	1.5	&		\citet{Schoier2013}		\\
S CMi	&	470	&	5000	&	2800	&	0.07	&	1.6	&		\cite{Knapp1998}		\\
R LMi	&	330	&	5500	&	2400	&	0.2	&	1.7	&		\cite{Whitelock2008}	$\ddagger$	\\
R Leo	&	130	&	4600	&	1800	&	0.10	&	1.6	&	$\dagger$	\citet{Schoier2013}		\\
S CrB	&	400	&	5400	&	2400	&	0.2	&	1.7	&		\cite{Knapp1998}		\\
V1111 Oph	&	750	&	7500	&	2000	&	0.75	&	2.0	&	$\dagger$	\citet{Schoier2013}		\\
RR Aql	&	530	&	7900	&	2000	&	0.70	&	2.1	&	$\dagger$	\citet{Schoier2013}		\\
V1943 Sgr	&	200	&	5000	&	2200&	0.05	&	1.6	&		\cite{van-Leeuwen2007}		\\
V1300 Aql	&	620	&	10600	&	2000	&	3.50	&	2.4	&	$\dagger$	\citet{Schoier2013}		\\
T Cep	&	190	&	5700	&	2400&	0.10	&	1.8	&		\cite{van-Leeuwen2007}		\\
\hline
\end{tabular}
\tablefoot{An * indicates that the luminosity is assumed (for semi-regular variables) rather than calculated. The references listed are for distances, $D$, and in some cases the luminosity, $L$, where no period is known. A $\dagger$ in the references indicates that all parameters including the effective temperature, $T_*$, and optical depth, $\tau_{10}$ were also taken from the referenced DUSTY modelling. The absence of a $\dagger$ indicates that temperature and optical depth were calculated as part of this work. $\ddagger$ indicates that the distance was calculated from the period-magnitude relation in the cited work.}
\end{table*}

\subsection{CO line modelling}\label{sec:comod}

\subsubsection{The circumstellar model}

In our radiative transfer modelling of the observed CO lines, we assumed a spherically symmetric CSE, formed through constant and isotropic mass loss with a smoothly accelerating wind. Some of our modelled stars, such as GY Cam and AFGL 292, show line profiles which we are not able to reproduce within our adopted circumstellar model. See Sect. \ref{bbyb} for further discussion of this.

The radial gas velocity law used in our modelling is given by
\begin{equation}\label{vel}
\upsilon(r) = \upsilon_\mathrm{min} + (\upsilon_\infty - \upsilon_\mathrm{min})
\left( 1 - \frac{R_\mathrm{in}}{r}\right) ^\beta
\end{equation}
where $\upsilon_\mathrm{min} = 3 \;\kms$ is the approximate sound speed at the dust condensation radius, $\upsilon_\infty$ is the observed terminal expansion velocity, $R_\mathrm{in}$ is the dust condensation radius as calculated in the corresponding SED model, and $\beta$ is a parameter used to adjust the acceleration in the inner part of the envelope. In general, we assume $\beta = 1.0$ but adjust it for the stars where this gives a noticeable improvement in the fits to the line shapes. The $\beta$ values used for each star are given in Table \ref{coresults}. For some stars we use a constant expansion velocity, either because the line width is especially low (such as for S CMi and L$_2$ Pup with $\upsilon_\infty = 2\;\kms$) or because the constant velocity model was a much better fit to the observations than a model with a velocity profile. The range of results is discussed in more detail in Sect. \ref{bbyb}. We also employ a constant turbulent gas velocity of $0.5~\kms$ for all the modelled stars.


A similar function to \eqn{vel} is used for the dust and drift velocities, where the drift velocity is the difference between the dust and gas velocities. The terminal drift velocity is assumed to be \citep[for details see][]{Kwok1975}:
\begin{equation}\label{driftvel}
\upsilon_{\mathrm{drift,}\infty}= \sqrt{\frac{L_*\upsilon_\infty Q}{\dot{M} c}}
\end{equation}
where $L_*$ is the stellar luminosity, $\dot{M}$ is the mass-loss rate, $Q$ is an efficiency factor assumed to be 0.03 \citep{Ramstedt2008}, and $c$ is the speed of light. The drift velocity profile is calculated from \eqn{vel} with this terminal velocity and with $\upsilon_\mathrm{min} = 1$. For each star, the same $R_\mathrm{in}$ and $\beta$ values are used for the drift (and hence dust) velocity as for the corresponding gas velocity. Where the gas velocity has $\beta_\mathrm{gas} = 0$, we assume that the drift velocity has $\beta_\mathrm{drift} = 1$. { Using this formulation, we find drift velocities in the range $2~\kms$ -- $18~\kms$ (with a median of $6~\kms$), which is below the sputtering-dominated threshold of $20~\kms$ calculated by \cite{Kwok1975}.}

\subsubsection{The modelling approach}\label{sec:modap}

We modelled the observed CO lines using a Monte Carlo program (MCP) which has been previously described in \citet{Bernes1979}, \citet{Schoier2001}, \citet{Schoier2002}, \citet{Ramstedt2008}, and \citet{Danilovich2014}. The MCP code takes basic stellar and molecular parameters, and the results of our SED model as input and calculates the molecular excitation by solving the statistical equilibrium equations using the Monte Carlo method. MCP also solves the energy balance equation to calculate the gas temperature as a function of radius throughout the CSE,

\begin{equation}
\frac{dT_\mathrm{kin}}{dr} = (2-2\gamma)\left(1+ \frac{r}{2 \upsilon(r)} \frac{d\upsilon}{dr} \right)\frac{T_\mathrm{kin}(r)}{r} + \frac{\gamma-1}{n_{\mathrm{H}_2} k_B \upsilon(r)} (H - C)
\end{equation}
where $T_\mathrm{kin}$ is the kinetic temperature of the gas, $n_{\mathrm{H}_2}$ is the hydrogen number density, $k_B$ is Boltzmann's constant, $\gamma$ is the adiabatic index with $\gamma = \frac{5}{3}$ for $T_\mathrm{kin} < 350$ K and $\gamma = \frac{7}{5}$ otherwise, and $H$ and $C$ are the sums of the heating and cooling terms, respectively. The adiabatic cooling is given by the first term on the right-hand side of the equation. Otherwise, the cooling terms include \h2 vibrational line cooling and CO rotational line cooling, and the heating terms include heating due to dust-gas collisions and photoelectric heating. CO line cooling can also act as heating in some circumstances \citep[for more details, see][]{Schoier2001}.

The most important heating term in the energy balance equation, the dust-gas collision term, includes a number of assumptions on the dust properties. These come in a multiplicative form and we combine them in the so called $h$-parameter defined as
\begin{equation}\label{hparam}
h = \left( \frac{\Psi}{0.01}\right)\left( \frac{2.0\;\mathrm{g}\;\mathrm{cm}^{-3}}{\rho_d}\right)\left( \frac{0.05\;\mic}{a_g}\right)
\end{equation}
where $\Psi$ is the dust-to-gas ratio, $\rho_d$ is the dust grain density and $a_g$ is the dust grain radius, assuming spherical grains. We assume $\rho_d = 2.2$ g cm$^{-3}$ for carbon dust and $\rho_d = 3.3$ g cm$^{-3}$ for silicate dust, and $a_g = 0.05~\mic$. The $h$ parameter is one of the two free parameters adjusted (in practice we vary $\Psi$) in our modelling to fit the observational data, the other being the mass-loss rate. The impact of the $h$-parameter is on the kinetic temperature structure of the circumstellar envelope, which tends to be close to a power law for part of the envelope. The molecular excitation analysis is complex and therefore the effect of changing $h$ is not always easily predictable, but the tuning of the $h$ parameter finally results in a kinetic temperature distribution that best matches the observed data, in particular the relative intensities of the different $J$ lines. However, since we may be missing terms in the energy balance equation, or the included terms do not fully capture the physics involved, it is clear that the resulting $\Psi$ of the best-fit model is not necessarily a good estimate of the circumstellar dust-to-gas ratio.

Our CO analyses assumed inner CO fractional abundances (with respect to \h2) of $1\e{-3}$ for C stars, $6\e{-4}$ for S stars and $3\e{-4}$ for M stars, in line with canonical results. The sizes of the CO envelopes, which are determined by photodissociation, were calculated using the results of \citet{Mamon1988}.

In our CO excitation analyses we included radiative transitions for the first 40 rotational energy levels in the ground and first vibrationally excited states, taken from \cite{Chandra1996}. The collisional rates, which were only available for the ground vibrational state, were taken from \citet{Yang2010} and weighted assuming an ortho-/para-\h2 ratio of 3 by \citet{Schoier2011}. They cover kinetic temperatures from 2 to 3000 K.

We calculated the best fit models for each star using a $\chi^2$ statistic, which we define as
\begin{equation}
\chi^2 = \sum^N_{i=1} \frac{(I_{\mathrm{mod},i} - I_{\mathrm{obs},i})^2}{\sigma_i^2}
\end{equation}
where $I$ is the integrated line intensity, $\sigma$ is the uncertainty in the observations (generally assumed to be 20\%), and $N$ is the number of lines being modelled. We also calculate a reduced $\chi^2$ value such that $\chi^2_\mathrm{red} = \chi^2/(N-p)$ where $p$ is the number of free parameters. We take the mass-loss rate and the $h$ parameter as the two free parameters, i.e. $p = 2$. (Although $\beta$ is also adjusted, it is done so based solely on the line widths and not the complete model results.)

\begin{figure}[t]
\begin{center}
\includegraphics[width=0.5\textwidth]{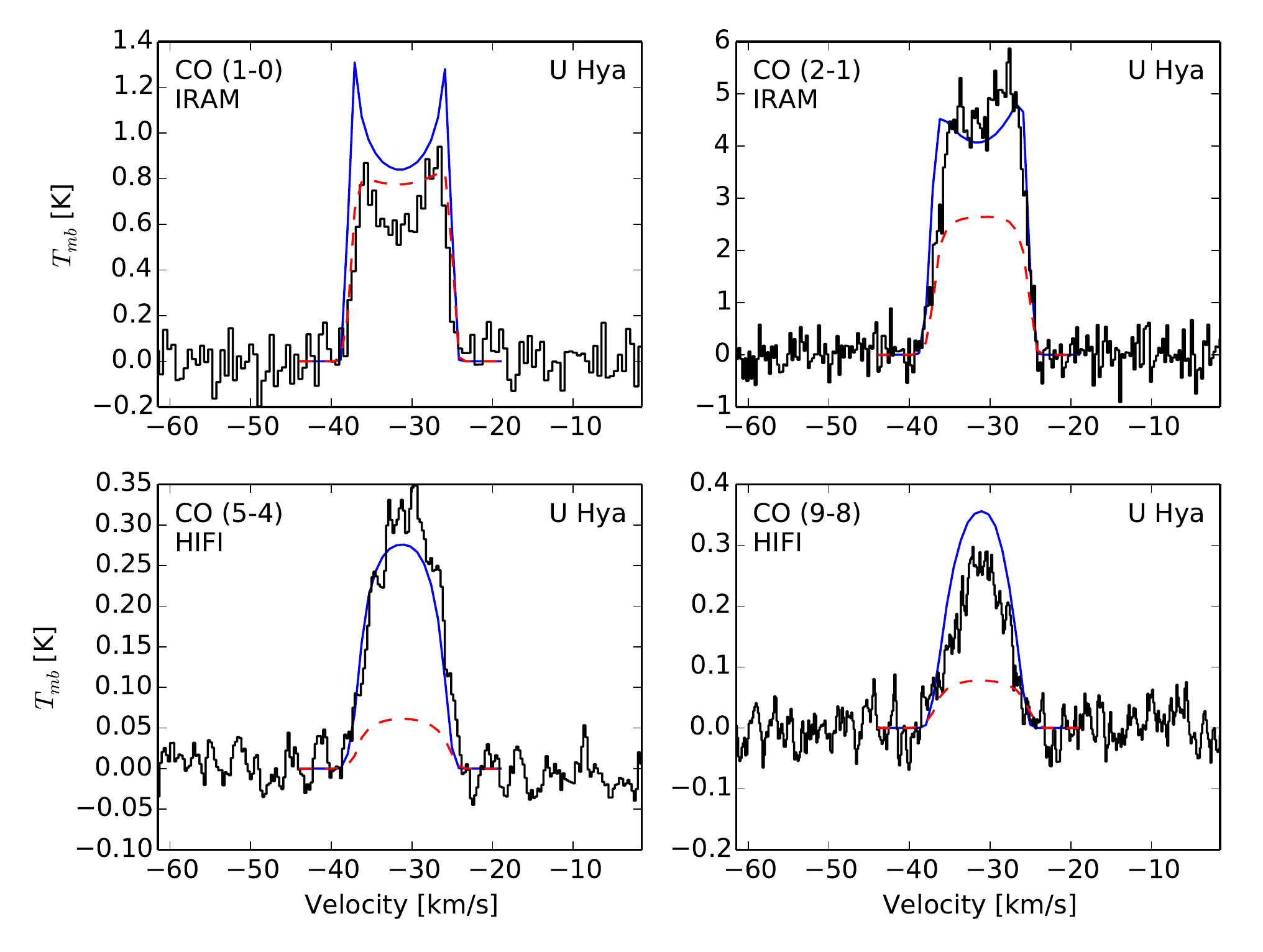}
\includegraphics[width=0.5\textwidth]{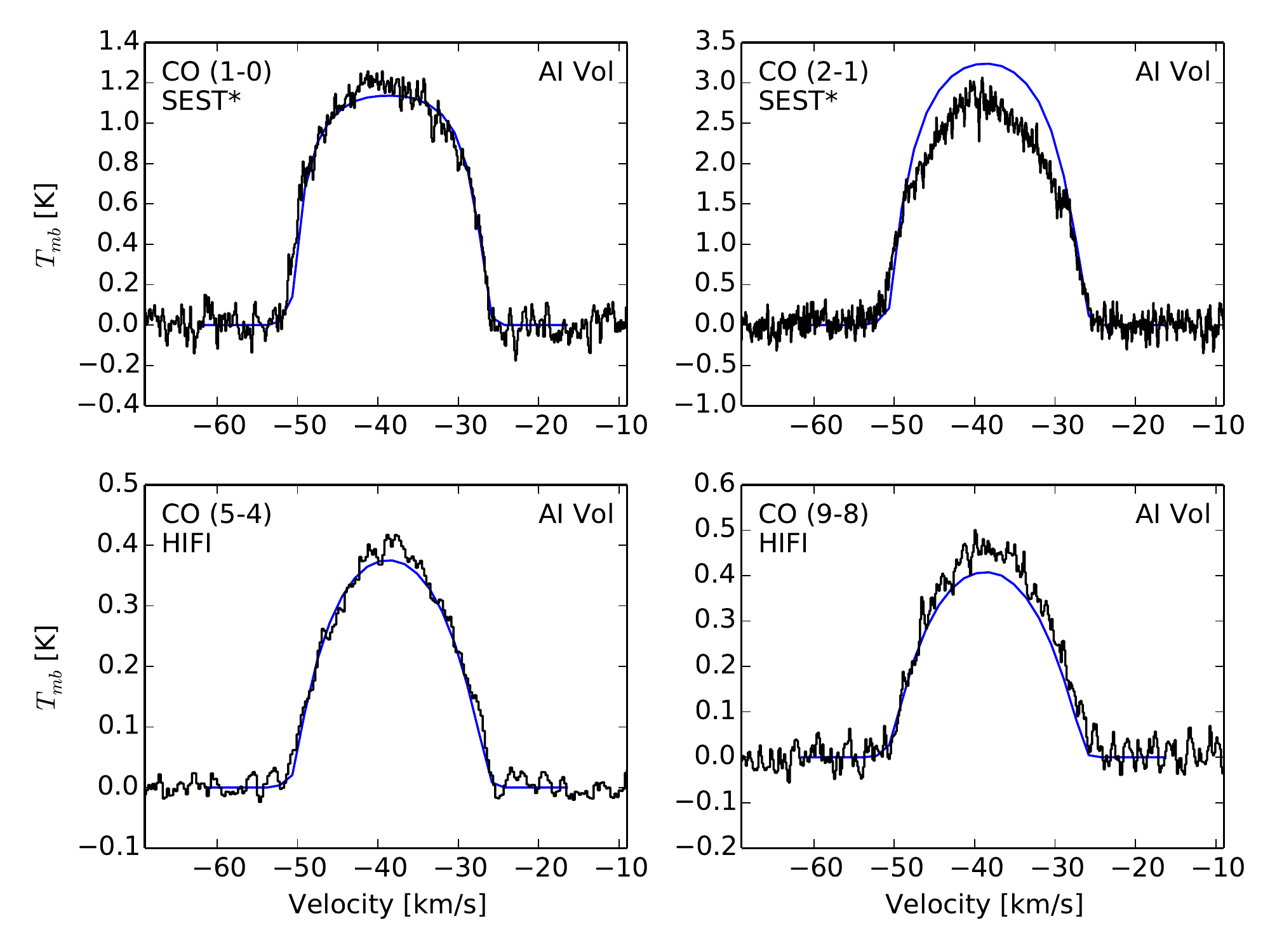}
\includegraphics[width=0.25\textwidth]{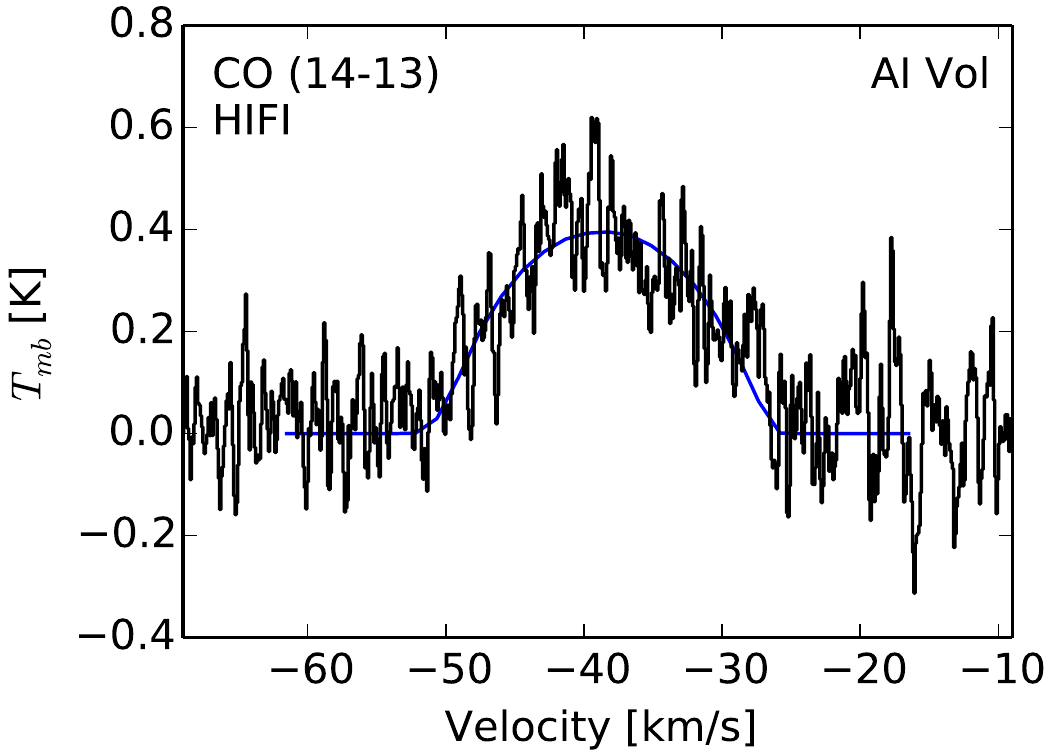}
\caption{Models (blue) and observed data (black) of two example C stars, plotted with respect to LSR velocity. An * next to the telescope name indicates that archival data is plotted. The dashed red lines on U Hya show the predicted model if a constant expansion velocity is used, keeping all parameters the same as in the blue model.}
\label{exmodsC}
\end{center}
\end{figure}

\begin{figure}[t]
\begin{center}
\includegraphics[width=0.5\textwidth]{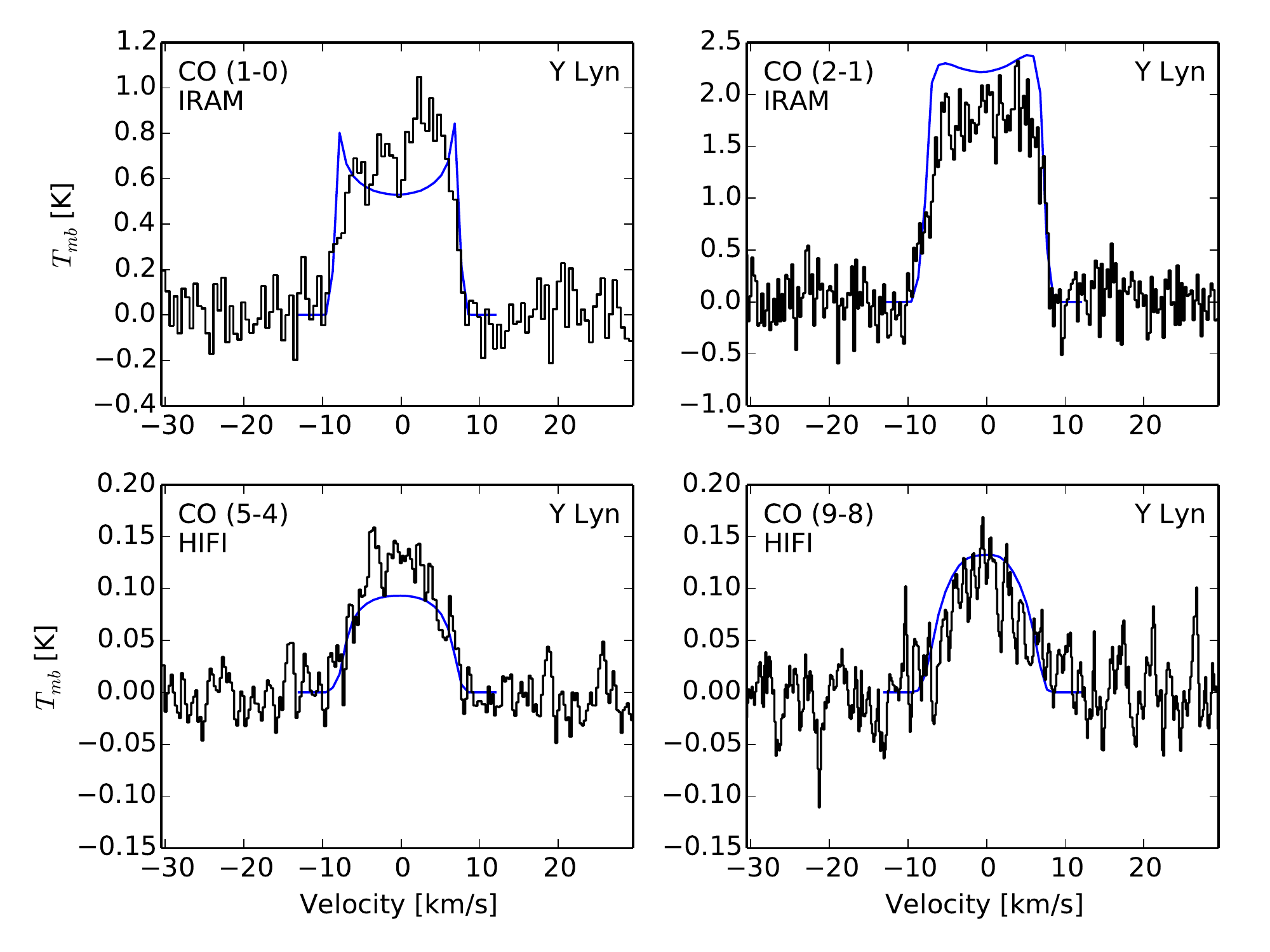}
\includegraphics[width=0.5\textwidth]{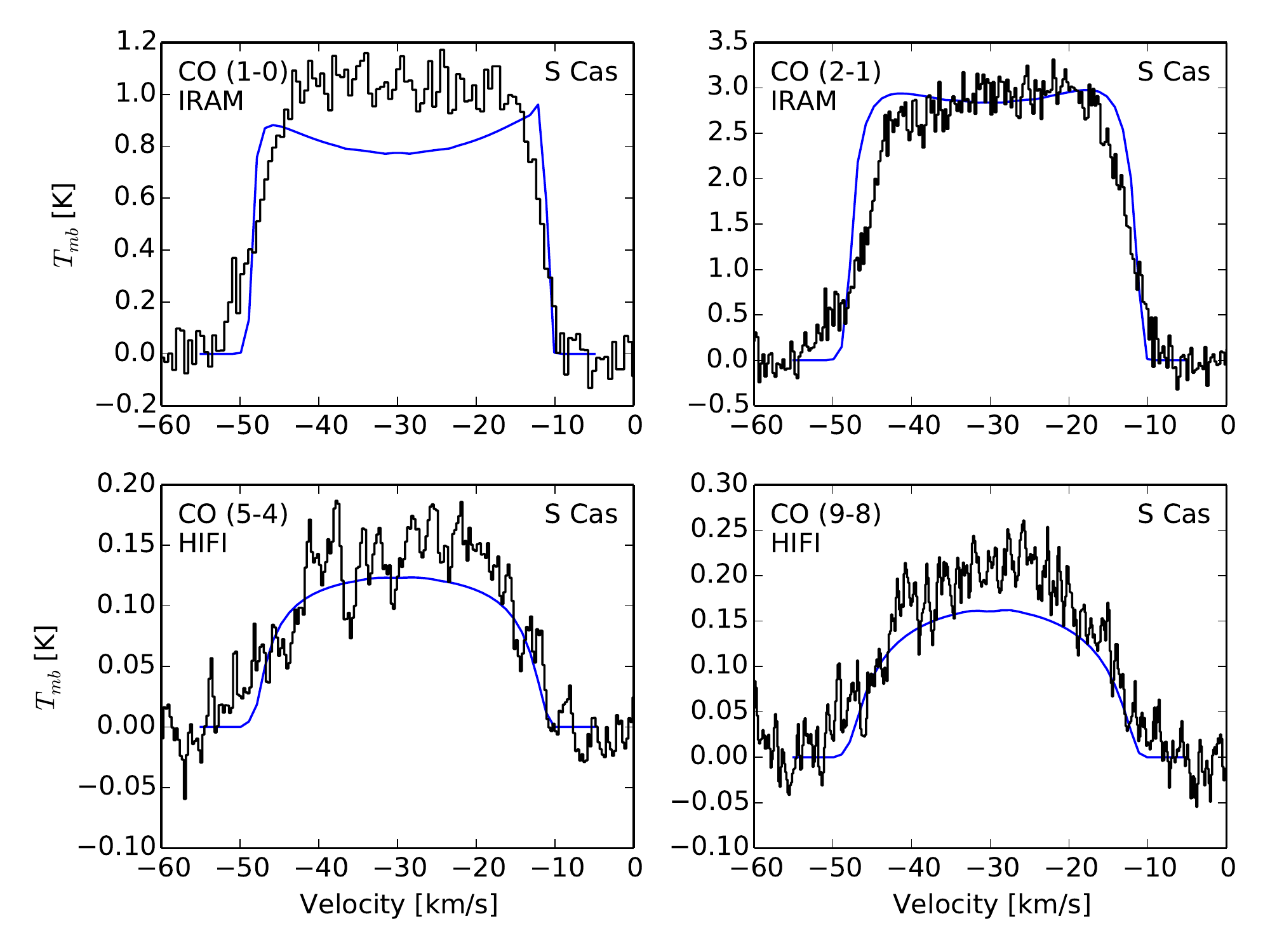}
\caption{Models (blue) and observed data (black) of two example S stars, plotted with respect to LSR velocity.}
\label{exmodsS}
\end{center}
\end{figure}

\begin{figure}[t]
\begin{center}
\includegraphics[width=0.5\textwidth]{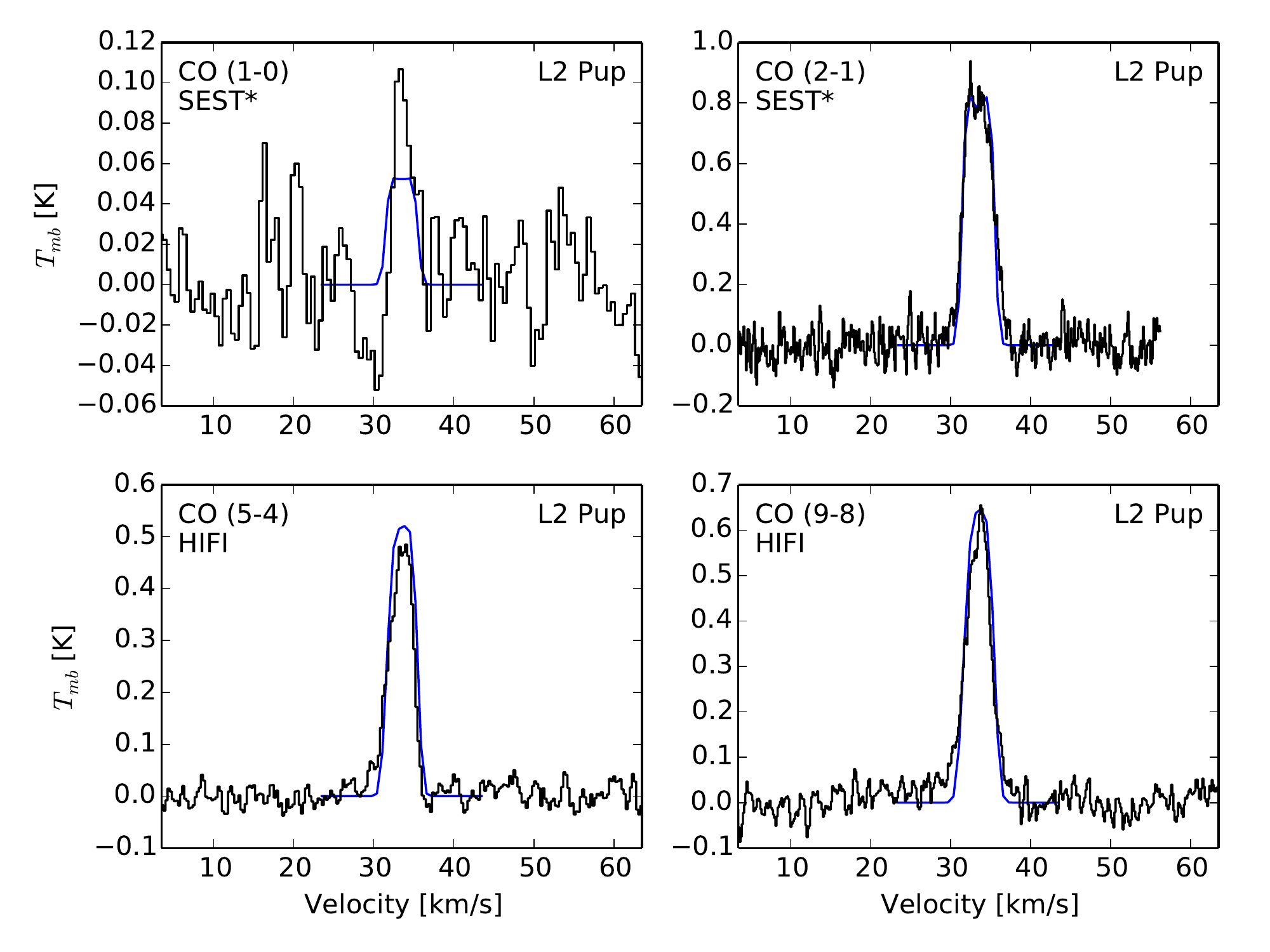}
\includegraphics[width=0.25\textwidth]{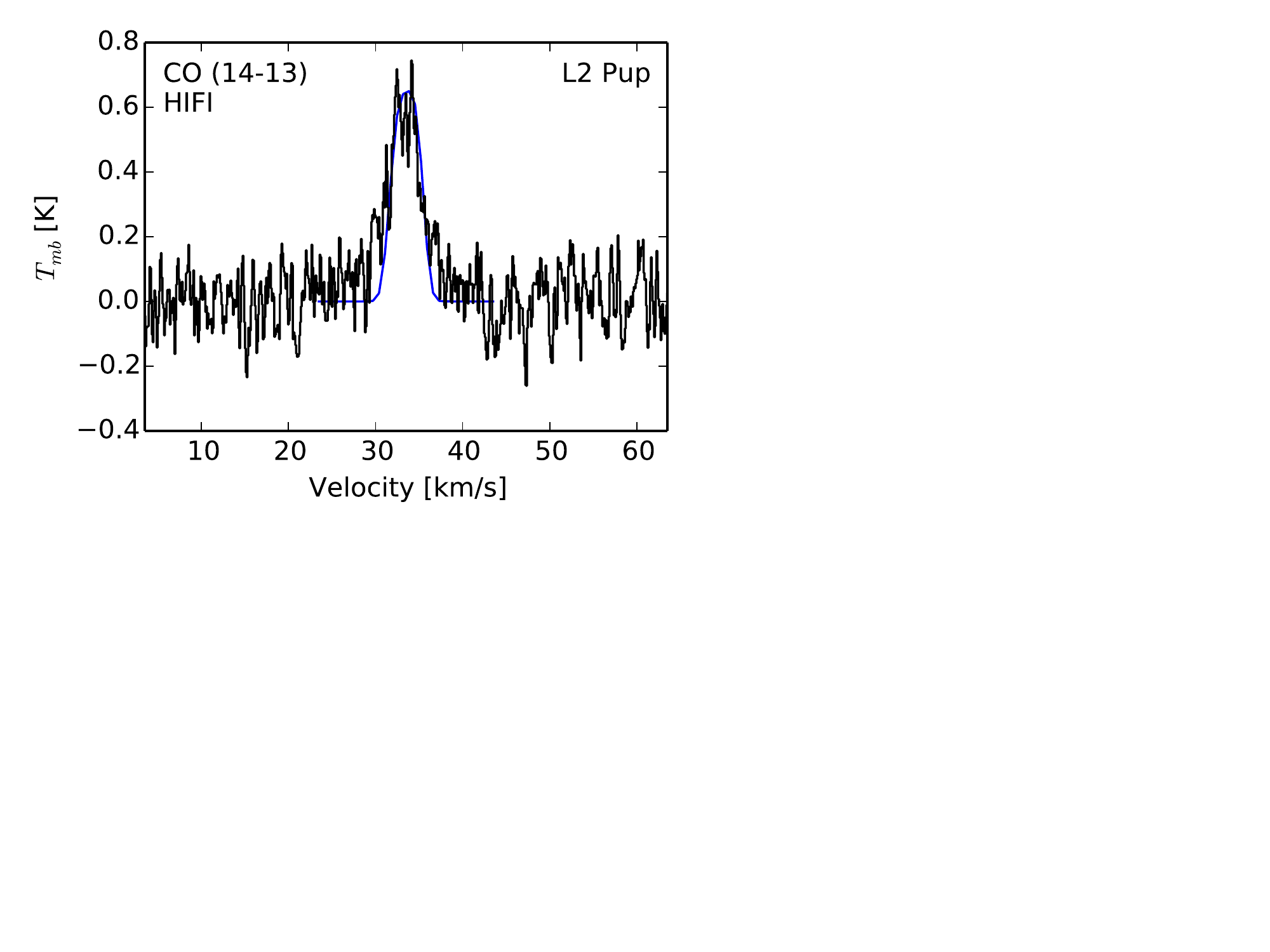}
\includegraphics[width=0.5\textwidth]{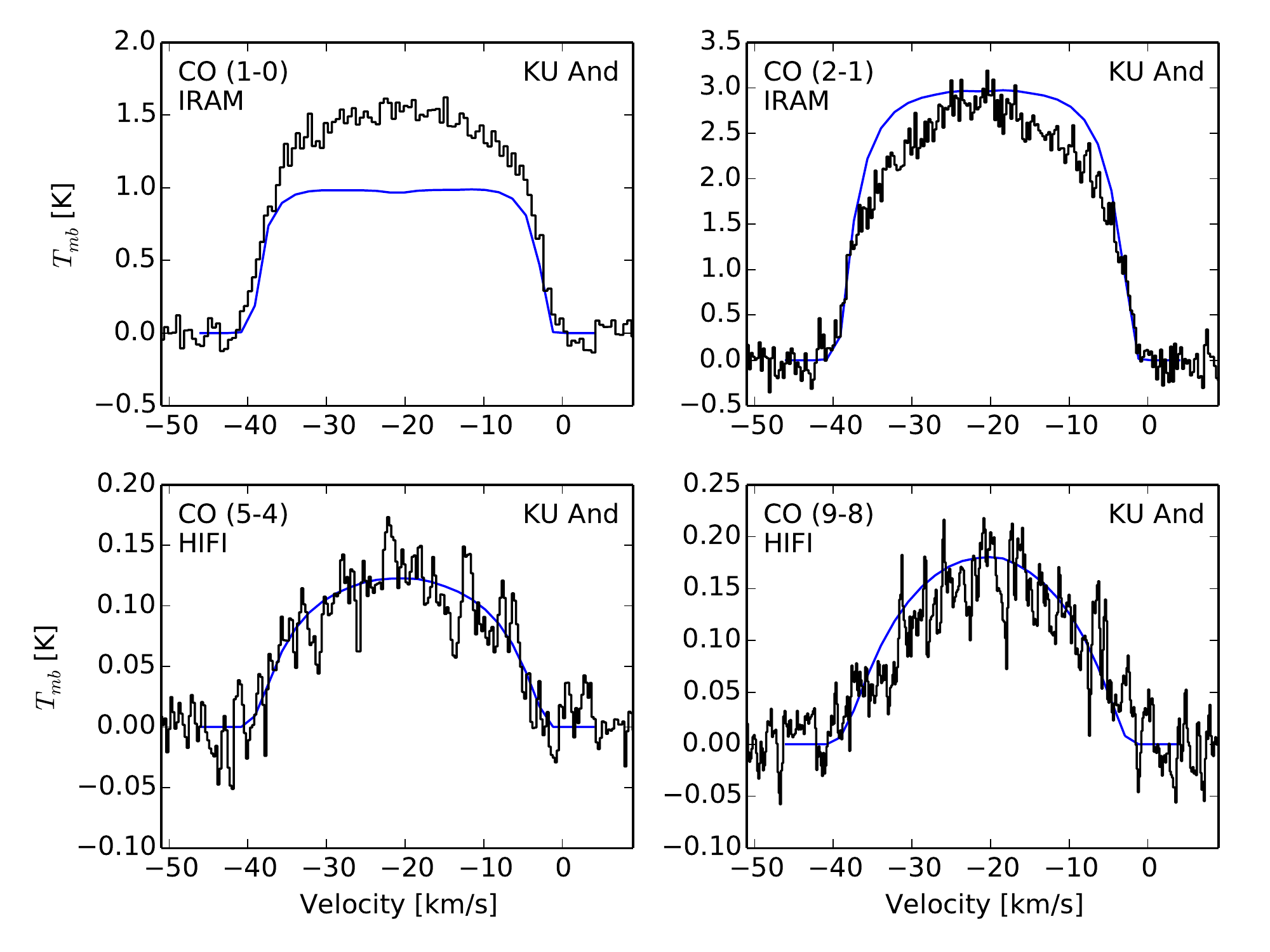}
\caption{Models (blue) and observed data (black) of two example M stars, plotted with respect to LSR velocity. An * next to the telescope name indicates that archival data is plotted.}
\label{exmodsM}
\end{center}
\end{figure}

In Fig. \ref{exmodsC} we show the resultant molecular emission line models plotted against observational data for two C stars: U Hya, which is a low mass-loss rate object, and AI Vol, which is a higher mass-loss rate object. Alongside our model results for U Hya (shown in blue) we have plotted the results of the same model with the only alteration being a constant velocity instead of a velocity law following Eq. \ref{vel} (red dashed line). As can be clearly seen, the constant velocity model severely under-predicts the higher-$J$ lines, compared with both the observations and the standard model with a velocity profile. This is a good example of both the importance of considering high-$J$ lines when constraining a model, and the effect of using a velocity profile. Figure \ref{exmodsS} shows model and observations for Y Lyn, a low mass-loss rate S star, and S Cas, a higher mass-loss rate S star. Fig. \ref{exmodsM} shows model and observations for L$_2$ Pup, a very low mass-loss rate (and low expansion velocity) M star, and KU And, a higher mass-loss rate M star.


The remainder of our results are plotted in Figures \ref{Cmods}, \ref{Smods}, and \ref{Mmods} and the model results are summarised in Table \ref{coresults}. We plot new HIFI and IRAM observations where they are available, and archival ($1\to0$) and ($2\to1$) lines where new low-$J$ observations are missing, to show as broad a range of lines with respect to $J$ as possible.

\begin{table}
\caption{CSE parameters from CO models}             
\label{coresults}      
\centering                          
\begin{tabular}{l c c c c c}
\hline \hline
Star	&	$\dot{M}$	&	$\beta$	&	$h$&	$\chi^2_\mathrm{red}$	&	N	\\
	&	[$\spy$]	&		&				&		\\
\hline					
\quad \it C stars\\						
V701 Cas	&	$	4.5	\e{	-6	}$	&	0	&	$	5	\e{	0	}$	&	5.1	&	4	\\
V384 Per	&	$	2.3	\e{	-6	}$	&	0	&	$	6	\e{	-1	}$	&	2.8	&	12	\\
GY Cam	&	$	3.7	\e{	-6	}$	&	2.0	&	$	9	\e{	-1	}$	&	2.0	&	5	\\
R Lep	&	$	8.7	\e{	-7	}$	&	2.0	&	$	5	\e{	-2	}$	&	1.4	&	9	\\
V1259 Ori	&	$	8.8	\e{	-6	}$	&	3.0	&	$	5	\e{	0	}$	&	1.7	&	9	\\
UU Aur	&	$	1.7	\e{	-7	}$	&	1.0	&	$	9	\e{	-3	}$	&	2.3	&	8	\\
V688 Mon	&	$	6.1	\e{	-6	}$	&	1.0	&	$	5	\e{	0	}$	&	0.58	&	5	\\
AI Vol	&	$	4.9	\e{	-6	}$	&	1.0	&	$	9	\e{	-1	}$	&	0.76	&	9	\\
U Hya	&	$	8.9	\e{	-8	}$	&	5.0	&	$	8	\e{	-2	}$	&	2.1	&	15	\\
X TrA	&	$	1.9	\e{	-7	}$	&	1.0	&	$	5	\e{	-1	}$	&	1.0	&	9	\\
II Lup	&	$	1.7	\e{	-5	}$	&	0	&	$	5	\e{	-1	}$	&	1.7	&	20	\\
V CrB	&	$	3.3	\e{	-7	}$	&	1.0	&	$	2	\e{	-1	}$	&	3.2	&	7	\\
V821 Her	&	$	3.0	\e{	-6	}$	&	1.0	&	$	7	\e{	-1	}$	&	2.7	&	14	\\
V Aql	&	$	1.3	\e{	-7	}$	&	0	&	$	5	\e{	-1	}$	&	2.0	&	11	\\
V1968 Cyg	&	$	7.5	\e{	-6	}$	&	1.0	&	$	8	\e{	0	}$	&	2.0	&	5	\\
RV Aqr	&	$	2.3	\e{	-6	}$	&	1.0	&	$	5	\e{	-1	}$	&	3.4	&	7	\\
\quad \it S stars\\
R And	&	$	5.3	\e{	-7	}$	&	1.5	&	$	6	\e{	-1	}$	&	1.8	&	6	\\
S Cas	&	$	2.8	\e{	-6	}$	&	1.0	&	$	5	\e{	-1	}$	&	2.4	&	6	\\
W And	&	$	2.8	\e{	-7	}$	&	3.0	&	$	4	\e{	-1	}$	&	0.93	&	5	\\
R Gem	&	$	4.3	\e{	-7	}$	&	2.0	&	$	6	\e{	-1	}$	&	2.2	&	5	\\
Y Lyn	&	$	1.7	\e{	-7	}$	&	1.5	&	$	5	\e{	-2	}$	&	0.92	&	5	\\
R Cyg	&	$	9.5	\e{	-7	}$	&	2.0	&	$	1	\e{	0	}$	&	1.9	&	5	\\
\quad \it M stars\\
KU And	&	$	9.4	\e{	-6	}$	&	2.0	&	$	3	\e{	-2	}$	&	1.7	&	7	\\
AFGL 292	&	$	2.1	\e{	-7	}$	&	1.0	&	$	5	\e{	-2	}$	&	3.5	&	4	\\
R Hor	&	$	5.9	\e{	-7	}$	&	1.0	&	$	4	\e{	-1	}$	&	0.19	&	6	\\
NV Aur	&	$	2.5	\e{	-5	}$	&	1.0	&	$	6	\e{	-1	}$	&	1.8	&	11	\\
BX Cam	&	$	4.4	\e{	-6	}$	&	1.0	&	$	5	\e{	-3	}$	&	3.1	&	7	\\
GX Mon	&	$	8.4	\e{	-6	}$	&	1.0	&	$	5	\e{	-1	}$	&	2.6	&	14	\\
L$_2$ Pup	&	$	1.4	\e{	-8	}$	&	0	&	$	2	\e{	-1	}$	&	1.3	&	10	\\
S CMi	&	$	4.9	\e{	-8	}$	&	0	&	$	6	\e{	-1	}$	&	1.2	&	4	\\
R LMi	&	$	2.6	\e{	-7	}$	&	2.0	&	$	2	\e{	-1	}$	&	2.2	&	6	\\
R Leo	&	$	1.1	\e{	-7	}$	&	5.0	&	$	5	\e{	-2	}$	&	5.7	&	15	\\
S CrB	&	$	2.3	\e{	-7	}$	&	5.0	&	$	2	\e{	-1	}$	&	3.9	&	10	\\
V1111 Oph	&	$	1.2	\e{	-5	}$	&	1.0	&	$	6	\e{	-1	}$	&	1.9	&	10	\\
RR Aql	&	$	2.4	\e{	-6	}$	&	2.0	&	$	5	\e{	-2	}$	&	2.7	&	7	\\
V1943 Sgr	&	$	9.9	\e{	-8	}$	&	1.0	&	$	5	\e{	-3	}$	&	3.1	&	4	\\
V1300 Aql	&	$	1.0	\e{	-5	}$	&	1.0	&	$	6	\e{	-2	}$	&	4.6	&	16	\\
T Cep	&	$	9.1	\e{	-8	}$	&	1.0	&	$	3	\e{	-1	}$	&	0.23	&	3	\\
\hline
\end{tabular}
\end{table}

\section{Analysis and discussion}

\subsection{Modelling target selection}\label{modselect}

Not all the stars in our observational sample were modelled in this paper. Some stars, such as the semi-regular variables (SRV) BK Vir, RX Boo, R Crt, Y UMa, SW Vir, RT Vir, X Her, and EP Aqr, will be modelled in a forthcoming paper (Alcolea et al., in prep.), while some were excluded because we knew {\it a priori} that our one-dimensional code, in which spherical symmetry is assumed, could not take their idiosyncrasies into account. 

The new observations for unmodelled carbon stars can be seen in Fig. \ref{Cplots}. R Scl was excluded because it is known to have a detached shell and a spiral structure in the gas indicative of an unseen companion \citep{Maercker2012}. See also \cite{Schoier2005a} for a discussion of difficulties modelling this particular star. Conversely, U Hya is also known to have a detached shell \citep{Waters1994}, but the shell is sufficiently distant that the CO in it is most likely photodissociated and hence it has no impact on the detected CO lines, enabling us to keep U Hya in our modelled sample. IRAS~06192+4657 was excluded because we had insufficient data for a robust model, particularly since the ($5\to4$) HIFI line was not convincingly detected. TX Psc was discounted because it has a two-component molecular wind, and is known to have an irregular structure as discussed in \cite{Heske1989}.

Two S stars were excluded from modelling, RS Cnc and $\pi^1$ Gru, both of which have two velocity components in their line profiles, as can be seen in Fig. \ref{Splots}. RS Cnc is known to have a bipolar outflow and a disc structure around the star \citep{Libert2010}, and $\pi^1$ Gru is known to have a bipolar outflow, a G0V binary companion, and evidence of a second hidden companion, as described by \citet{Mayer2014}.

The unmodelled M-type star sample contains the SRVs listed above and R Hya and V370 And, which were excluded for having double component winds.

\begin{figure}[t]
\begin{center}
\includegraphics[width=0.5\textwidth]{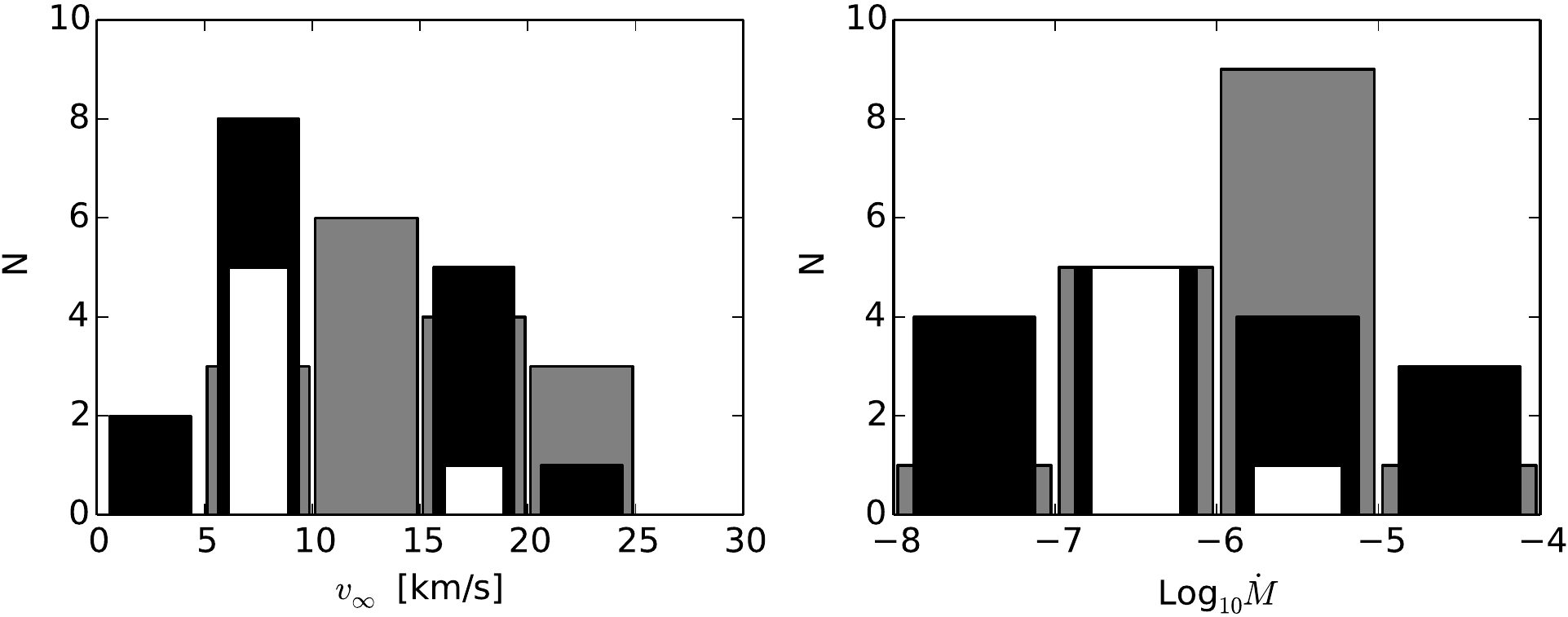}
\caption{The numbers of stars with the indicated expansion velocities and mass-loss rates. C stars are grey, S stars are white and M stars are black.}
\label{distributions}
\end{center}
\end{figure}

These exclusions have left us with a sample for which we have plotted histograms of mass-loss rates and expansion velocities by chemical type in Fig. \ref{distributions}. Note in particular the gap in expansion velocity for M stars with no modelled stars in the range $10 \;\kms \leq \upsilon_\infty < 15\;\kms$. There is a similar gap in the S star distribution, however this is more a result of having only one S star of high mass-loss rate and high velocity. The C stars are fairly evenly distributed in velocity.

\subsection{Goodness of fit}\label{fitgoods}

The goodness of fit of our models has been primarily estimated using the $\chi^2$ method discussed in Sect. \ref{sec:comod}. To further visualise the goodness of fit we have plotted the values of modelled line intensity divided by the observed line intensity ($I_\mathrm{mod}/I_\mathrm{obs}$) for each observed line in each star. The resulting plots can be seen in Fig. \ref{Cfits} for C stars, Fig. \ref{Sfits} for S stars and Fig. \ref{Mfits} for M stars. For the large majority of sources, we can model the CO lines with upper energy levels below 250 K (and below 580 K for sources with observed $J = 14\to13$ lines) with reasonably accuracy using a constant mass-loss rate. In particular, there is no trend with $J$-number.

\begin{figure}[t]
\begin{center}
\includegraphics[width=0.5\textwidth]{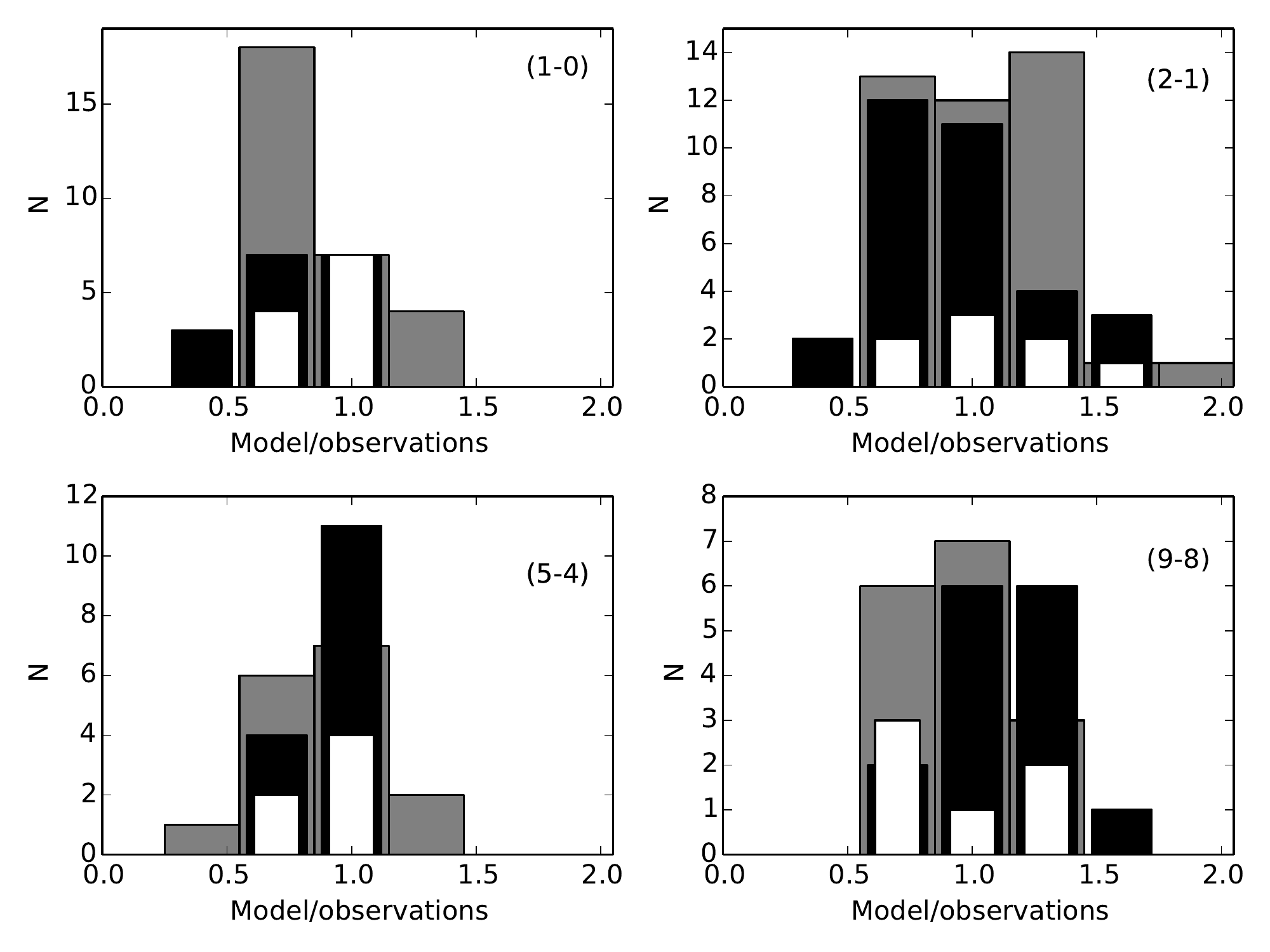}
\caption{Goodness of fit as defined by model/observed integrated line intensity of all stars per transition line for C (grey), S (white) and M (black) stars. $N$ is the number of lines of each chemical type in each bin.}
\label{transgood}
\end{center}
\end{figure}

We have also combined the same quantity for all stars grouped by line, for the ($1\to0$), ($2\to1$), ($5\to4$), and ($9\to8$) lines. The resultant plots can be seen in Fig. \ref{transgood}. The distributions are reasonably symmetric, except for the ($1\to0$) that seems to be under-predicted in the models for all three chemical types. The ($1\to0$) line intensity is particularly sensitive to the size of the CO envelope and the former discrepancy can be (at least partly) remedied if a larger CO envelope is used than the size predicted by the \citet{Mamon1988} model.

However, this line, and to some extent the ($2\to1$) line also have a tendency to towards double-peaked profiles rather than the flat or slightly rounded observed line profiles. This happens in about 5 out of 38 cases and is a well-known fact in CO line modelling, for example see \citet{Olofsson2002}. An increase in the size of the CO envelope would further enhance the double-peaked nature of the line profiles. In addition to the CO envelope size, there are several possible reasons for the double-peaked ($1\to0$) line profiles: a too-warm outer CSE, a too-distant source, or maser action in the inner CSE (which was in evidence in some of our models but is known to be produced in nature). Since we have studied a large sample of sources in this paper, we have preferred to treat all sources the same and, therefore, make no adjustments for individual sources. consequently, we conclude that we produce overall good fits to the CO line intensities for the majority of our sources. while there remain some discrepancies (some known from previous studies, such as the tendency for double-peaked model line profiles), we believe they have no major impact on the conclusions in this paper.

\subsection{Trends with mass-loss rate and chemical type}

\begin{figure*}[t]
\begin{center}
\includegraphics[width=0.49\textwidth]{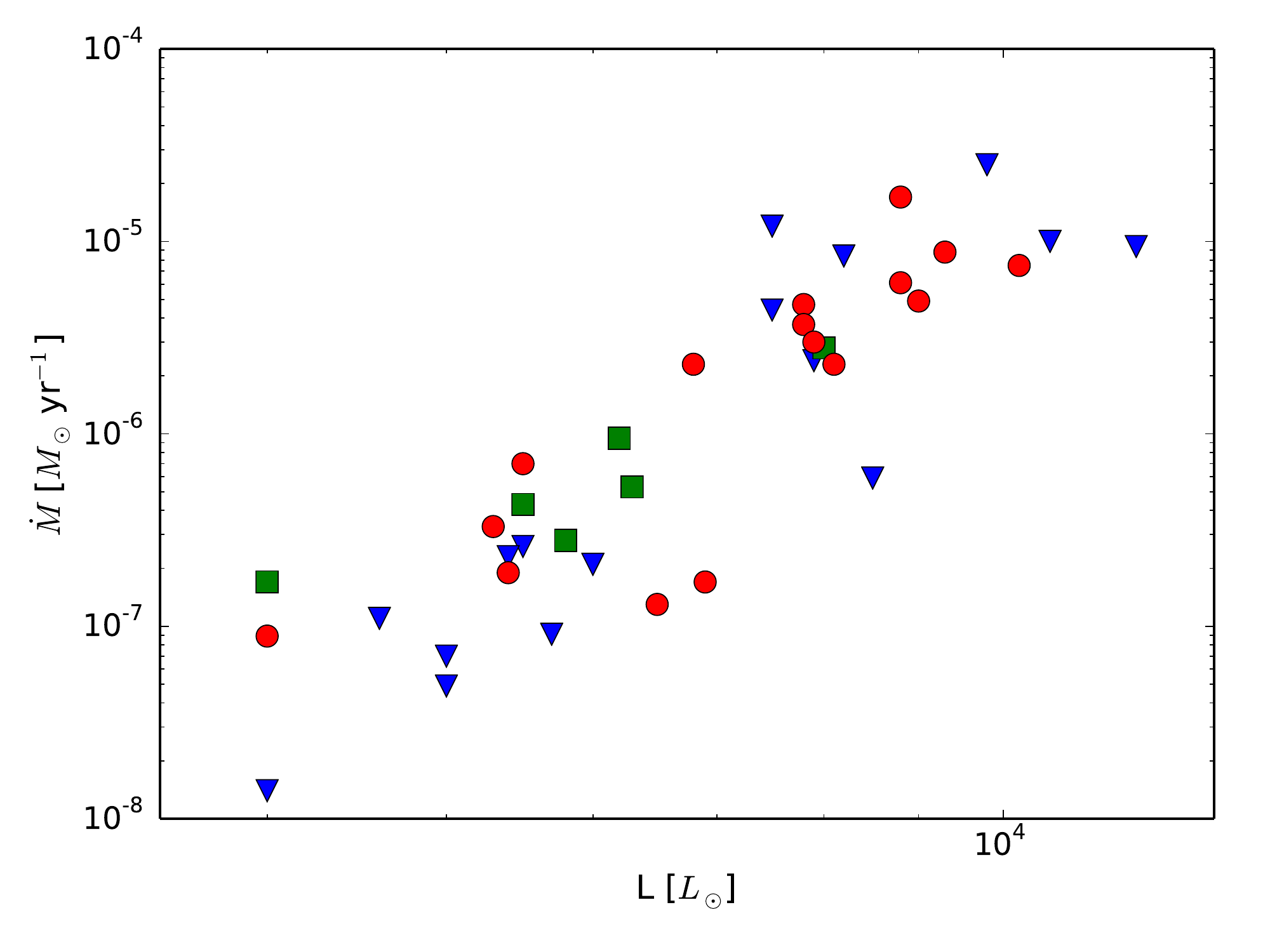}
\includegraphics[width=0.48\textwidth]{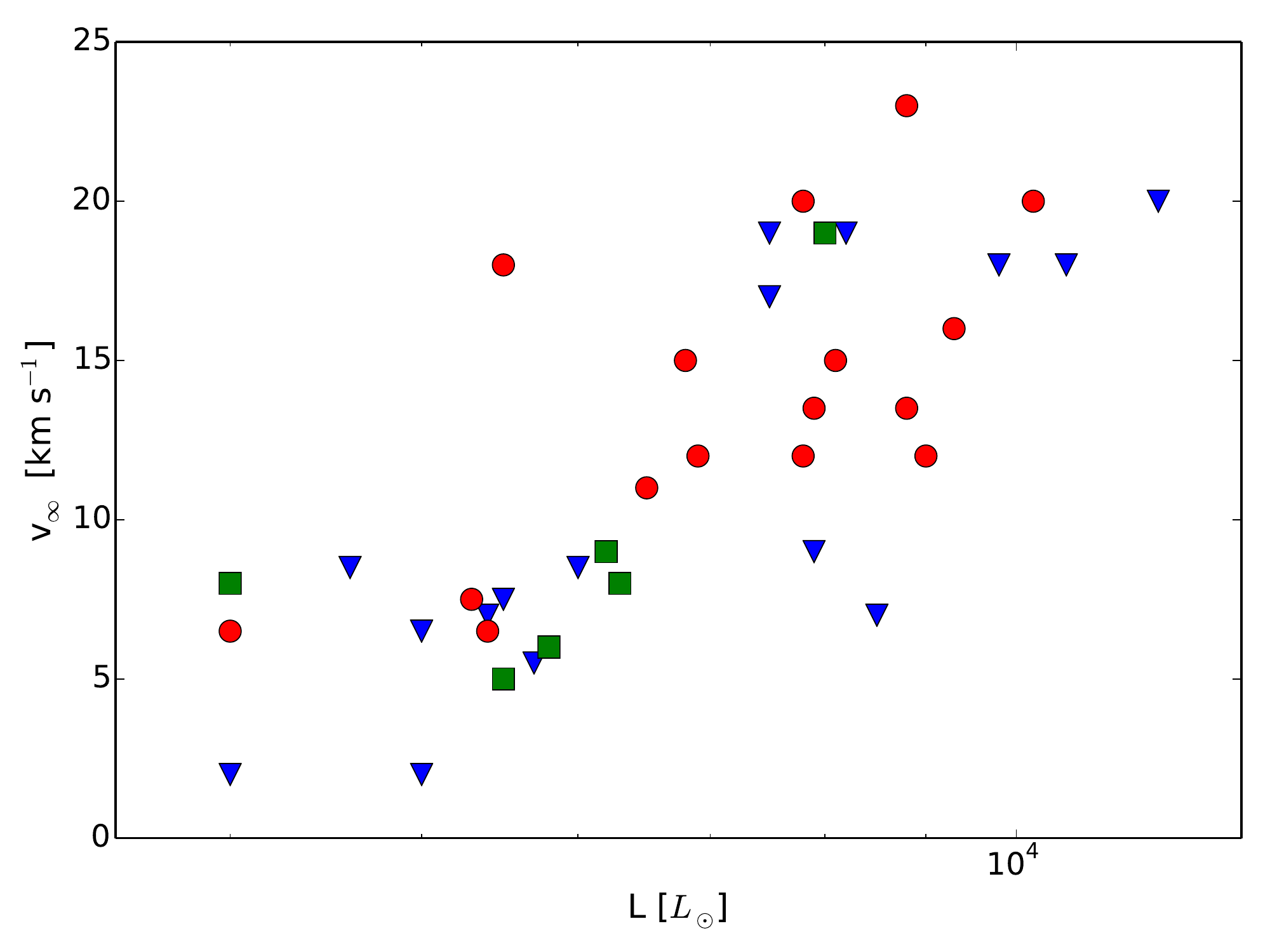}
\includegraphics[width=0.49\textwidth]{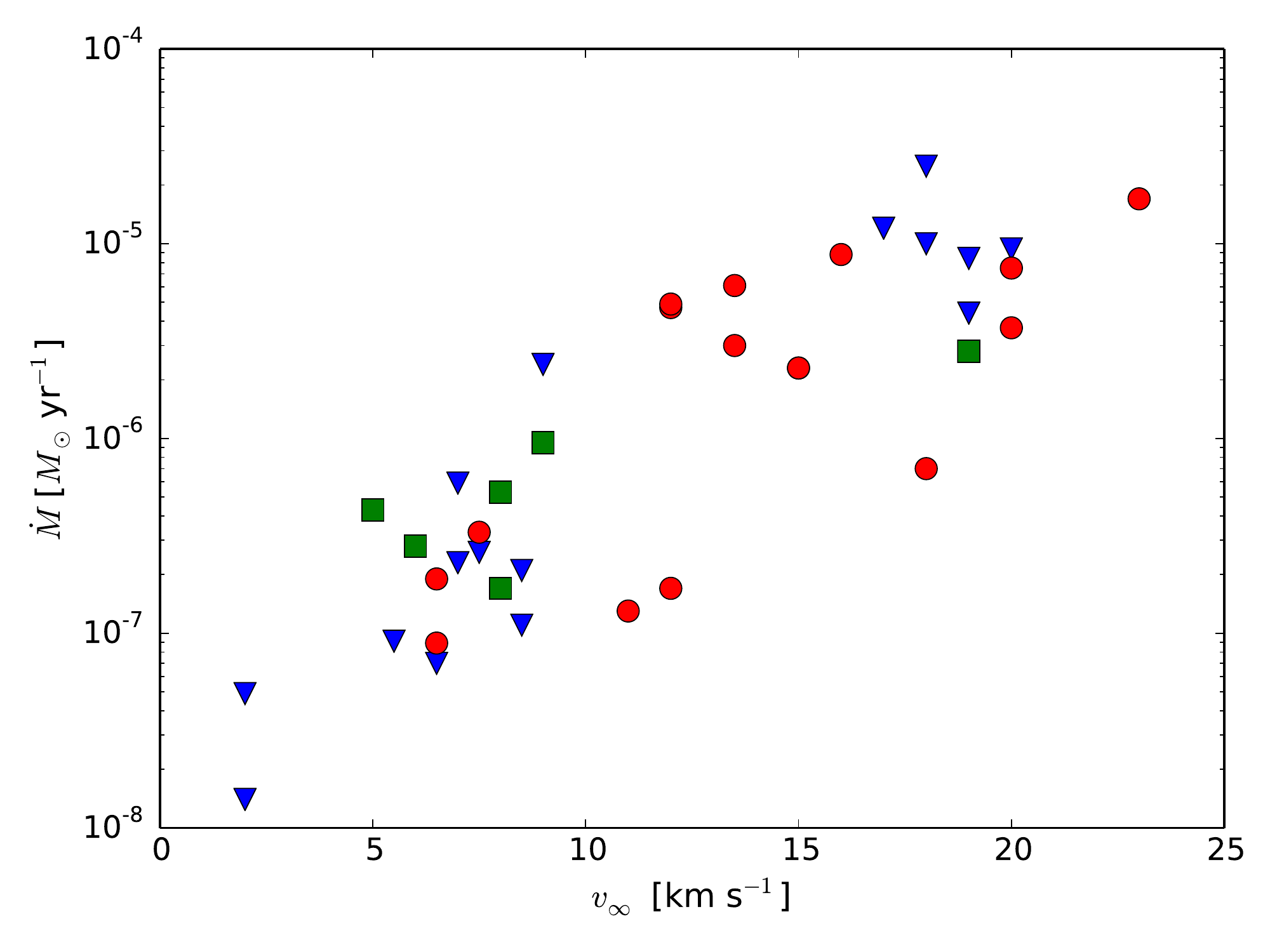}
\includegraphics[width=0.49\textwidth]{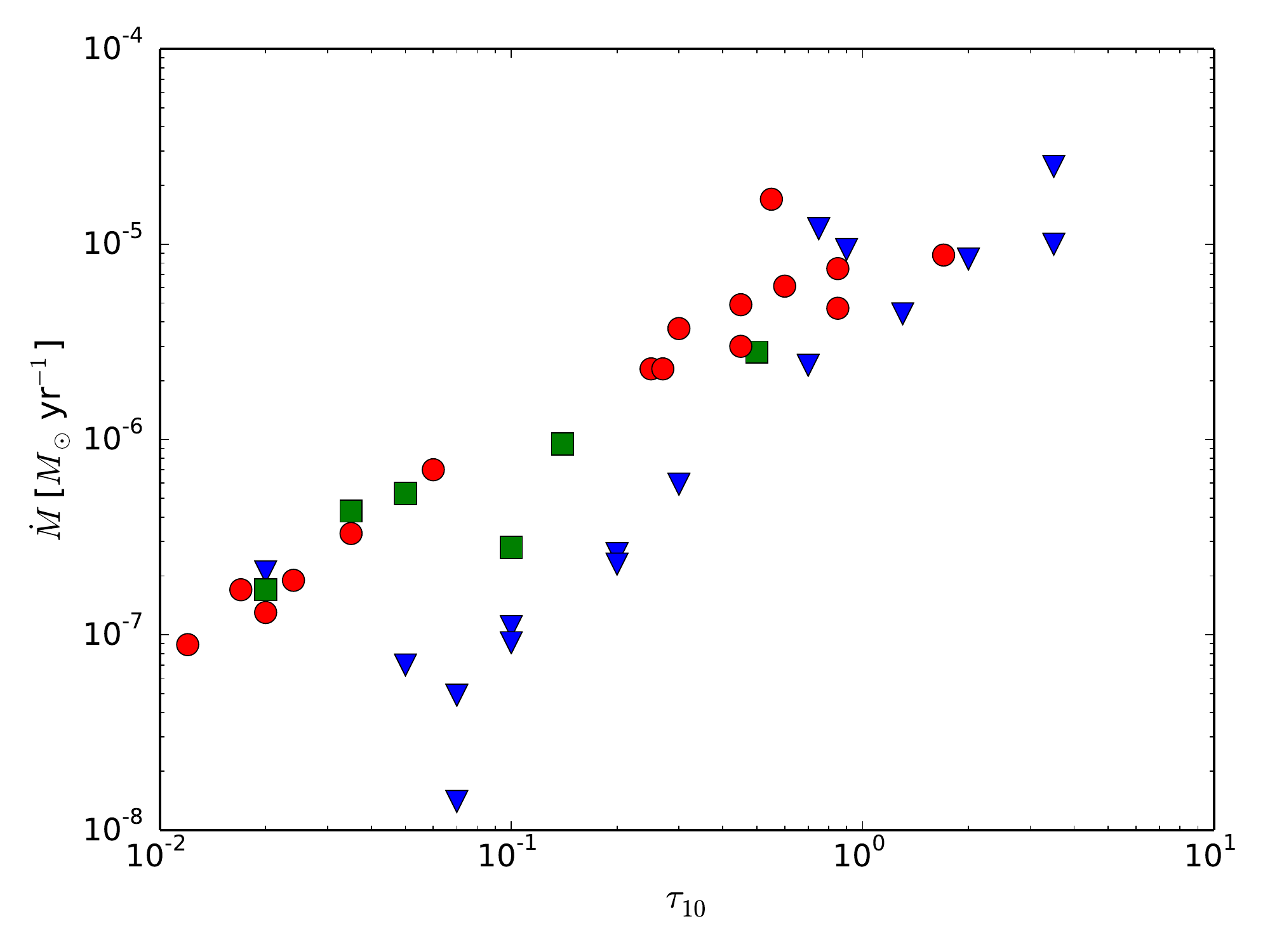}
\includegraphics[width=0.49\textwidth]{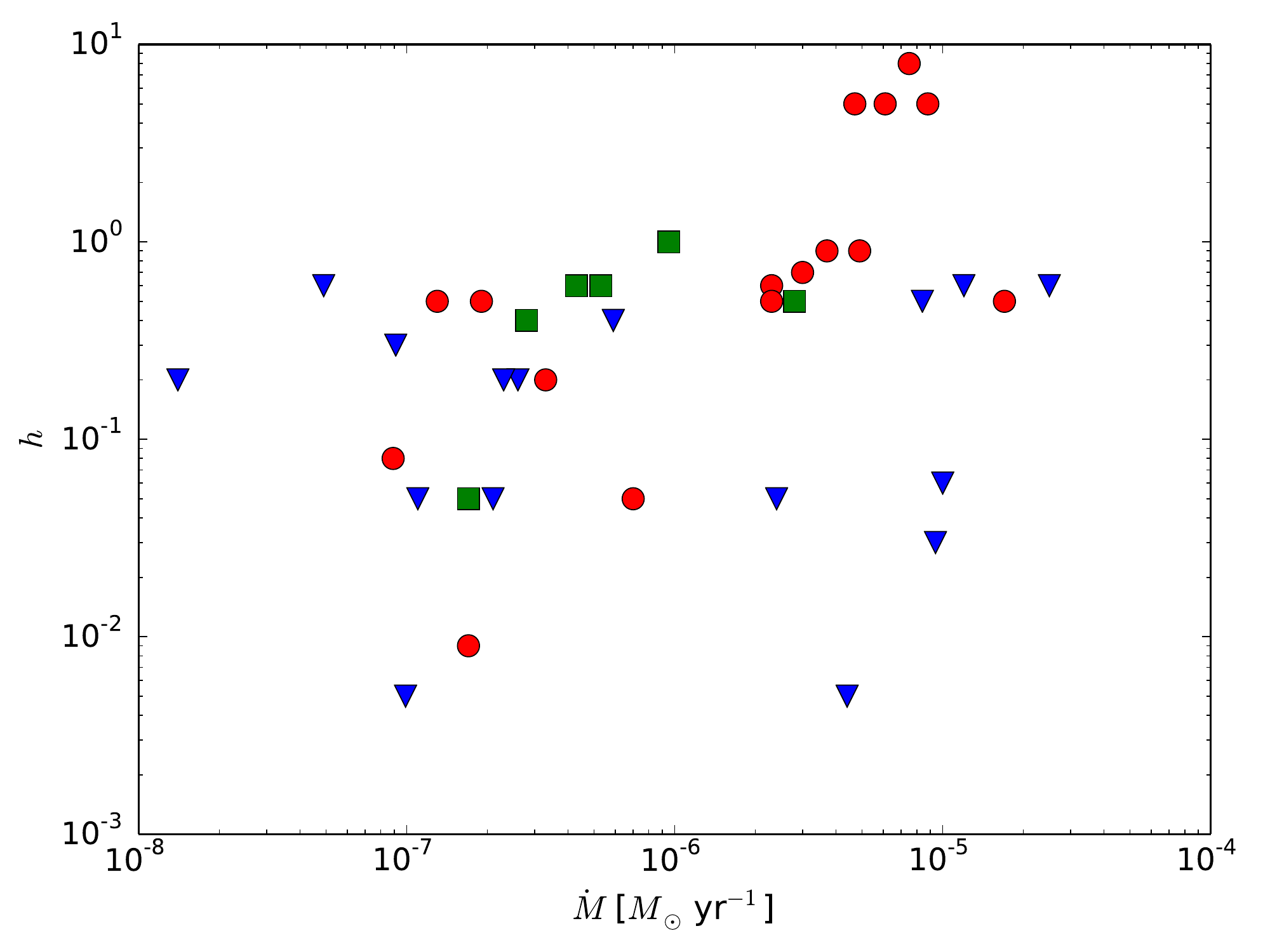}
\includegraphics[width=0.49\textwidth]{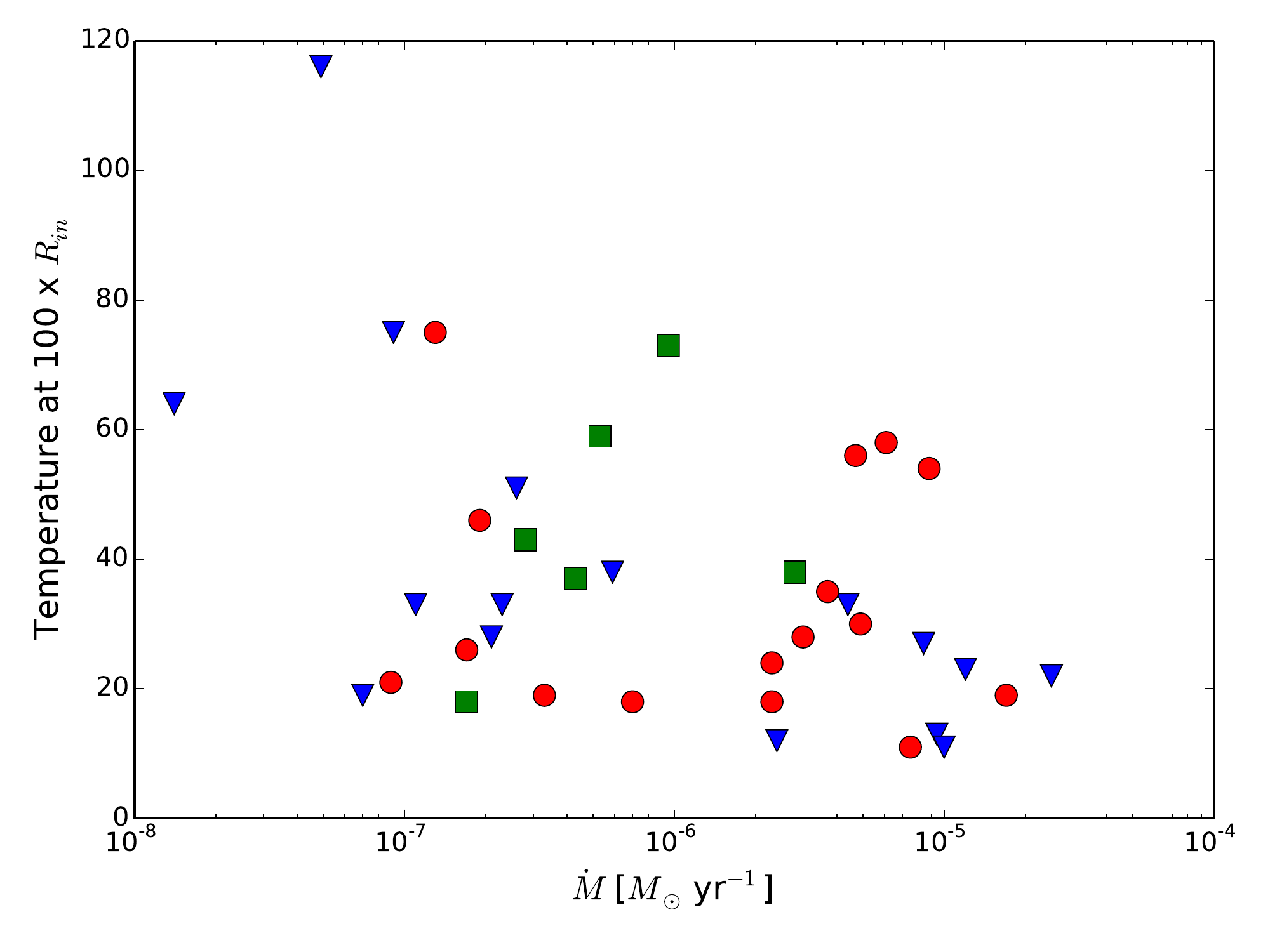}
\caption{Mass-loss rate and expansion velocity plotted against luminosity; mass-loss rate plotted against expansion velocity, optical depth and the $h$-parameter; temperature in the CSE at a radius of $100\times R_\mathrm{in}$ plotted against mass-loss rate. Carbon stars are red circles, S stars are green squares, M stars are blue triangles.}
\label{trends}
\end{center}
\end{figure*}

To investigate any trends that appear in our results, we plotted mass-loss rate against luminosity, expansion velocity, optical depth at $10~\mic$, and the $h$ parameter in Fig. \ref{trends}. There is a clear trend between mass-loss rate and luminosity, which does not seem to depend on chemical type. This is in line with the expectation that the higher luminosity stars should have higher mass-loss rates, either due to having a higher mass, or due to a more advanced age on the AGB. There is also a correlation between expansion velocity and stellar luminosity, although it is less tight than the mass-loss rate and luminosity correlation. This suggests that the gas expansion velocity is tied to the stellar luminosity.
It is not surprising, then to find a correlation between the mass-loss rate and gas expansion velocity. The relation between these two mass-loss characteristics puts constraints on any viable mass-loss mechanism.
Although there is some segregation in both of these plots, this is due to gaps in our sample rather than intrinsic trends. 

There is also a correlation between mass-loss rate and optical depth at $10~\mic$, as is expected if the gas and dust mass-loss rates are correlated. The correlation has slightly different slopes for the C and M stars. For C stars we find the best fit slope is $1.0 \pm 0.3$ and for M stars $1.3 \pm 0.6$. There are not enough S stars in our sample for a meaningful determination. { It should also be noted that the optical depth is taken at $10~\mic$, near a strong silicate feature in the M stars and an SiC feature in the C stars. This probably contributes to the difference in slope, along with differences in wind-driving efficiency between the two chemical types}.

The plot of mass-loss rate against $h$ parameter does not show any obvious trends, indicating that the dust properties embedded in the $h$ parameter do not directly depend on the mass-loss rate. 

We have modelled CO lines covering a relatively broad range in energies, which means that the kinetic temperature distribution is well constrained, except perhaps for the very inner part. In Fig. \ref{trends} (bottom right) we plot the kinetic temperature in the CSE at a radius of $100\times R_\mathrm{in}$ against mass-loss rate. There is no obvious correlation, but the trend is that the CSEs become, on average, cooler the higher the mass-loss rate. This is certainly expected, since, with all else being equal, a high mass-loss rate leads to more efficient CO line cooling and a lower drift velocity contributing to less efficient heating of the gas.



\subsection{Comments on individual stars}\label{bbyb}

For the discussion of individual stars below, note that the modelled and observed lines plotted in Figures \ref{exmodsC}, \ref{exmodsS}, and \ref{exmodsM}, and in the appendices in Figures \ref{Cmods}, \ref{Smods}, and \ref{Mmods}, show a maximum of five emission lines. In most cases there are several archival observations which were included in the modelling, and are listed in Table \ref{lowjobs}, but are not plotted with the new observations. These archival observations can have a significant impact on the choice of best fit model. The full collection of line observations for each star is included in the goodness of fit plots in Figures \ref{Cfits}, \ref{Sfits}, and \ref{Mfits}. The number of lines included in the radiative transfer modelling for each star is noted in Table \ref{coresults}.

\subsubsection{C stars}\label{Cbbyb}

V1259 Ori, V688 Mon, AI Vol, V821 Her, V1968 Cyg, and RV Aqr all have higher mass-loss rates (above $10^{-6} \spy$) and expansion velocities in excess of $10 \;\kms$. Their CO line profiles can all be described by models that include a velocity profile as described in Eq. \ref{vel} with $\beta =1$, except for in the case of V1259 Ori for which $\beta = 3$. V688 Mon has ISM contamination in the low-$J$ data for which we compensated when calculating the integrated line intensities.

UU Aur, U Hya, X TrA, and V CrB are the low mass-loss rate carbon stars, all having mass-loss rates lower than $10^{-6} \spy$ and expansion velocities $\leq 10\;\kms$. Their CO line profiles are all reasonably well-described by our models with $\beta =1$, except for in the case of U Hya for which $\beta = 5$. U Hya also has the lowest mass-loss rate ($8.9\e{-8}\spy$) of the carbon stars in this sample.

There were a handful of stars for which a constant velocity model ($\beta = 0$) gave a markedly better fit than the standard velocity profile model. The stars which fell into this group were V701 Cas, V384 Per, V Aql, and II Lup. The first two objects have moderate mass-loss rates and V Aql has a relatively low mass-loss rate for a C star. The models of these three objects all suffer from maser emission in the low-$J$ lines if a velocity profile is used (which does not entirely disappear in V Aql even with a constant velocity as can be seen in Fig. \ref{Cmods}). The model maser emission arises in the inner regions of the CSE. It may be that, for various reasons,  such maser action is not produced in nature. V701 Cas also suffers from ISM contamination in the low-$J$ lines which was compensated for.
II Lup is both a high mass-loss rate and high expansion velocity object. The line intensities of a $\beta > 0$ model do not match the observed data as well as a constant velocity model by a significant amount: the best fit $\beta = 1$ model has $\chi^2_\mathrm{red} = 3.3$ compared with $\chi^2_\mathrm{red} = 1.7$ for the constant velocity model. 

GY Cam has a narrow component ($\sim 8~\kms$ wide) lying approximately at the systemic velocity of $\upsilon_\mathrm{lsr} =-49\;\kms$ that is clearly visible in the ($1\to0$), ($2\to1$) and ($5\to4$) lines. Note that it is absent in the ($9\to8$) line, which suggests that it may be cooler in temperature than the gas contributing to the broader component of the line profile. Our CO model was not able to take the narrow component in these lines into account and we did not compensate for it in any way. Since our best-fit models are determined from total line intensity, this had a slight effect on our result. However, it does not seem to have been as large an issue as in some stars such as AFGL 292 (see Sect. \ref{Mbbyb} below).

\subsubsection{S stars}

The majority of the S stars were straightforward to model. Although R And, R Gem, and Y Lyn are known to have binary companions \citep{Proust1981,Pourbaix2003}, this did not seem to have any significant impact on the observed line profiles when compared to the line models. The most problematic of these five stars was R Gem, which has the noisiest data.

The only S star which we found problematic to model was S~Cas. It has the largest terminal expansion velocity (more than a factor of two higher than any of the other S~stars) and we found it to also have the highest mass-loss rate. It was also the only S~star for which we detected any CN ``bonus" lines (see Sect. \ref{a:bonus}). In these ways it bears many similarities to some of the carbon stars.

\subsubsection{M stars}\label{Mbbyb}

KU And, NV Aur, BX Cam, GX Mon, V1111 Oph, and V1300~Aql are the oxygen-rich stars in our sample with the highest terminal expansion velocities ($\upsilon_\infty > 15 \;\kms$). They can all be described well by models which include a velocity profile with $\beta=1$ except for NV Aur, for which $\beta=2$. It should also be noted that the NV Aur lines exhibit some ISM contamination, especially in the low-$J$ lines.

R Hor, R LMi, S CrB, RR Aql, V1943 Sgr, and T Cep have relatively low terminal expansion velocities, with $\upsilon_\infty < 10\;\kms$. Their CO profiles are described well by our models and have velocity profiles with $\beta$ in the range $1.0 \leq \beta \leq 2.0$.

{ L$_2$ Pup and S CMi both appear to have very low terminal expansion velocities, with $\upsilon_\infty = 2 \;\kms$, and we have therefore modelled the CO line emission in both cases assuming a constant velocity. The fits to the observed lines are in general very good, although it is unclear whether the S CMi $(1\to0)$ line is really as strongly double-peaked as it appears, in which case this behaviour is not reproduced in our model. However, there are indications that L$_2$ Pup has a central toroidal or disc structure \citep{Lykou2015,Kervella2014}, with a possible close companion and bipolar outflow \citep{Kervella2015}. In this case, our spherically symmetric model most likely does not represent a realistic view of L$_2$ PupÕs circumstellar environment, and the narrow line is not produced in a slowly, spherically expanding envelope. Similarly, \cite{Ragland2006} have found evidence of asymmetry in S CMi, suggesting non-spherical geometry.}


AFGL 292 appears to have a two-component line profile.  The new IRAM observations in particular suggest an additional narrow component centred at the stellar systemic velocity. This component is not visible in the high-$J$ lines observed with HIFI, which could be an effect of cooler gas or simply because the signal-to-noise ratio of the HIFI observations is insufficient. As a result, the model we present in Fig. \ref{Mmods} is the best fit based on the integrated intensities, but does not match the line profiles, particularly not for the low-$J$ lines.

R Leo shows asymmetric line profiles, with the HIFI lines being less bright on the red-shifted side, as can be seen in Fig. \ref{Mmods}. Conversely, some of the low-$J$ lines, most notably the $(2\to1)$ IRAM line and the $(3\to2)$ CSO line, have significant peaks on the red side of the line profiles. Also, the $(1\to0)$ IRAM line seems to have a central peak as well as two outer peaks. This could be indicative of localised clumps of gas at different temperatures.

\subsection{Comparison with other studies}

\begin{figure}[t]
\begin{center}
\includegraphics[width=0.5\textwidth]{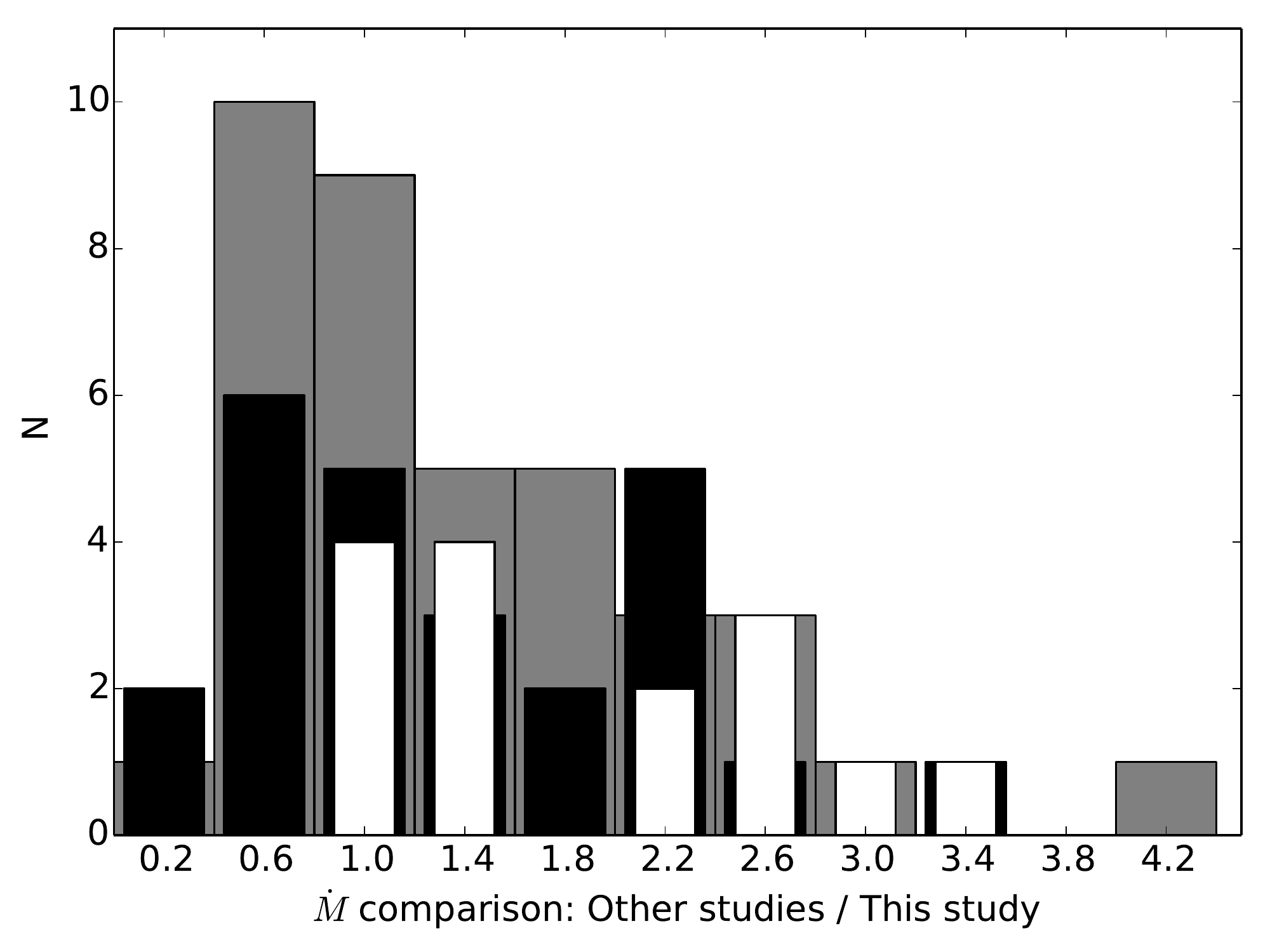}
\caption{Comparisons with mass-loss rates from past studies: $\dot{M}_\mathrm{previous}$/$\dot{M}_\mathrm{new}$, grouped by chemical type. C stars are grey, M stars are black and S stars are white.}
\label{mdotcomparisons}
\end{center}
\end{figure}

We compared our modelling results with those of other studies of stars in our sample. Correcting the mass-loss rates for both different distances to the object and different CO abundance assumptions, we compared results using the metric $\dot{M}_\mathrm{previous}$/$\dot{M}_\mathrm{new}$, where ``previous" indicates the mass-loss rate found in an earlier study and ``new" refers to our mass-loss rate results. The previous studies with the largest overlap with our sample were those of \cite{Schoier2001}, \cite{Knapp1998}, \cite{De-Beck2012}, and \cite{Ramstedt2014}. Where possible we also included an extra data point for each star, coming from other overlapping studies such as \cite{Olofsson1993}, \cite{Woods2003}, \cite{Neri1998}, \cite{Ramstedt2008}, \cite{Teyssier2006}, \cite{Young1995}, and \cite{Olofsson2002}. Grouping all the comparison ratios into chemical type, we find the mean ratio and standard deviation for C stars is $1.4 \pm 0.9$, for M stars is $1.4 \pm 0.8$ and for S stars is $1.9 \pm 0.8$. The collected ratios are also plotted in histograms in Fig. \ref{mdotcomparisons}. 

In general, our models tend to give mass-loss rates on average 40\% lower than previous studies, even after correcting for distances and and CO abundance assumptions. Apart from this offset, the spread in the estimates is about $\pm 50\%$. We present the first study where high-$J$ lines up to  $(9\to8)$ --- or in some cases $(14\to13)$ --- are used in the analysis of a large sample of stars. It could hence be inferred that taking high-$J$ lines properly into account when modelling mass-loss rates results in lower predicted mass-loss rates. Another possible factor is in our use of the velocity profile described in Eq. \ref{vel}. As discussed in \ref{sec:modap} and shown in Fig. \ref{exmodsC}, the inclusion of an accelerating wind model compared with a constant expansion velocity wind can have a significant effect on the resultant line intensities (and intensity ratios between different lines). 

\section{Conclusions}

We have presented new \textit{Herschel}/HIFI and IRAM CO line data for a sample of 53 AGB stars, which cover a wide range of transitions from $J=1\to0$ up to $J=9\to8$ (and $J=14\to13$ in a few cases). For 38 of the stars, we used radiative transfer models to determine their mass-loss rates. Our results cover a wide range of mass-loss rates, from $\sim 10^{-8} - 2\e{-5}\spy$, and include all three chemical types of AGB stars (M, S, and C). In general, we find it possible to get a very good fit to the observed CO lines which cover up to 250 K in energy (in a few cases up to 600 K) without the need to invoke mass-loss rate changes with time. Our model results are also in reasonable agreement with past studies (within a factor two for 63\% of the stars), although they generally predict slightly lower mass-loss rates than past studies based primarily on lower-$J$ lines. We found that our models under-predict the CO ($1\to0$) line intensity more often than other lines, although the precise reason for this is not clear.

In analysing our results we found correlations between mass-loss rate and stellar luminosity, gas terminal expansion velocity, and dust optical depth. The latter correlation seems to show slightly different slopes for the M and C stars, with the S stars falling in between the two. We also found that there is a tendency for the CSEs to have a lower kinetic temperature the higher the mass-loss rate, an effect expected due to less efficient dust-gas collision heating and stronger CO line cooling in the denser objects.

We also found that the inclusion of an expansion velocity profile in our models can have a significant effect on the results. In particular, the use of a constant velocity instead of an accelerating wind can severely under-predict the high-$J$ line intensities, such as those observed by \textit{Herschel}/HIFI. The exact shape of the velocity profile is obtained by fitting the shapes of, in particular, the higher-$J$ lines.

Our observations also included a series of ``bonus" lines, which fell within the observing ranges of our CO observations. \up{13}CO was the most commonly detected ``bonus" line in all three chemical types of AGB stars. It was detected in eight C stars, five S stars and eight M stars. CN and HC$_3$N were detected in C stars, with one S star (S Cas) also being detected in some of the observed CN lines. SiO was frequently detected in M stars (17 detections), as opposed to the C stars where only two were detected. SiS was detected in the higher mass-loss rate C and M stars, although not in all of the highest mass-loss rate stars. There were seven detections in C stars and five in M stars.

\begin{acknowledgements}
TD and KJ acknowledge funding from the SNSB. DT acknowledges support from the Faculty of the European Space Astronomy Centre (ESAC). JC, VB and JA thank spanish MINECO for funding under grants AYA2009-07304, AYA2012-32032, CSD2009-00038, and
ERC under ERC-2013-SyG, G.A. 610256 NANOCOSMOS

HIFI has been designed and built by a consortium of institutes and university departments from across Europe, Canada and the United States under the leadership of SRON Netherlands Institute for Space Research, Groningen, The Netherlands and with major contributions from Germany, France and the US. Consortium members are: Canada: CSA, U.Waterloo; France: CESR, LAB, LERMA, IRAM; Germany: KOSMA, MPIfR, MPS; Ireland, NUI Maynooth; Italy: ASI, IFSI-INAF, Osservatorio Astrofisico di Arcetri-INAF; Netherlands: SRON, TUD; Poland: CAMK, CBK; Spain: Observatorio Astron\'omico Nacional (IGN), Centro de Astrobiolog\'ia (CSIC-INTA). Sweden: Chalmers University of Technology - MC2, RSS \& GARD; Onsala Space Observatory; Swedish National Space Board, Stockholm University - Stockholm Observatory; Switzerland: ETH Zurich, FHNW; USA: Caltech, JPL, NHSC.

This publication is based on data acquired with the Atacama Pathfinder Experiment (APEX). APEX is a collaboration between the Max-Planck-Institut f\"ur Radioastronomie, the European Southern Observatory, and the Onsala Space Observatory.

This research has made use of the International Variable Star Index (VSX) database, operated at AAVSO, Cambridge, Massachusetts, USA.

\end{acknowledgements}

%

\bibliographystyle{aa}
\bibliography{SUCCESS.bbl}

\Online
\appendix

\section{Modelled stars}\label{modplots}

Here we present the observations and models for the stars not included in the body of the paper. In each instance, we present all new data from IRAM and HIFI --- black histograms --- overplotted with model results for the parameters given in Table \ref{coresults}. For those stars that were not observed with IRAM, we include archival CO $(1\to0)$ and $(2\to1)$ lines from various telescopes as available. These archival lines are indicated by an * next to the telescope name in the plot and allow us to present an overview of our models from low- to high-$J$. R Hor is the only star for which these low-$J$ lines were not available. Our model for R Hor still incorporates some low-$J$ lines as noted in Table \ref{lowjobs}.

The plots for C stars are shown in Fig. \ref{Cmods}, for S stars in Fig. \ref{Smods} and for M stars in Fig. \ref{Mmods}.

As well as plotting the model and observed CO lines, we have calculated ``goodness of fit" per line in each star. This gives us an indication of which lines may be outliers or whether there are any trends across lines.

Goodness of fit for C stars is shown in Fig. \ref{Cfits}, for S stars in Fig. \ref{Sfits}, and for M stars in Fig. \ref{Mfits}. See Fig. \ref{transgood} and Sect. \ref{fitgoods} for a discussion of goodness of fit across the entire sample. A list of archival lines included in our models is given in Table \ref{lowjobs}.

\begin{figure*}[!p]
\begin{center}
\includegraphics[width=0.49\textwidth]{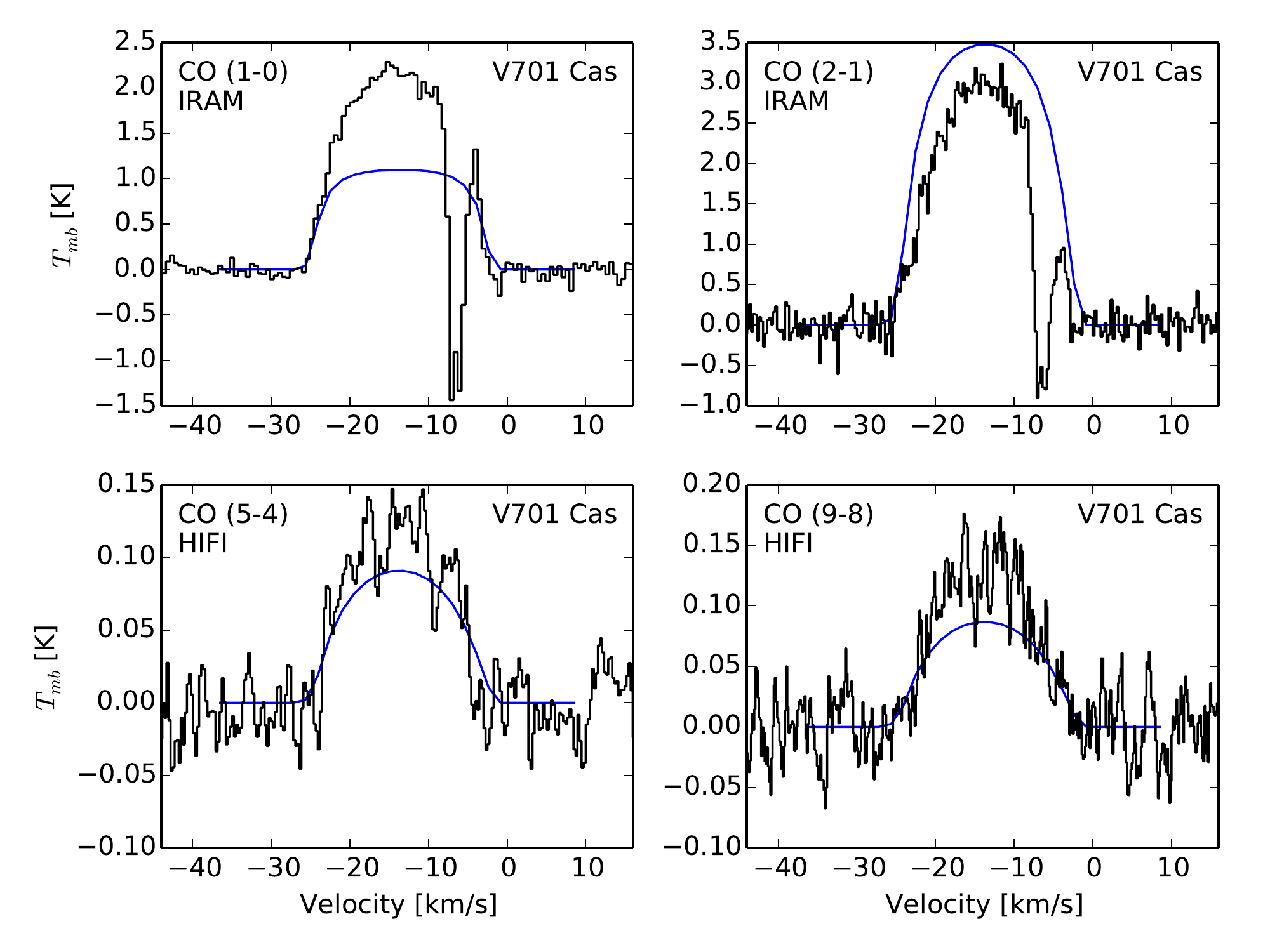}
\includegraphics[width=0.49\textwidth]{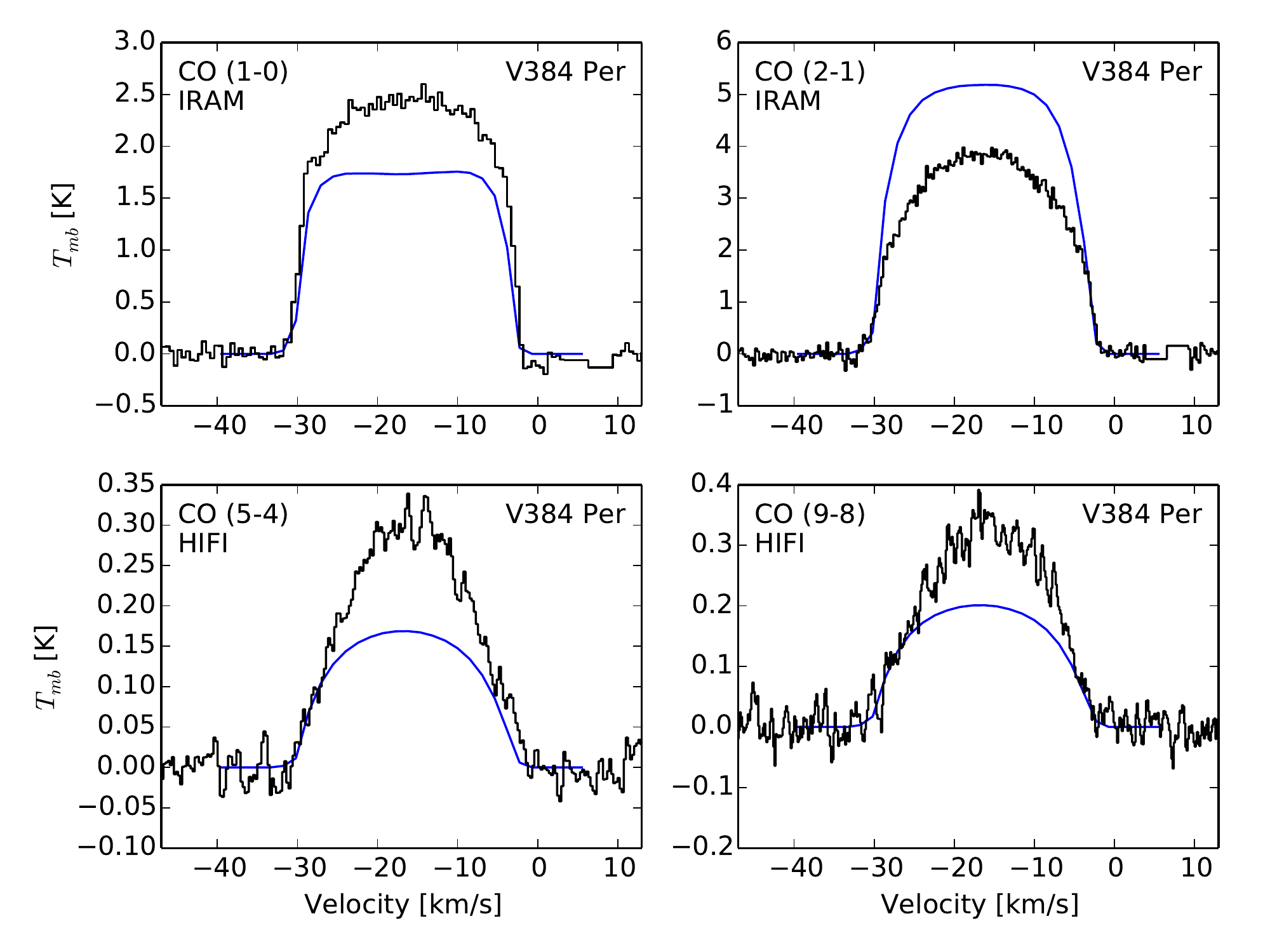}
\includegraphics[width=0.49\textwidth]{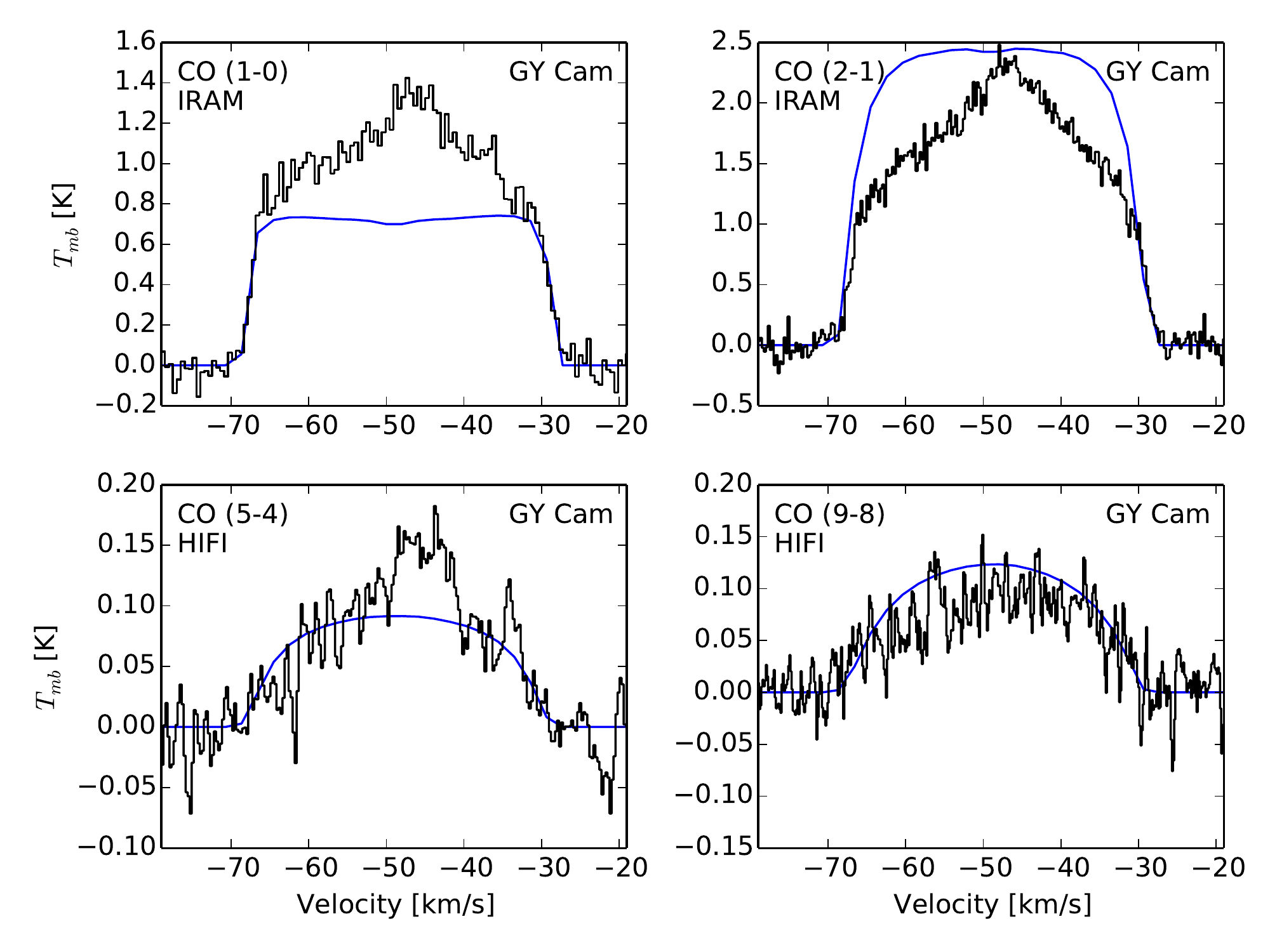}
\includegraphics[width=0.49\textwidth]{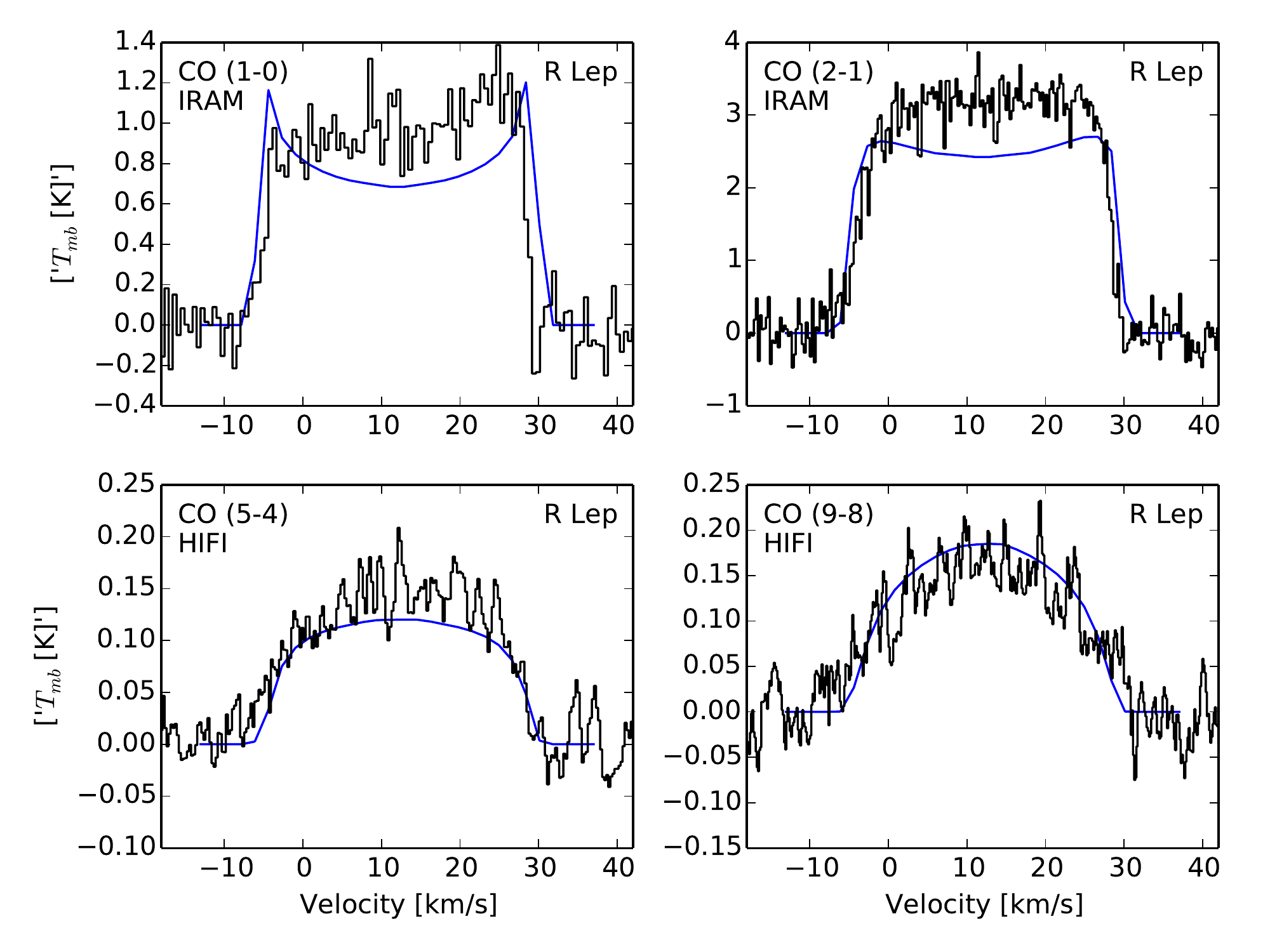}
\includegraphics[width=0.49\textwidth]{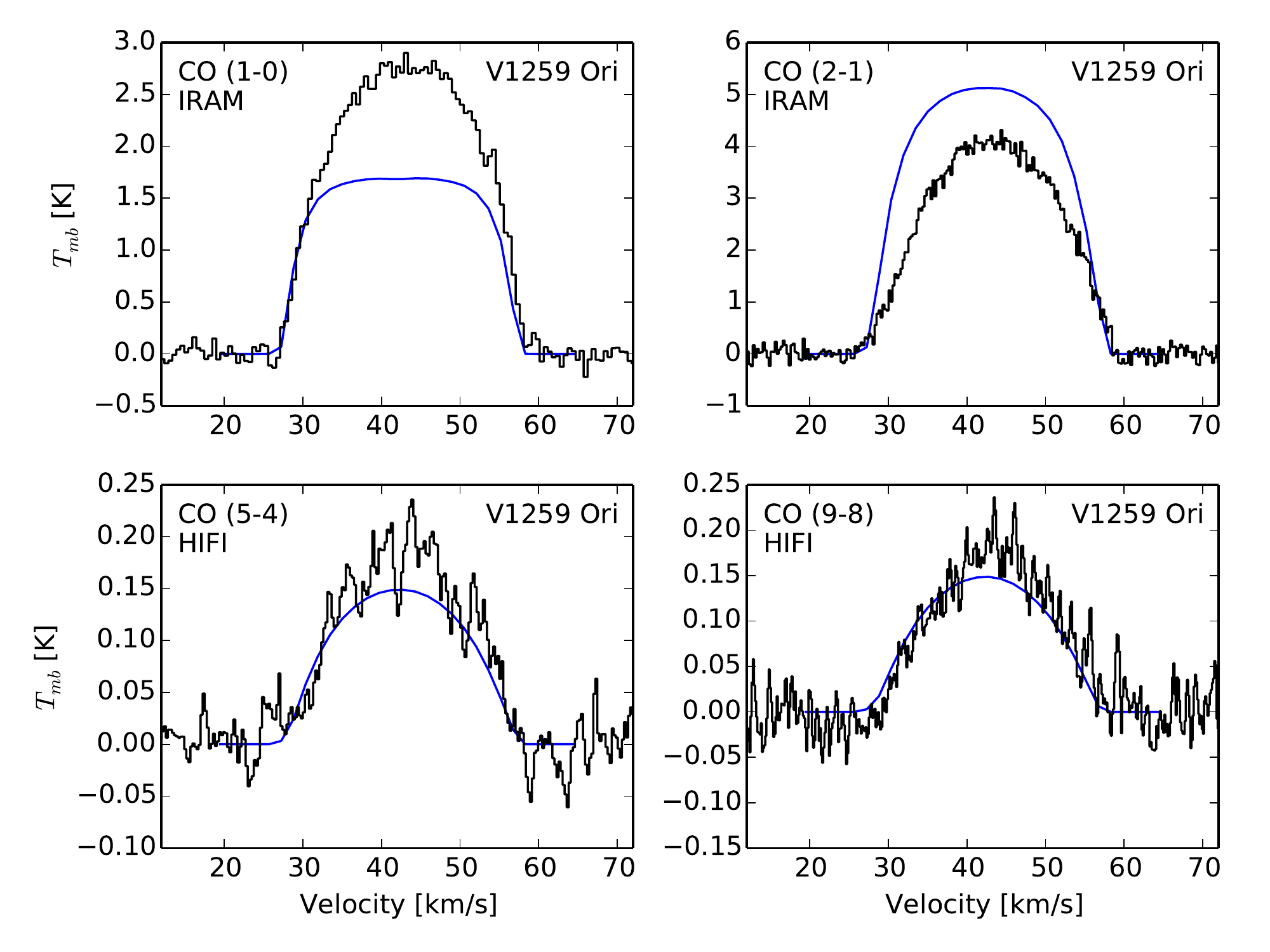}
\includegraphics[width=0.49\textwidth]{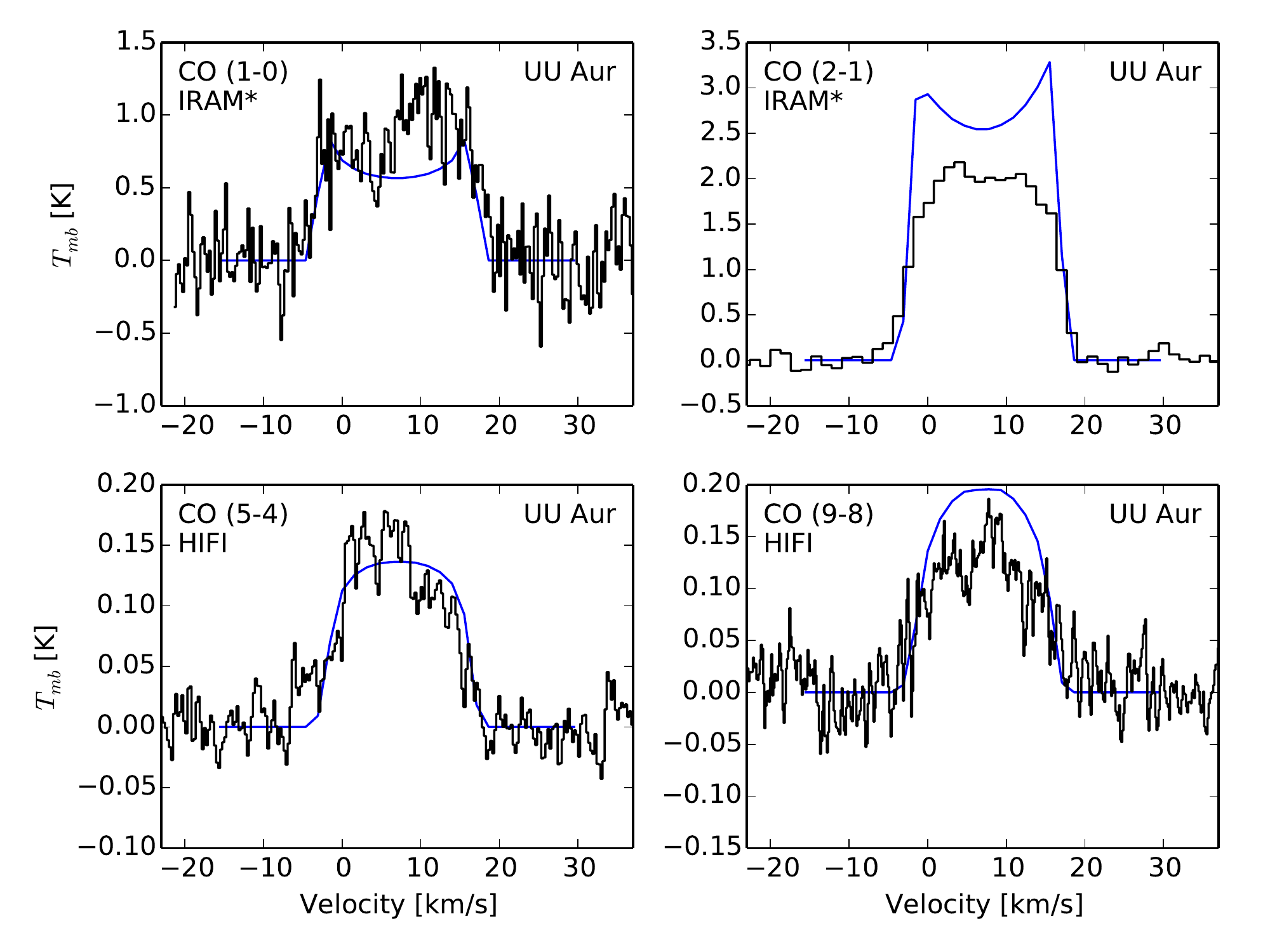}
\caption{Models (blue) and observed data (black) of C stars, plotted with respect to LSR velocity. An * next to the telescope name indicates that archival data is plotted.}
\label{Cmods}
\end{center}
\end{figure*}
\addtocounter{figure}{-1}
\begin{figure*}[!p]
\begin{center}
\includegraphics[width=0.49\textwidth]{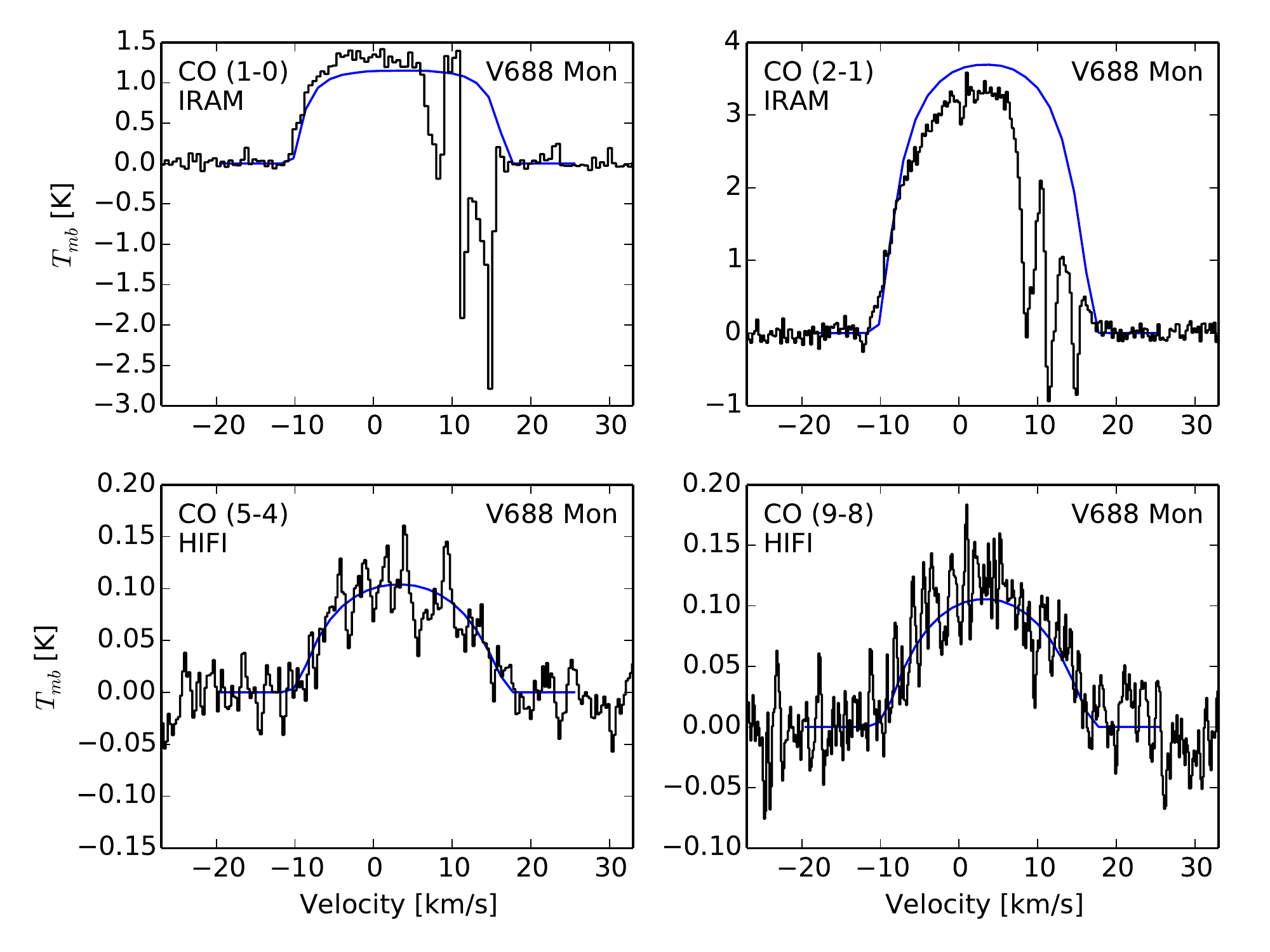}
\includegraphics[width=0.49\textwidth]{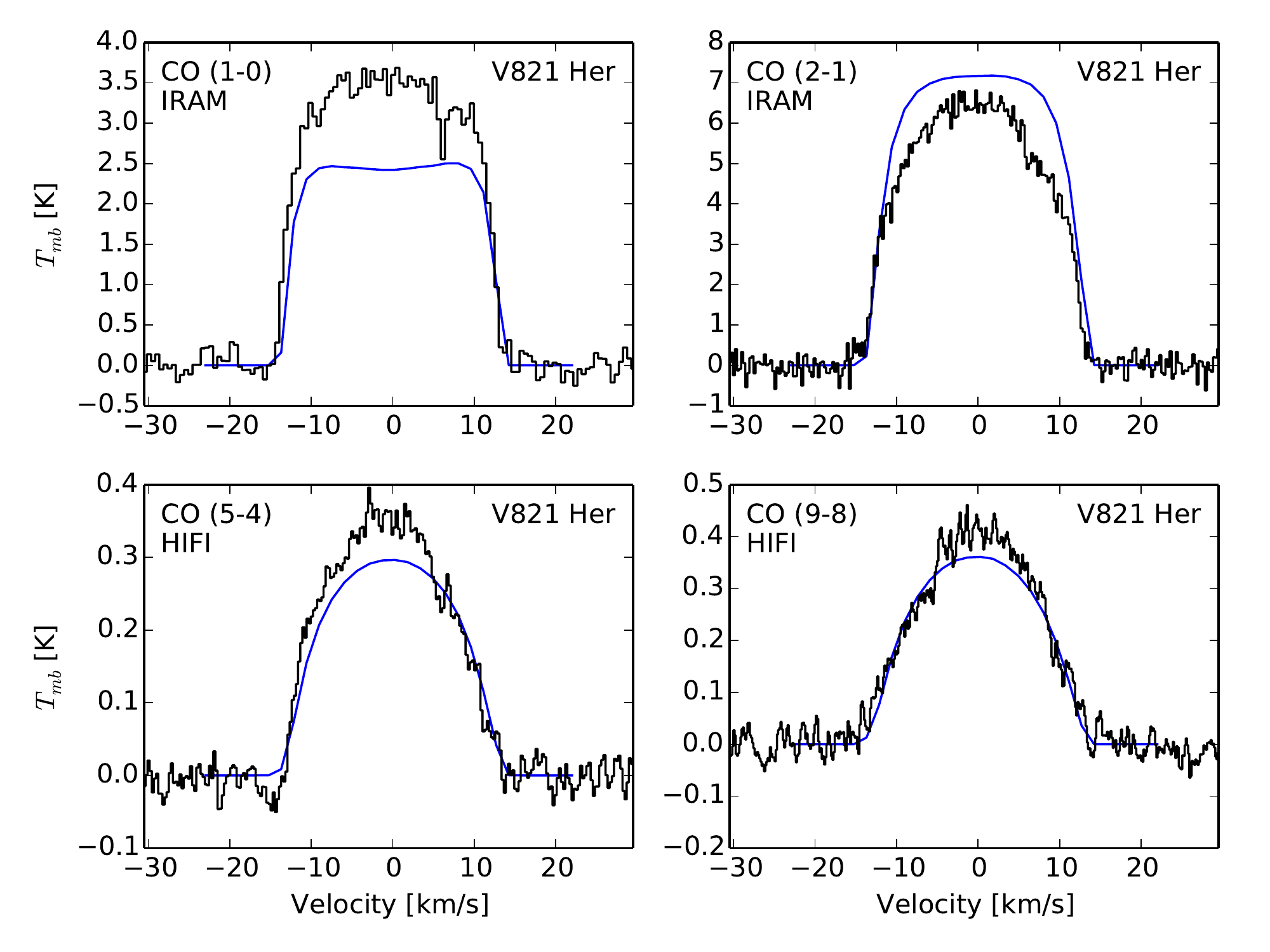}
	\includegraphics[width=0.49\textwidth]{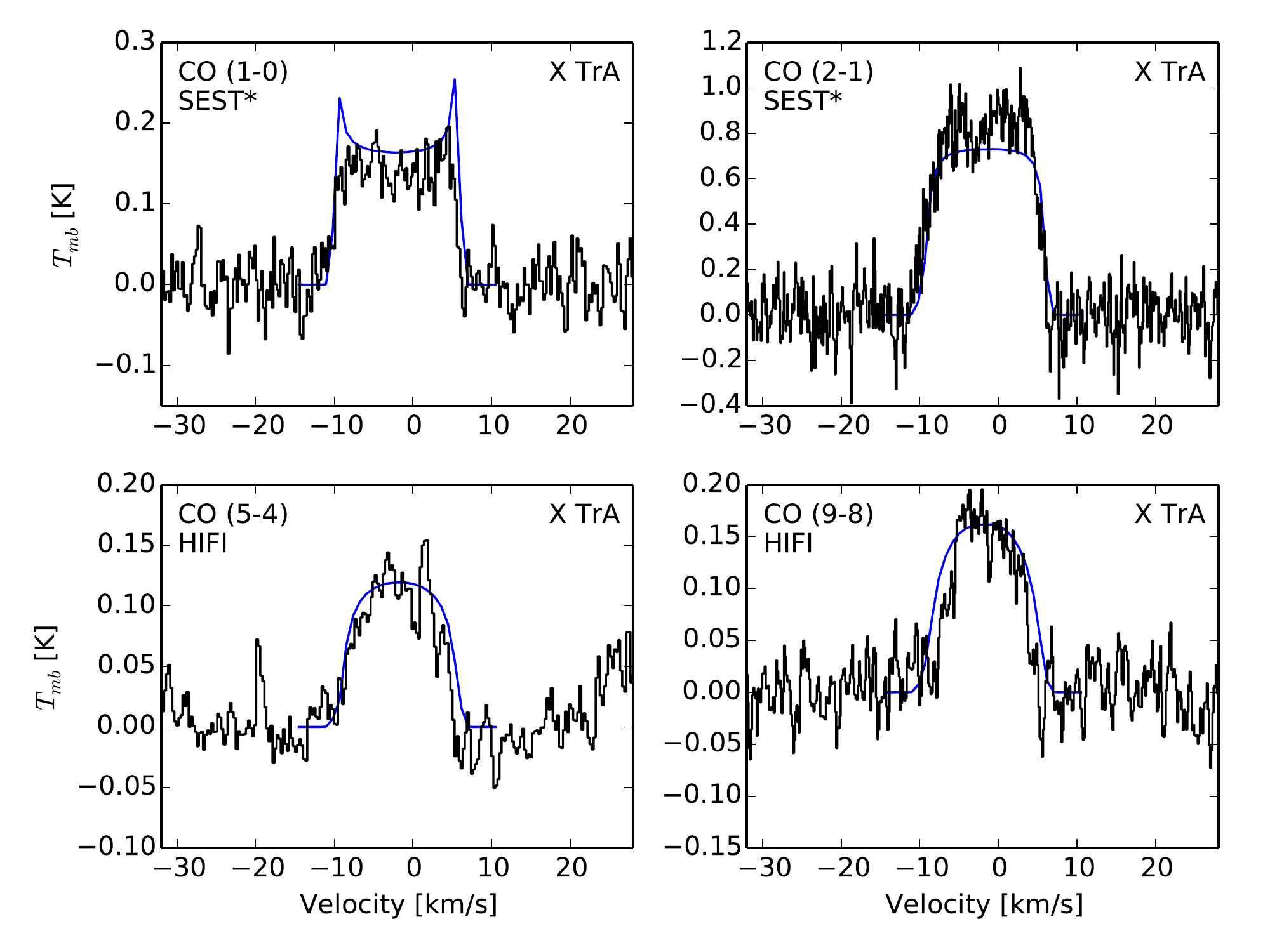}
	\includegraphics[width=0.49\textwidth]{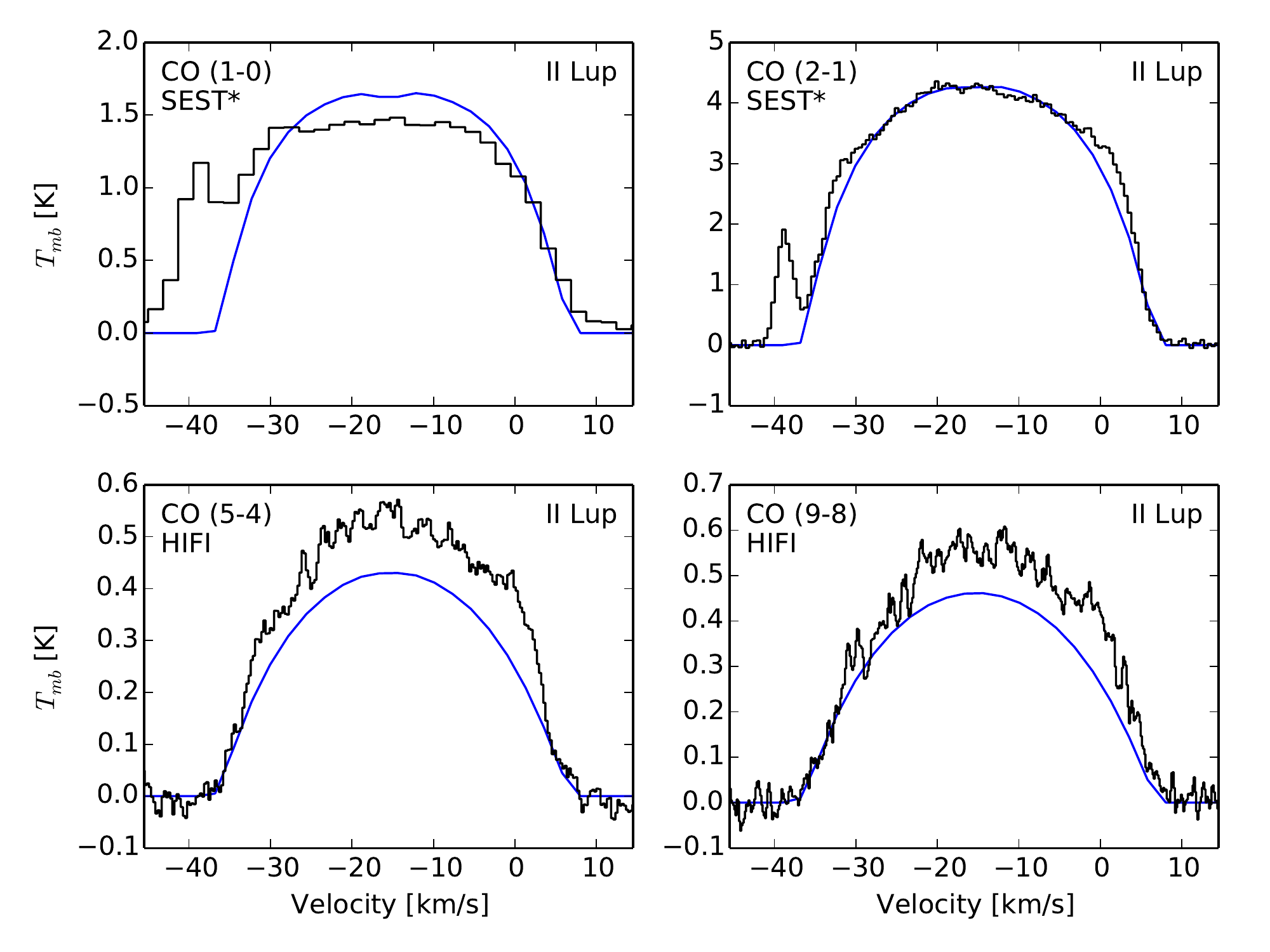}
	\includegraphics[width=0.49\textwidth]{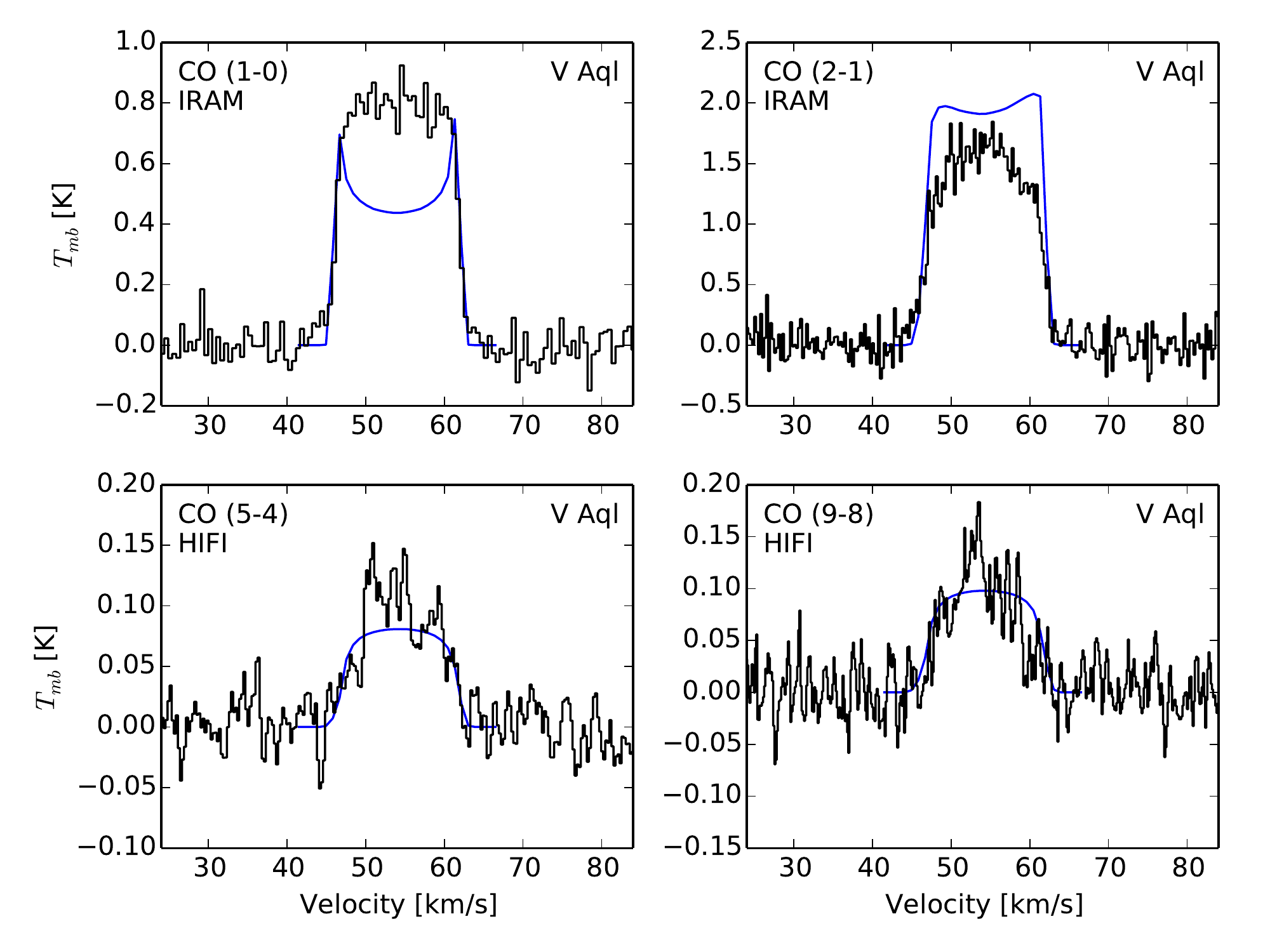}
	\includegraphics[width=0.49\textwidth]{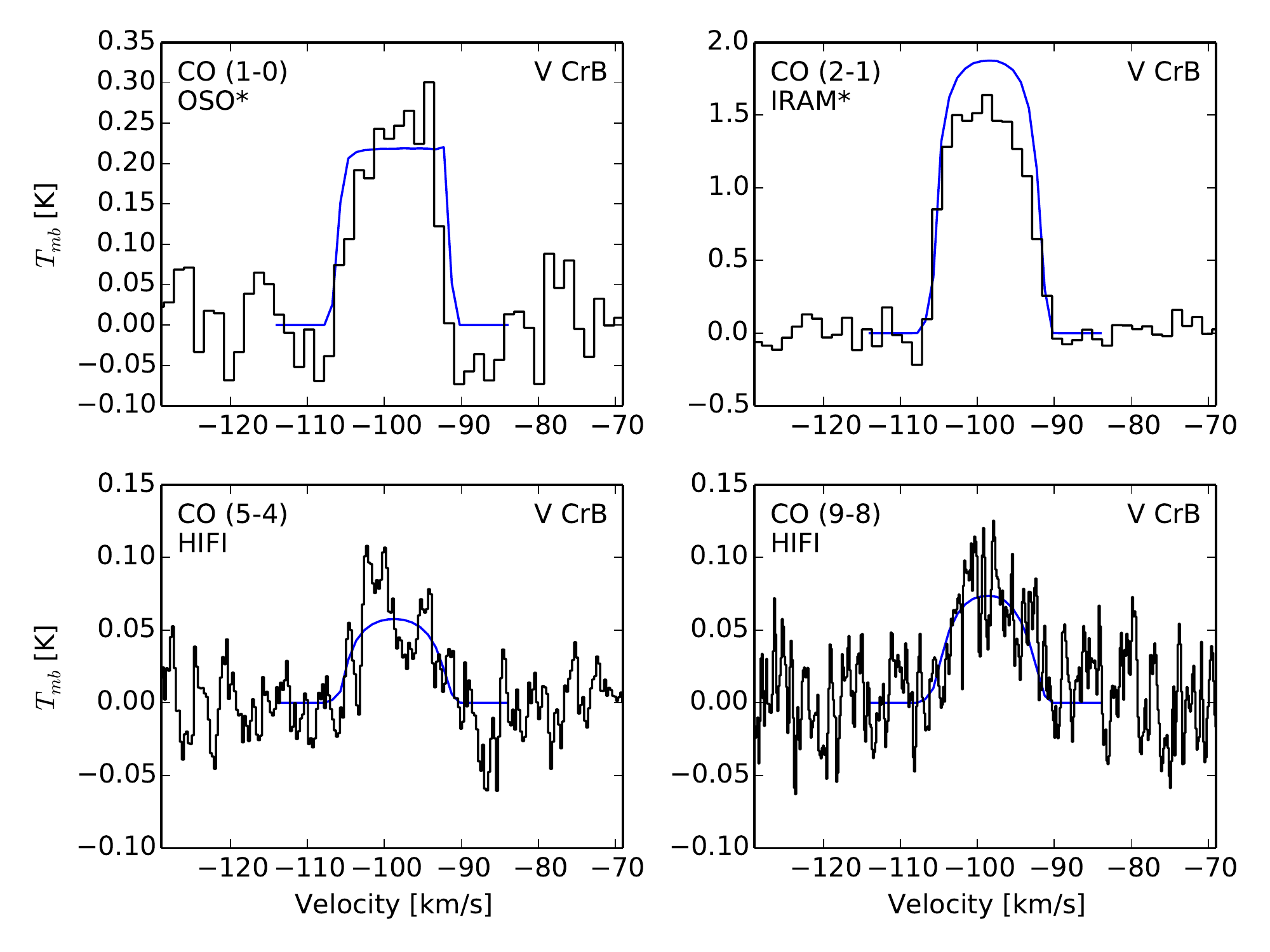}
\caption{{\bf cont.} Models (blue) and observed data (black) of C stars, plotted with respect to LSR velocity. An * next to the telescope name indicates that archival data is plotted.}
\end{center}
\end{figure*}
\addtocounter{figure}{-1}
\begin{figure*}[t]
\begin{center}
	\includegraphics[width=0.49\textwidth]{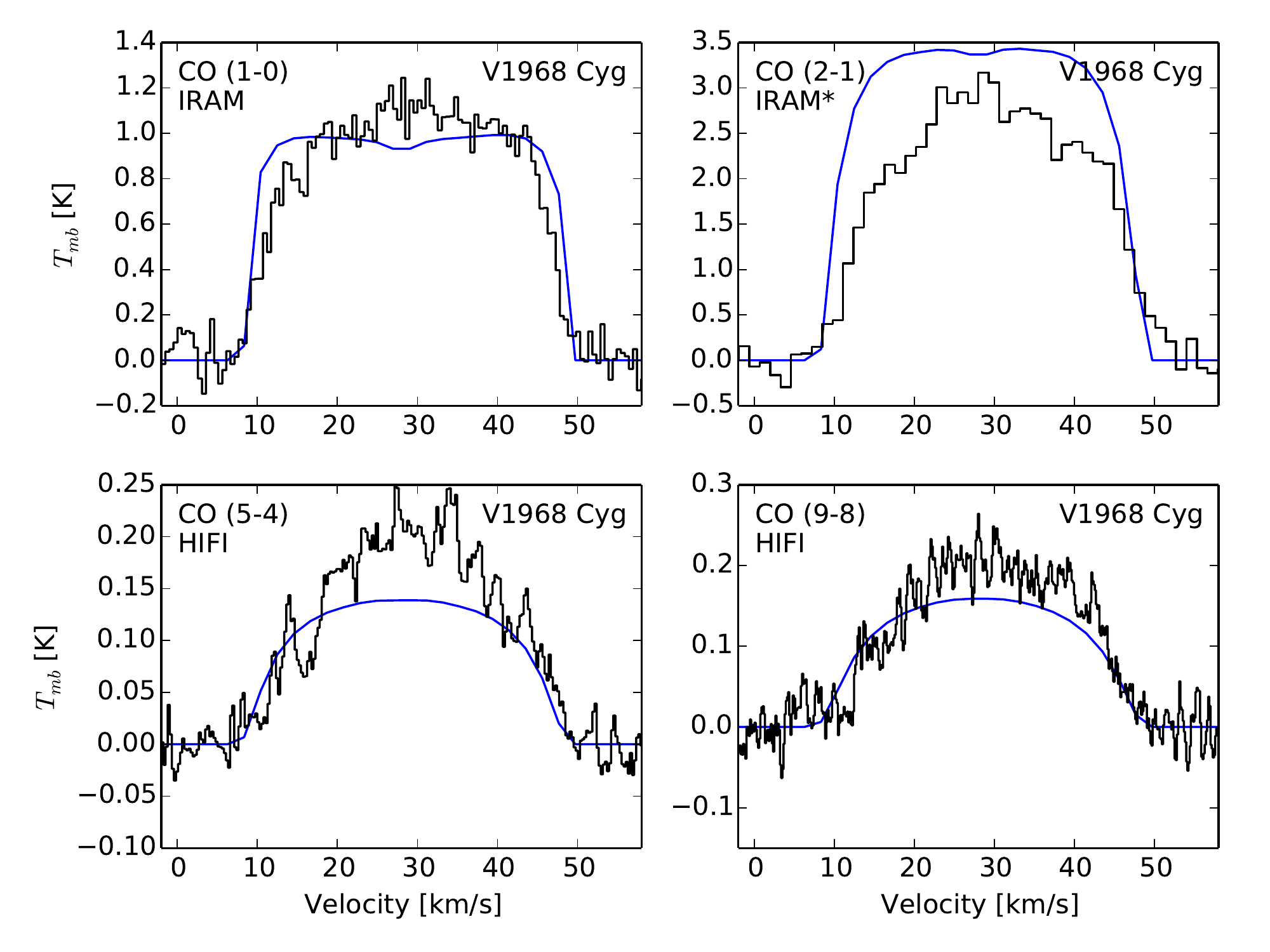}
	\includegraphics[width=0.49\textwidth]{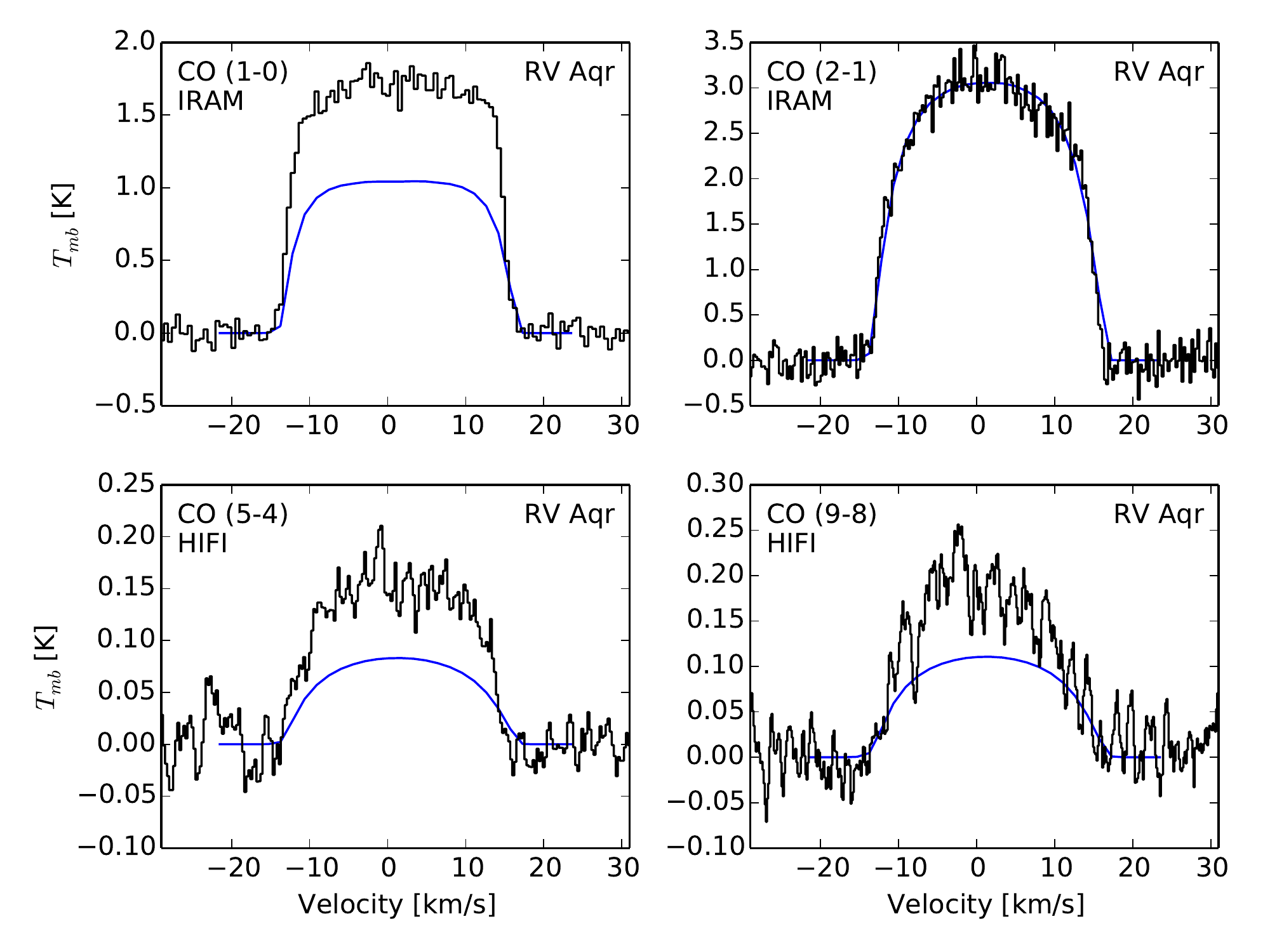}
\caption{{\bf cont.} Models (blue) and observed data (black) of C stars, plotted with respect to LSR velocity. An * next to the telescope name indicates that archival data is plotted.}
\end{center}
\end{figure*}

\begin{figure*}[t]
\begin{center}
\includegraphics[width=0.49\textwidth]{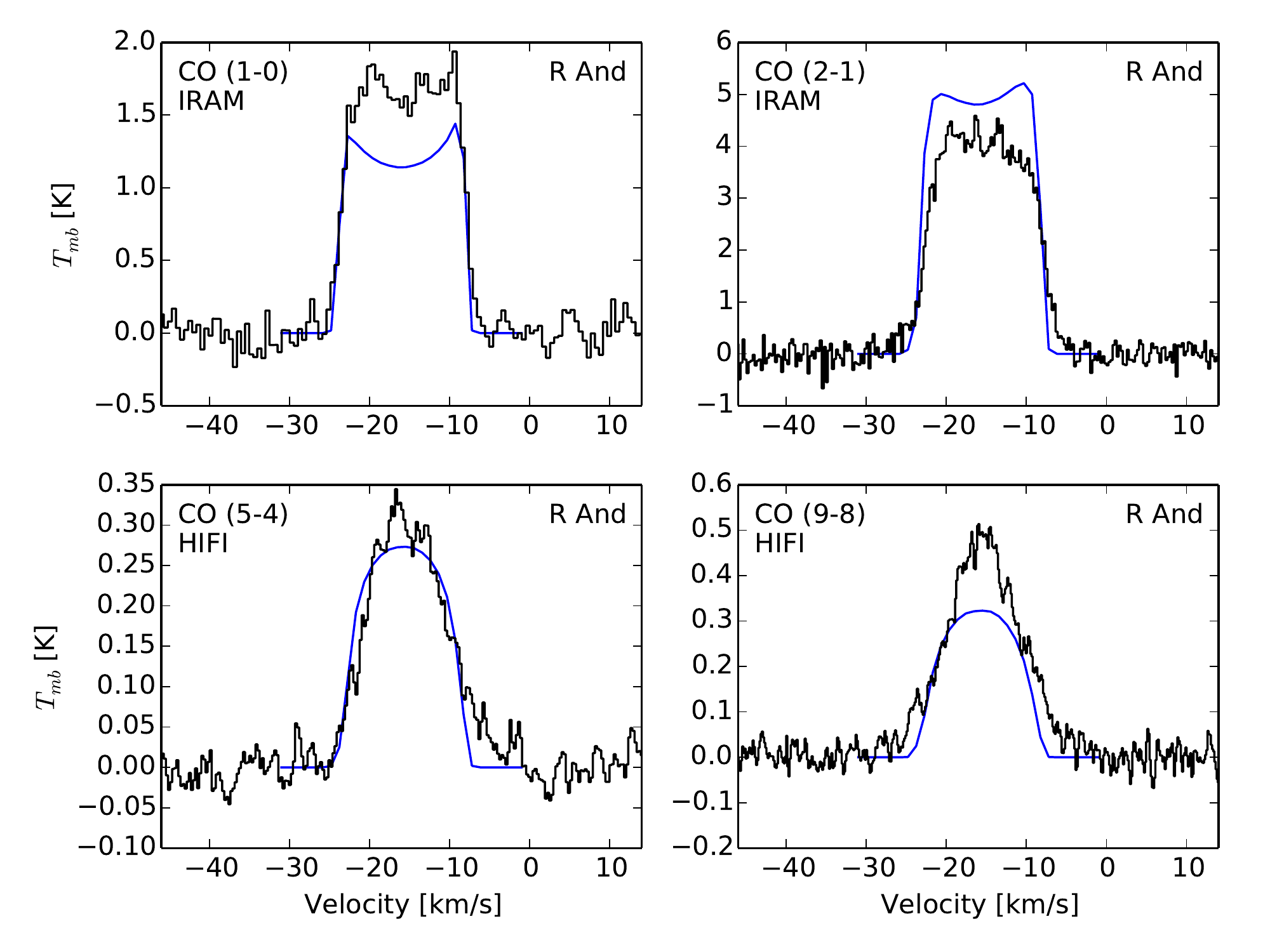}
\includegraphics[width=0.49\textwidth]{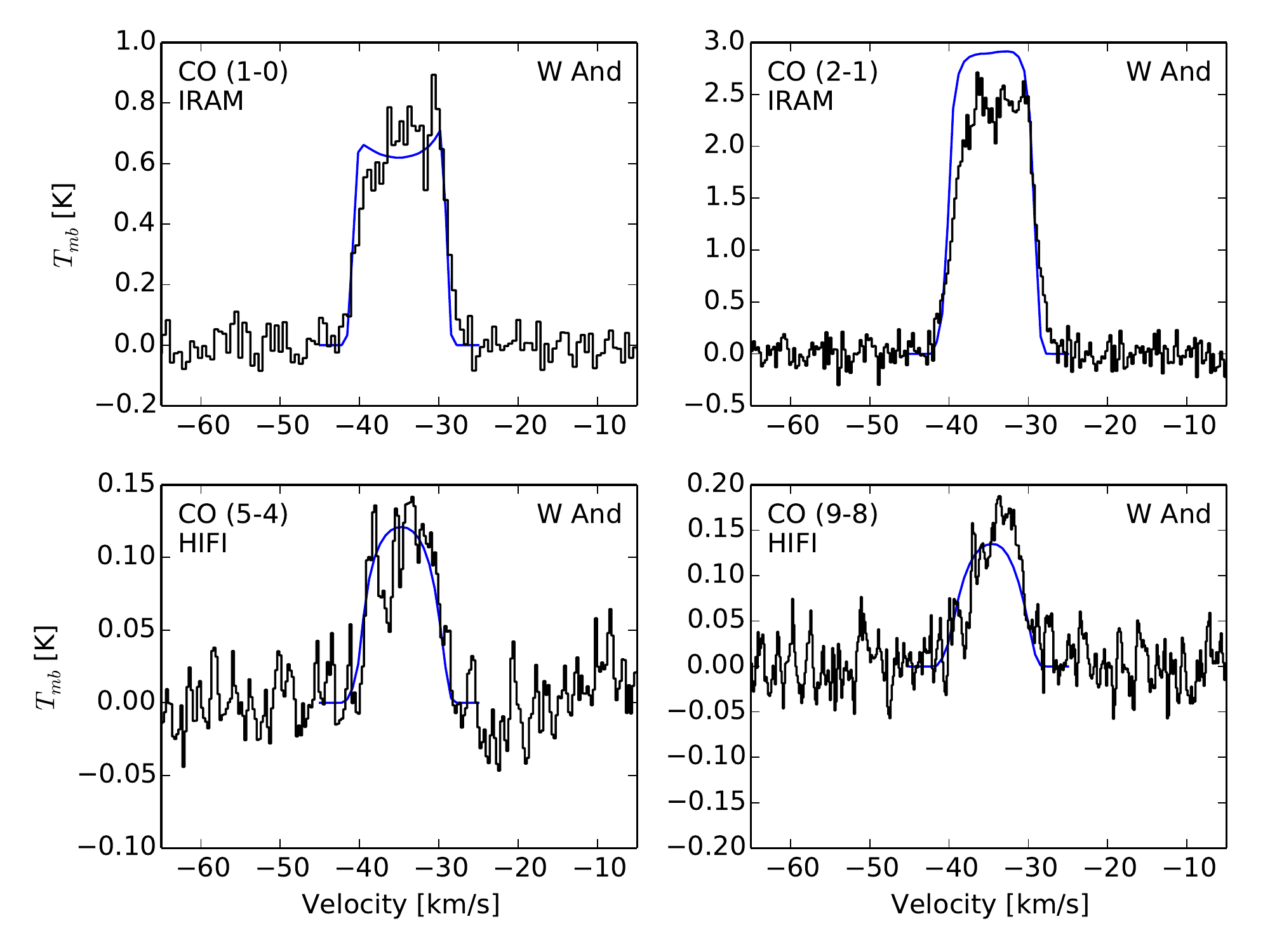}
\includegraphics[width=0.49\textwidth]{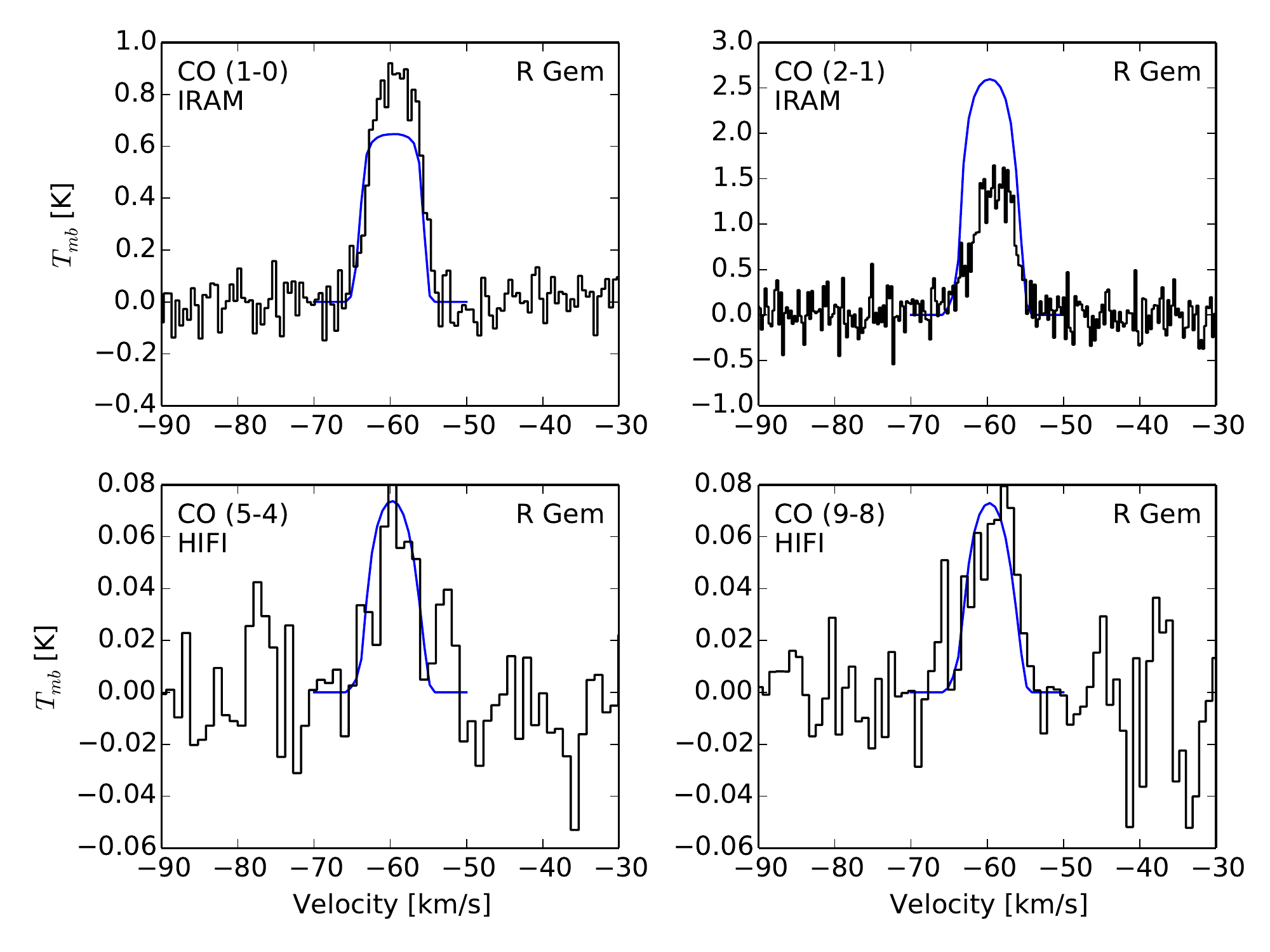}
\includegraphics[width=0.49\textwidth]{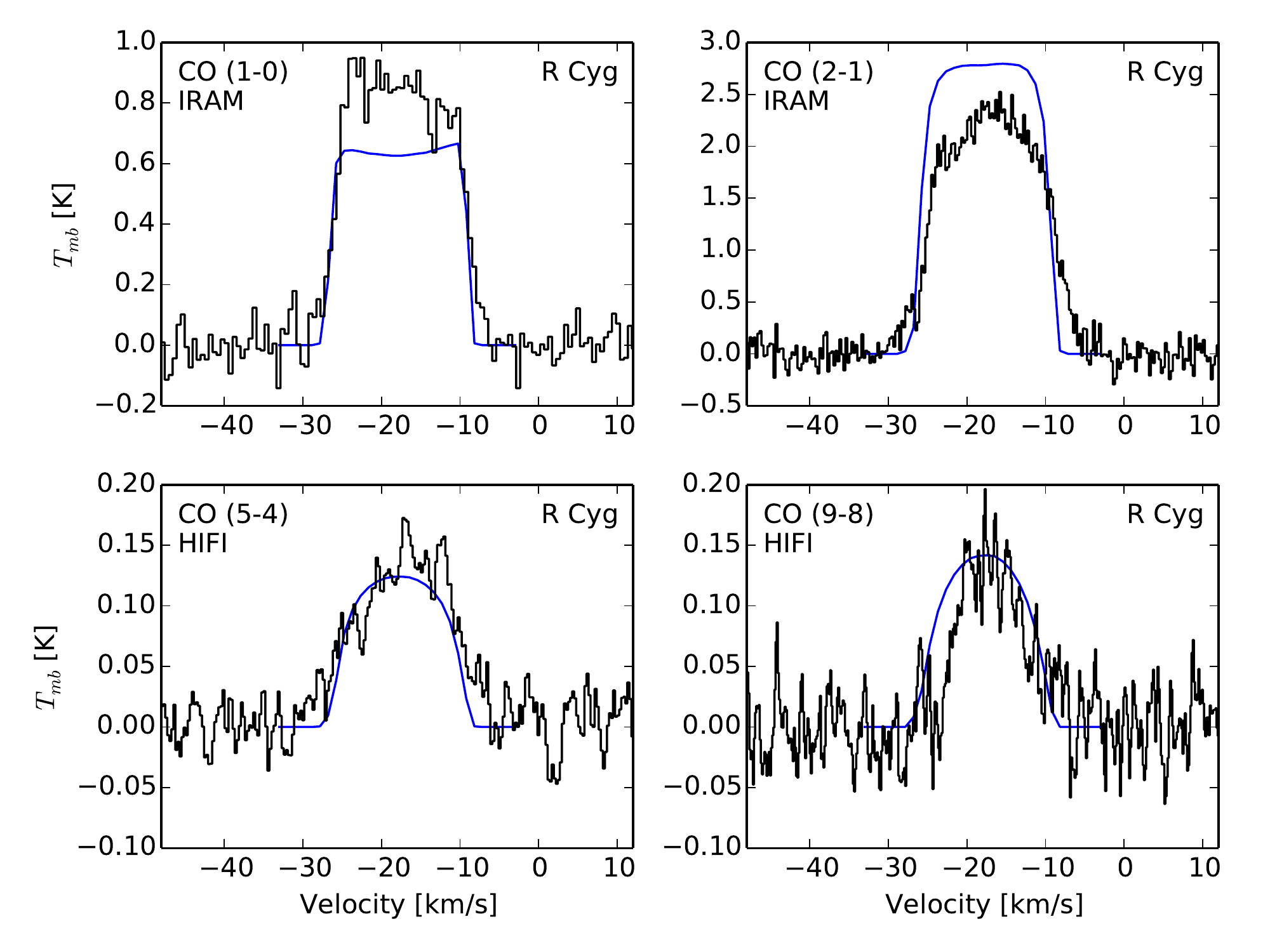}
\caption{Models (blue) and observed data (black) of S stars, plotted with respect to LSR velocity.}
\label{Smods}
\end{center}
\end{figure*}

\begin{figure*}[!p]
\begin{center}
\includegraphics[width=0.49\textwidth]{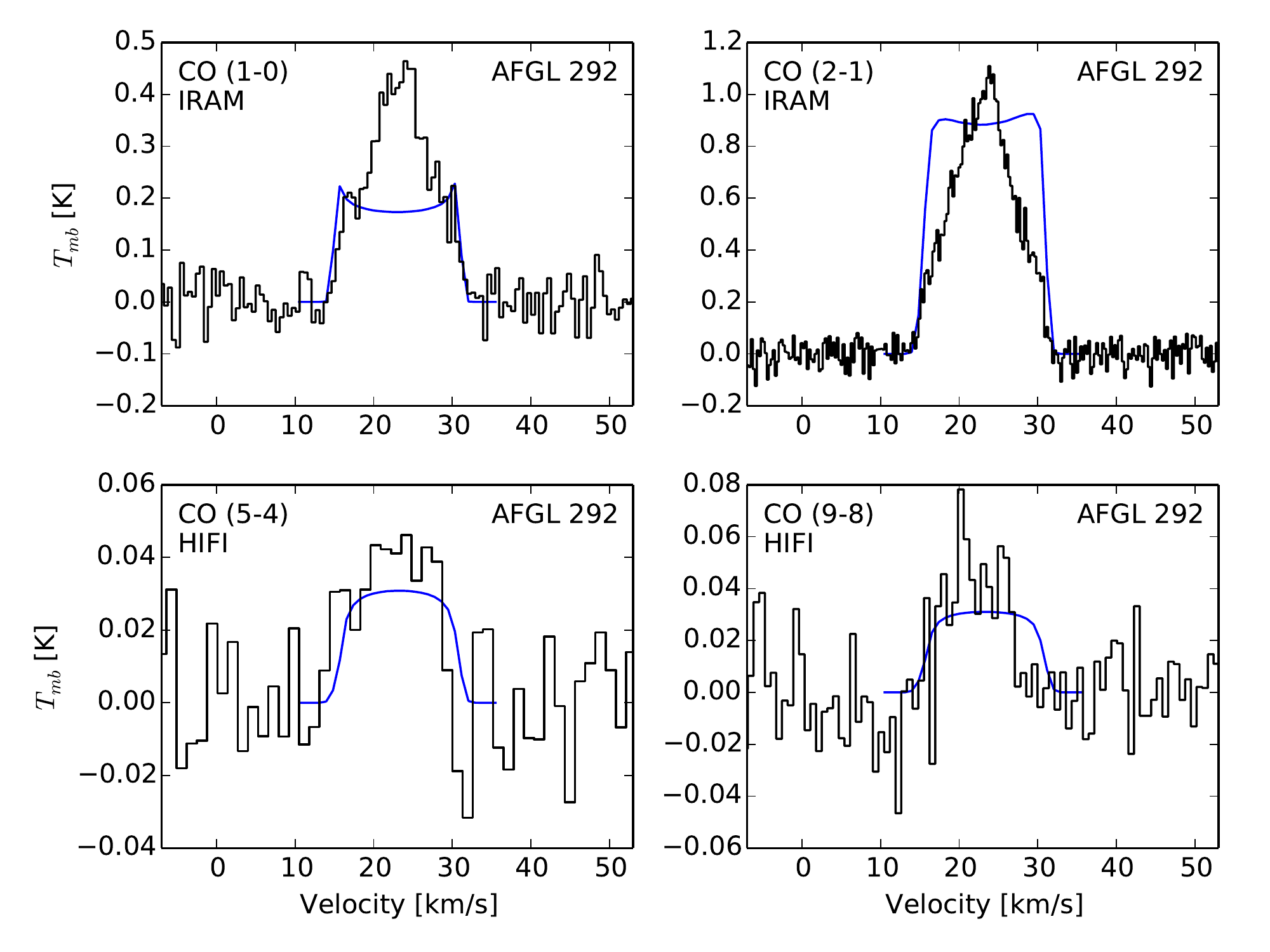}
\includegraphics[width=0.49\textwidth]{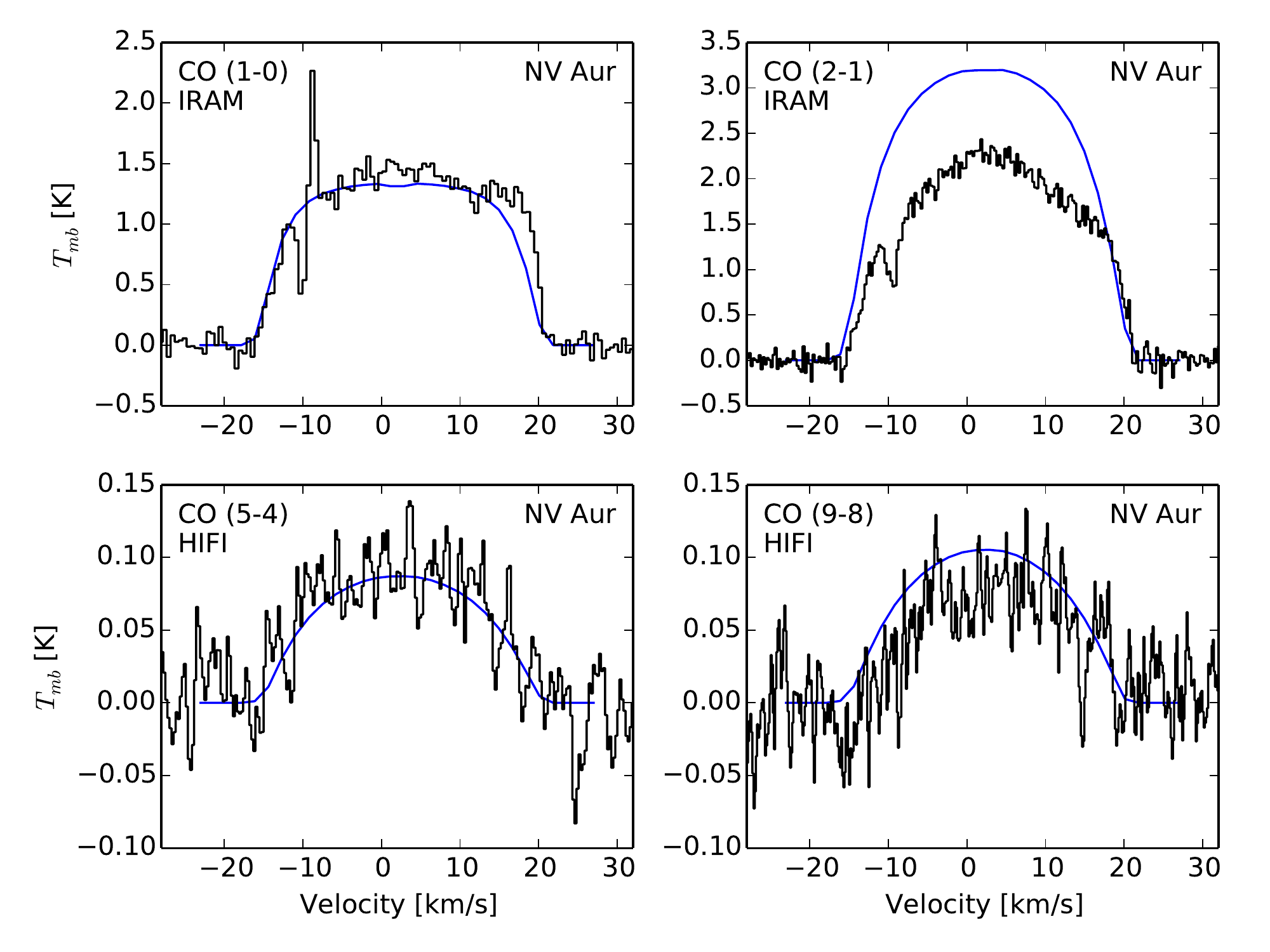}
\includegraphics[width=0.49\textwidth]{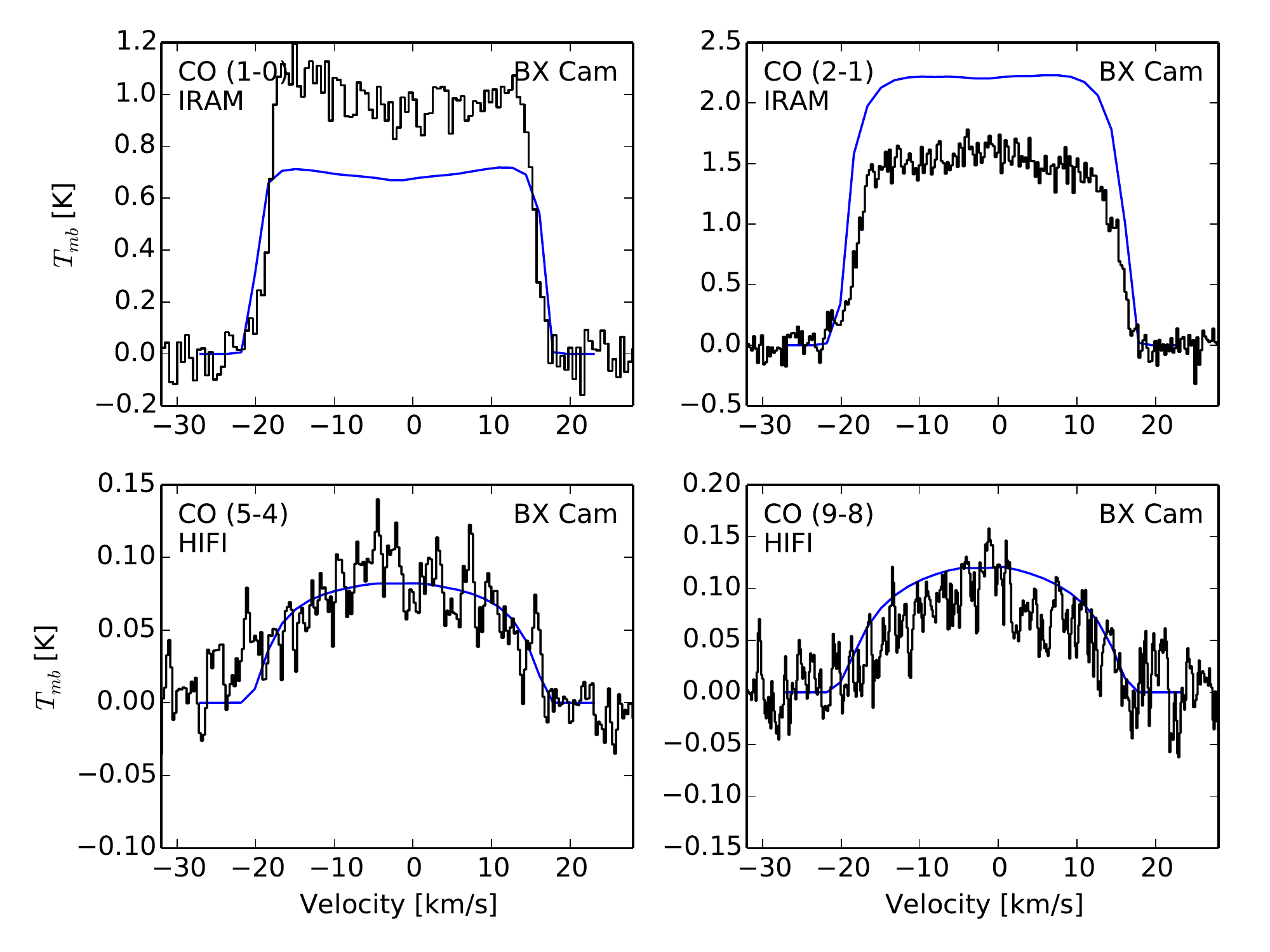}
	\includegraphics[width=0.49\textwidth]{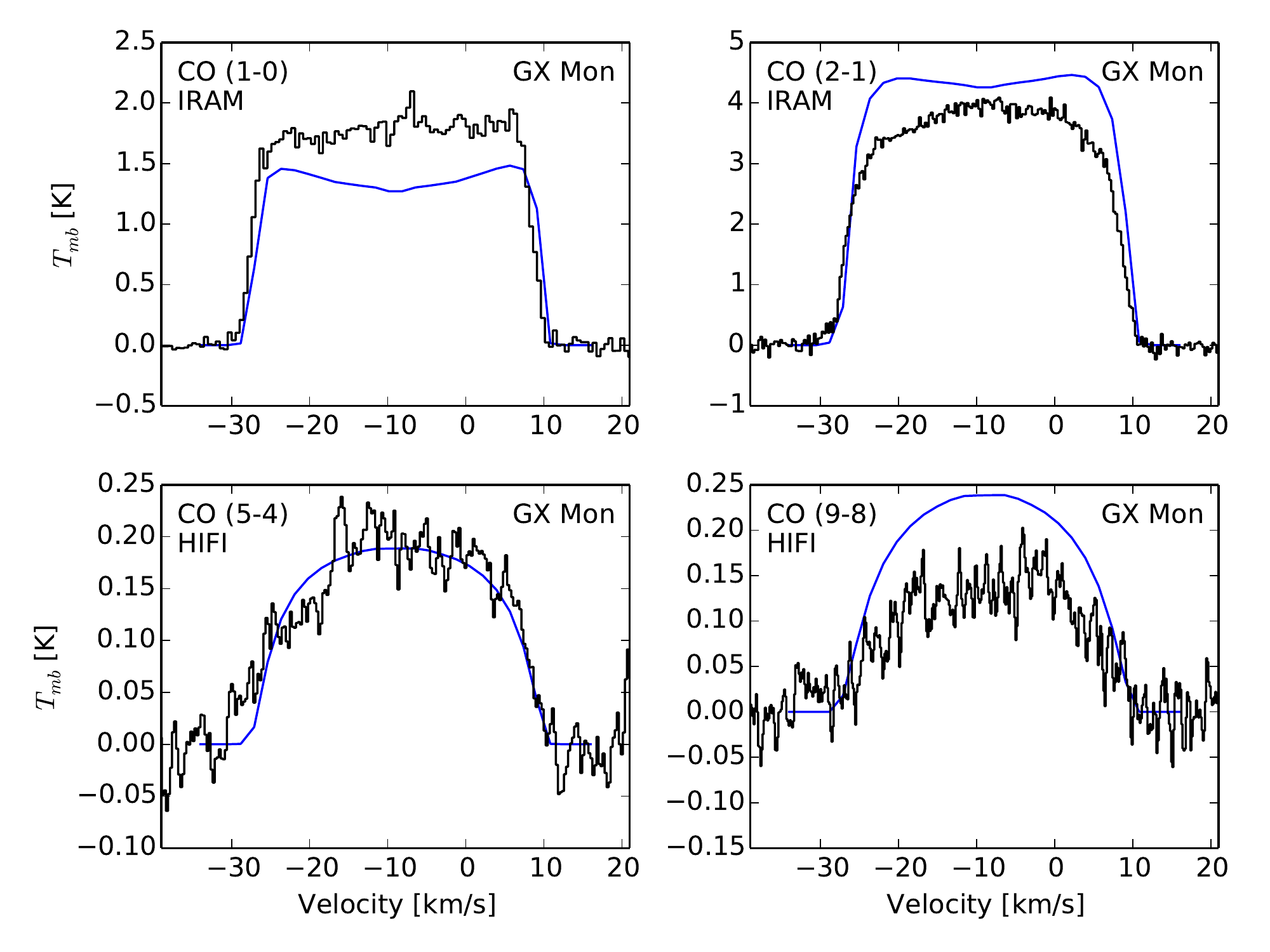}
	\includegraphics[width=0.49\textwidth]{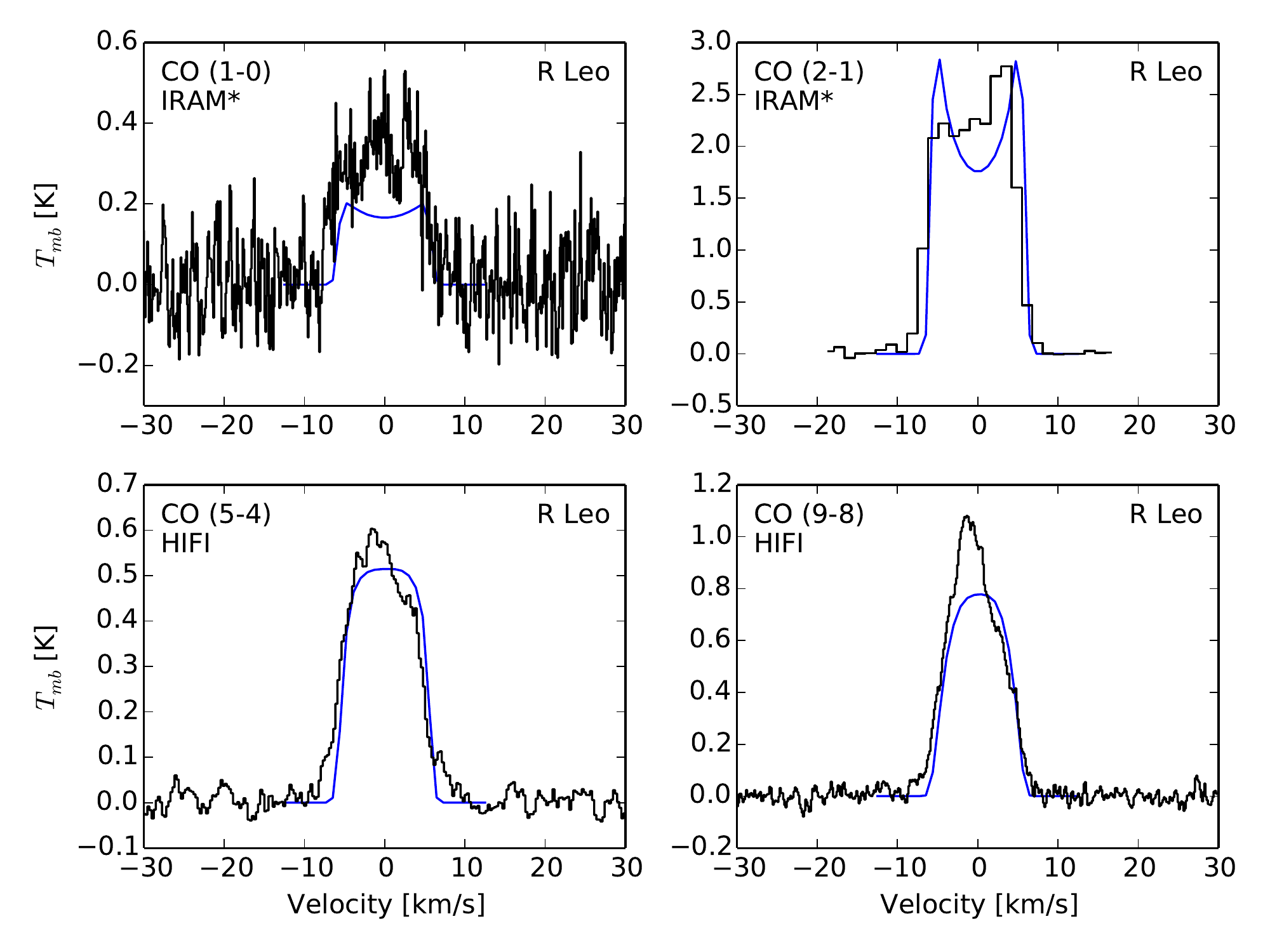}
	\includegraphics[width=0.49\textwidth]{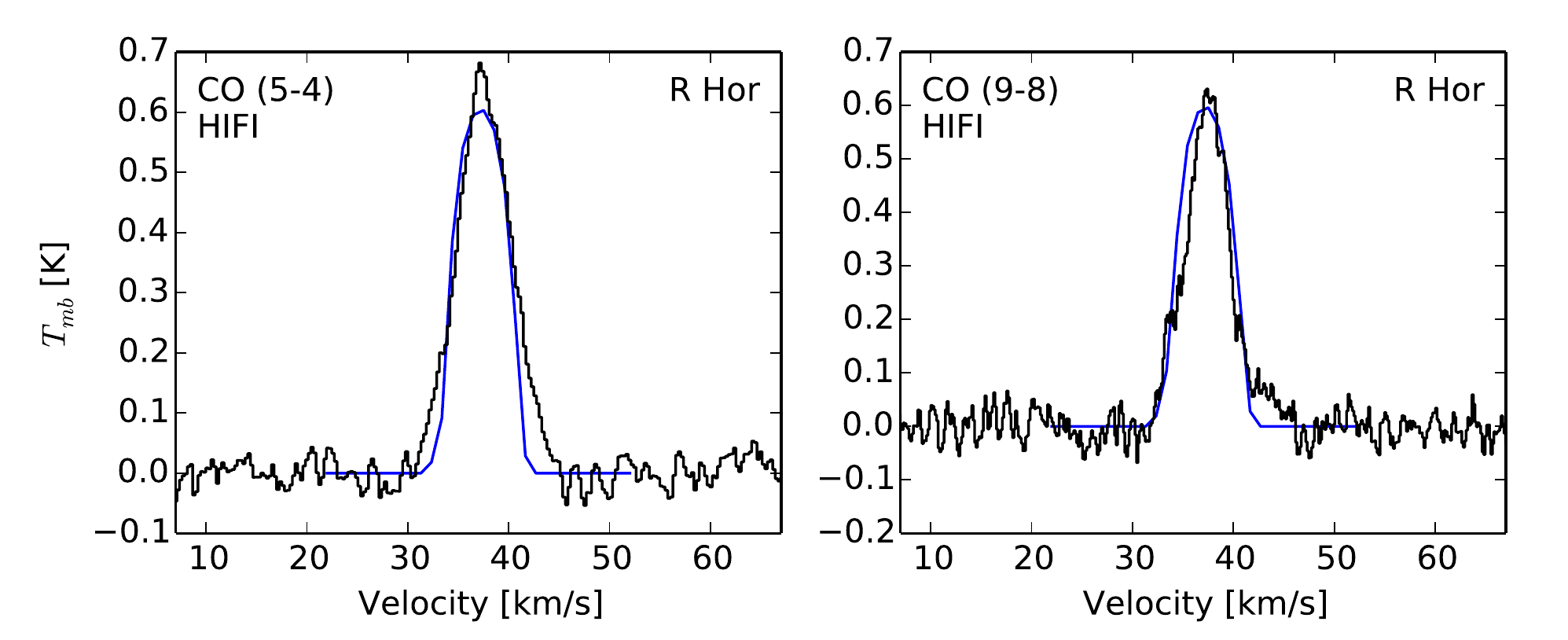}
\caption{Models (blue) and observed data (black) of M stars, plotted with respect to LSR velocity. An * next to the telescope name indicates archival data is plotted.}
\label{Mmods}
\end{center}
\end{figure*}
\addtocounter{figure}{-1}
\begin{figure*}[!p]
\begin{center}
\includegraphics[width=0.49\textwidth]{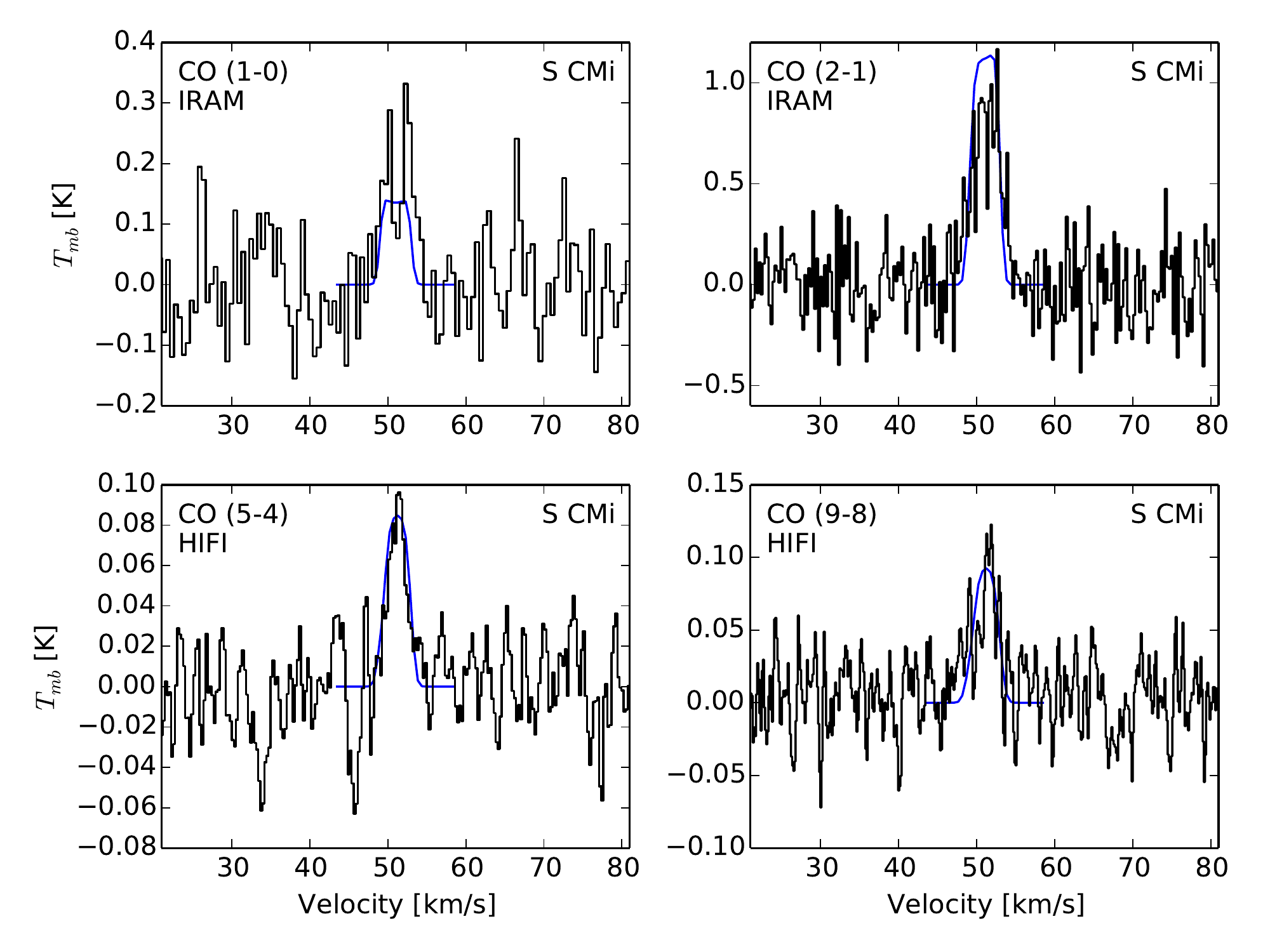}
\includegraphics[width=0.49\textwidth]{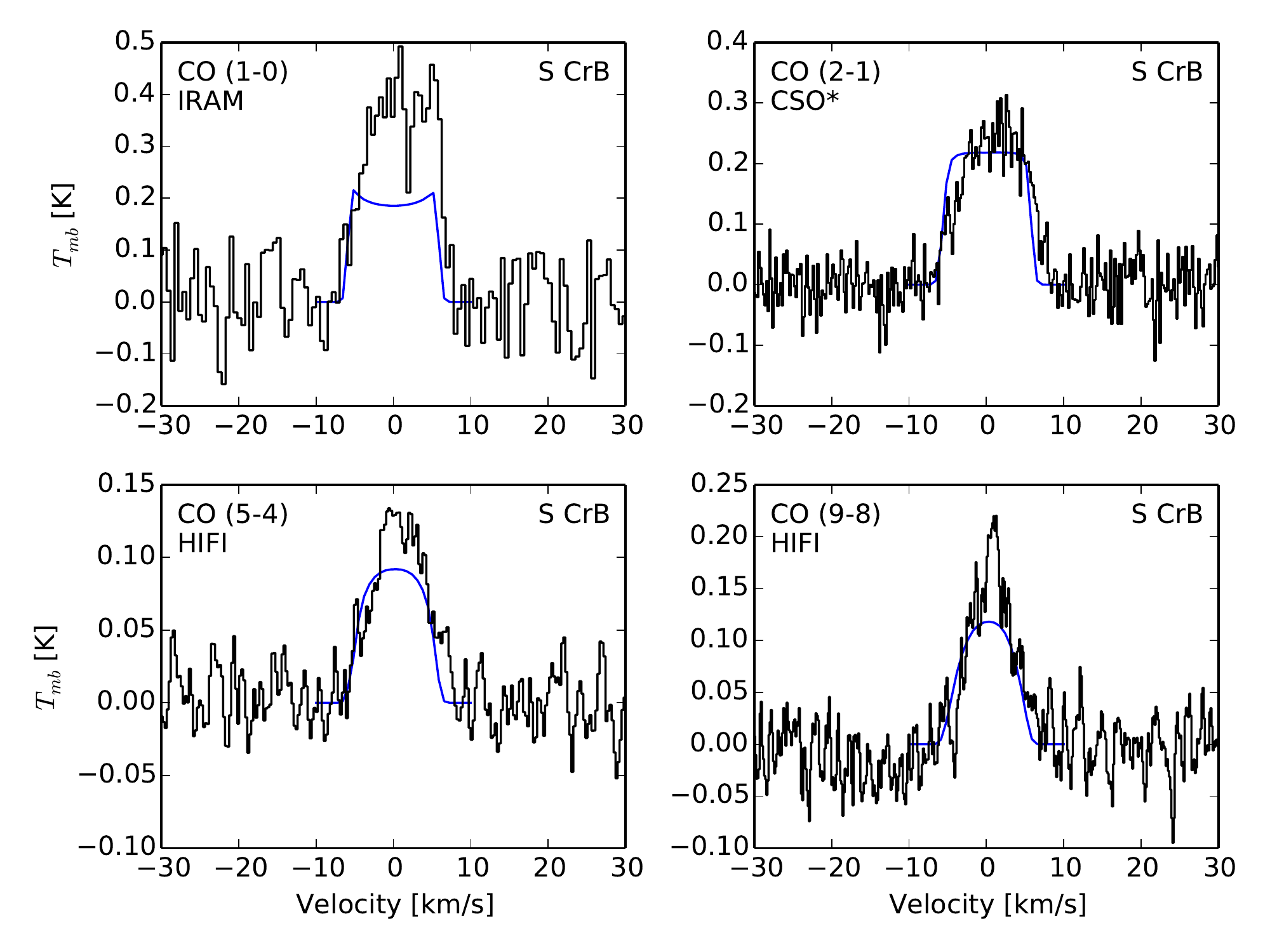}
\includegraphics[width=0.49\textwidth]{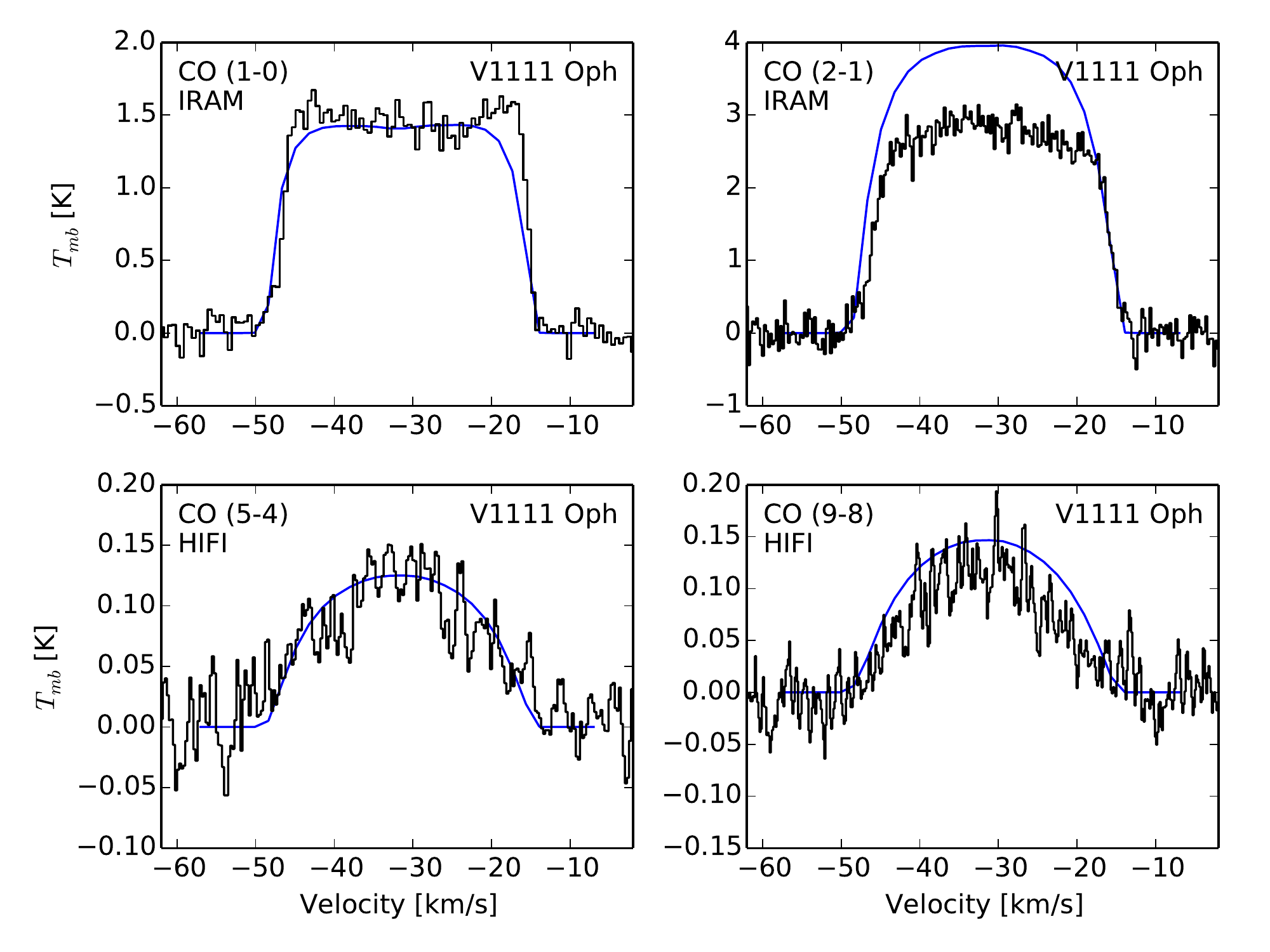}
\includegraphics[width=0.49\textwidth]{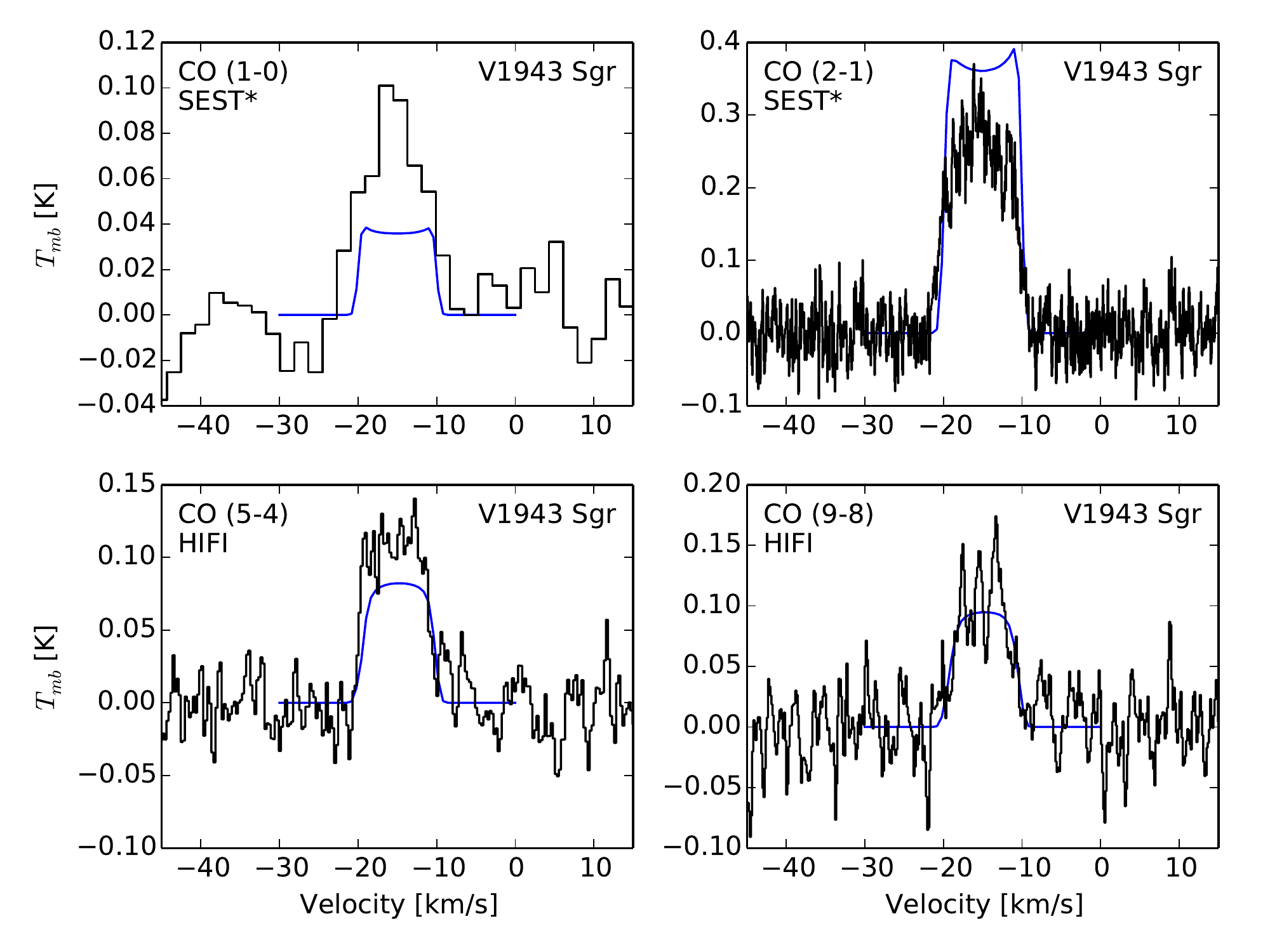}
\includegraphics[width=0.49\textwidth]{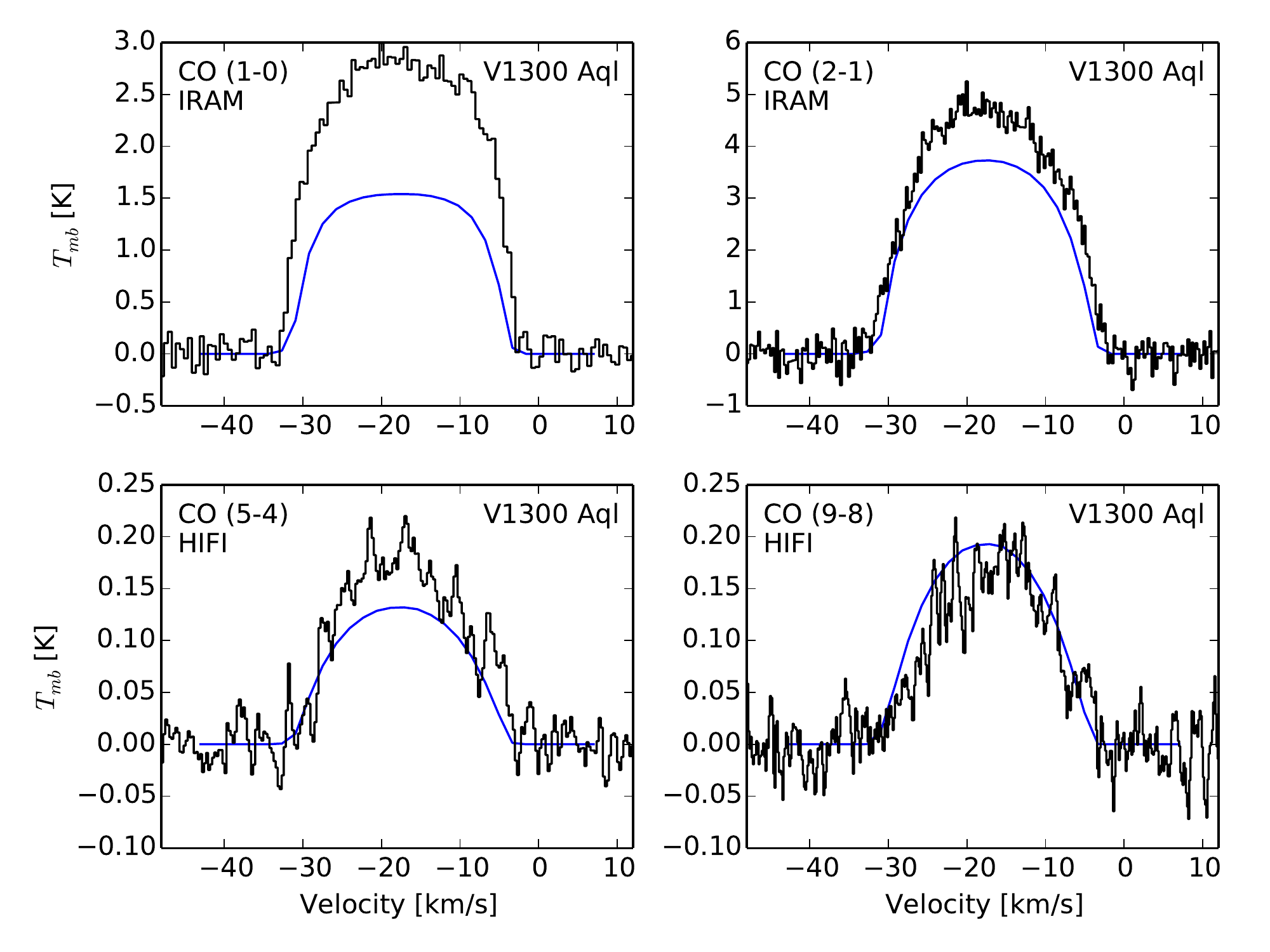}
\includegraphics[width=0.49\textwidth]{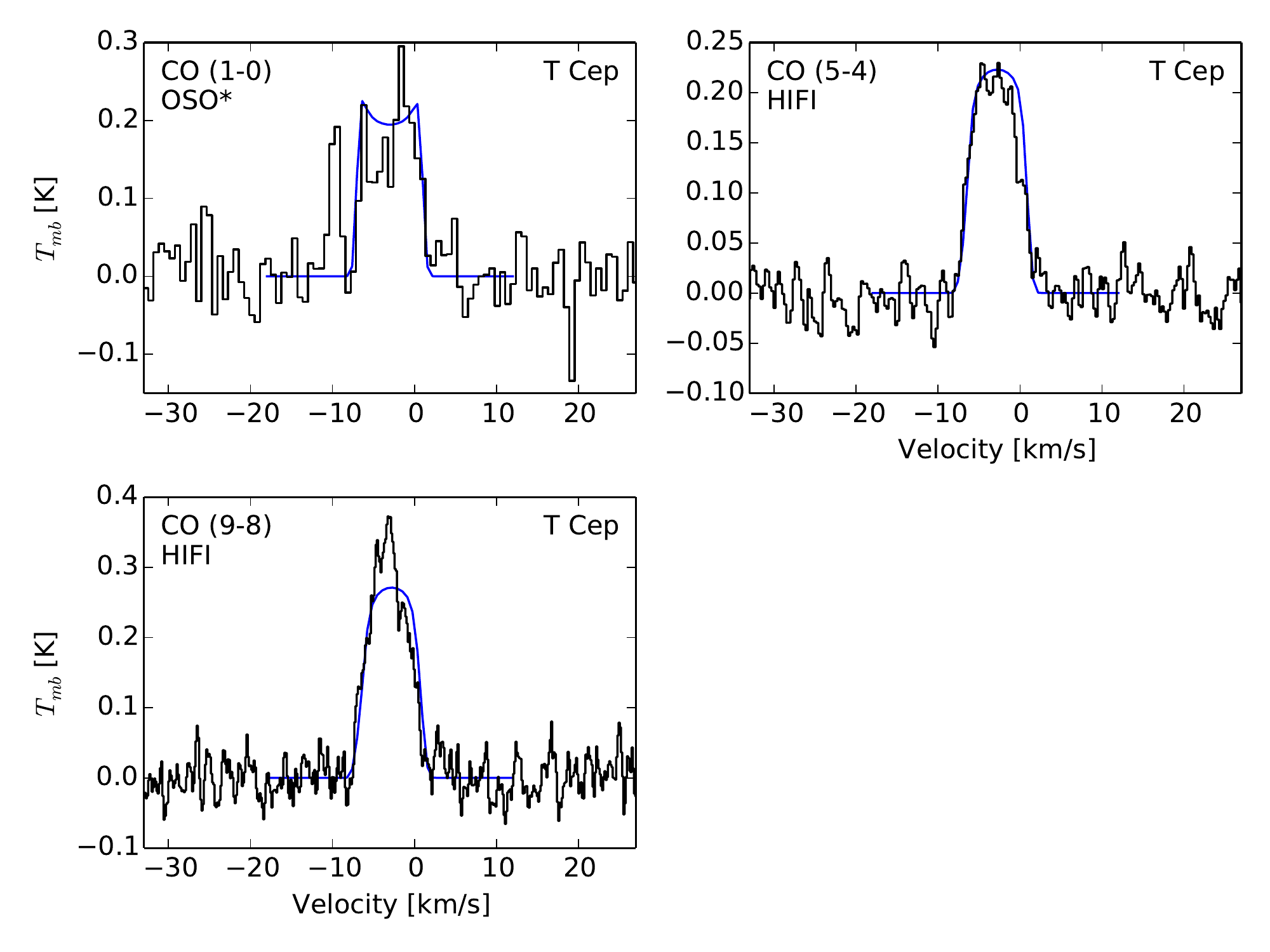}
\caption{{\bf cont.} Models (blue) and observed data (black) of M stars, plotted with respect to LSR velocity. An * next to the telescope name indicates archival data is plotted.}
\end{center}
\end{figure*}
\addtocounter{figure}{-1}
\begin{figure*}[t]
\begin{center}
\includegraphics[width=0.49\textwidth]{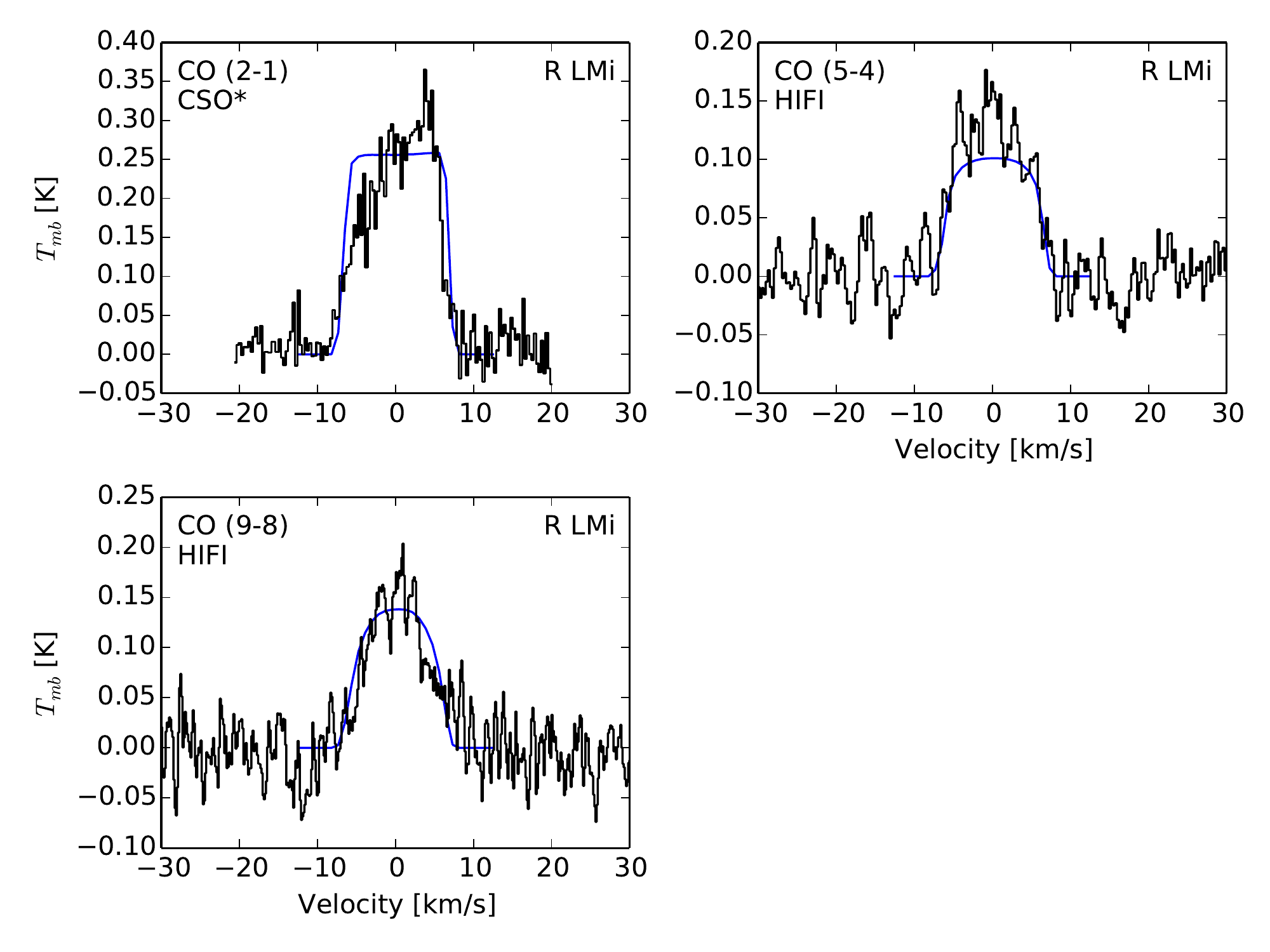}
\includegraphics[width=0.49\textwidth]{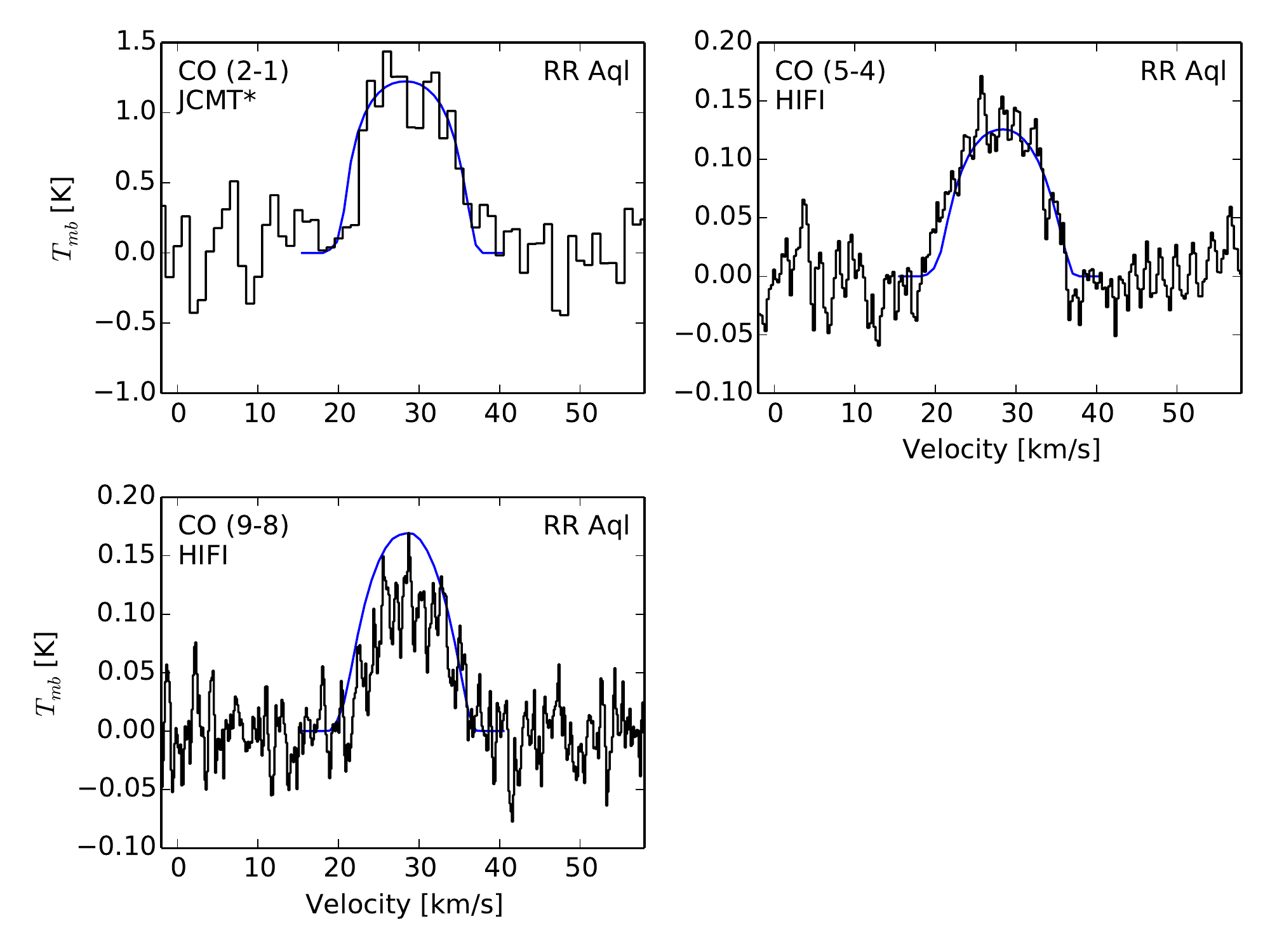}
\caption{{\bf cont.} Models (blue) and observed data (black) of M stars, plotted with respect to LSR velocity. An * next to the telescope name indicates archival data is plotted.}
\end{center}
\end{figure*}

\begin{figure*}[t]
\begin{center}
\includegraphics[width=0.19\textwidth]{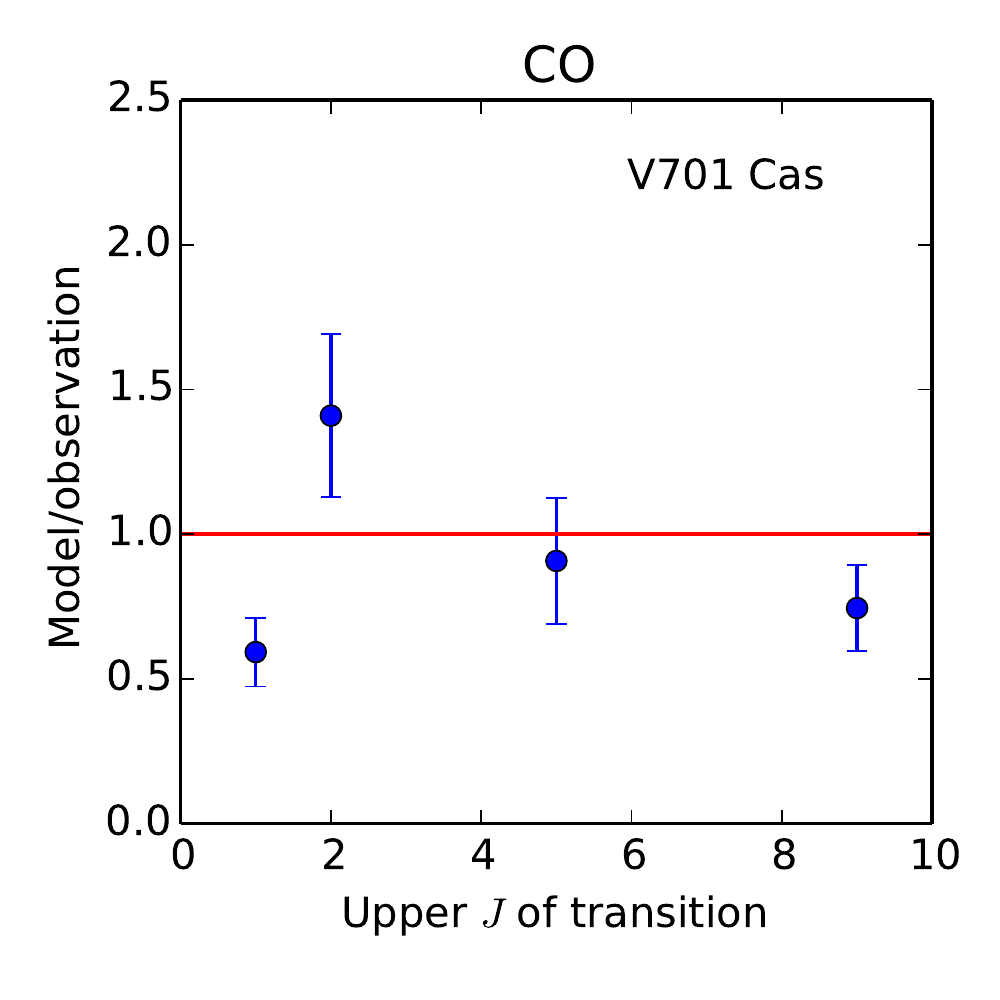}
\includegraphics[width=0.19\textwidth]{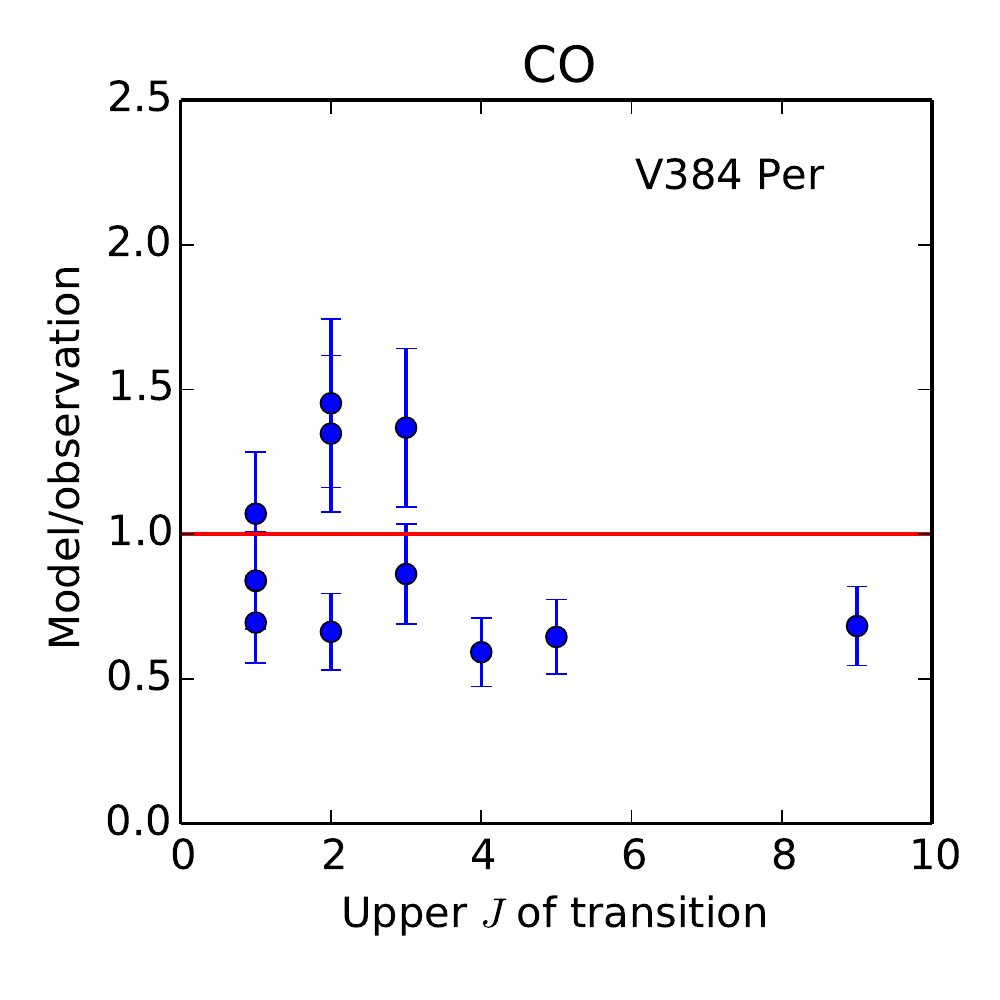}
\includegraphics[width=0.19\textwidth]{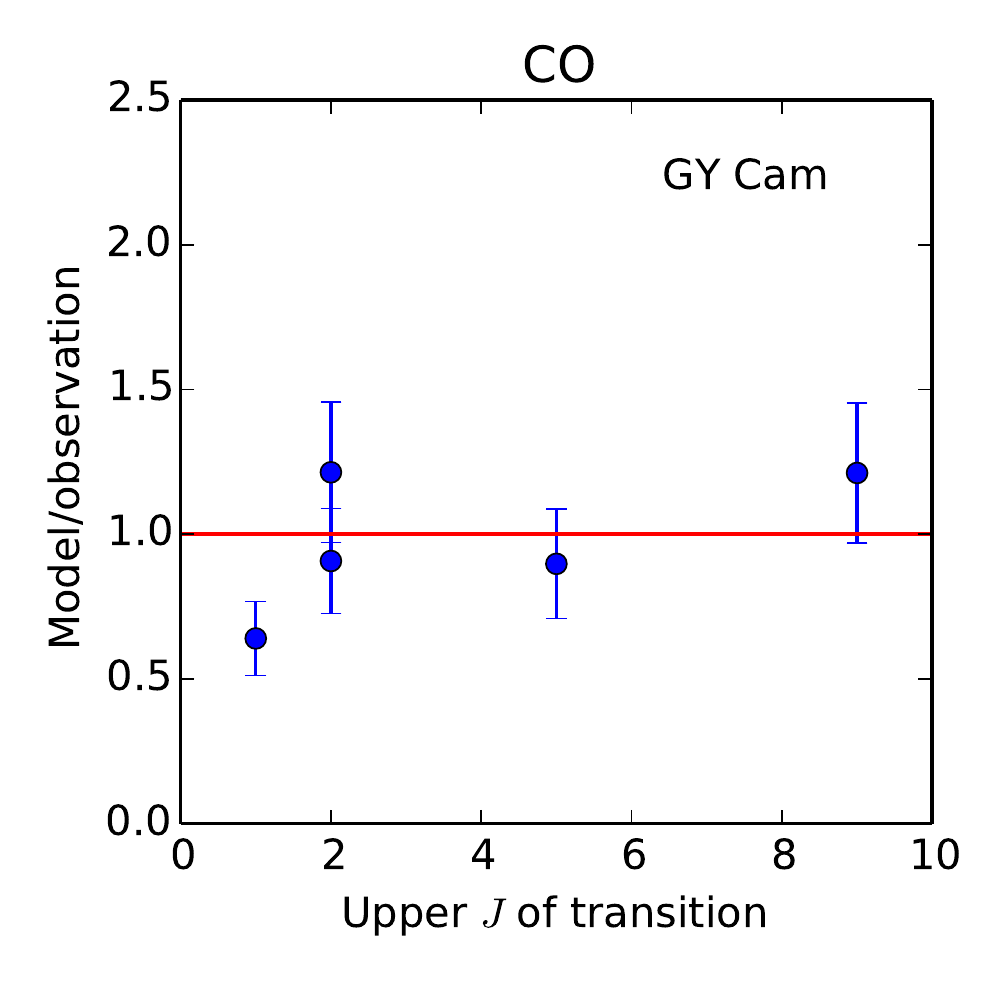}
\includegraphics[width=0.19\textwidth]{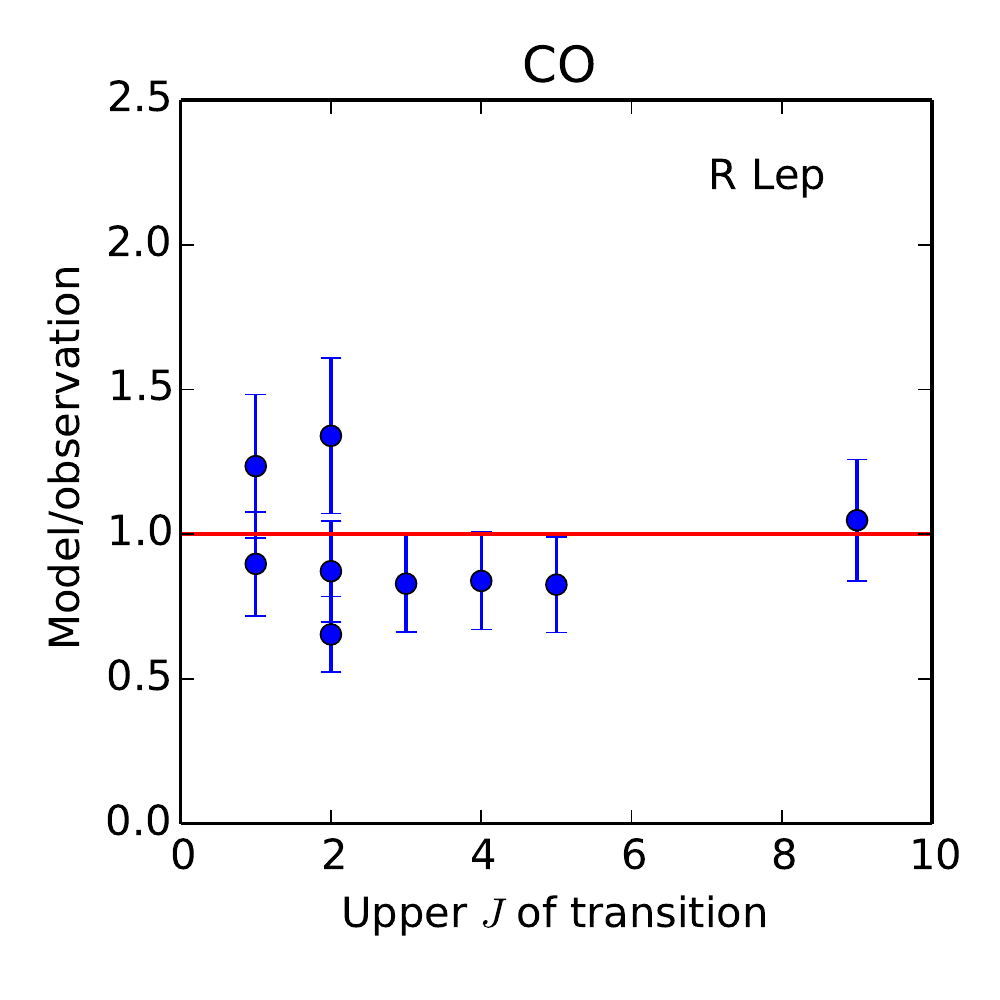}
\includegraphics[width=0.19\textwidth]{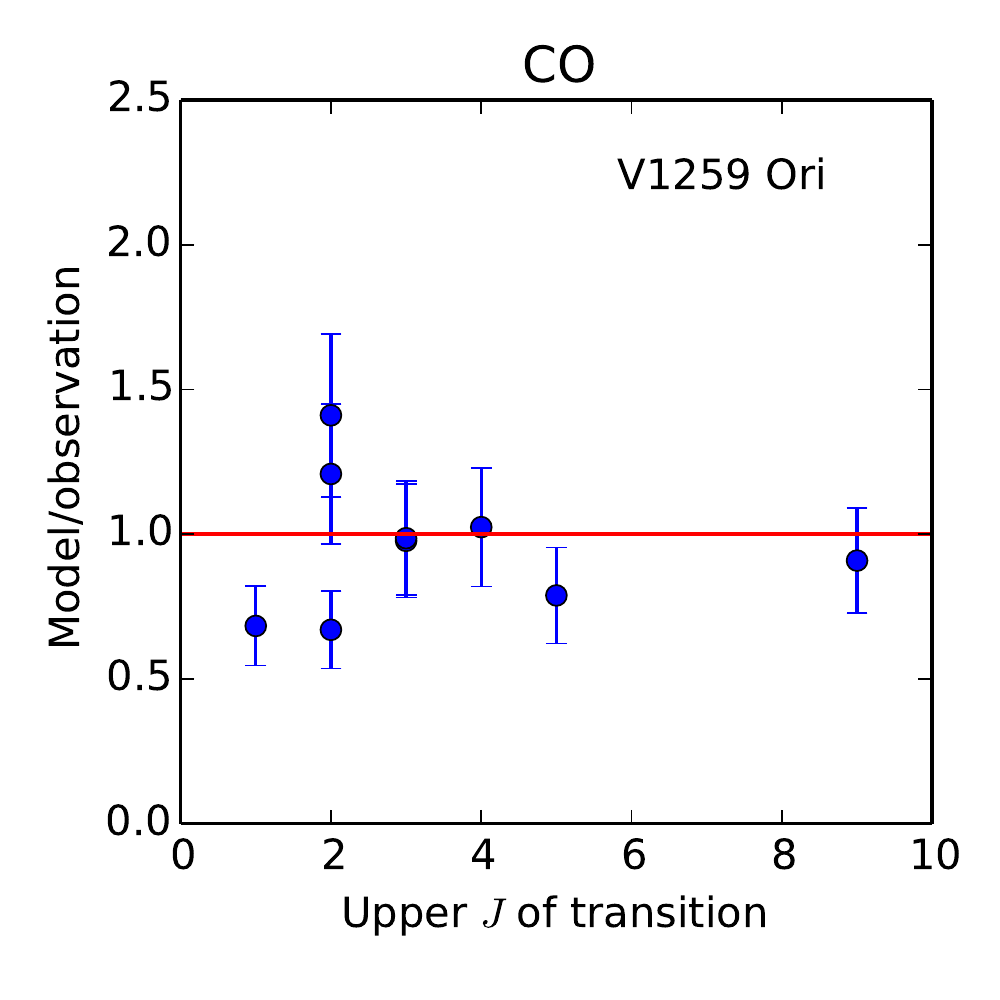}
\includegraphics[width=0.19\textwidth]{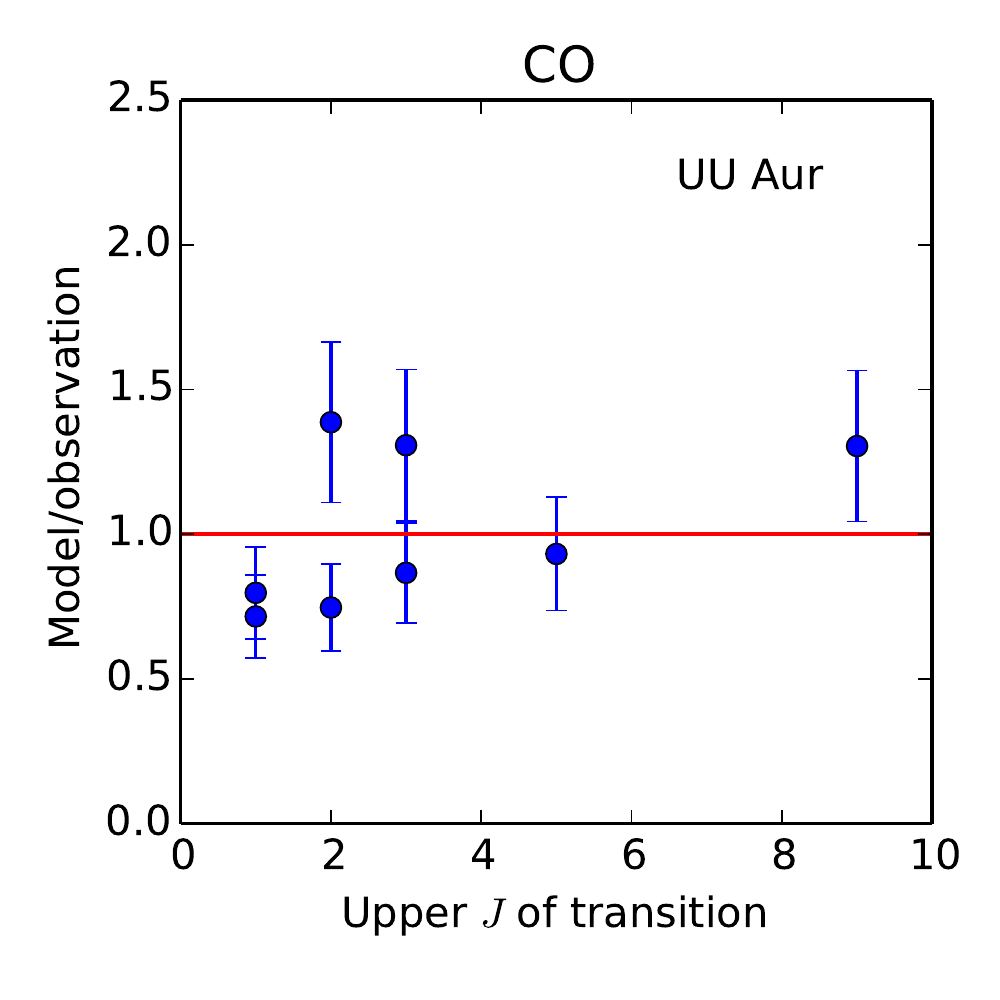}
\includegraphics[width=0.19\textwidth]{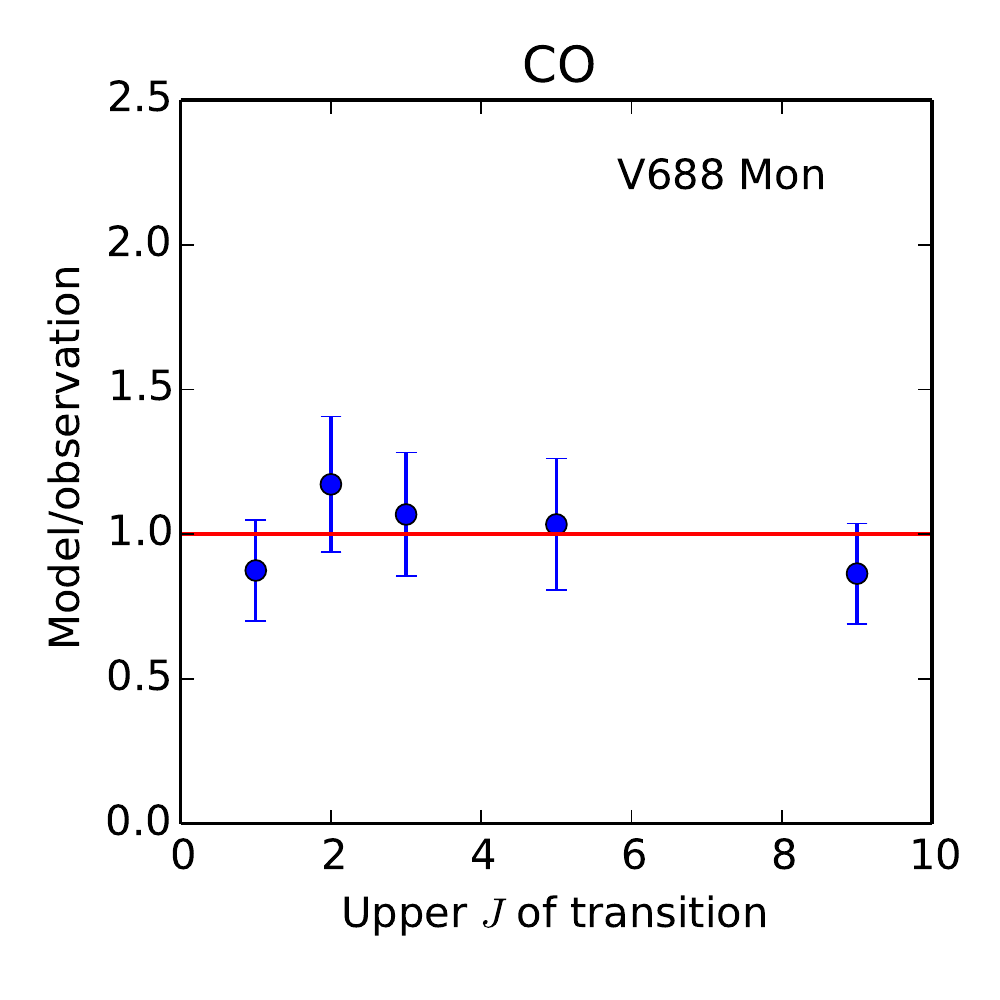}
\includegraphics[width=0.19\textwidth]{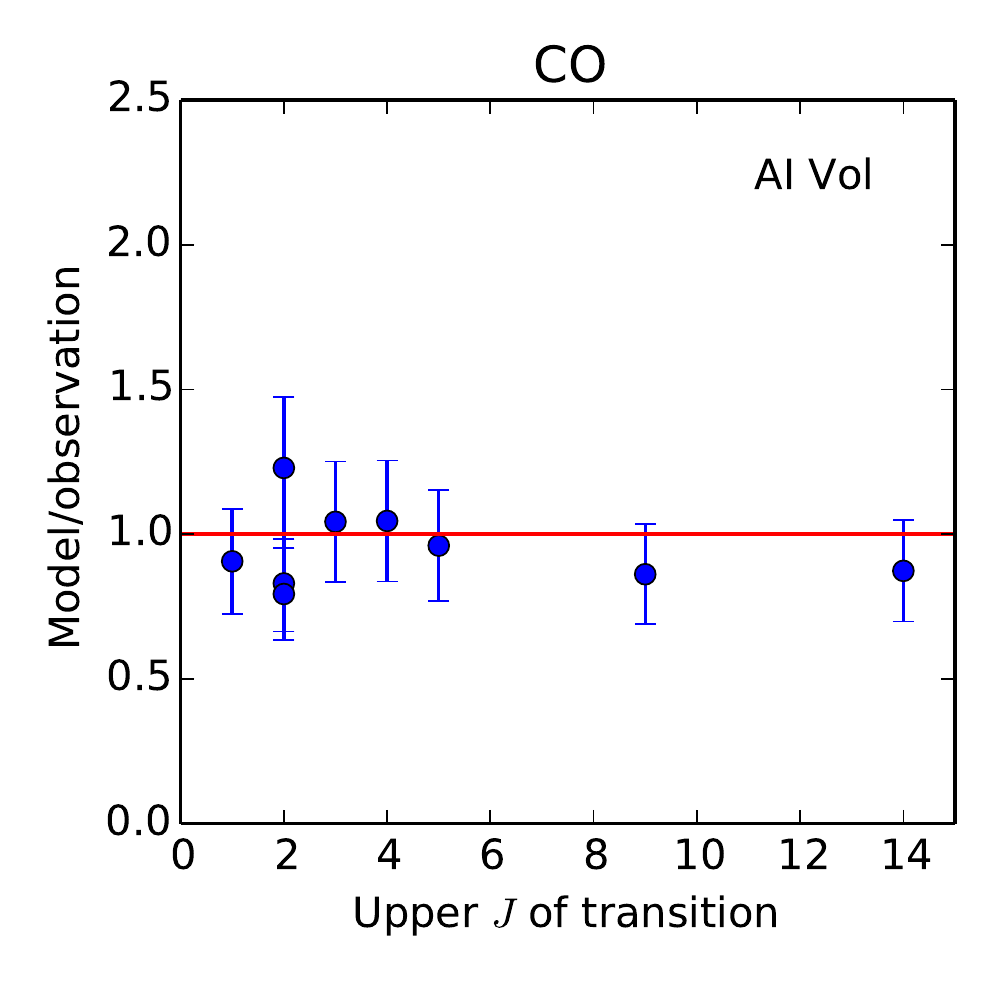}
\includegraphics[width=0.19\textwidth]{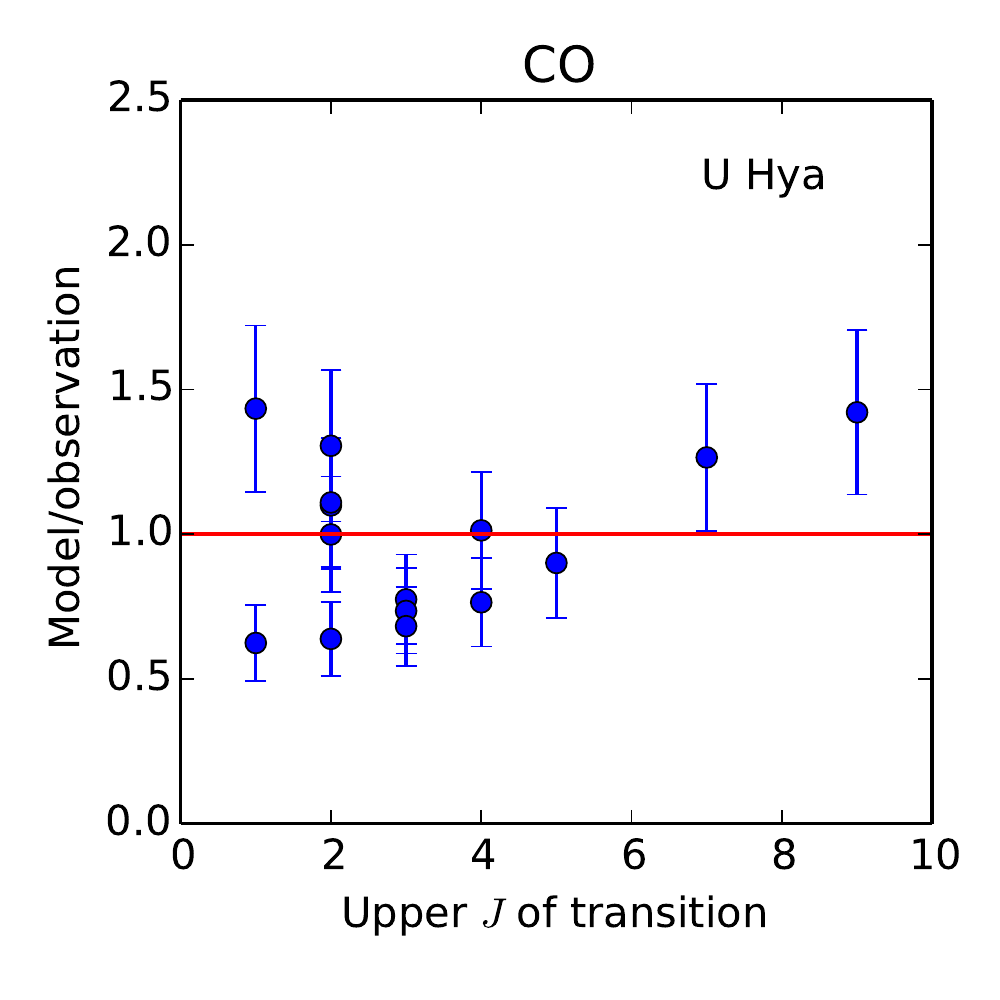}
\includegraphics[width=0.19\textwidth]{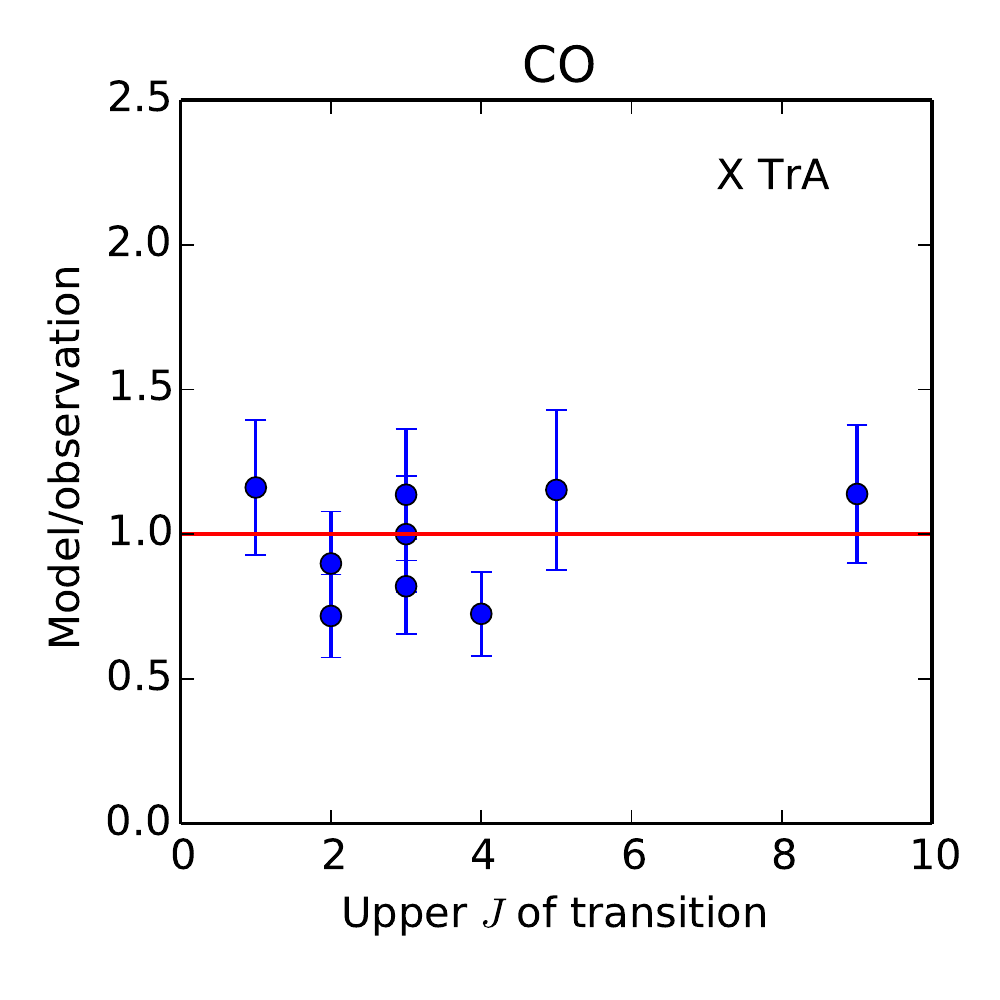}
\includegraphics[width=0.19\textwidth]{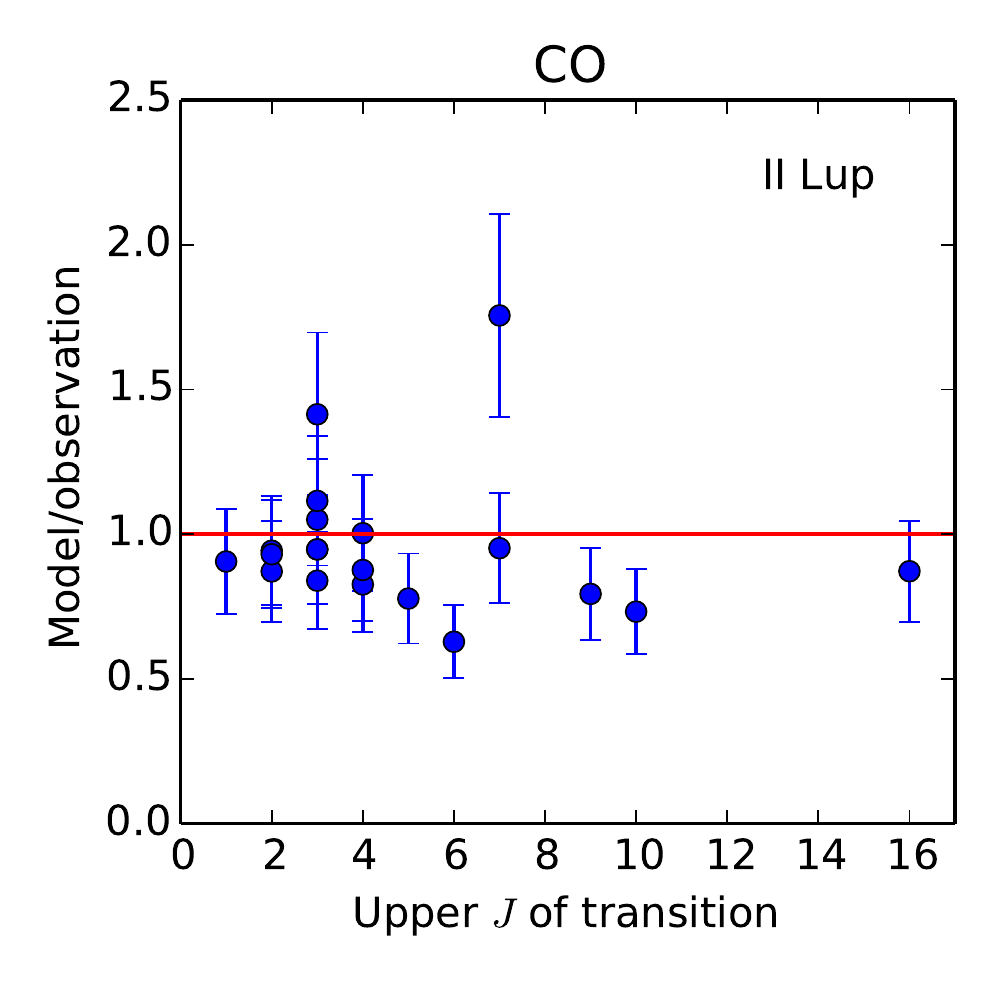}
\includegraphics[width=0.19\textwidth]{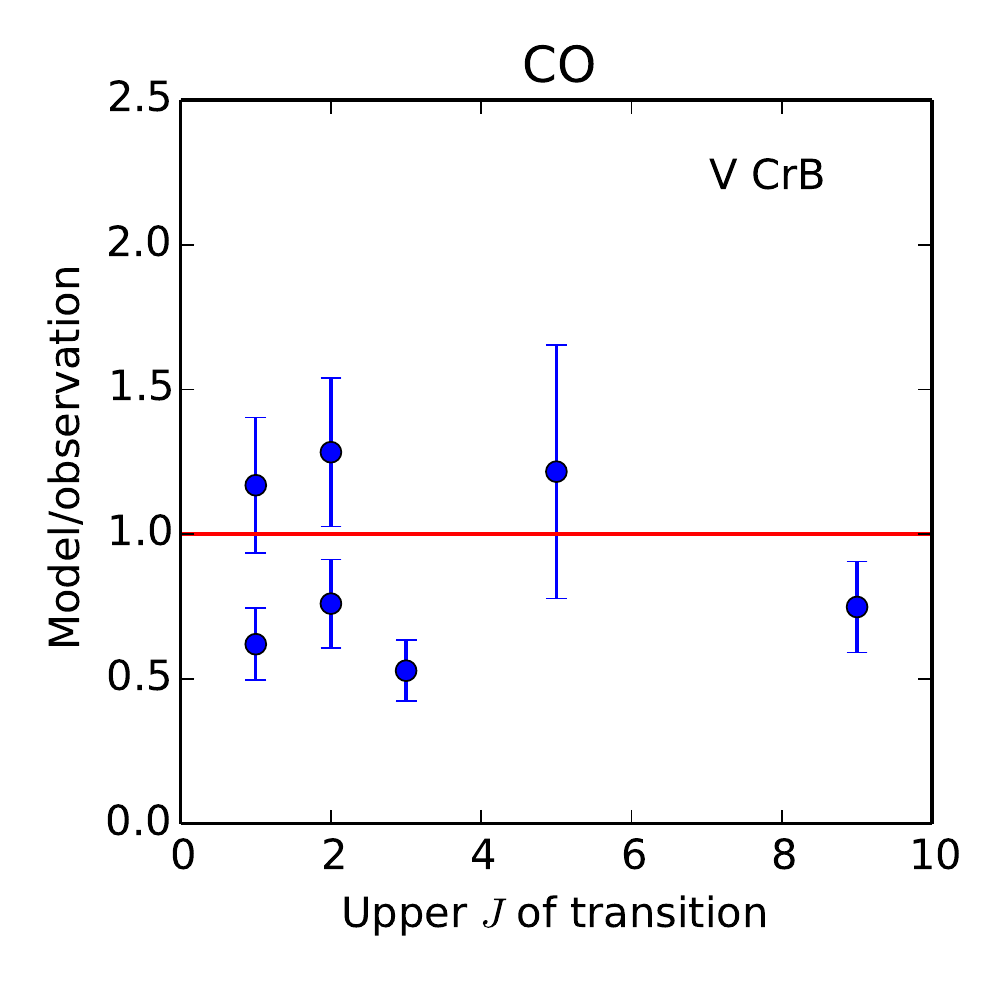}
\includegraphics[width=0.19\textwidth]{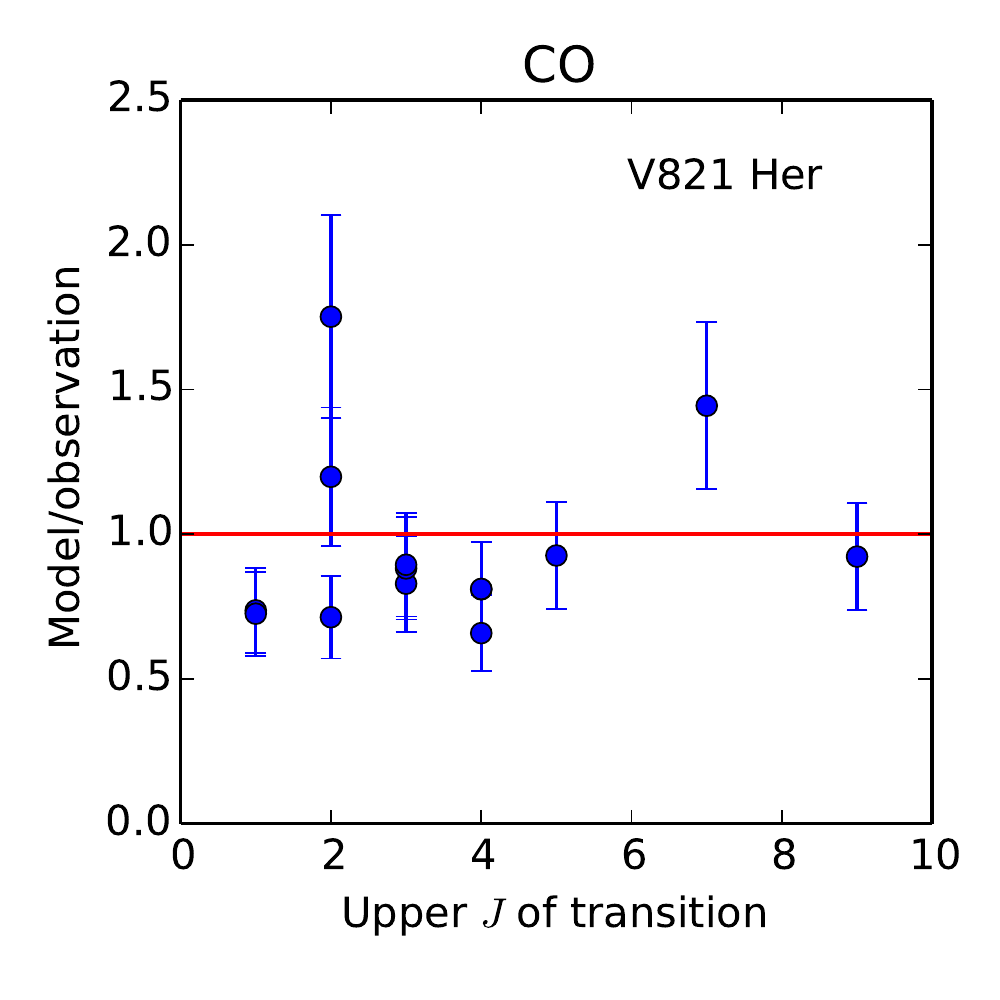}
\includegraphics[width=0.19\textwidth]{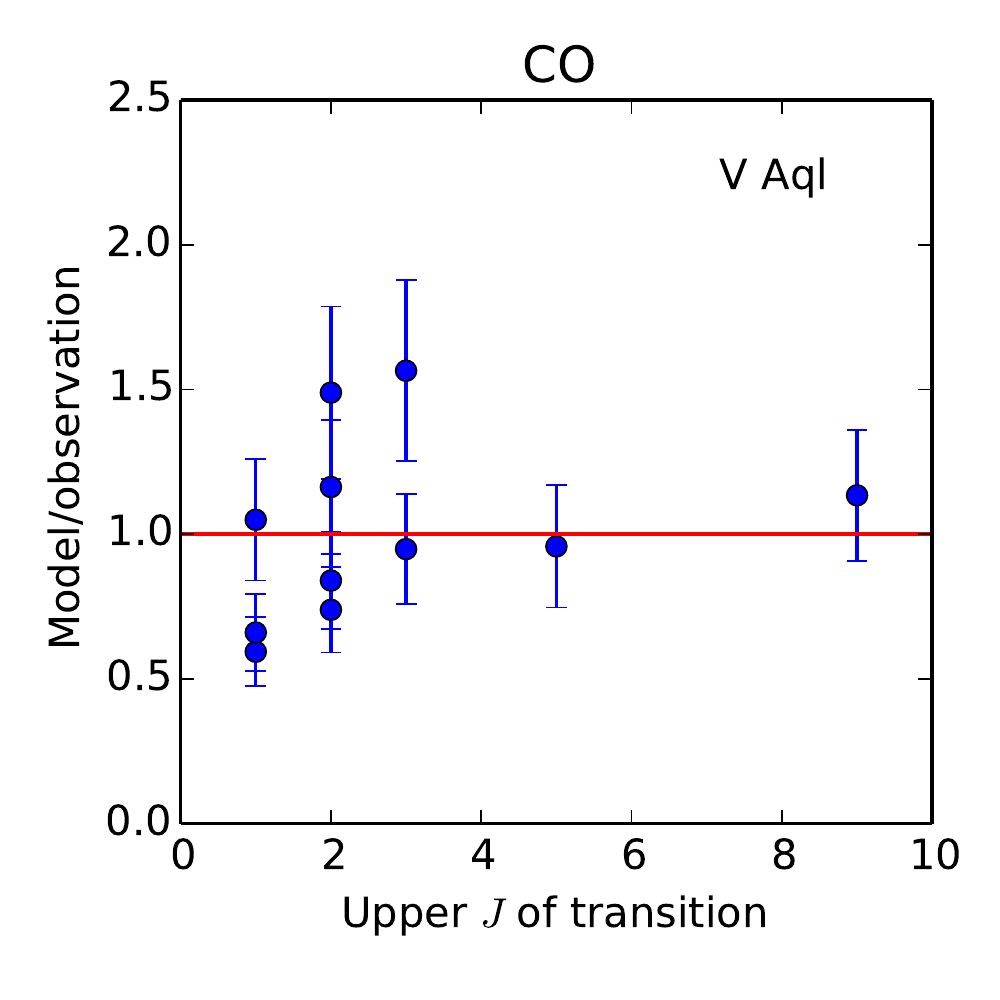}
\includegraphics[width=0.19\textwidth]{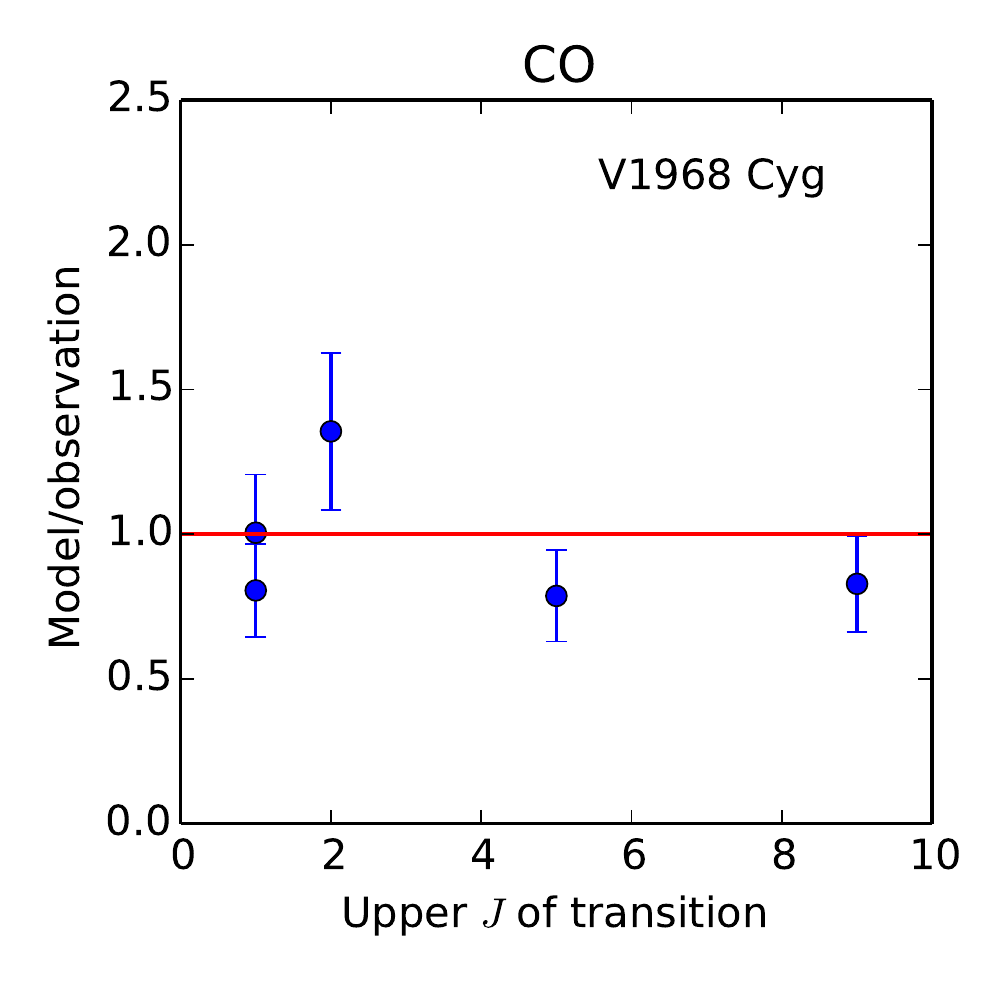}
\includegraphics[width=0.19\textwidth]{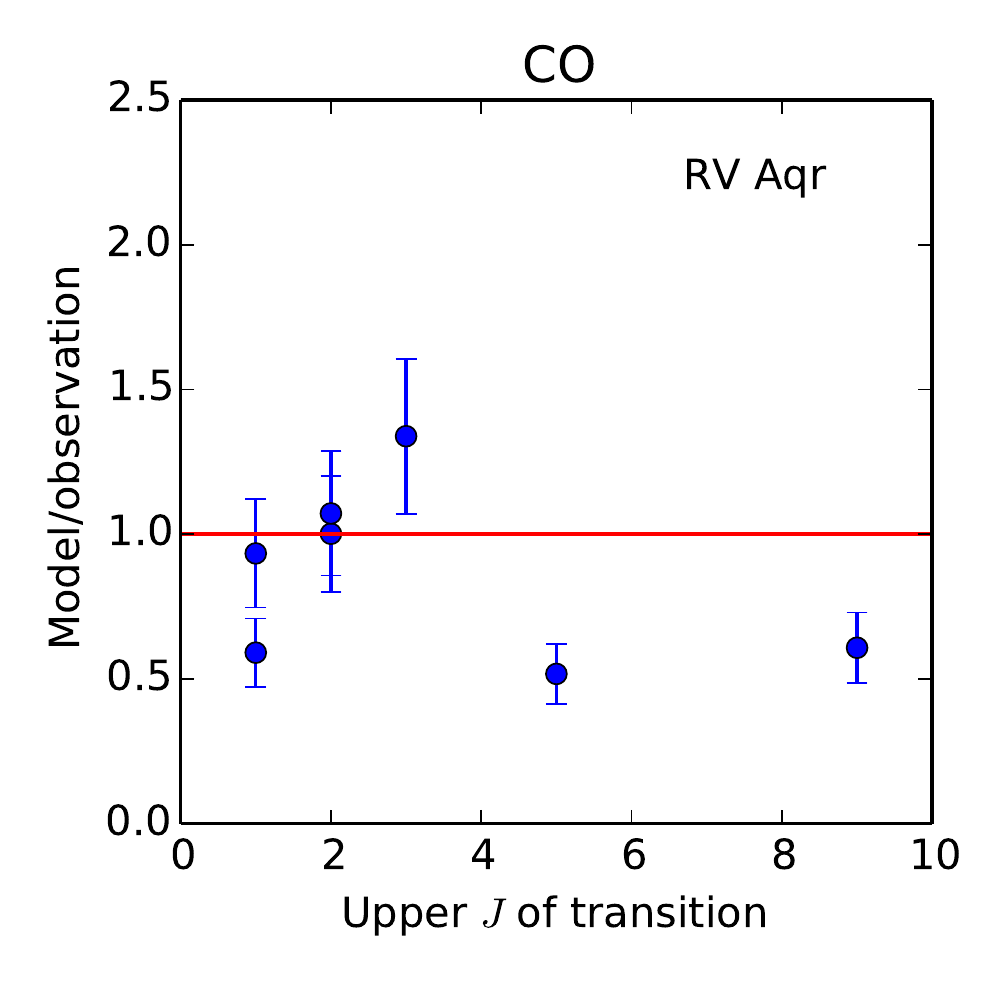}
\caption{Goodness of fit as defined by model/observed intensity for C stars.}
\label{Cfits}
\end{center}
\end{figure*}

\begin{figure*}[t]
\begin{center}
\includegraphics[width=0.19\textwidth]{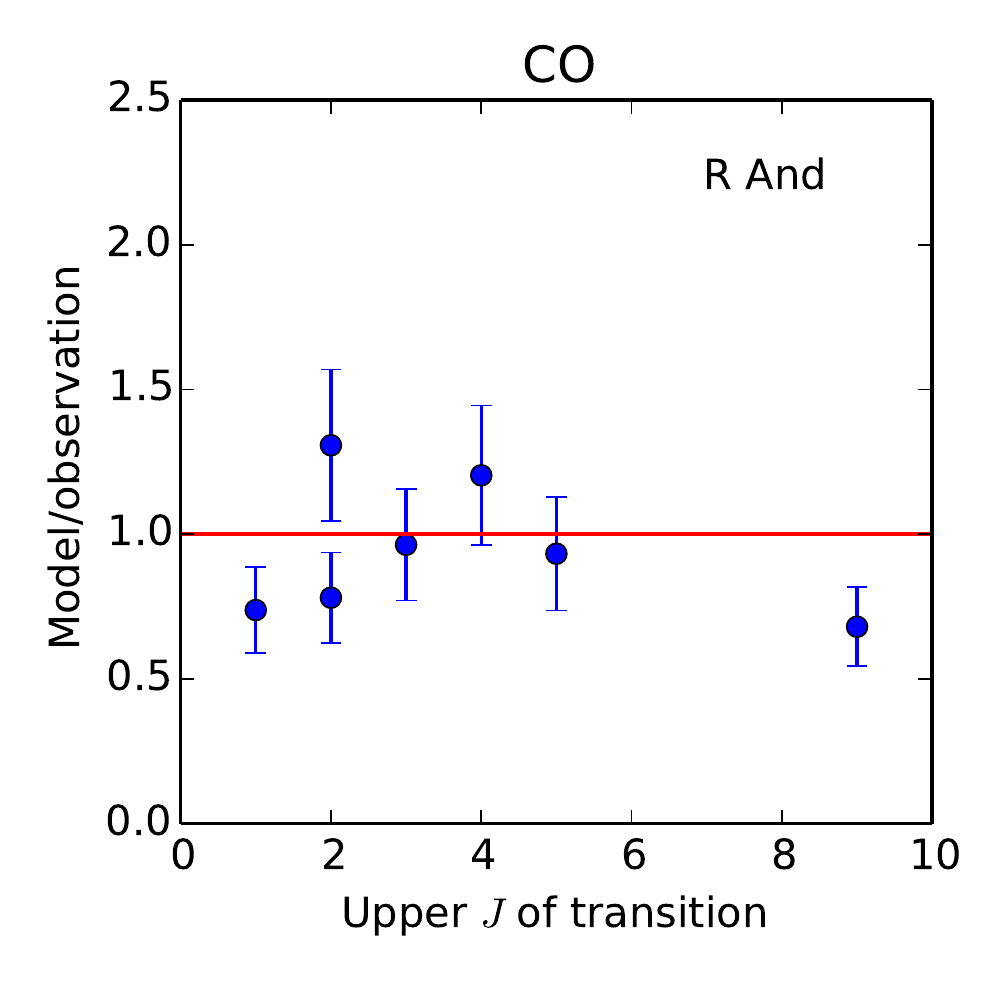}
\includegraphics[width=0.19\textwidth]{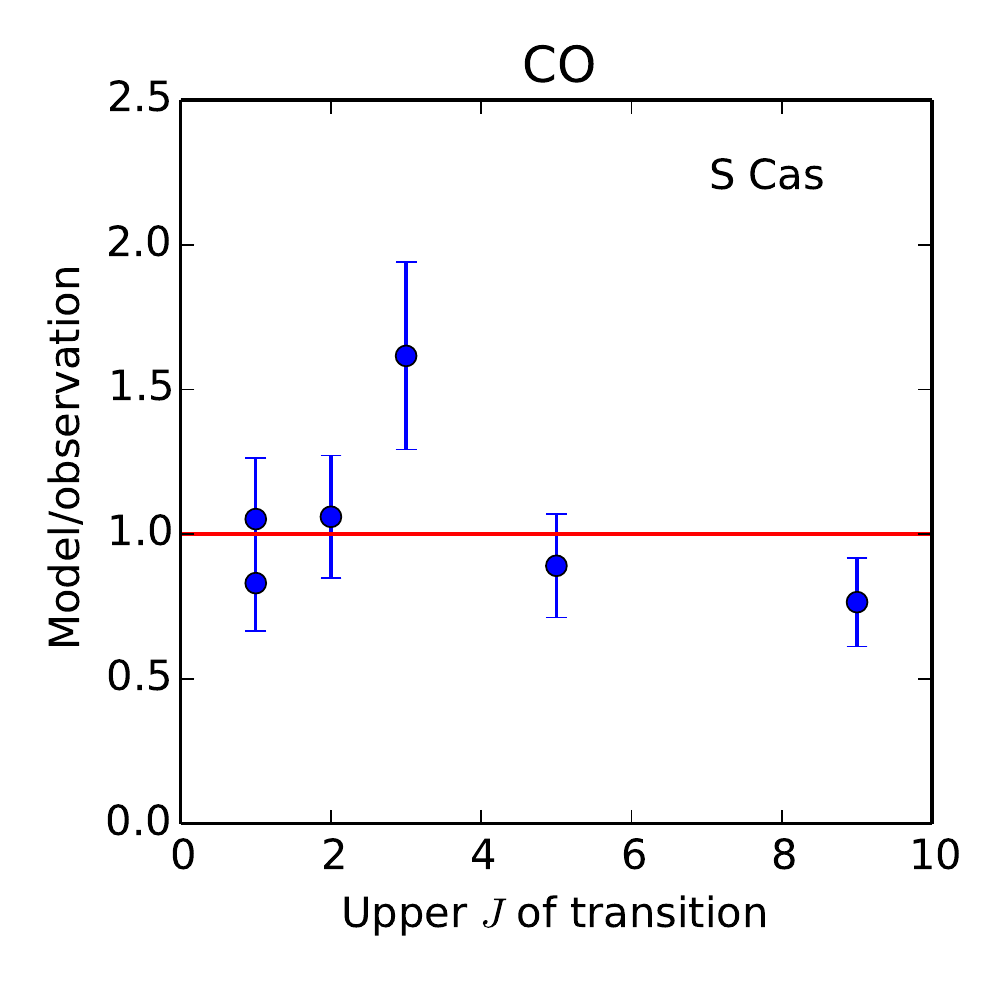}
\includegraphics[width=0.19\textwidth]{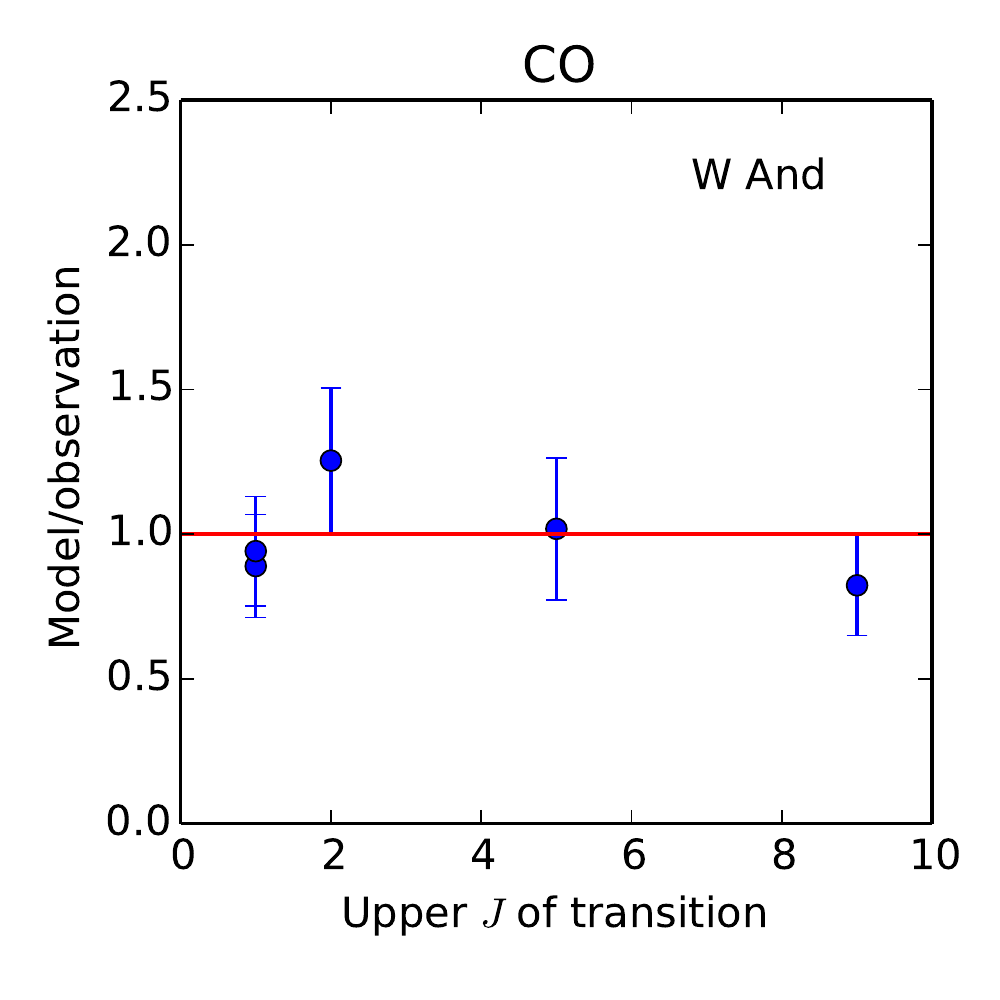}
\includegraphics[width=0.19\textwidth]{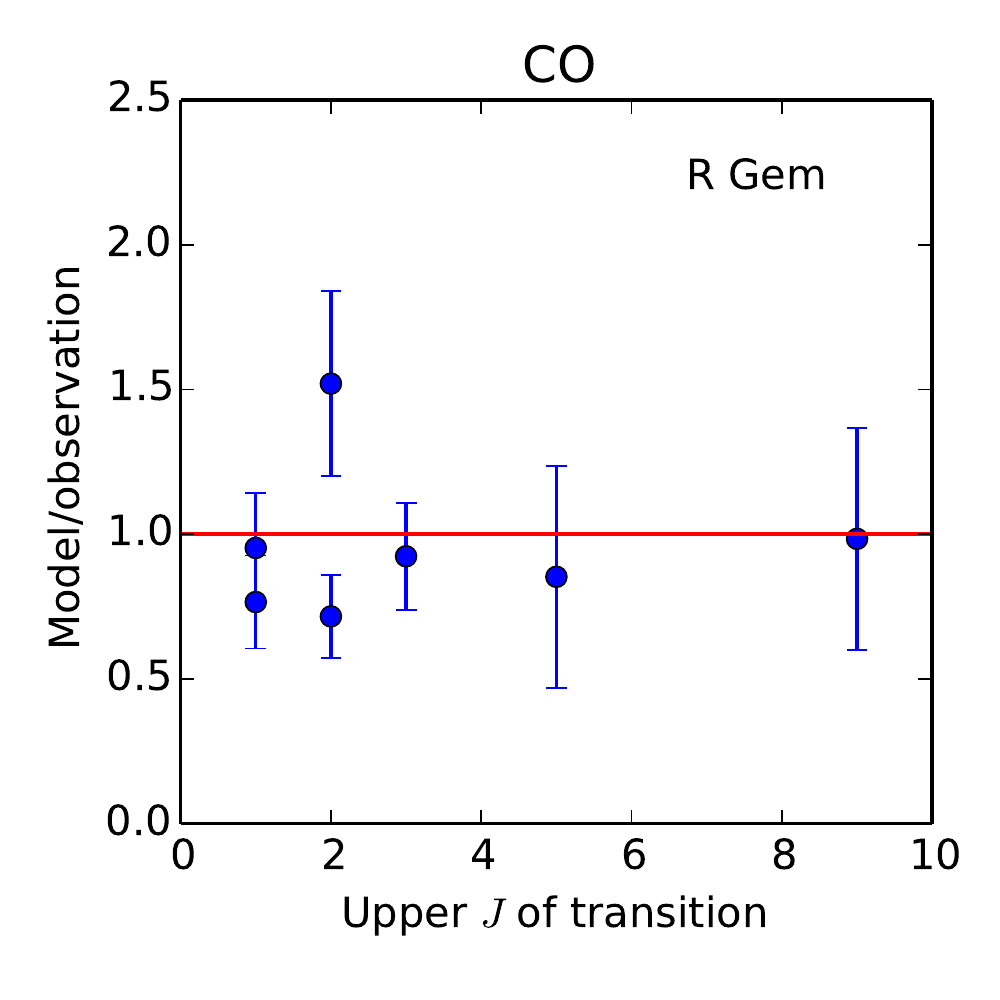}
\includegraphics[width=0.19\textwidth]{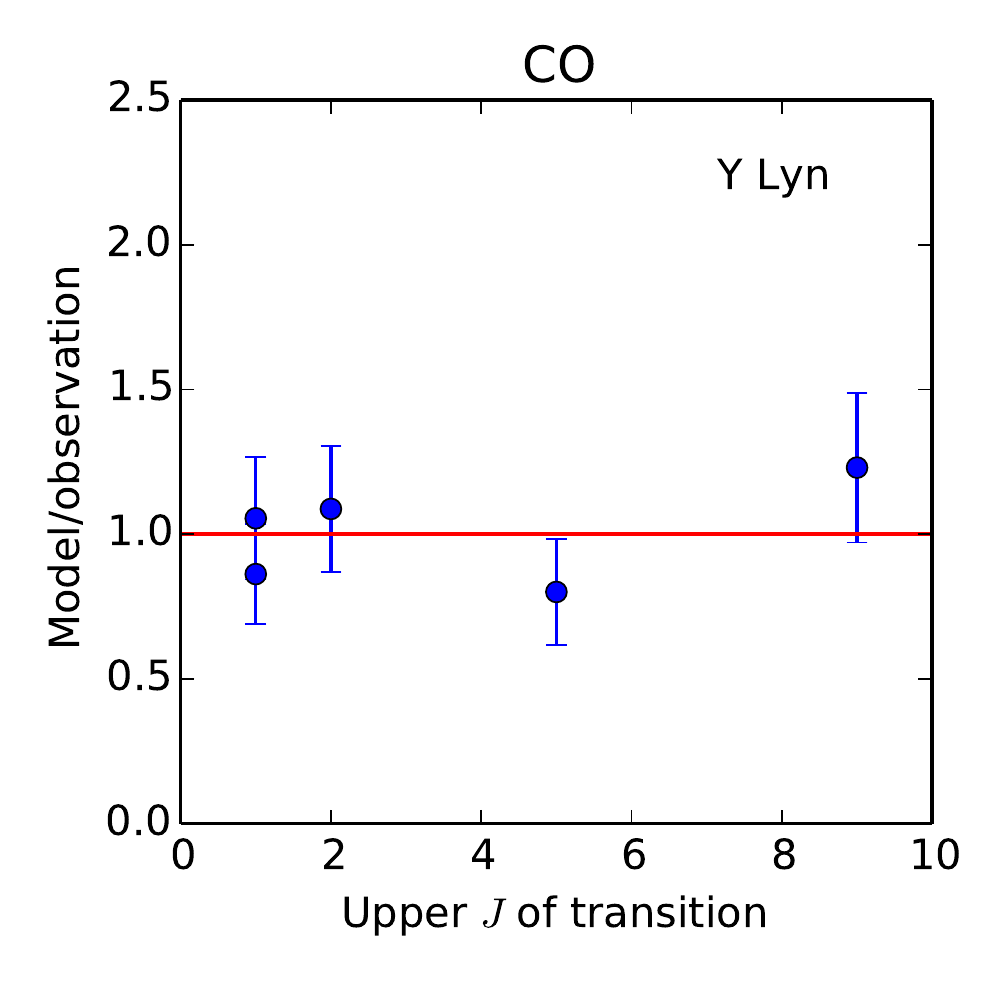}
\includegraphics[width=0.19\textwidth]{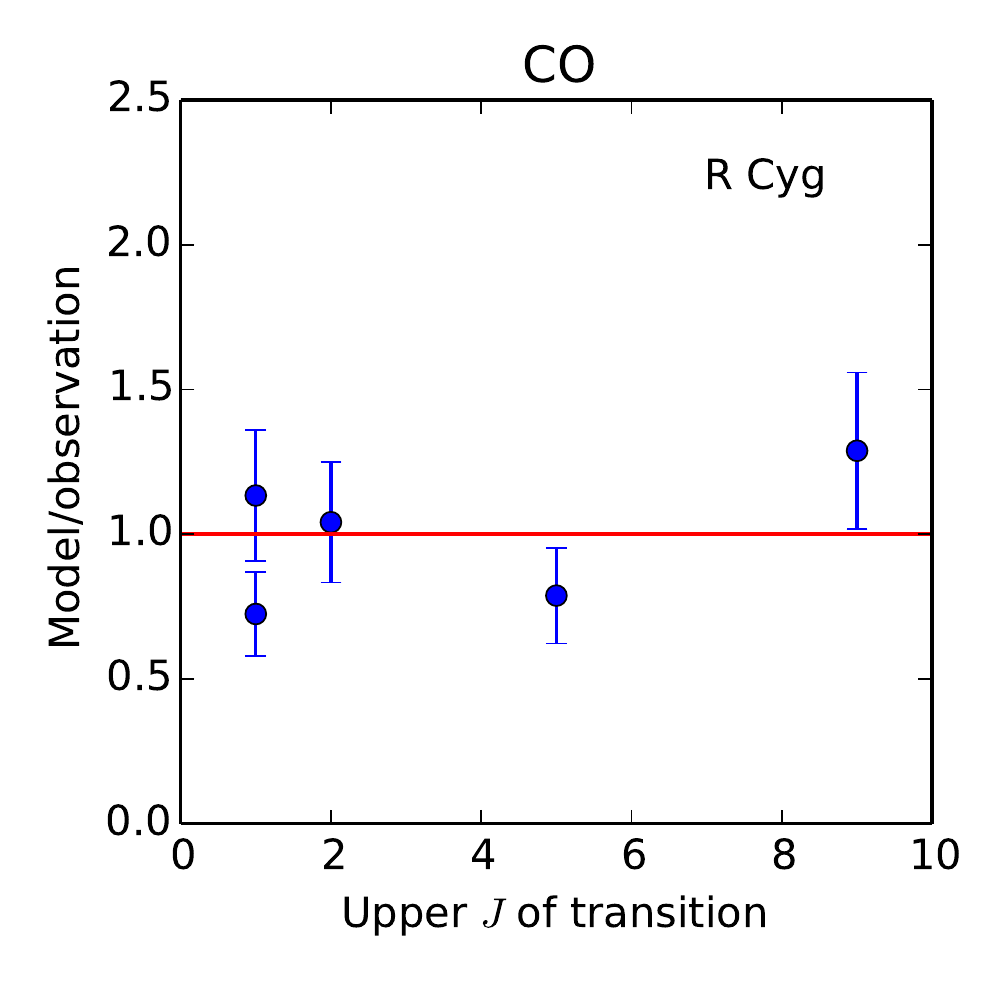}
\caption{Goodness of fit as defined by model/observed intensity for S stars.}
\label{Sfits}
\end{center}
\end{figure*}

\begin{figure*}[t]
\begin{center}
\includegraphics[width=0.19\textwidth]{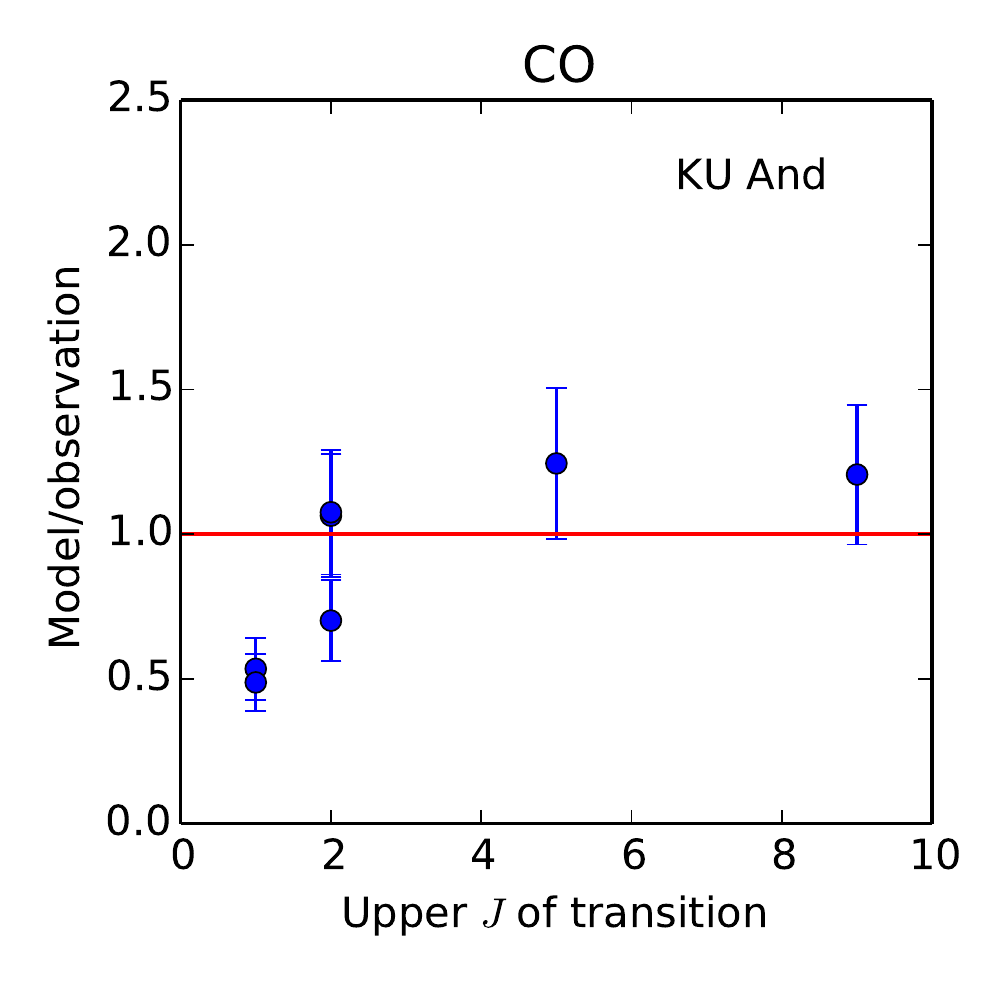}
\includegraphics[width=0.19\textwidth]{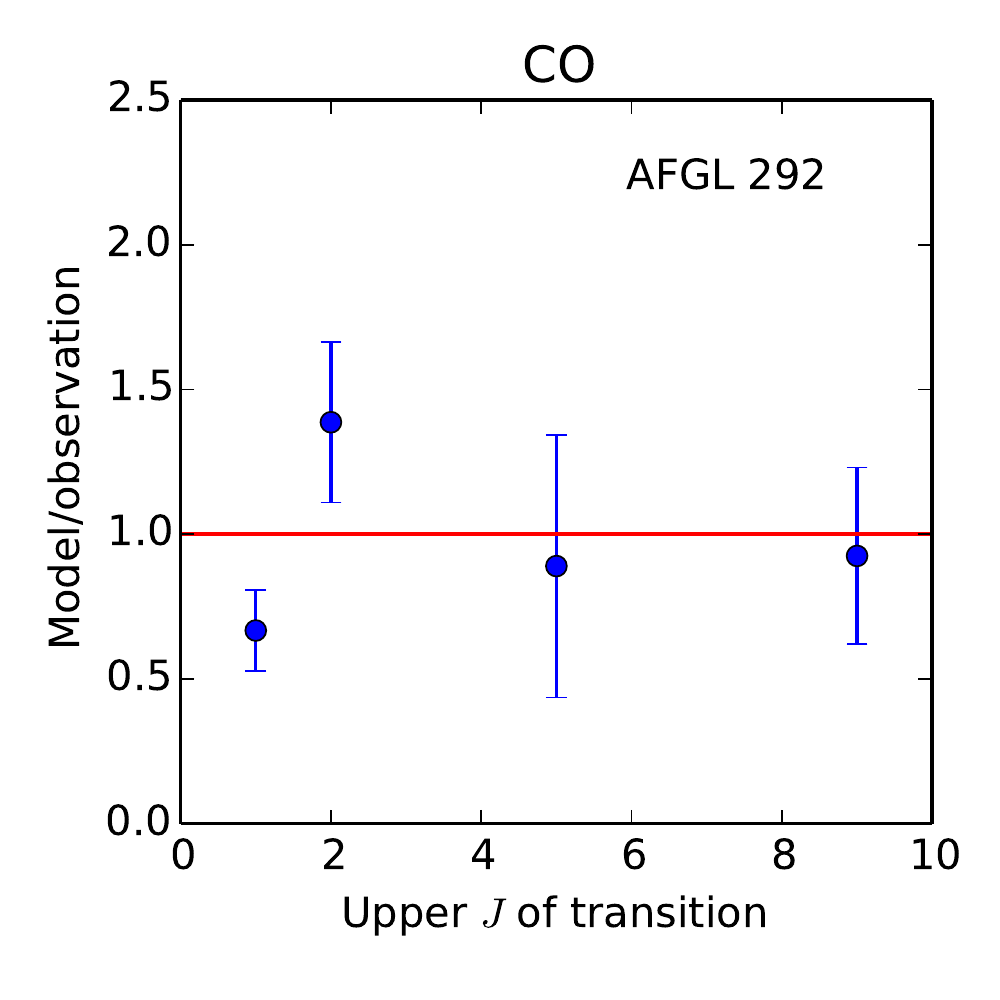}
\includegraphics[width=0.19\textwidth]{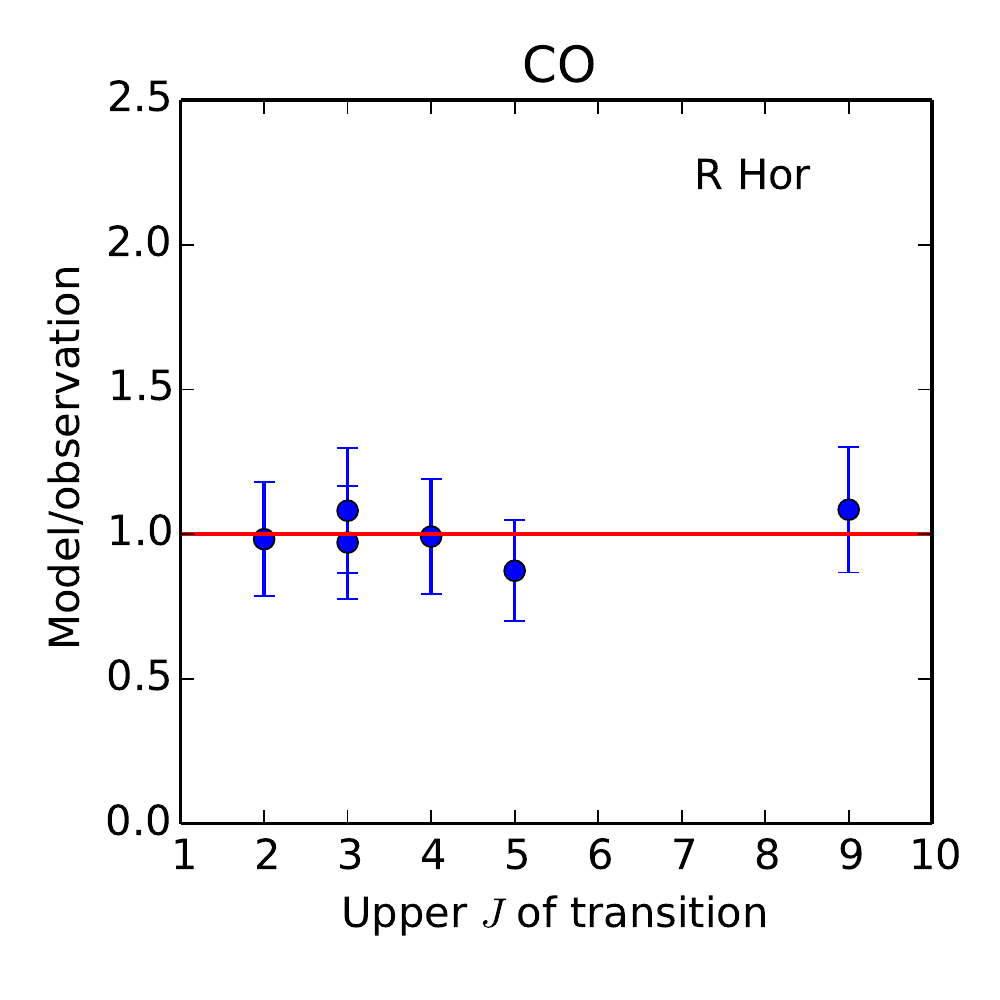}
\includegraphics[width=0.19\textwidth]{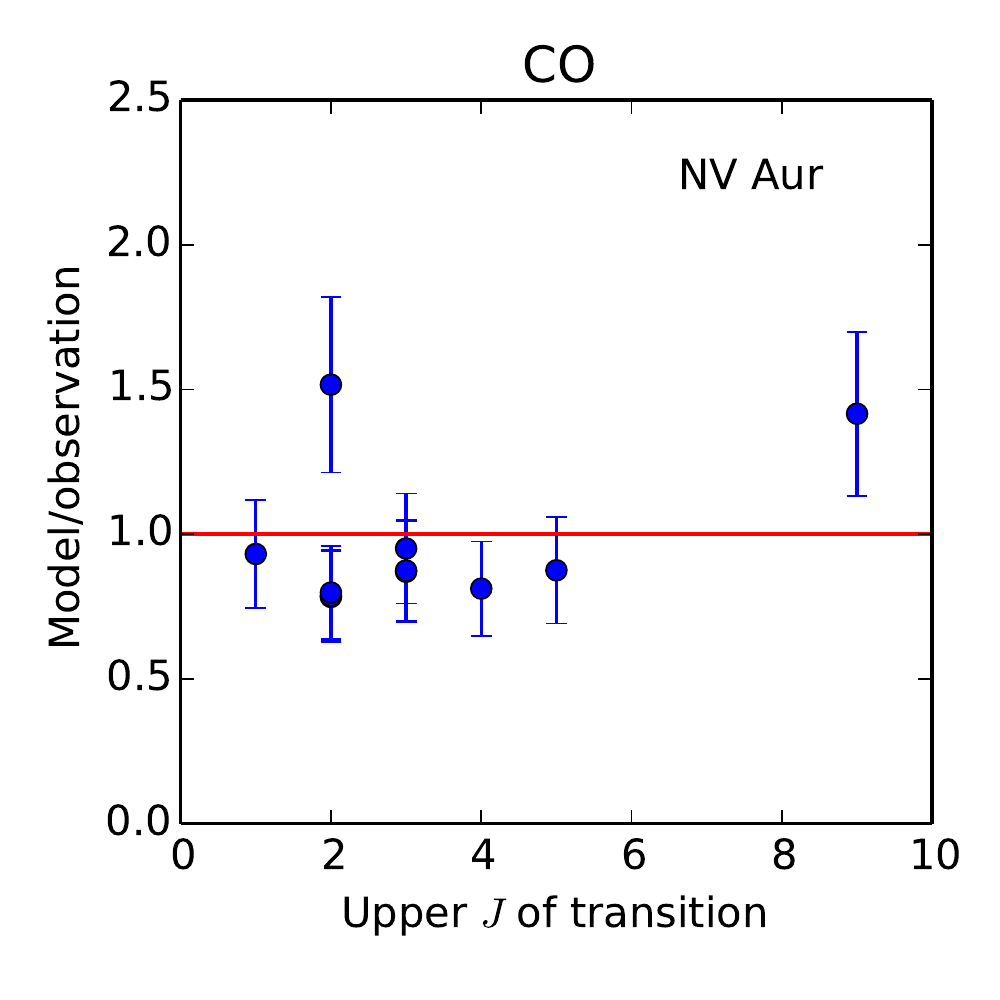}
\includegraphics[width=0.19\textwidth]{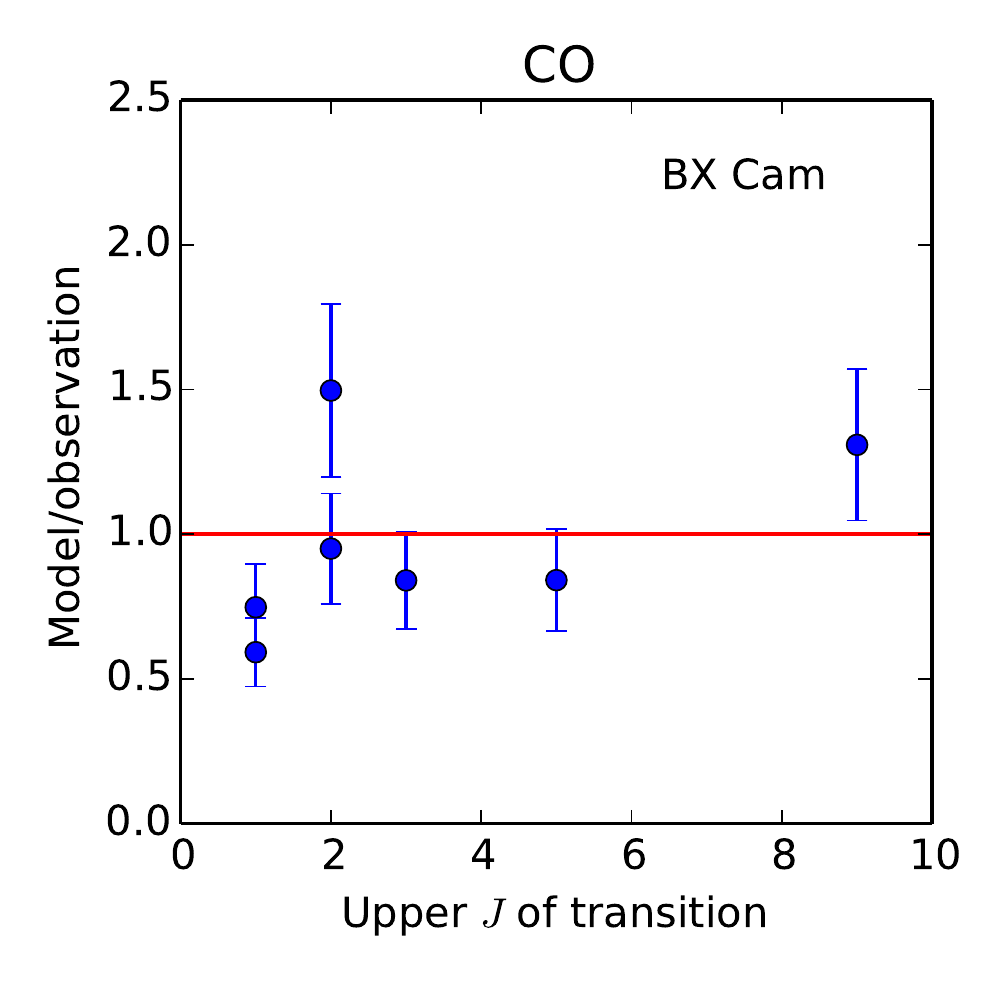}
\includegraphics[width=0.19\textwidth]{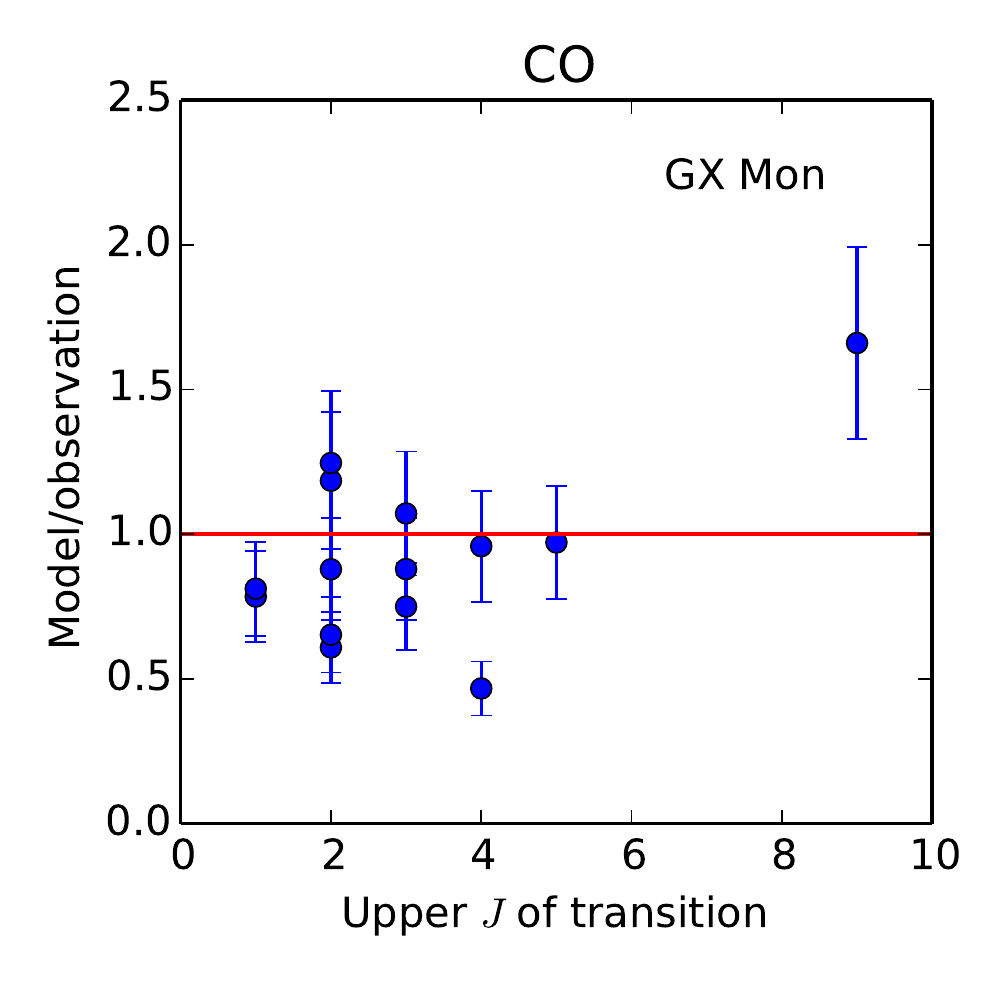}
\includegraphics[width=0.19\textwidth]{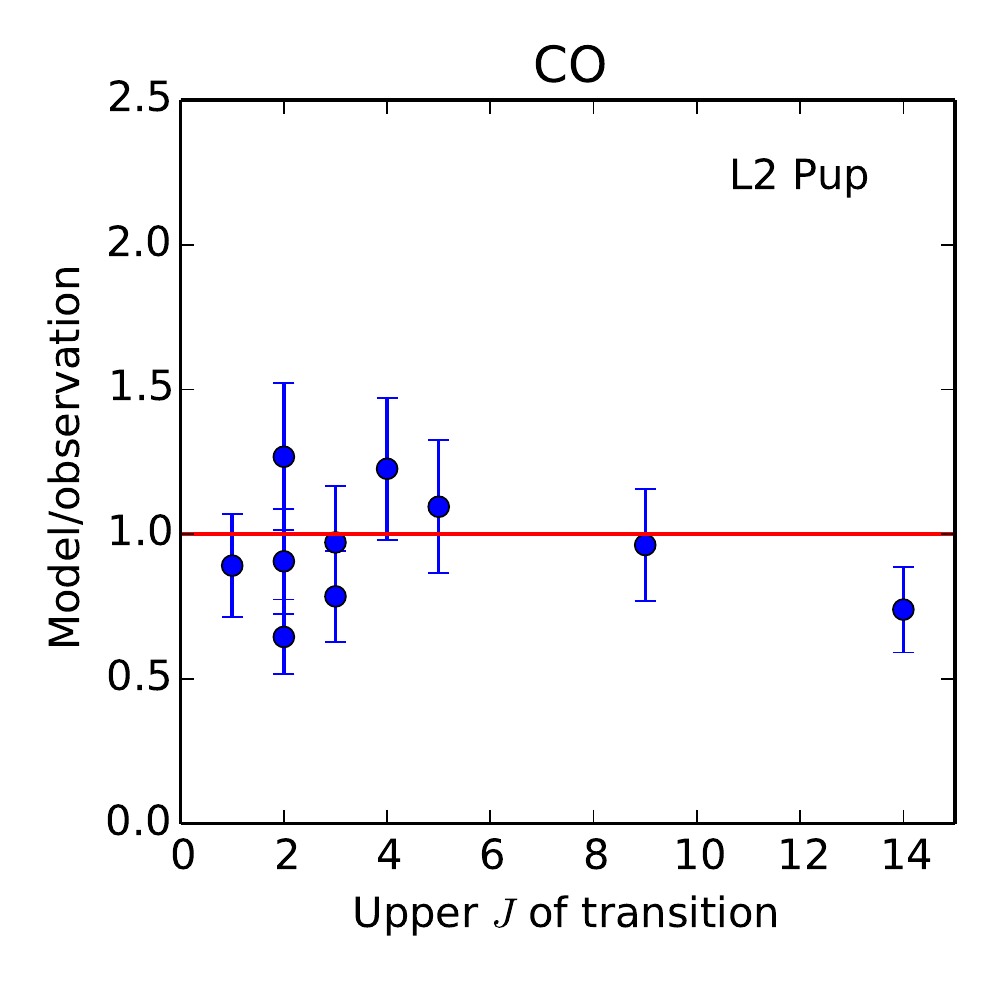}
\includegraphics[width=0.19\textwidth]{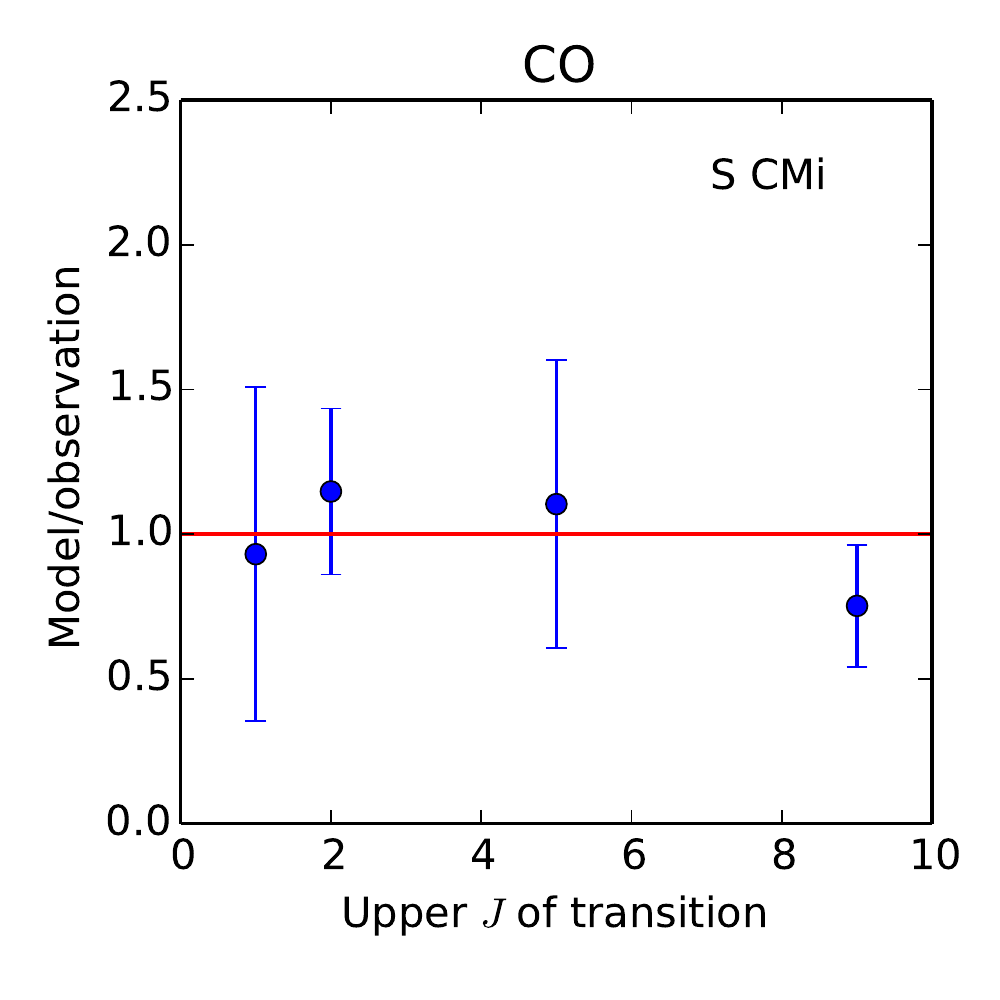}
\includegraphics[width=0.19\textwidth]{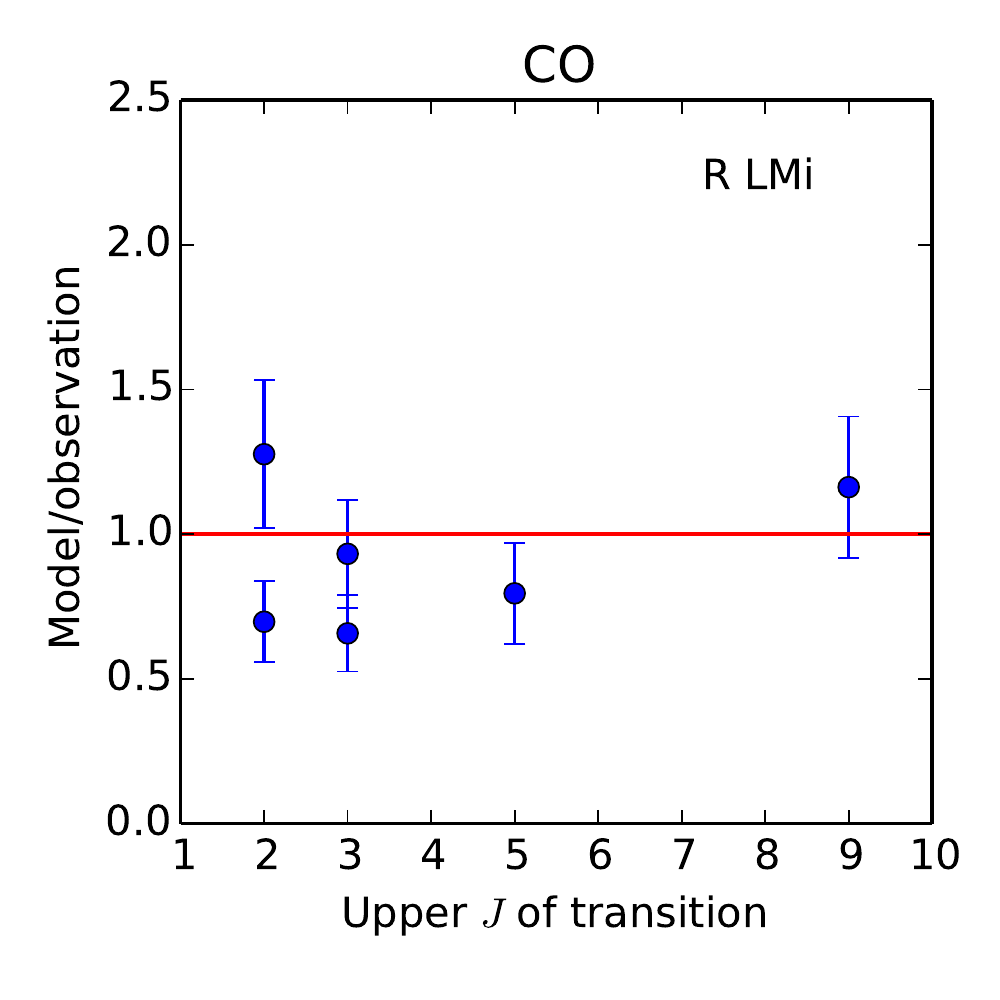}
\includegraphics[width=0.19\textwidth]{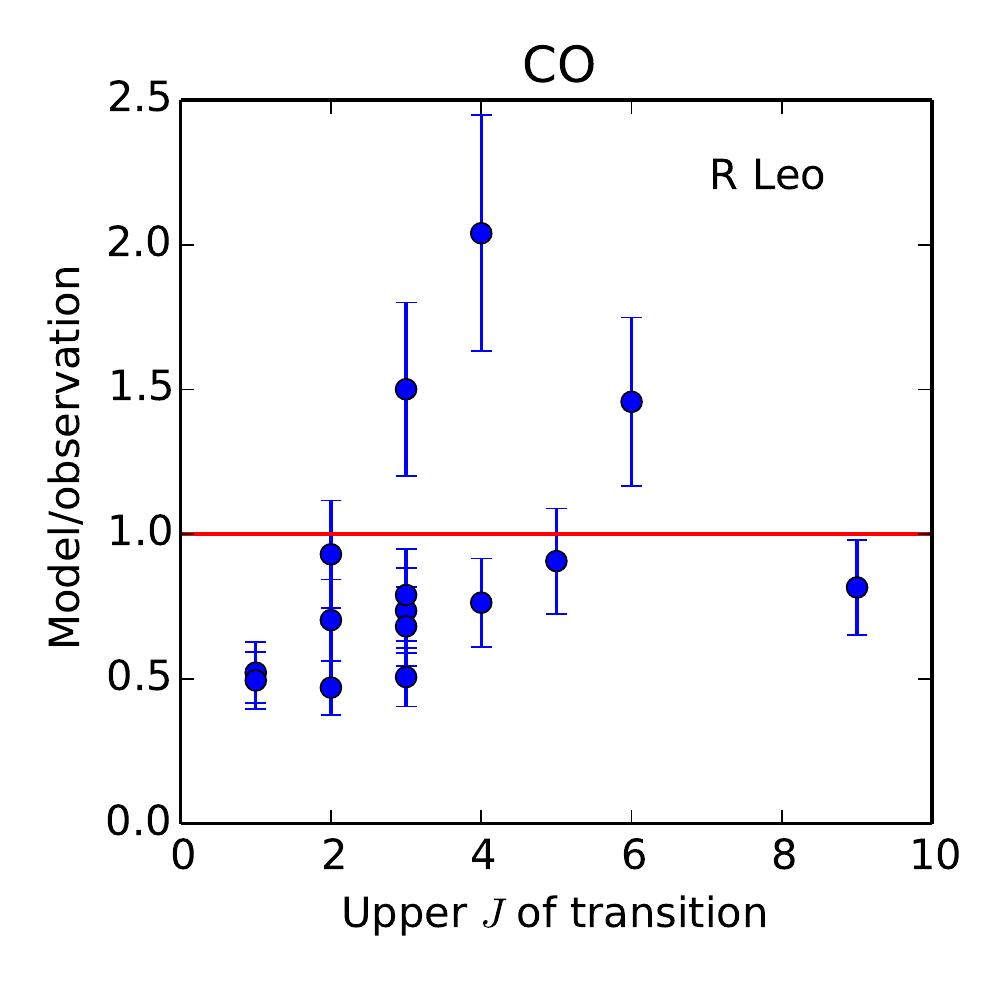}
\includegraphics[width=0.19\textwidth]{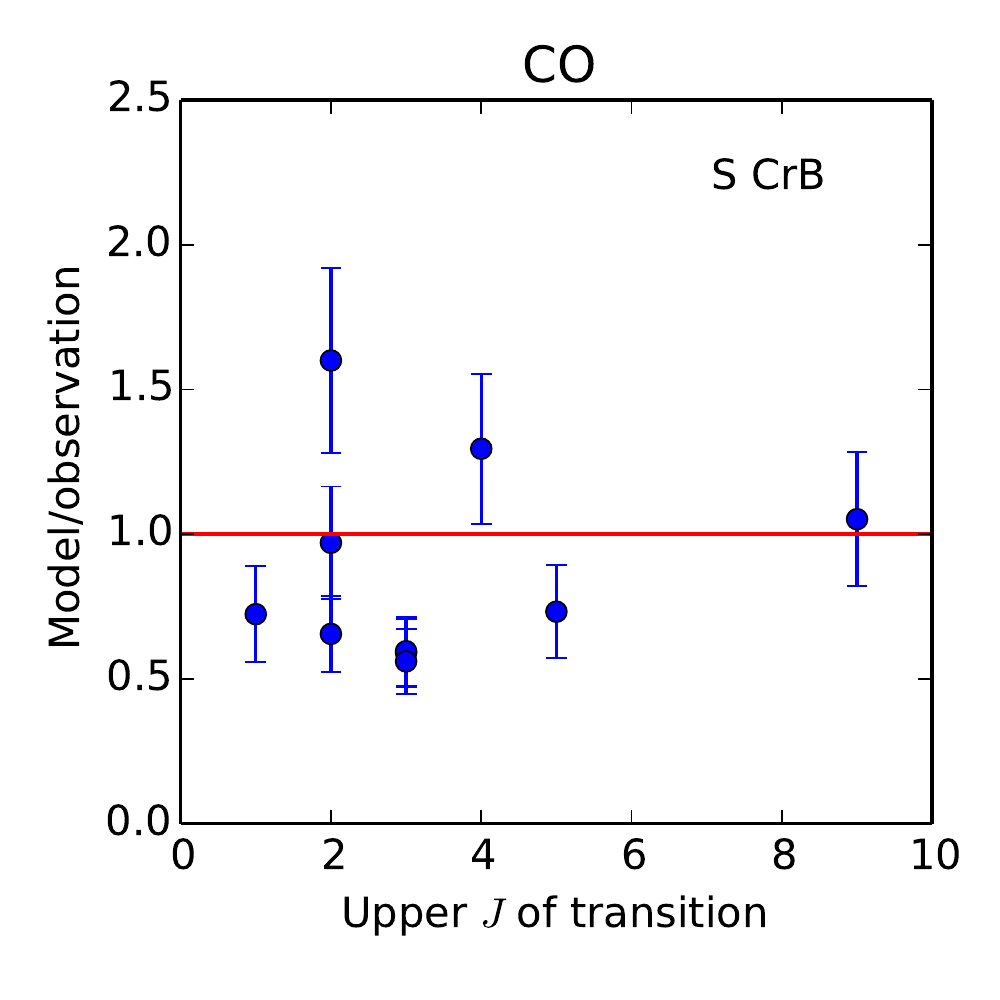}
\includegraphics[width=0.19\textwidth]{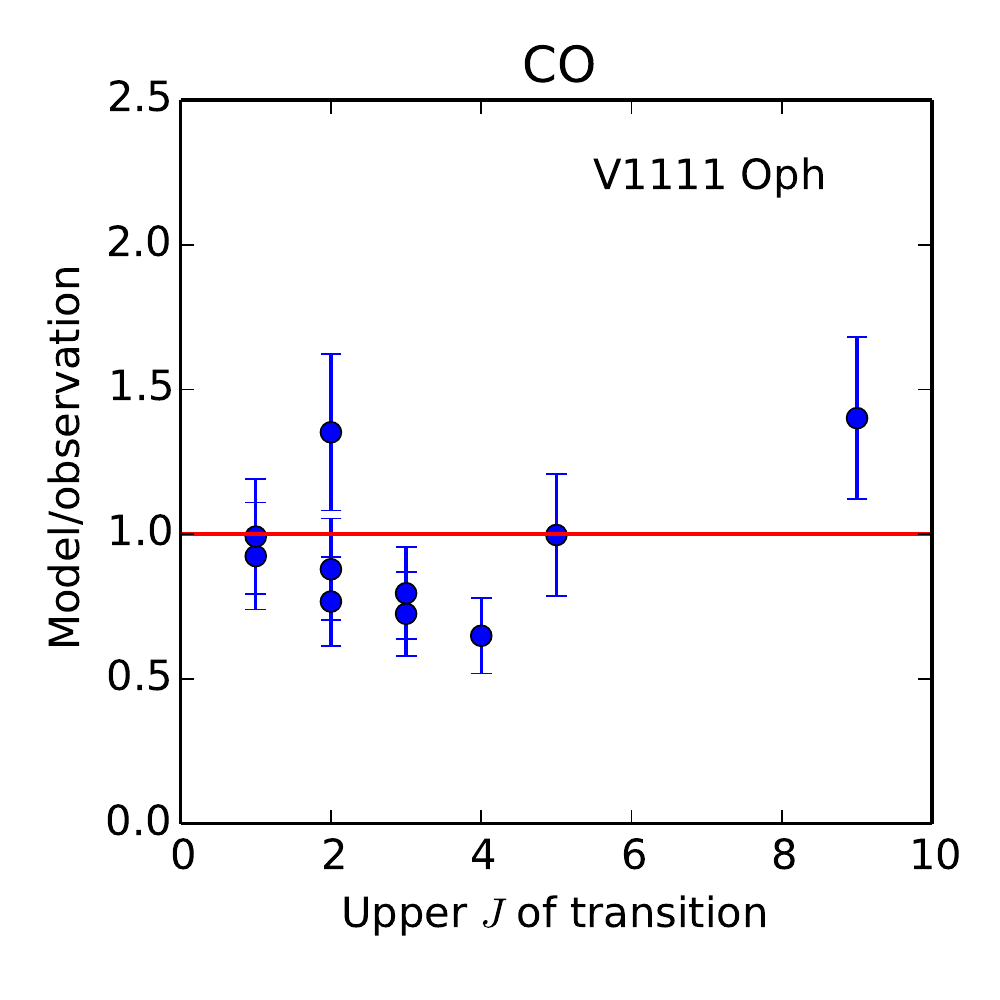}
\includegraphics[width=0.19\textwidth]{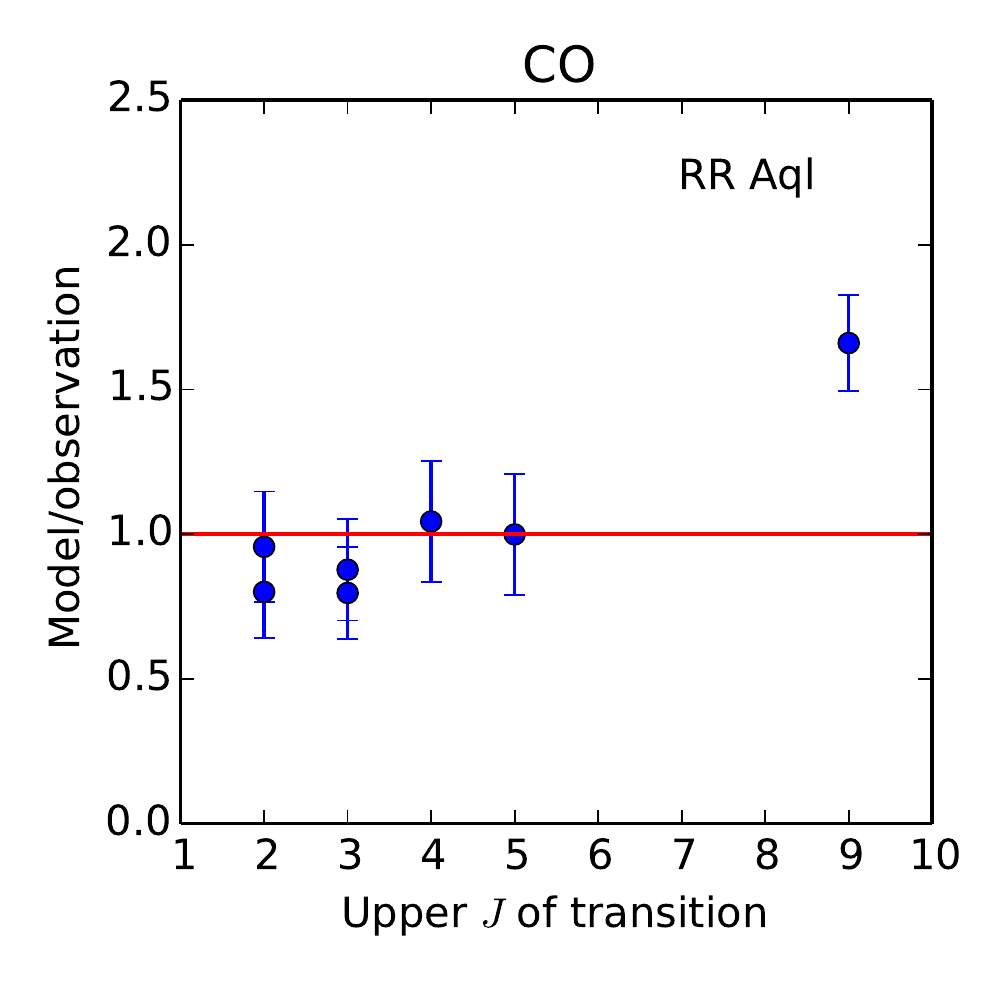}
\includegraphics[width=0.19\textwidth]{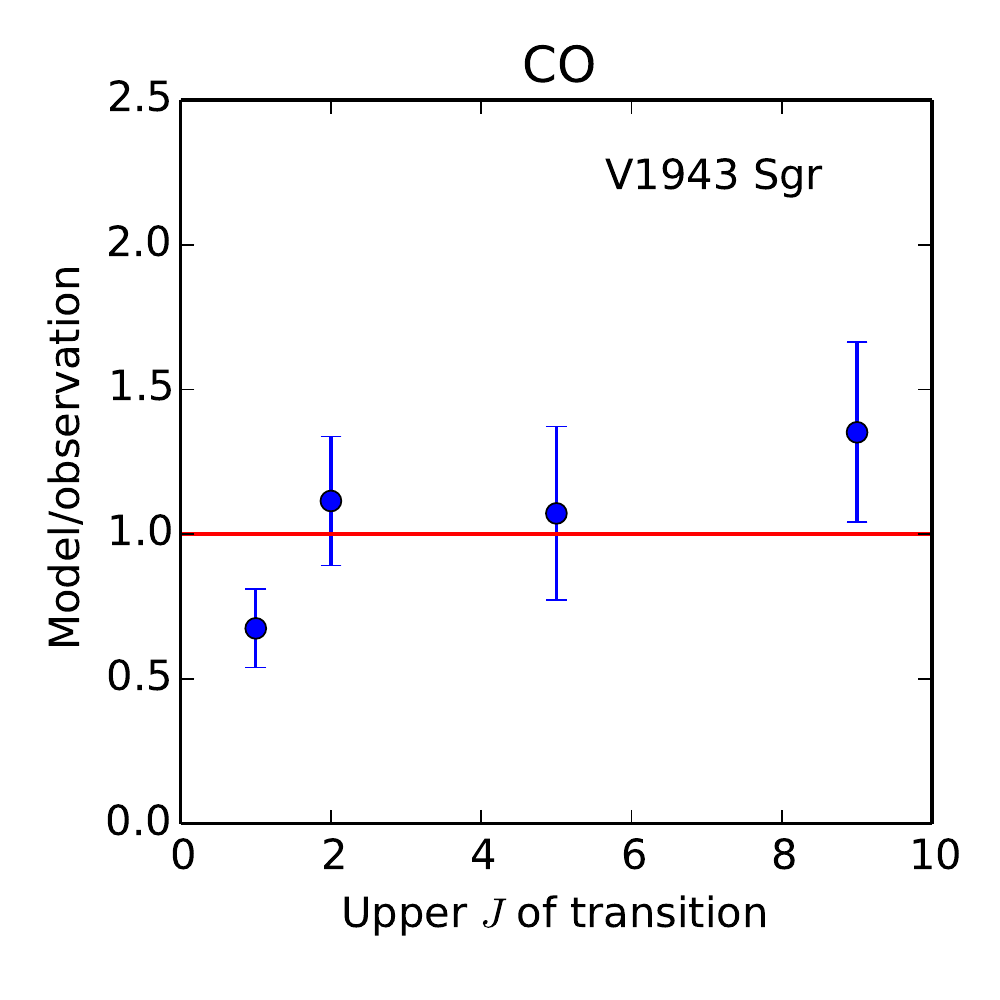}
\includegraphics[width=0.19\textwidth]{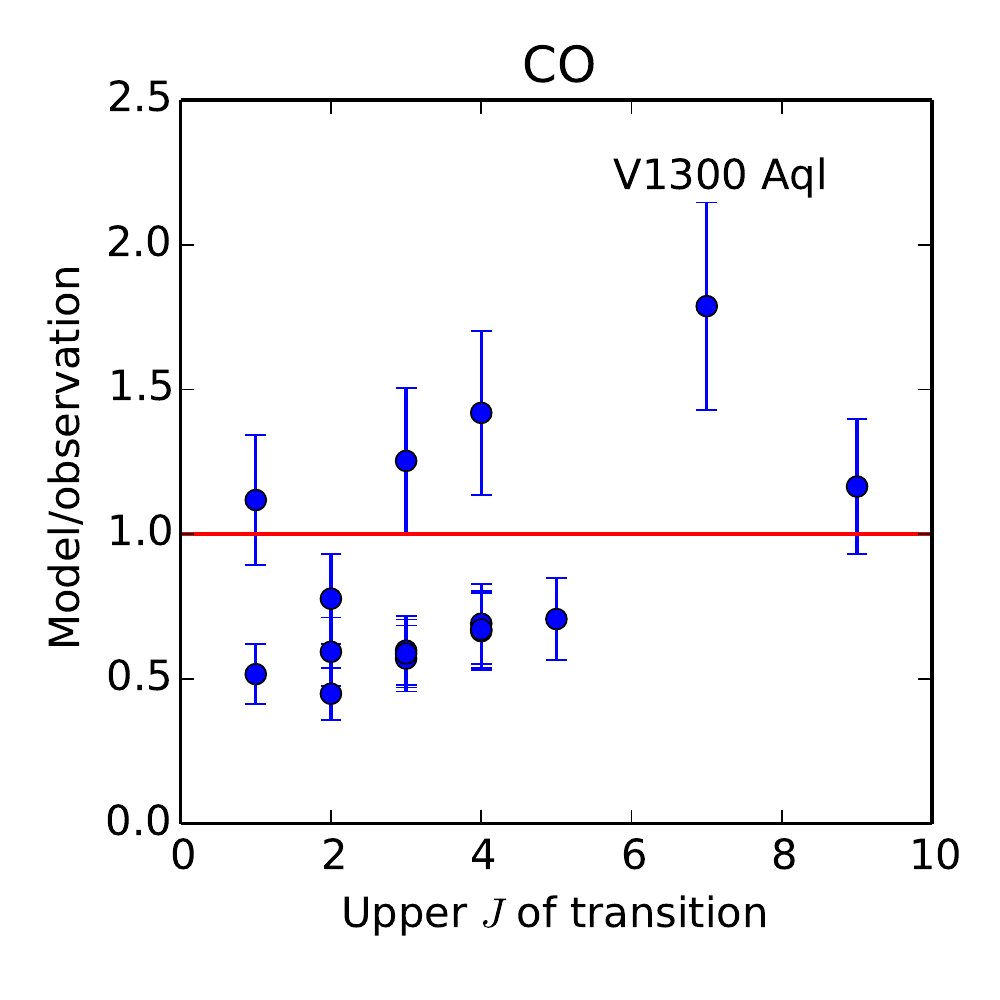}
\includegraphics[width=0.19\textwidth]{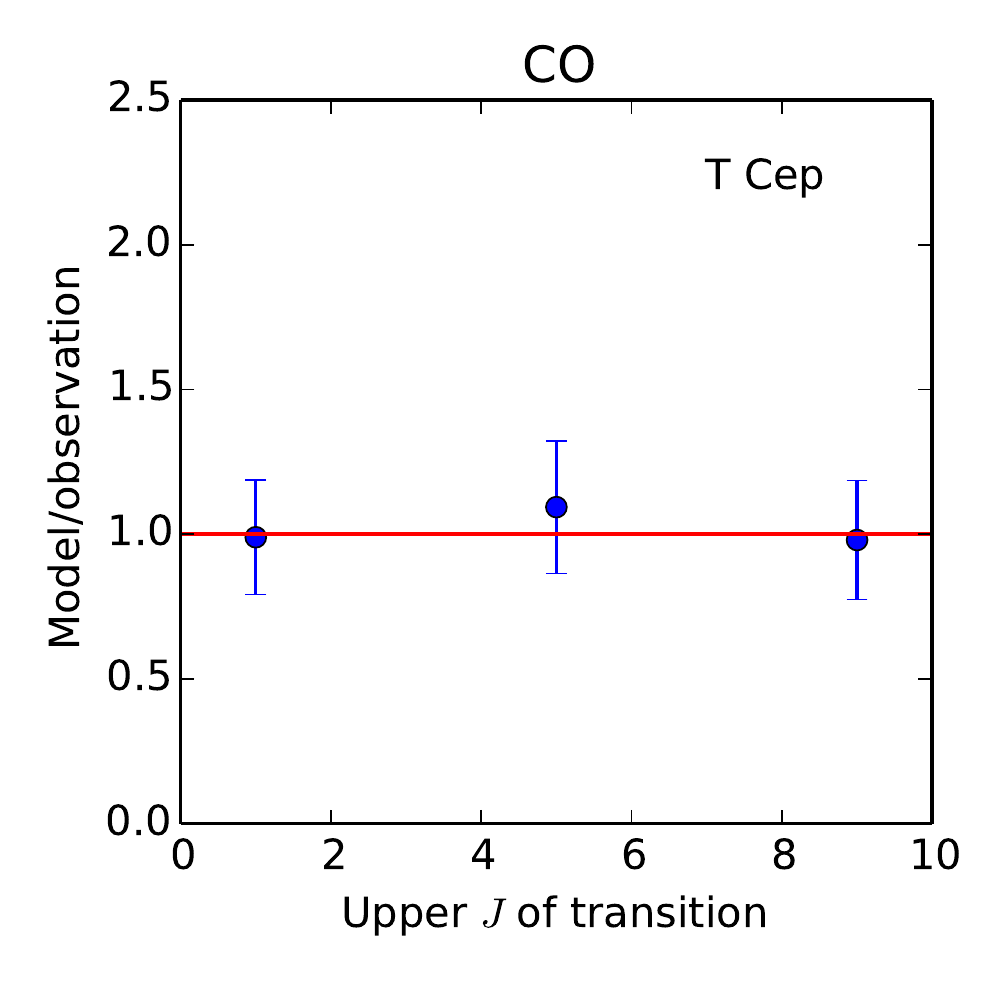}
\caption{Goodness of fit as defined by model/observed intensity for M stars.}
\label{Mfits}
\end{center}
\end{figure*}

\clearpage
\section{New observations}
\subsection{CO lines}

\begin{figure*}[t]
\begin{center}
\includegraphics[width=\textwidth]{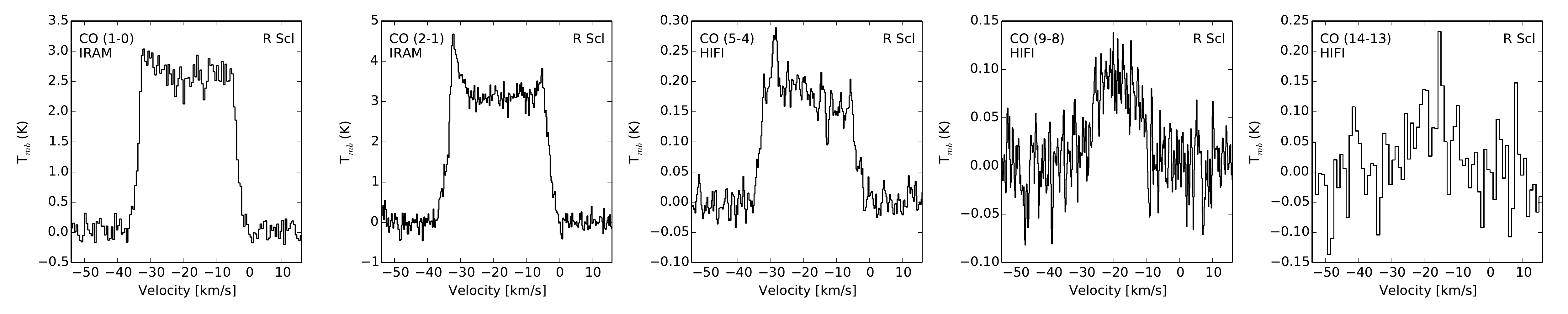}
\includegraphics[width=\textwidth]{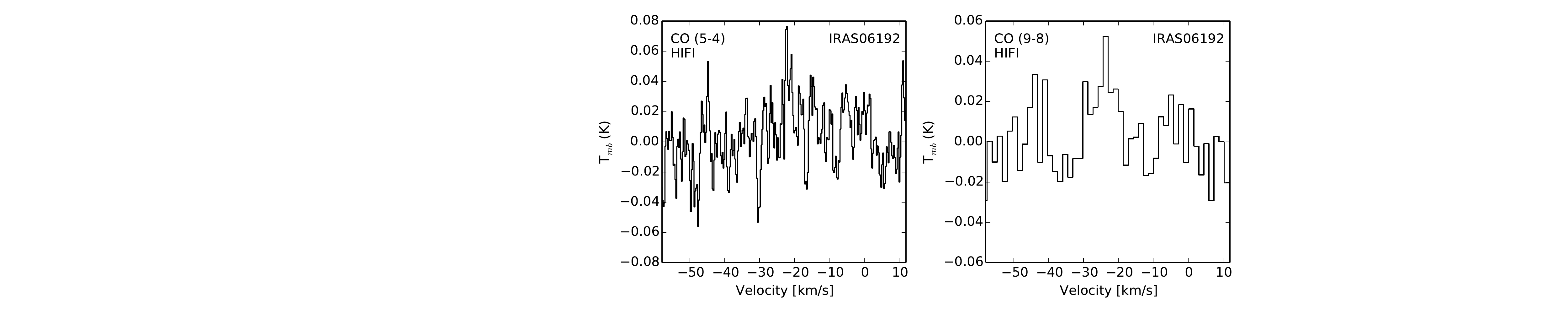}
\includegraphics[width=\textwidth]{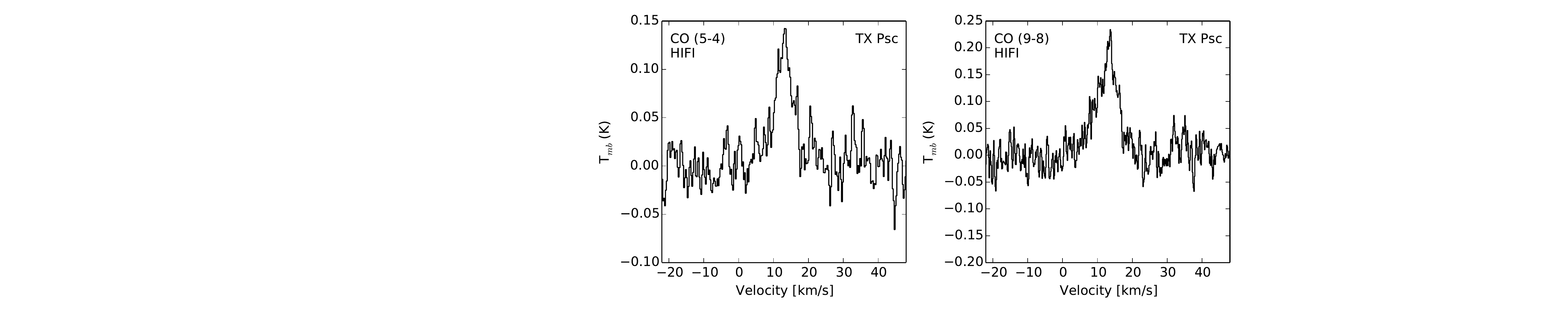}
\caption{New data from HIFI and IRAM for C stars not modelled in this paper, plotted with respect to LSR velocity.}
\label{Cplots}
\end{center}
\end{figure*}

\begin{figure*}[t]
\begin{center}
\includegraphics[width=\textwidth]{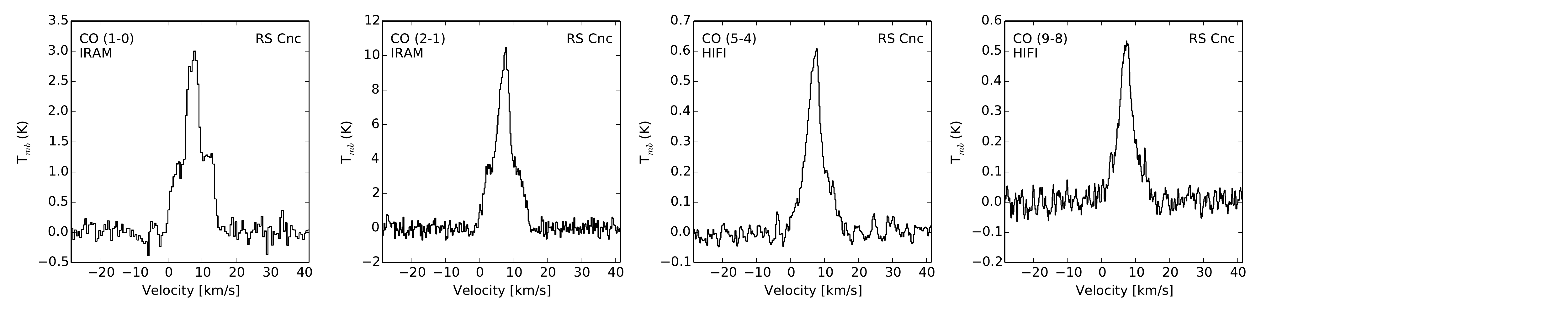}
\includegraphics[width=\textwidth]{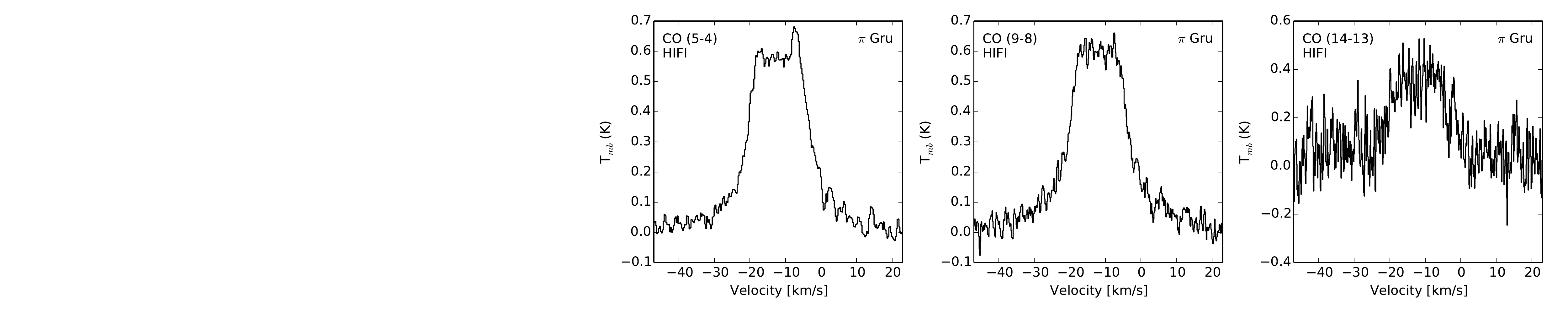}
\caption{New data from HIFI and IRAM for S stars not modelled in this paper, plotted with respect to LSR velocity.}
\label{Splots}
\end{center}
\end{figure*}

Of the data described in Table \ref{obstbl}, those stars for which we ran radiative transfer models were plotted in the body of the paper and in Appendix \ref{modplots}. The remaining lines, which were excluded from modelling for various reasons (see discussion in Sect. \ref{modselect}) are now presented here. The C stars are plotted in Fig. \ref{Cplots}, the S stars are plotted in Fig. \ref{Splots} and the M stars are plotted in Fig. \ref{Mplots}.

In particular, the unusual line profile due to the presence of a detached shell can be seen in the C star R Scl and double-component winds are clearly evident in C star TX Psc and S stars RS Cnc and $\pi^1$ Gru.

The observation identifiers (ObsIDs) for our \textit{Herschel} observations are listed in Table \ref{obsids}.

\begin{figure*}[t]
\begin{center}
\includegraphics[width=\textwidth]{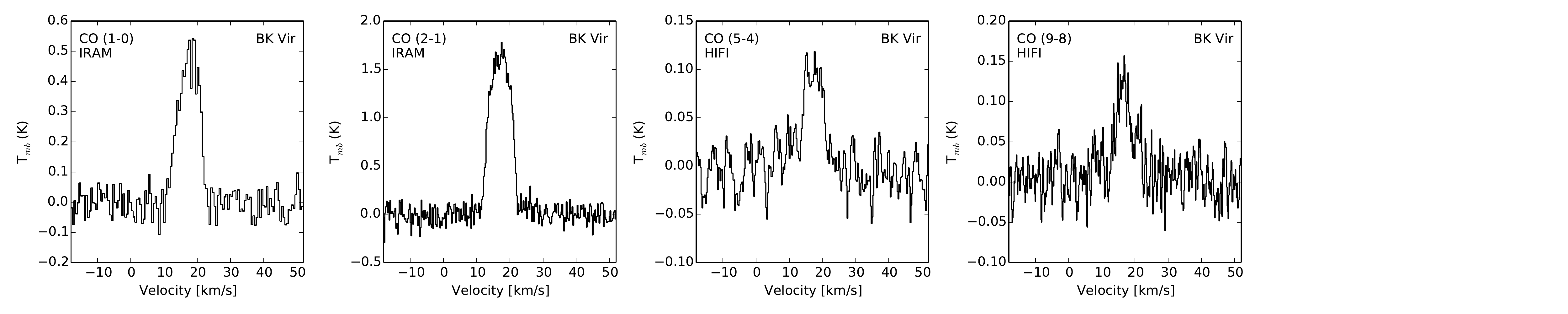}
\includegraphics[width=\textwidth]{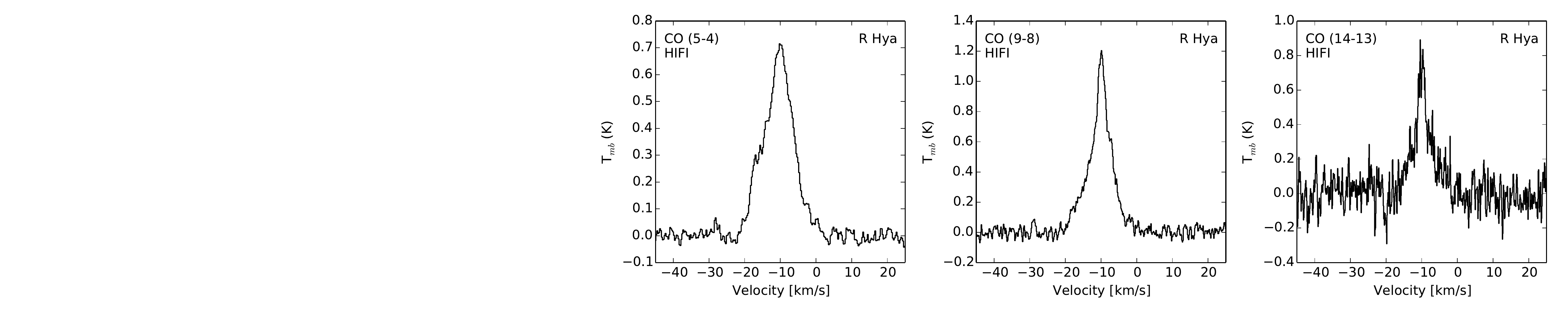}
\includegraphics[width=\textwidth]{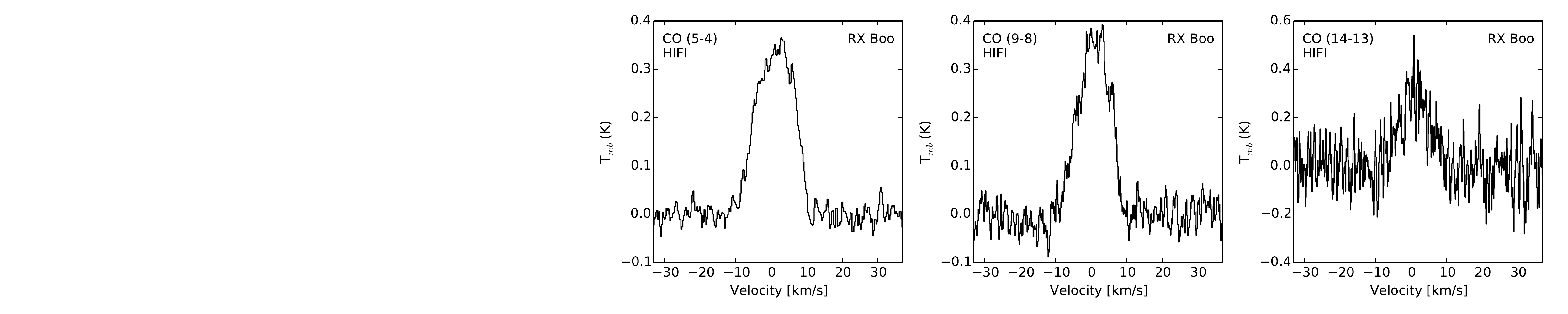}
\includegraphics[width=0.4\textwidth]{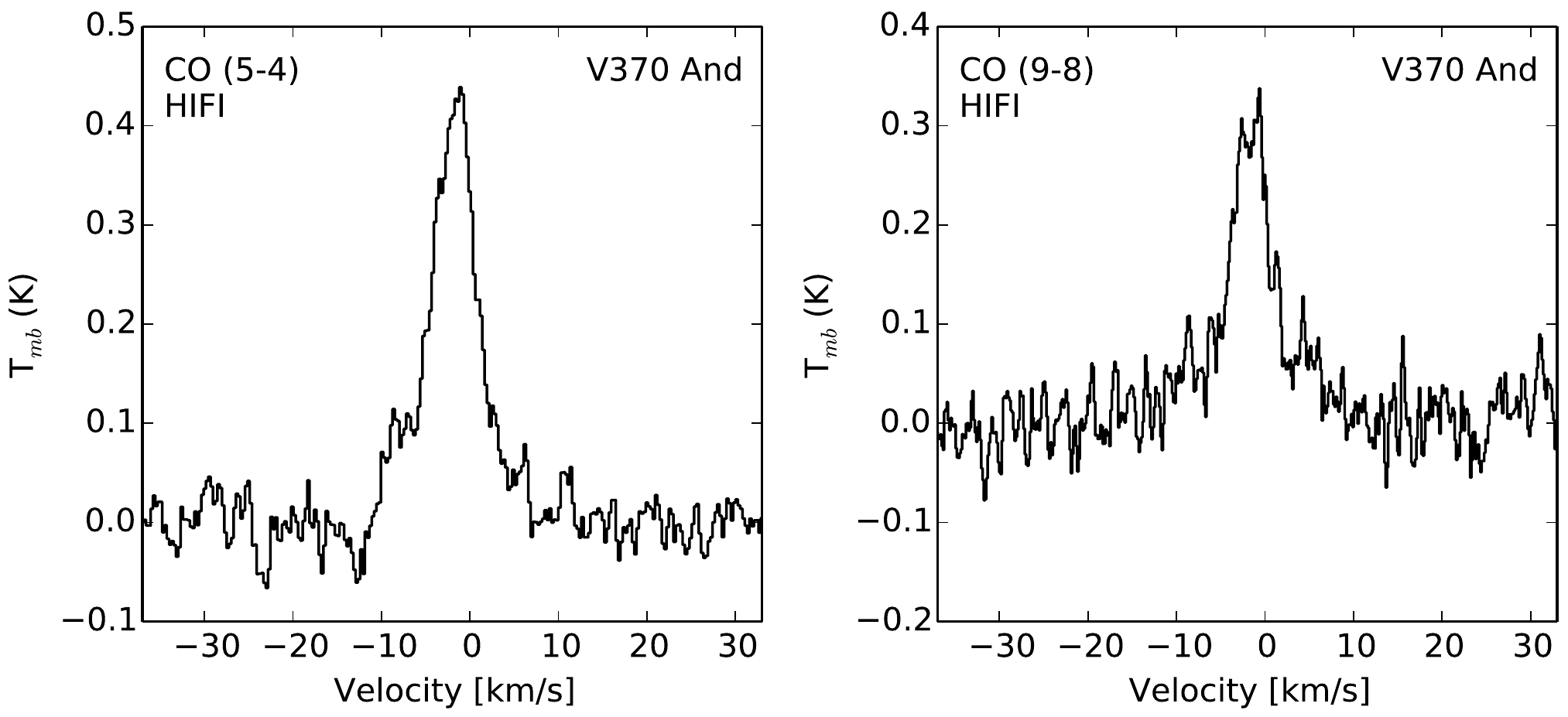}
\hspace{1cm}
\includegraphics[width=0.4\textwidth]{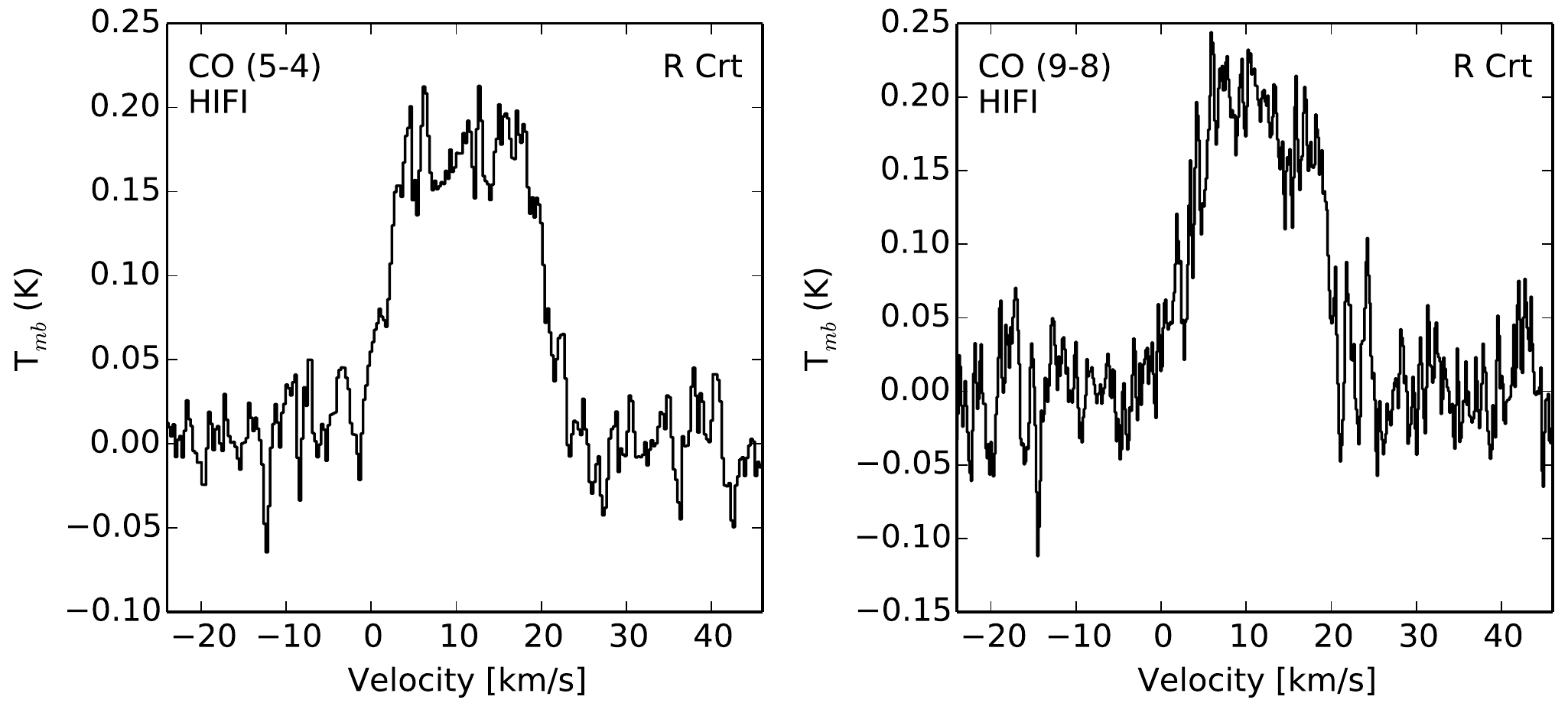}
\includegraphics[width=0.4\textwidth]{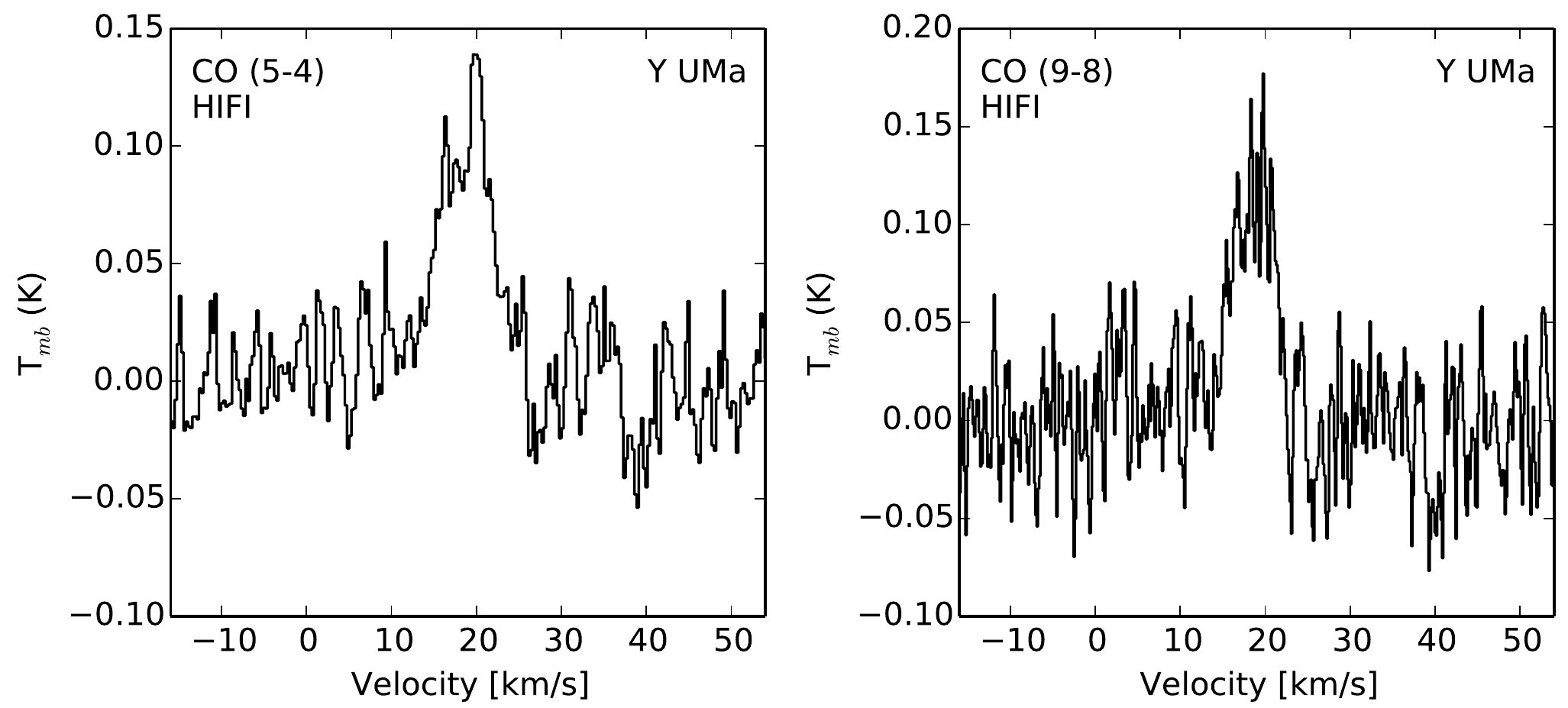}
\hspace{1cm}
\includegraphics[width=0.4\textwidth]{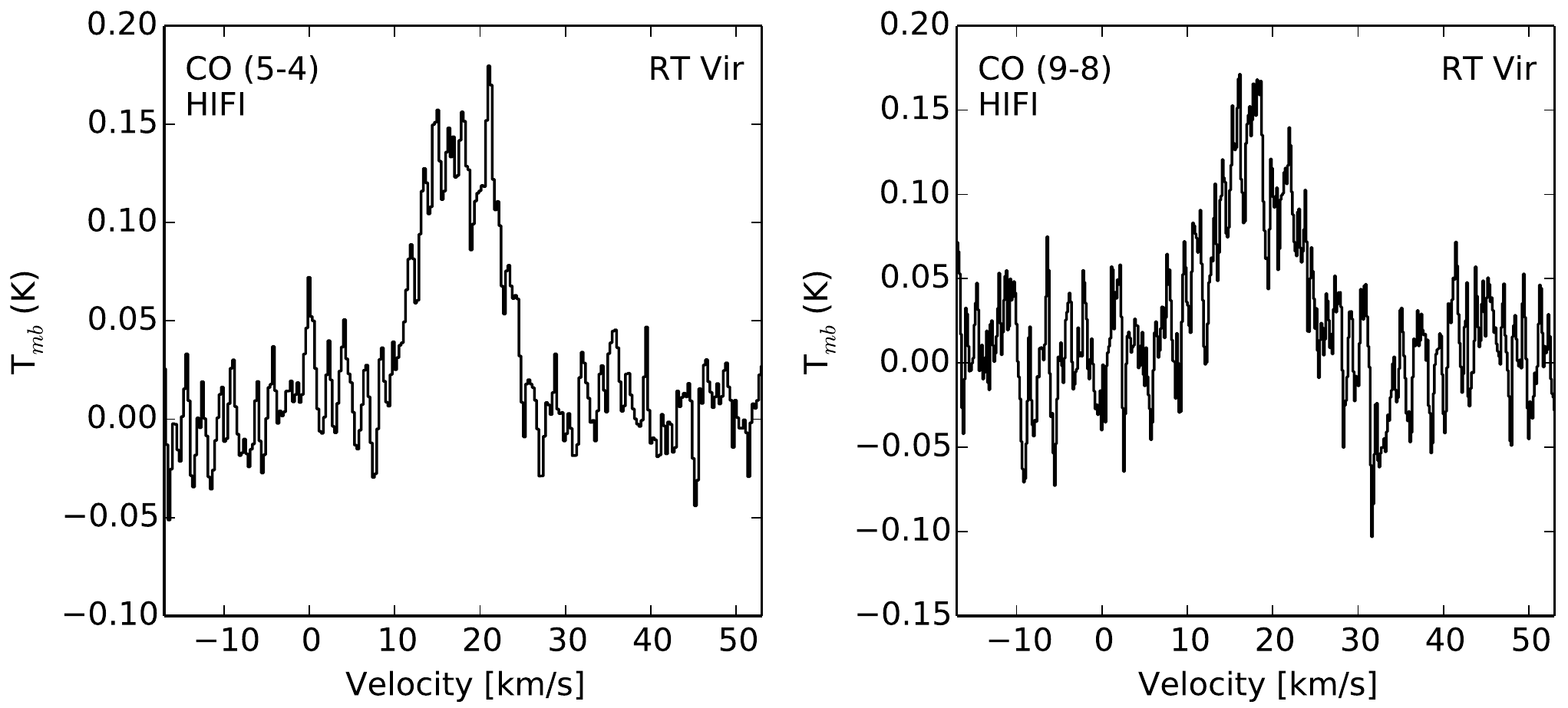}
\caption{New data from HIFI and IRAM for M stars not modelled in this paper, plotted with respect to LSR velocity.}
\label{Mplots}
\end{center}
\end{figure*}
\addtocounter{figure}{-1}
\begin{figure*}[t]
\begin{center}
\includegraphics[width=0.4\textwidth]{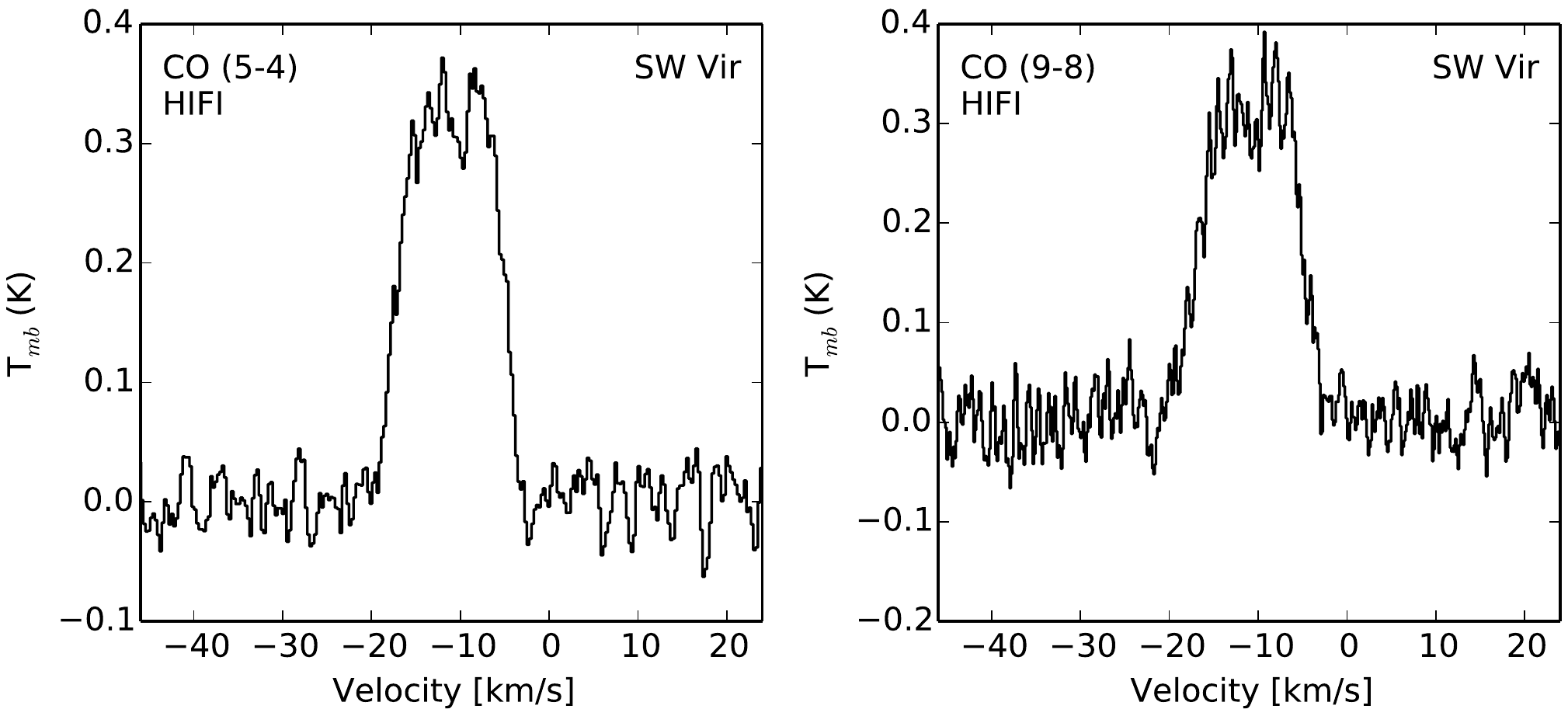}
\hspace{1cm}
\includegraphics[width=0.4\textwidth]{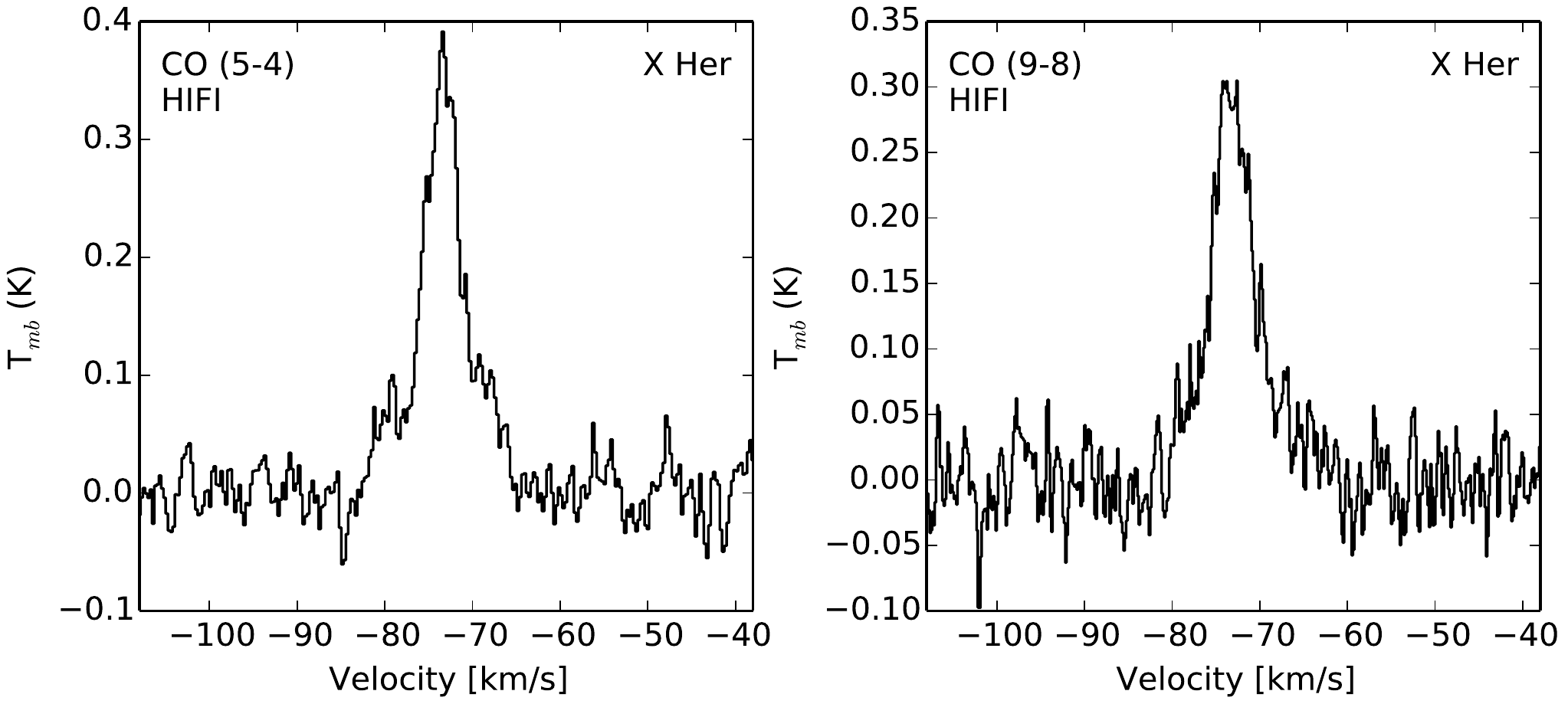}
\includegraphics[width=0.4\textwidth]{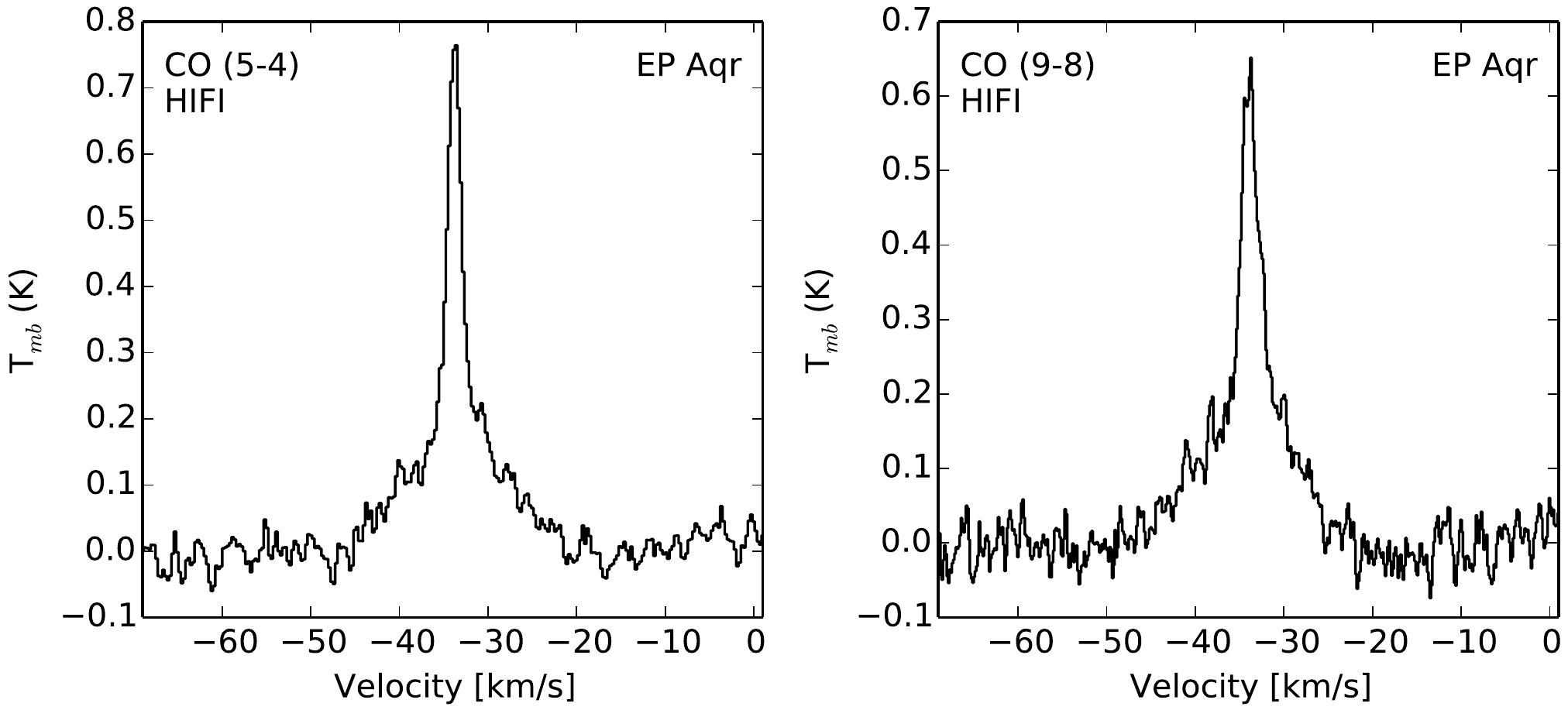}
\caption{{\bf Cont.} New data from HIFI and IRAM for M stars not modelled in this paper, plotted with respect to LSR velocity.}
\end{center}
\end{figure*}

\begin{table*}
\caption{ObsIDs for \textit{Herschel} observations.\label{obsids}}             
\centering                          
\begin{tabular}{l c c c}
\hline \hline
Source &  ObsID CO($5\to4$) &  ObsID CO($9\to8$) &  ObsID CO($14\to13$) \\ 
\hline
V1968 Cyg  &  1342210073  &  1342221424  &  ... \\
AFGL 292  &  1342200903  &  1342213353  &  ... \\
V701 Cas  &  1342201532  &  1342227533  &  ... \\
V1259 Ori  &  1342218903  &  1342227530  &  ... \\
V688 Mon  &  1342218902  &  1342229929  &  ... \\
BK Vir  &  1342210683  &  1342212125  &  ... \\
BX Cam  &  1342230363  &  1342227532  &  ... \\
EP Aqr  &  1342210079  &  1342220501  &  ... \\
GX Mon  &  1342228562  &  1342228602  &  ... \\
IRAS 06192+4657  &  1342230362  &  1342227531  &  ... \\
AI Vol  &  1342204015  &  1342210044  &  1342229787 \\
 &  1342200911  & ... & ... \\
II Lup  &  1342202049  &  1342227540  &  ... \\
V1111 Oph  &  1342230369  &  1342229924  &  ... \\
KU And  &  1342200923  &  1342213361  &  ... \\
V370 And  &  1342201528  &  1342213354  &  ... \\
V384 Per  &  1342204006  &  1342227534  &  ... \\
  &  1342204007  &  1342216333  &  ... \\
NV Aur  &  1342217689  &  1342218627  &  ... \\
GY Cam  &  1342214328  &  1342218626  &  ... \\
V1300 Aql  &  1342210080  &  1342216809  &  ... \\
L$_2$ Pup  &  1342202055  &  1342210049  &  1342231773 \\
$\pi$ Gru  &  1342210084  &  1342210043  &  1342220504 \\
R And  &  1342200924  &  1342213362  &  ... \\
R Crt  &  1342210087  &  1342212121  &  ... \\
R Cyg  &  1342200920  &  1342210034  &  ... \\
R Gem  &  1342228561  &  1342228601  &  ... \\
R Hor  &  1342200910  &  1342213350  &  ... \\
R Hya  &  1342200913  &  1342212120  &  1342223432 \\
R LMi  &  1342220512  &  1342220498  &  ... \\
R Leo  &  1342210684  &  1342220493  &  ... \\
R Lep  &  1342214326  &  1342216330  &  ... \\
R Scl  &  1342210695  &  1342213347  &  1342221454 \\
 &  1342200909  & ... & ... \\
RR Aql  &  1342216806  &  1342216808  &  ... \\
RS Cnc  &  1342220513  &  1342220497  &  ... \\
RT Vir  &  1342200915  &  1342212126  &  ... \\
RV Aqr  &  1342210078  &  1342218413  &  ... \\
RX Boo  &  1342200916  &  1342212129  &  1342223433 \\
S Cas  &  1342204009  &  1342213358  &  ... \\
 &  1342201533  & ... & ... \\
S CMi  &  1342220514  &  1342220496  &  ... \\
S CrB  &  1342200917  &  1342212128  &  ... \\
SW Vir  &  1342200914  &  1342212124  &  ... \\
T Cep  &  1342201535  &  1342210666  &  ... \\
TX Psc  &  1342210686  &  1342222348  &  ... \\
U Hya  &  1342210088  &  1342212122  &  ... \\
UU Aur  &  1342230361  &  1342229926  &  ... \\
V Aql  &  1342230383  &  1342229921  &  ... \\
V CrB  &  1342200918  &  1342214438  &  ... \\
V1943 Sgr  &  1342216804  &  1342216812  &  ... \\
V821 Her  &  1342230368  &  1342215900  &  ... \\
W And  &  1342201527  &  1342213355  &  ... \\
X Her  &  1342200919  &  1342210033  &  ... \\
X TrA  &  1342202053  &  1342216329  &  ... \\
Y Lyn  &  1342230360  &  1342229925  &  ... \\
Y Uma  &  1342210067  &  1342212127  &  ... \\
\hline
\end{tabular}
\end{table*}

\subsection{Bonus lines}\label{a:bonus}

As mentioned in Sect. \ref{sec:bonus}, we acquired some ``bonus" line spectra for molecules that were observable within our target frequency ranges. In HIFI, in the same range as the CO $(5\to4)$ line, we detected SiO $(13\to12)$ at 564.249 GHz. As can be seen in Table \ref{bonustbl}, it was mostly detected in M stars, especially those of lower mass-loss rates, which is in agreement with the trend found by \cite{Gonzalez-Delgado2003} and the calculations performed by \cite{Schoier2004}. 

\begin{figure*}[t]
\begin{center}
\includegraphics[width=\textwidth]{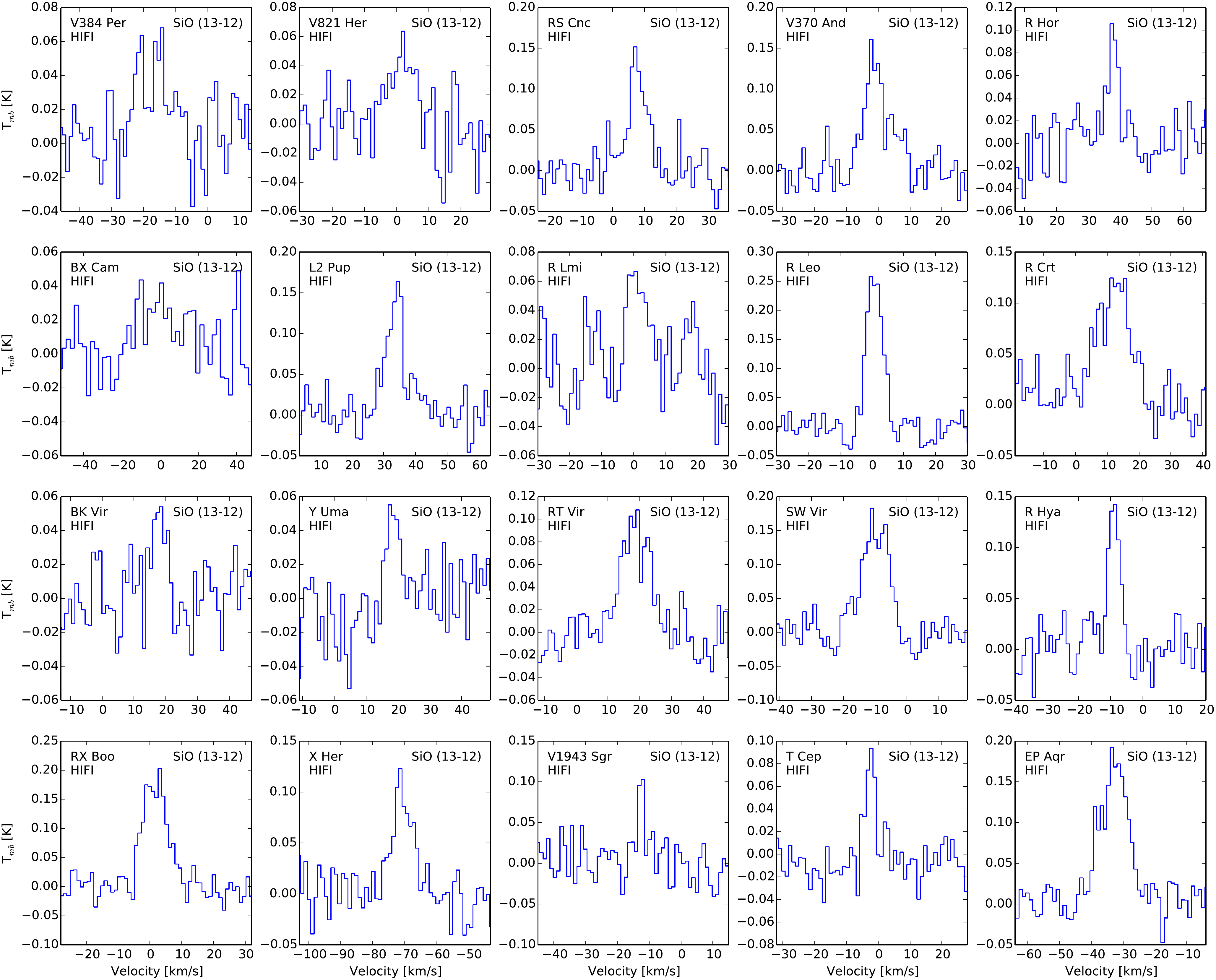}
\caption{New SiO data from HIFI, plotted with respect to LSR velocity.}
\label{SiOlines}
\end{center}
\end{figure*}

Our detections are plotted in Fig. \ref{SiOlines}. There was one detection in an S star, RS Cnc, which is the most ``M-like" S star in our sample, based on optical classifications. There were also two detections in C stars: V384 Per and V821 Her. They both have mass-loss rates in the range $\sim$ 2--3 $\e{-6}\spy$, putting them in the mid-to-high mass-loss rate range. They are located at 560 and 600 pc respectively, making them two of the nearest C stars in the higher mass-loss rate range ($> 10^{-6} \spy$). This could be why they had (weak) detections, while there were no detections in other C stars. The two C stars are among the sample modelled in SiO by \citet{Schoier2006}. These authors used observations of SiO lines from $J = 8\to7$ down to $J = 2\to1$ and there are six overlapping stars between their sample and the one in this paper, leaving four stars (AI Vol, II Lup, RV Aqr, and R Lep) detected in the lower-$J$ SiO lines but not in the higher-$J$ HIFI line.

\begin{figure*}[t]
\begin{center}
\includegraphics[width=\textwidth]{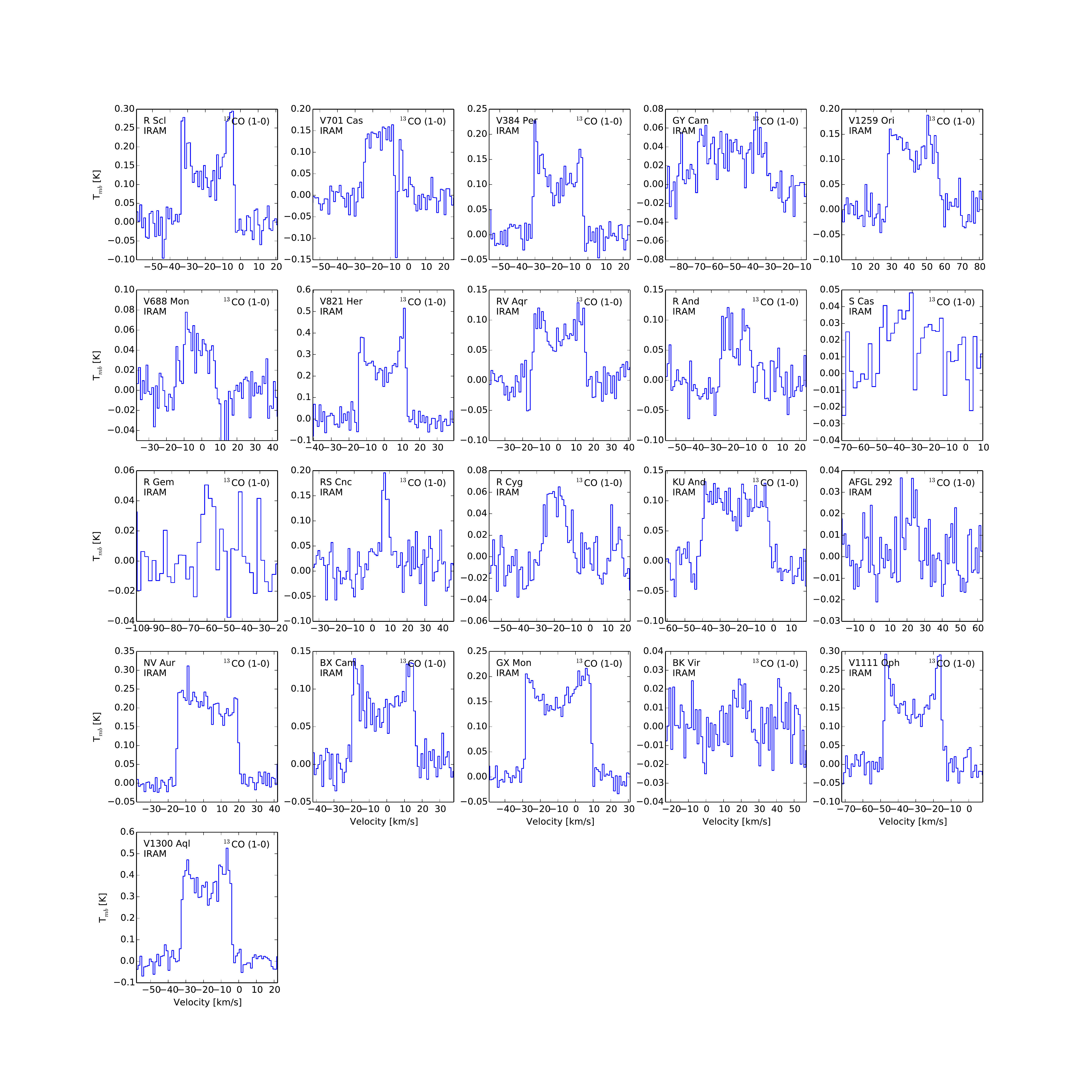}
\caption{New \up{13}CO data from IRAM, plotted with respect to LSR velocity.}
\label{13COlines}
\end{center}
\end{figure*}

Covered by our IRAM observations was the \up{13}CO $(1\to0)$ line at 110.201 GHz. The integrated intensities for these detections are given in Table \ref{bonustbl} and the observations are plotted in Fig. \ref{13COlines}. The \up{13}CO $(1\to0)$ line seems to have been most reliably detected in higher mass-loss rate sources across the three chemical types. It was not detected at all in stars with mass-loss rates $\sim 10^{-8}$--$10^{-7}\spy$ (note, however, that we are only dealing with two stars in this range) and was detected increasingly often for increasing mass-loss rates across chemical types.

\begin{figure*}[t]
\begin{center}
\includegraphics[width=\textwidth]{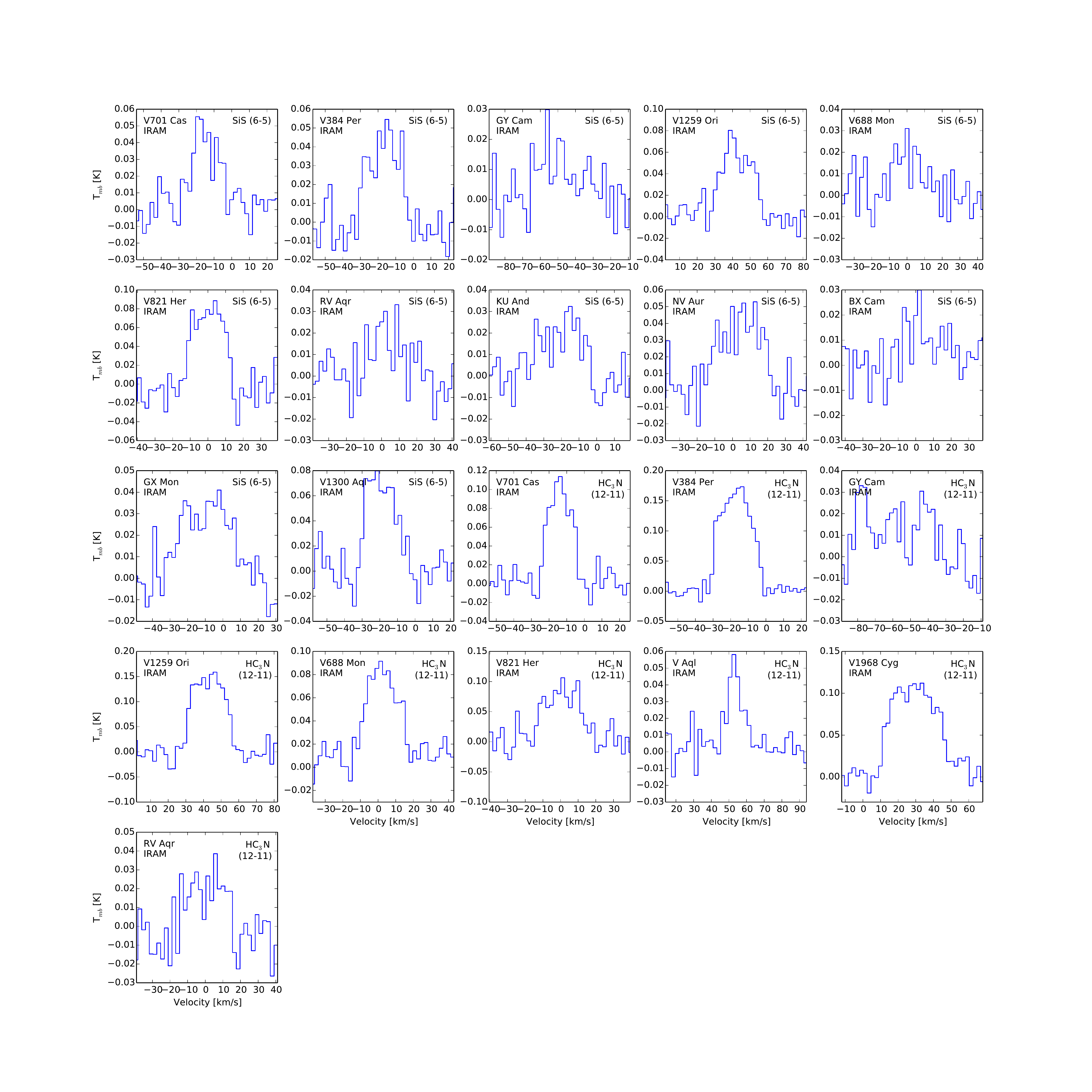}
\caption{New SiS and HC$_3$N data from IRAM, plotted with respect to LSR velocity. Note that the peak at $\sim-80~\kms$ in the GY Cam H$_{3}$CN spectrum is an artefact and not part of the H$_{3}$CN line.}
\label{SiSlines}
\end{center}
\end{figure*}

The SiS $(6\to5)$ line at 108.924 GHz was also detected in seven C stars, five M stars and no S stars. The integrated intensities are listed in Table \ref{bonustbl} and the spectra are plotted in \ref{SiSlines}. The C stars with detections were all in the mass-loss rate range $\sim 10^{-6}$ to $10^{-5}\;\spy$, with no detections for lower mass-loss rate objects and only one detection out of the two highest mass-loss rate C stars observed. Of the M stars, SiS was also detected in the higher mass-loss rate objects, but not in the highest mass-loss rate star, V1111 Oph. This trend suggests that SiS is more readily formed --- or at least more readily detectable --- in sources of intermediate mass-loss rate, around the range $\sim 10^{-6}$ to $10^{-5}\;\spy$.

CN lines were covered by both HIFI and IRAM observations. The CN $(5_{9/2}\to4_{7/2})$ and $(5_{11/2}\to4_{9/2})$ line groups with rest frequencies taken as 566.693 GHz and 566.947 GHz were covered in the observing range for CO ($5\to4$) and were detected in a handful of C stars. Our IRAM observations covered the CN $N=1\to0$ lines at 113.123 GHz and 113.488 GHz for the $(1_{1/2}-0_{1/2})$ and $(1_{3/2}-0_{1/2})$ line groups, respectively, and the CN $N=2\to1$ lines at 226.617 GHz, 226.874 GHz and 226.360 GHz for the $(2_{3/2}-1_{1/2})$, $(2_{3/2}-1_{3/2})$, and $(2_{5/2}-1_{3/2})$ line groups, respectively. The hyperfine structure of the low-$N$ CN lines can be seen particularly clearly. The integrated intensities of each line group are given in Table \ref{CNbonus} and the observations themselves are plotted in Fig. \ref{CNlines}.

Low-$N$ CN lines were detected in all of the observed C stars. Not all lines were detected in all stars, however. The lowest mass-loss rate star, U Hya, did not yield a clear detection of the $(1_{1/2}-0_{1/2})$ or $(2_{3/2}-1_{3/2})$ groups, although the remaining lines, including the $(5_{9/2}\to4_{7/2})$ and $(5_{11/2}\to4_{9/2})$ groups were clearly seen. The $(2_{3/2}-1_{3/2})$ was also not detected in V701 Cas, V1259 Ori, V688 Mon, or V821 Her, all of which are relatively high mass-loss rate objects with $\dot{M} \sim 10^{-6} - 10^{-5}\spy$. One S star, S Cas, was also detected in CN, in the $(1_{3/2}-0_{1/2})$, $(2_{3/2}-1_{1/2})$, $(2_{3/2}-1_{3/2})$, and $(2_{5/2}-1_{3/2})$ line groups. S Cas is the highest mass-loss rate and expansion velocity S star and, from its optical classification of S4/6e, is on the higher C/O end of the S star scale.

The last bonus line we detected was HC$_3$N ($12\to11$) at 109.174 GHz. The integrated intensities for the detections are listed in Table \ref{CNbonus} and the spectra are plotted in Fig. \ref{SiSlines}. HC$_3$N was only detected in C stars and not in the three lowest mass-loss rate objects with mass-loss rates below $10^{-6}\spy$. This is probably due to a higher density of available carbon to form this (simple) carbon-chain molecule in the higher mass-loss rate C stars.

\begin{table}
\caption{IRAM and HIFI \up{13}CO, SiO and SiS line observations}             
\label{bonustbl}      
\centering                          
\begin{tabular}{l c c c}
\hline \hline
Star	&	\up{13}CO ($1\to0$)	&	 SiO ($13\to12$)	& SiS ($6\to5$)	\\
	&	[K km s$^{-1}$]	&	[K km s$^{-1}$]	&	[K km s$^{-1}$]	\\
\hline	
\quad \it C Stars	\\
R Scl	&	4.7	(0.4)	&	$<$		0.1		&	$<$		0.4		\\
V701 Cas	&*	2.7	(0.2)	&	$<$		0.22		&	0.94	(0.15)	\\
V384 Per	&	3.9	(0.2)	&	0.73	(0.18)	&	0.77	(0.14)	\\
GY Cam	&	2.0	(0.2)	&	$<$		0.20		&	0.38	(0.15)	\\
V1259 Ori	&	3.5	(0.2)	&	$<$		0.21		&	1.40	(0.13)	\\
V688 Mon	&*	1.5	(0.2)	&	$<$		0.19		&	0.41	(0.12)	\\
V821 Her	&	7.7	(0.4)	&	0.29	(0.16)	&	1.33	(0.22)	\\
RV Aqr	&	2.2	(0.2)	&	$<$		0.13		&	0.39	(0.14)	\\
\quad \it S Stars	\\															
R And	&	0.83	(0.29)	&	$<$		0.16		&	$<$		0.3		\\
S Cas	&	1.3	(0.2)	&	$<$		0.18		&	$<$		0.2		\\
R Gem	&	0.34	(0.26)	&	$<$		0.23		&	$<$		0.3		\\
RS Cnc	&	1.0	(0.3)	&	0.90	(0.16)	&	$<$		0.3		\\
R Cyg	&	0.41	(0.21)	&	$<$		0.16		&	$<$		0.2		\\
\quad \it M Stars	\\															
KU And	&	3.7	(0.2)	&	$<$		0.16		&	0.55	(0.15)	\\
V370 And	&	...				&	1.07	(0.22)	&	...				\\
AFGL 292	&	0.16	(0.10)	&	$<$		0.23		&	$<$		0.1		\\
R Hor	&	...				&	0.41	(0.18)	&	...				\\
NV Aur	&	7.9	(0.2)	&	$<$		0.13		&	1.25	(0.12)	\\
BX Cam	&	2.9	(0.2)	&	0.85	(0.22)	&	0.40	(0.11)	\\
GX Mon	&	6.2	(0.2)	&	$<$		0.18		&	1.04	(0.12)	\\
L$_2$ Pup	&	...				&	1.13	(0.19)	&	...				\\
R LMi	&	...				&	0.54	(0.22)	&	...				\\
R Leo	&	...				&	1.46	(0.18)	&	...				\\
R Crt	&	...				&	1.73	(0.17)	&	...				\\
BK Vir	&	0.04	(0.13)	&	0.29	(0.17)	&	$<$		0.1		\\
Y UMa	&	...				&	0.10	(0.18)	&	...				\\
RT Vir	&	...				&	0.87	(0.18)	&	...				\\
SW Vir	&	...				&	1.50	(0.22)	&	...				\\
R Hya	&	...				&	0.72	(0.18)	&	...				\\
RX Boo	&	...				&	1.63	(0.20)	&	...				\\
X Her	&	...				&	0.79	(0.27)	&	...				\\
V1111 Oph	&	5.6	(0.3)	&	$<$		0.19		&	$<$		0.3		\\
V1943 Sgr	&	...				&	0.37	(0.17)	&	...				\\
V1300 Aql	&	11.0	(0.4)	&	$<$		0.15		&	1.53	(0.22)	\\
T Cep	&	...				&	0.32	(0.17)	&	...				\\
EP Aqr	&	...				&	2.00	(0.21)	&	...				\\
\hline
\end{tabular}
\tablefoot{The value in brackets after the flux gives the integrated noise RMS. An ellipsis (...) indicates that the line was not observed for the indicated star. * indicates that flux has been corrected for ISM emission.}
\end{table}

\begin{table*}
\caption{HIFI and IRAM CN and H\down{3}CN line group observations}             
\label{CNbonus}      
\centering                          
\begin{tabular}{l c c c c c c c c}
\hline \hline
Star	& \multicolumn{7}{c}{CN} & HC$_3$N\\
&	$1_{1/2}\to0_{1/2}$	&	$1_{3/2}\to0_{1/2}$	&	$2_{3/2}\to1_{1/2}$	&	$2_{3/2}\to1_{3/2}$	&	$2_{5/2}\to1_{3/2}$	&	$5_{9/2}\to4_{7/2}$	& $5_{11/2}\to4_{9/2}$	& $12\to11$\\
	&	[K km s$^{-1}$]	&	[K km s$^{-1}$]	&	[K km s$^{-1}$]	&	[K km s$^{-1}$]	&	[K km s$^{-1}$]	&	[K km s$^{-1}$]&	[K km s$^{-1}$]&	[K km s$^{-1}$]	\\
\hline	
R Scl	&	15.6	(1.0)	&	13.2	(1.0)	&	31.4	(0.9)	&	7.1	(0.9)	&	25.2	(0.9)	&	3.0	(0.1)	&	1.7	(0.1)	&	$<$		1.0		\\
V701 Cas	&	3.7	(0.3)	&	8.5	(0.3)	&	5.0	(0.4)	&	$<$		0.4		&	6.3	(0.4)	&	$<$		0.2		&	$<$		0.2		&	1.76	(0.17)	\\
V384 Per	&	21.6	(0.2)	&	28.4	(0.2)	&	21.9	(0.3)	&	7.0	(0.3)	&	23.3	(0.3)	&	$<$		0.2		&	$<$		0.2		&	3.72	(0.12)	\\
GY Cam	&	13.8	(0.2)	&	10.9	(0.2)	&	16.2	(0.3)	&	3.7	(0.3)	&	12.9	(0.3)	&	$<$		0.2		&	$<$		0.2		&	0.69	(0.17)	\\
R Lep	&	11.8	(0.4)	&	11.7	(0.4)	&	37.2	(0.6)	&	6.8	(0.6)	&	29.5	(0.6)	&	2.0	(0.1)	&	1.2	(0.1)	&	$<$		0.4		\\
V1259 Ori	&	6.1	(0.2)	&	12.6	(0.2)	&	4.6	(0.3)	&	$<$		0.3		&	8.9	(0.3)	&	$<$		0.2		&	$<$		0.2		&	3.13	(0.16)	\\
V688 Mon	&	4.2	(0.3)	&	8.7	(0.3)	&	8.1	(0.2)	&	$<$		0.2		&	11.6	(0.2)	&	$<$		0.2		&	$<$		0.2		&	2.40	(0.15)	\\
U Hya	&	$<$		0.4		&	1.4	(0.4)	&	17.1	(1.0)	&	$<$		1.0		&	11.6	(1.0)	&	1.8	(0.2)	&	1.1	(0.2)	&	$<$		0.4		\\
X TrA	&	...				&	...				&	...				&	...				&	...				&	1.0	(0.2)	&	0.7	(0.2)	&	...				\\
V821 Her	&	15.8	(0.5)	&	22.6	(0.5)	&	17.1	(0.7)	&	$<$		0.7		&	25.2	(0.7)	&	$<$		0.2		&	$<$		0.2		&	2.26	(0.24)	\\
V Aql	&	4.4	(0.2)	&	4.9	(0.2)	&	10.6	(0.3)	&	2.4	(0.3)	&	13.0	(0.3)	&	$<$		0.1		&	$<$		0.1		&	0.26	(0.08)	\\
V1968 Cyg	&	3.0	(0.3)	&	5.6	(0.3)	&	...				&	...				&	...				&	$<$		0.2		&	$<$		0.2		&	1.06	(0.14)	\\
RV Aqr	&	14.7	(0.2)	&	15.7	(0.2)	&	22.2	(0.4)	&	4.7	(0.4)	&	20.2	(0.4)	&	$<$		0.1		&	$<$		0.1		&	0.28	(0.15)	\\
S Cas	&	$<$		0.3		&	1.4	(0.3)	&	11.8	(0.5)	&	2.4	(0.5)	&	11.6	(0.5)	&	$<$		0.2		&	$<$		0.2		&	$<$		0.3		\\
\hline
\end{tabular}
\tablefoot{The value in brackets after the flux gives the integrated noise RMS. An ellipsis (...) indicates that the line was not observed for the indicated star. An * indicates the line was detected but contaminated by an artefact and the integrated intensity cannot be relied upon.}
\end{table*}

\begin{figure*}[t]
\begin{center}
\includegraphics[width=\textwidth]{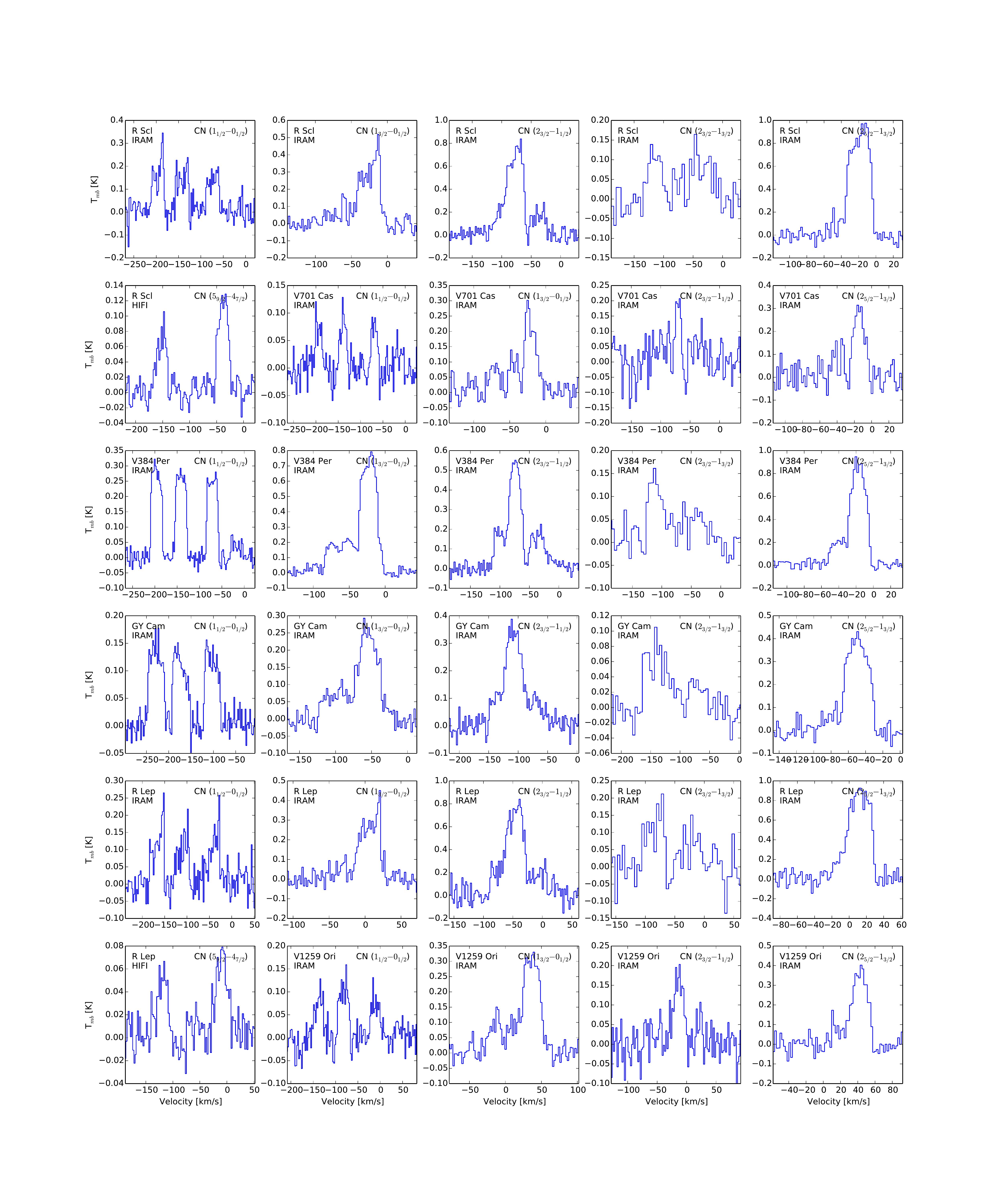}
\caption{New CN data from HIFI and IRAM, plotted with respect to LSR velocity of the reddest component. In the case of the HIFI lines, both the ($5_{9/2}\to4_{7/2}$) and ($5_{11/2}\to4_{9/2}$) lines are plotted together at the rest frequency of the former.}
\label{CNlines}
\end{center}
\end{figure*}
\addtocounter{figure}{-1}
\begin{figure*}[t]
\begin{center}
\includegraphics[width=\textwidth]{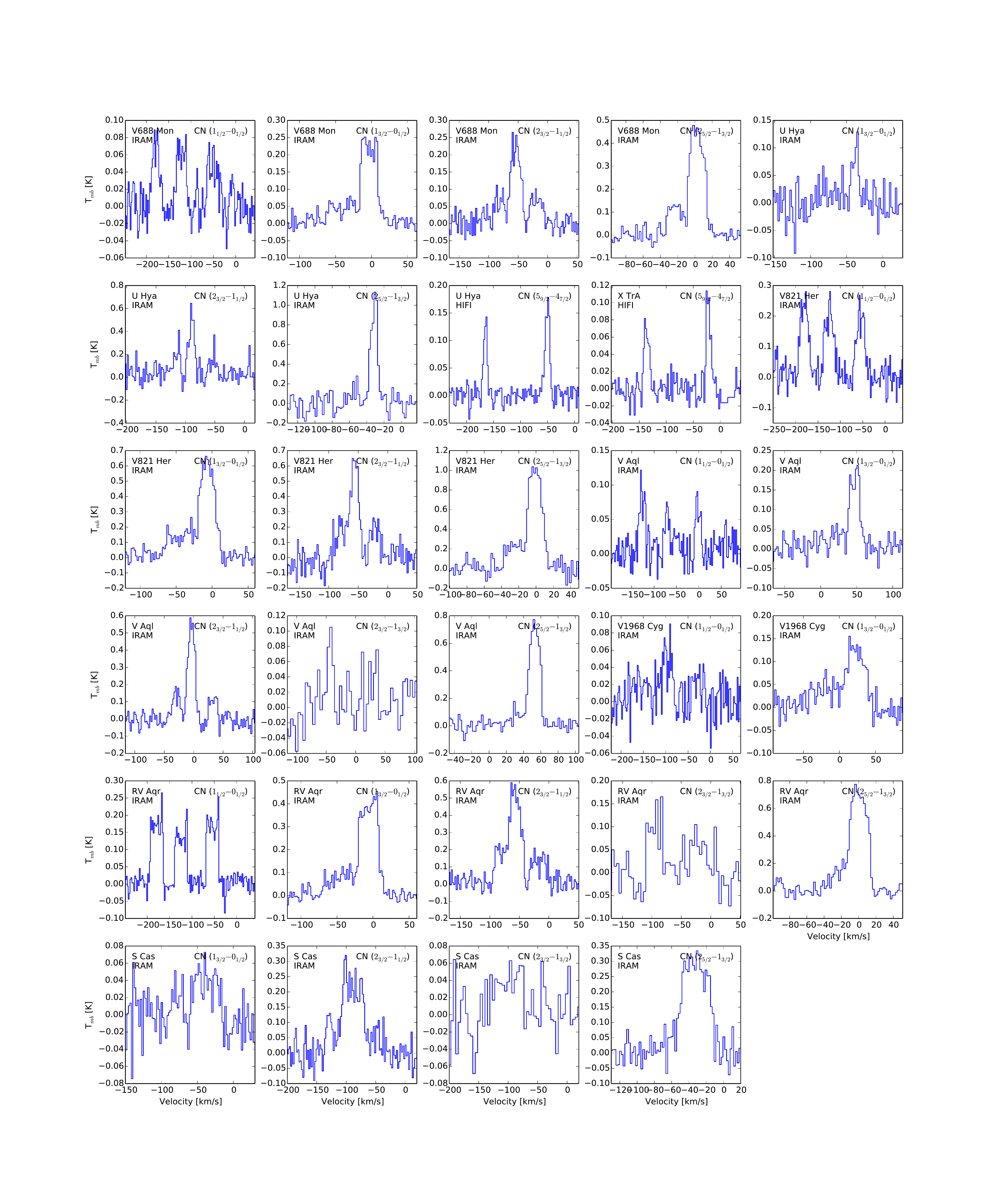}
\caption{{\bf Cont.} New CN data from HIFI and IRAM, plotted with respect to LSR velocity of the reddest component. In the case of the HIFI lines, both the ($5_{9/2}\to4_{7/2}$) and ($5_{11/2}\to4_{9/2}$) lines are plotted together at the rest frequency of the former.}
\end{center}
\end{figure*}

\clearpage
\section{Supplementary line data}

As discussed in Sect. \ref{suppobs}, we included substantial archival data in our modelling procedure to find the models which best fit the widest range of data possible. The archival data we used to constrain our models is listed in Table \ref{lowjobs}.

\clearpage

\addtocounter{table}{1}

\begin{longtab}{1}{
\begin{longtable}{lcccl}
\caption{Archival data of other CO observations of the stars used in our modelling.\label{lowjobs}}\\          
\hline\hline
Star	&		Transition		&	$I_\mathrm{mb}$	&	Telescope	&	Reference	\\
	&				&	[K km s$^{-1}$]	&		&		\\
\hline		
\quad \it C Stars	\\
\endfirsthead
\caption{continued...}\\
\hline\hline
Star	&		Transition		&	$I_\mathrm{mb}$	&	Telescope	&	Reference	\\
	&				&	[K km s$^{-1}$]	&		&		\\
\hline
\endhead
\hline
\multicolumn{4}{l}{* indicates uncorrected ISM contamination} \\
\endfoot
\multicolumn{4}{l}{* indicates uncorrected ISM contamination} \\
\endlastfoot
V384 Per	&	1	$\rightarrow$	0	&	52.7	&	IRAM	&	\cite{Olofsson1993}	\\
	&	1	$\rightarrow$	0	&	7.8*	&	NRAO	&	\cite{Schoier2001}	\\
	&	1	$\rightarrow$	0	&	25.5*	&	OSO	&	\cite{Olofsson1993}	\\
	&	2	$\rightarrow$	1	&	83.6	&	IRAM	&	\cite{Olofsson1993}	\\
	&	2	$\rightarrow$	1	&	61.9	&	JCMT	&	\cite{Schoier2001}	\\
	&	3	$\rightarrow$	2	&	32.8	&	CSO	&	\cite{Knapp1998}	\\
	&	3	$\rightarrow$	2	&	40.1	&	JCMT	&	\cite{Schoier2001}	\\
	&	4	$\rightarrow$	3	&	78.7	&	JCMT	&	H. Olofsson, private communication	\\
GY Cam	&	2	$\rightarrow$	1	&	14.3	&	CSO	&	\cite{Knapp1998}	\\
R Lep	&	1	$\rightarrow$	0	&	6.2	&	SEST	&	\cite{Olofsson1993}	\\
	&	2	$\rightarrow$	1	&	27.6	&	APEX	&	Pointing Catalogue	\\
	&	2	$\rightarrow$	1	&	18.1	&	SEST	&	\cite{Olofsson1993}	\\
	&	3	$\rightarrow$	2	&	32.3	&	APEX	&	Pointing Catalogue	\\
	&	4	$\rightarrow$	3	&	36.6	&	APEX	&	Pointing Catalogue	\\
V1259 Ori	&	2	$\rightarrow$	1	&	36.3	&	APEX	&	Pointing Catalogue	\\
	&	2	$\rightarrow$	1	&	16.5	&	CSO	&	\cite{Knapp1998}	\\
	&	3	$\rightarrow$	2	&	32.5	&	APEX	&	Pointing Catalogue	\\
	&	3	$\rightarrow$	2	&	32.2	&	APEX	&	Archive	\\
	&	4	$\rightarrow$	3	&	30.4	&	APEX	&	Pointing Catalogue	\\
UU Aur	&	1	$\rightarrow$	0	&	18.8	&	IRAM	&	\cite{Olofsson1993}	\\
	&	1	$\rightarrow$	0	&	7.9	&	OSO	&	\cite{Olofsson1993}	\\
	&	2	$\rightarrow$	1	&	15.0	&	APEX	&	Pointing Catalogue	\\
	&	2	$\rightarrow$	1	&	39.0	&	IRAM	&	\cite{Olofsson1993}	\\
	&	3	$\rightarrow$	2	&	21.6	&	APEX	&	Pointing Catalogue	\\
	&	3	$\rightarrow$	2	&	22.7	&	JCMT	&	H. Olofsson, private communication	\\
V688 Mon	&	3	$\rightarrow$	2	&	15.4*	&	CSO	&	\cite{Knapp1998}	\\
AI Vol	&	1	$\rightarrow$	0	&	25.57	&	SEST	&	\cite{Woods2003}	\\
	&	2	$\rightarrow$	1	&	56.5	&	APEX	&	Pointing Catalogue	\\
	&	2	$\rightarrow$	1	&	59.1	&	APEX	&	Hans pointing data	\\
	&	2	$\rightarrow$	1	&	50.00	&	SEST	&	\cite{Woods2003}	\\
	&	3	$\rightarrow$	2	&	57.8	&	APEX	&	Pointing Catalogue	\\
	&	4	$\rightarrow$	3	&	59.1	&	APEX	&	Pointing Catalogue	\\
U Hya	&	1	$\rightarrow$	0	&	5.4	&	SEST	&	\cite{Olofsson1993}	\\
	&	2	$\rightarrow$	1	&	17.9	&	APEX	&	Pointing Catalogue	\\
	&	2	$\rightarrow$	1	&	48.8	&	IRAM	&	\cite{Schoier2001}	\\
	&	2	$\rightarrow$	1	&	20.2	&	JCMT	&	\cite{Schoier2001}	\\
	&	2	$\rightarrow$	1	&	13.8	&	SEST	&	\cite{Schoier2001}	\\
	&	3	$\rightarrow$	2	&	25.8	&	APEX	&	Pointing Catalogue	\\
	&	3	$\rightarrow$	2	&	27.2	&	APEX	&	Archive	\\
	&	3	$\rightarrow$	2	&	29.3	&	APEX	&	\cite{De-Beck2010}	\\
	&	4	$\rightarrow$	3	&	23.1	&	APEX	&	Pointing Catalogue	\\
	&	4	$\rightarrow$	3	&	30.6	&	APEX	&	\cite{De-Beck2010}	\\
	&	7	$\rightarrow$	6	&	25.1	&	APEX	&	\cite{De-Beck2010}	\\
X TrA	&	1	$\rightarrow$	0	&	2.5	&	SEST	&	\cite{Olofsson1993}	\\
	&	2	$\rightarrow$	1	&	10.9	&	APEX	&	Pointing Catalogue	\\
	&	2	$\rightarrow$	1	&	11.8	&	SEST	&	\cite{Olofsson1993}	\\
	&	3	$\rightarrow$	2	&	15.0	&	APEX	&	Pointing Catalogue	\\
	&	3	$\rightarrow$	2	&	12.3	&	APEX	&	Archive	\\
	&	3	$\rightarrow$	2	&	15.3	&	SEST	&	\cite{Schoier2001}	\\
	&	4	$\rightarrow$	3	&	19.5	&	APEX	&	Pointing Catalogue	\\
II Lup	&	1	$\rightarrow$	0	&	61	&	SEST	&	\cite{Ryde1999}	\\
	&	2	$\rightarrow$	1	&	118.4	&	APEX	&	Pointing Catalogue	\\
	&	2	$\rightarrow$	1	&	128.12	&	APEX	&	Archive	\\
	&	2	$\rightarrow$	1	&	151	&	SEST	&	\cite{Ryde1999}	\\
	&	3	$\rightarrow$	2	&	163.0	&	APEX	&	\cite{Ramstedt2014}	\\
	&	3	$\rightarrow$	2	&	144.4	&	APEX	&	\cite{De-Beck2010}	\\
	&	3	$\rightarrow$	2	&	144.4	&	APEX	&	\cite{De-Beck2010}	\\
	&	3	$\rightarrow$	2	&	130.3	&	APEX	&	Pointing Catalogue	\\
	&	3	$\rightarrow$	2	&	122.71	&	APEX	&	Archive	\\
	&	3	$\rightarrow$	2	&	129.3	&	SEST	&	\cite{Ryde1999}	\\
	&	4	$\rightarrow$	3	&	155.9	&	APEX	&	\cite{De-Beck2010}	\\
	&	4	$\rightarrow$	3	&	128.4	&	APEX	&	\cite{De-Beck2010}	\\
	&	4	$\rightarrow$	3	&	147.0	&	APEX	&	Pointing Catalogue	\\
	&	6	$\rightarrow$	5	&	20.2	&	HIFI	&	\textit{Herschel} Science Archive (HIFISTARS)	\\
	&	7	$\rightarrow$	6	&	151.43	&	APEX	&	\cite{De-Beck2010}	\\
	&	7	$\rightarrow$	6	&	82.1	&	APEX	&	\cite{De-Beck2010}	\\
	&	10	$\rightarrow$	9	&	19.4	&	HIFI	&	\textit{Herschel} Science Archive (HIFISTARS)	\\
	&	16	$\rightarrow$	15	&	17.2	&	HIFI	&	\textit{Herschel} Science Archive (HIFISTARS)	\\
V CrB	&	1	$\rightarrow$	0	&	11.7	&	IRAM	&	\cite{Schoier2001}	\\
	&	1	$\rightarrow$	0	&	2.7	&	OSO	&	\cite{Olofsson1993}	\\
	&	2	$\rightarrow$	1	&	5.4	&	APEX	&	Pointing Catalogue	\\
	&	2	$\rightarrow$	1	&	18.4	&	IRAM	&	\cite{Olofsson1993}	\\
	&	3	$\rightarrow$	2	&	10.9	&	APEX	&	Archive	\\
V821 Her	&	1	$\rightarrow$	0	&	84.1	&	IRAM	&	\cite{Neri1998}	\\
	&	2	$\rightarrow$	1	&	52.1	&	APEX	&	Pointing Catalogue	\\
	&	2	$\rightarrow$	1	&	93.4	&	IRAM	&	\cite{Neri1998}	\\
	&	3	$\rightarrow$	2	&	60.3	&	APEX	&	\cite{De-Beck2010}	\\
	&	3	$\rightarrow$	2	&	56.7	&	APEX	&	Pointing Catalogue	\\
	&	3	$\rightarrow$	2	&	55.9	&	APEX	&	Archive	\\
	&	4	$\rightarrow$	3	&	66.1	&	APEX	&	\cite{De-Beck2010}	\\
	&	4	$\rightarrow$	3	&	66.0	&	APEX	&	Pointing Catalogue	\\
	&	4	$\rightarrow$	3	&	81.3	&	APEX	&	Archive	\\
	&	7	$\rightarrow$	6	&	49.8	&	APEX	&	\cite{De-Beck2010}	\\
V Aql	&	1	$\rightarrow$	0	&	3.2	&	OSO	&	\cite{Olofsson1993}	\\
	&	1	$\rightarrow$	0	&	2.8	&	SEST	&	\cite{Olofsson1993}	\\
	&	2	$\rightarrow$	1	&	8.8	&	APEX	&	Pointing Catalogue	\\
	&	2	$\rightarrow$	1	&	9.0	&	JCMT	&	\cite{Schoier2001}	\\
	&	2	$\rightarrow$	1	&	8.1	&	SEST	&	\cite{Olofsson1993}	\\
	&	3	$\rightarrow$	2	&	11.5	&	APEX	&	Pointing Catalogue	\\
	&	3	$\rightarrow$	2	&	9	&	JCMT	&	\cite{Schoier2001}	\\
V1968 Cyg	&	1	$\rightarrow$	0	&	46.4	&	IRAM	&	\cite{Neri1998}	\\
	&	2	$\rightarrow$	1	&	88.6	&	IRAM	&	\cite{Neri1998}	\\
RV Aqr	&	1	$\rightarrow$	0	&	7.5	&	SEST	&	\cite{Olofsson1993}	\\
	&	2	$\rightarrow$	1	&	18.1	&	SEST	&	\cite{Olofsson1993}	\\
	&	3	$\rightarrow$	2	&	18.6	&	SEST	&	\cite{Schoier2001}	\\
\quad \it S Stars	\\							
R And	&	2	$\rightarrow$	1	&	32.0	&	JCMT	&	\citet{Ramstedt2009}	\\
	&	3	$\rightarrow$	2	&	43.0	&	JCMT	&	\citet{Ramstedt2009}	\\
	&	4	$\rightarrow$	3	&	25.6	&	APEX	&	Pointing Catalogue	\\
S Cas	&	1	$\rightarrow$	0	&	13.8	&	OSO	&	\citet{Ramstedt2009}	\\
	&	3	$\rightarrow$	2	&	31.0	&	JCMT	&	\citet{Ramstedt2009}	\\
W And	&	1	$\rightarrow$	0	&	3.8	&	OSO	&	\citet{Ramstedt2009}	\\
R Gem	&	1	$\rightarrow$	0	&	2.4	&	OSO	&	\citet{Ramstedt2009}	\\
	&	2	$\rightarrow$	1	&	4.5	&	APEX	&	Pointing Catalogue	\\
	&	3	$\rightarrow$	2	&	5.2	&	APEX	&	Pointing Catalogue	\\
Y Lyn	&	1	$\rightarrow$	0	&	4.1	&	OSO	&	\citet{Ramstedt2009}	\\
R Cyg	&	1	$\rightarrow$	0	&	4.4	&	OSO	&	\citet{Ramstedt2009}	\\
\quad \it M Stars	\\
KU And	&	1	$\rightarrow$	0	&	24.8	&	OSO	&	\cite{Gonzalez-Delgado2003}	\\
	&	2	$\rightarrow$	1	&	14.0	&	CSO	&	\cite{Knapp1998}	\\
	&	2	$\rightarrow$	1	&	41.5	&	JCMT	&	\cite{Gonzalez-Delgado2003}	\\
R Hor	&	2	$\rightarrow$	1	&	19.0	&	APEX	&	Pointing Catalogue	\\
	&	3	$\rightarrow$	2	&	30.8	&	APEX	&	Pointing Catalogue	\\
	&	3	$\rightarrow$	2	&	23	&	CSO	&	\cite{Young1995}	\\
	&	4	$\rightarrow$	3	&	33.5	&	APEX	&	Pointing Catalogue	\\
NV Aur	&	2	$\rightarrow$	1	&	37	&	JCMT	&	\cite{Kemper2003}	\\
	&	2	$\rightarrow$	1	&	36.8	&	JCMT	&	\cite{Gonzalez-Delgado2003}	\\
	&	2	$\rightarrow$	1	&	36.3	&	JCMT	&	\cite{De-Beck2010}	\\
	&	3	$\rightarrow$	2	&	39	&	JCMT	&	\cite{Kemper2003}	\\
	&	3	$\rightarrow$	2	&	35.7	&	JCMT	&	\cite{Gonzalez-Delgado2003}	\\
	&	3	$\rightarrow$	2	&	38.8	&	JCMT	&	\cite{De-Beck2010}	\\
	&	4	$\rightarrow$	3	&	35.4	&	JCMT	&	\cite{De-Beck2010}	\\
BX Cam	&	1	$\rightarrow$	0	&	20.0	&	OSO	&	\cite{Olofsson1998}	\\
	&	2	$\rightarrow$	1	&	25.5	&	JCMT	&	\citet{Ramstedt2014}	\\
	&	3	$\rightarrow$	2	&	39.4	&	JCMT	&	\citet{Ramstedt2014}	\\
GX Mon	&	1	$\rightarrow$	0	&	31	&	OSO	&	\citet{Ramstedt2008}	\\
	&	2	$\rightarrow$	1	&	56.7	&	APEX	&	Pointing Catalogue	\\
	&	2	$\rightarrow$	1	&	52.9	&	APEX	&	Archive	\\
	&	2	$\rightarrow$	1	&	22.9	&	CSO	&	\cite{Knapp1998}	\\
	&	2	$\rightarrow$	1	&	61	&	JCMT	&	\citet{Ramstedt2008}	\\
	&	3	$\rightarrow$	2	&	64.5	&	APEX	&	Pointing Catalogue	\\
	&	3	$\rightarrow$	2	&	55.0	&	APEX	&	Archive	\\
	&	3	$\rightarrow$	2	&	71	&	JCMT	&	\citet{Ramstedt2008}	\\
	&	4	$\rightarrow$	3	&	54.3	&	APEX	&	Pointing Catalogue	\\
	&	4	$\rightarrow$	3	&	149	&	JCMT	&	\citet{Ramstedt2008}	\\
L$_2$ Pup	&	1	$\rightarrow$	0	&	0.24	&	SEST	&	\cite{Kerschbaum1996}	\\
	&	2	$\rightarrow$	1	&	3.9	&	APEX	&	Pointing Catalogue	\\
	&	2	$\rightarrow$	1	&	3.1	&	JCMT	&	H. Olofsson, private communication	\\
	&	2	$\rightarrow$	1	&	3.7	&	SEST	&	\cite{Olofsson2002}	\\
	&	3	$\rightarrow$	2	&	8.8	&	APEX	&	Pointing Catalogue	\\
	&	3	$\rightarrow$	2	&	16.6	&	JCMT	&	\cite{Olofsson2002}	\\
	&	4	$\rightarrow$	3	&	11.5	&	APEX	&	Pointing Catalogue	\\
R LMi	&	2	$\rightarrow$	1	&	6.1	&	APEX	&	Pointing Catalogue	\\
	&	2	$\rightarrow$	1	&	2.72	&	CSO	&	\cite{Knapp1998}	\\
	&	3	$\rightarrow$	2	&	9.1	&	APEX	&	Pointing Catalogue	\\
	&	3	$\rightarrow$	2	&	12.9	&	APEX	&	Archive	\\
R Leo	&	1	$\rightarrow$	0	&	4.1	&	IRAM	&	\cite{Teyssier2006}	\\
	&	1	$\rightarrow$	0	&	2.1	&	OSO	&	\cite{Gonzalez-Delgado2003}	\\
	&	2	$\rightarrow$	1	&	14.8	&	APEX	&	Pointing Catalogue	\\
	&	2	$\rightarrow$	1	&	28.5	&	IRAM	&	\cite{Teyssier2006}	\\
	&	2	$\rightarrow$	1	&	15.0	&	JCMT	&	\cite{Gonzalez-Delgado2003}	\\
	&	3	$\rightarrow$	2	&	30.3	&	APEX	&	Pointing Catalogue	\\
	&	3	$\rightarrow$	2	&	32.7	&	APEX	&	Archive	\\
	&	3	$\rightarrow$	2	&	21.9	&	CSO	&	\cite{Knapp1998}	\\
	&	3	$\rightarrow$	2	&	41.6	&	JCMT	&	\cite{Gonzalez-Delgado2003}	\\
	&	4	$\rightarrow$	3	&	39.5	&	APEX	&	Pointing Catalogue	\\
	&	4	$\rightarrow$	3	&	28.1	&	CSO	&	\cite{Young1995}	\\
	&	6	$\rightarrow$	5	&	31.0	&	CSO	&	\cite{Teyssier2006}	\\
S CrB	&	2	$\rightarrow$	1	&	6.4	&	APEX	&	Pointing Catalogue	\\
	&	2	$\rightarrow$	1	&	2.53	&	CSO	&	\cite{Knapp1998}	\\
	&	2	$\rightarrow$	1	&	12.7	&	IRAM	&	\cite{Neri1998}	\\
	&	3	$\rightarrow$	2	&	10.6	&	APEX	&	Pointing Catalogue	\\
	&	3	$\rightarrow$	2	&	10.5	&	APEX	&	Archive	\\
	&	3	$\rightarrow$	2	&	9.1	&	CSO	&	\cite{Young1995}	\\
	&	4	$\rightarrow$	3	&	4.9	&	CSO	&	\cite{Young1995}	\\
V1111 Oph	&	1	$\rightarrow$	0	&	21.5	&	OSO	&	\cite{Gonzalez-Delgado2003}	\\
	&	2	$\rightarrow$	1	&	30.6	&	APEX	&	Pointing Catalogue	\\
	&	2	$\rightarrow$	1	&	42.0	&	JCMT	&	\cite{Gonzalez-Delgado2003}	\\
	&	3	$\rightarrow$	2	&	40.5	&	APEX	&	Pointing Catalogue	\\
	&	3	$\rightarrow$	2	&	59.1	&	JCMT	&	\cite{Ramstedt2014}	\\
	&	4	$\rightarrow$	3	&	46.1	&	APEX	&	Pointing Catalogue	\\
RR Aql	&	2	$\rightarrow$	1	&	12.5	&	APEX	&	Pointing Catalogue	\\
	&	2	$\rightarrow$	1	&	16.6	&	JCMT	&	H. Olofsson, private communication	\\
	&	3	$\rightarrow$	2	&	16.1	&	APEX	&	Pointing Catalogue	\\
	&	3	$\rightarrow$	2	&	23.5	&	JCMT	&	H. Olofsson, private communication	\\
	&	4	$\rightarrow$	3	&	14.3	&	APEX	&	Pointing Catalogue	\\
V1943 Sgr	&	1	$\rightarrow$	0	&	1.25	&	SEST	&	\cite{Kerschbaum1999}	\\
	&	2	$\rightarrow$	1	&	5.17	&	SEST	&	\cite{Kerschbaum1999}	\\
V1300 Aql	&	1	$\rightarrow$	0	&	16	&	OSO	&	\citet{Ramstedt2008}	\\
	&	2	$\rightarrow$	1	&	38.1	&	APEX	&	Pointing Catalogue	\\
	&	2	$\rightarrow$	1	&	45	&	JCMT	&	\citet{Ramstedt2008}	\\
	&	3	$\rightarrow$	2	&	37.0	&	APEX	&	\cite{De-Beck2010}	\\
	&	3	$\rightarrow$	2	&	35.3	&	APEX	&	Pointing Catalogue	\\
	&	3	$\rightarrow$	2	&	35.9	&	APEX	&	Archive	\\
	&	3	$\rightarrow$	2	&	27	&	JCMT	&	\citet{Ramstedt2008}	\\
	&	4	$\rightarrow$	3	&	33.5	&	APEX	&	\cite{De-Beck2010}	\\
	&	4	$\rightarrow$	3	&	34.8	&	APEX	&	Pointing Catalogue	\\
	&	4	$\rightarrow$	3	&	34.5	&	APEX	&	Archive	\\
	&	4	$\rightarrow$	3	&	22	&	JCMT	&	\citet{Ramstedt2008}	\\
	&	7	$\rightarrow$	6	&	19.7	&	APEX	&	\cite{De-Beck2010}	\\
T Cep	&	1	$\rightarrow$	0	&	1.7	&	OSO	&	\cite{Olofsson1998}	\\
\hline
\end{longtable}
} \end{longtab}

\end{document}